\documentclass{article}

\usepackage[utf8]{inputenc}

\usepackage[margin=1in]{geometry}

\usepackage[section]{placeins}

\usepackage{amsfonts, amssymb, amsmath, amsthm,esint,multicol}
\usepackage[colorlinks=true, pdfstartview=FitV, linkcolor=blue,
            citecolor=blue, urlcolor=blue]{hyperref}
\usepackage[usenames,dvipsnames,table]{xcolor}
\definecolor{Red}{rgb}{0.7,0,0.1}
\definecolor{Green}{rgb}{0,0.7,0}
\usepackage{accents}
\usepackage{comment}
\usepackage{graphicx}
\usepackage[capitalize,nameinlink,noabbrev]{cleveref}
\usepackage{enumerate}
\usepackage{enumitem}

\DeclareFontFamily{U}{mathx}{}
\DeclareFontShape{U}{mathx}{m}{n}{<-> mathx10}{}
\DeclareSymbolFont{mathx}{U}{mathx}{m}{n}
\DeclareMathAccent{\widehat}{0}{mathx}{"70}
\DeclareMathAccent{\widecheck}{0}{mathx}{"71}

\DeclareSymbolFont{extraup}{U}{zavm}{m}{n}
\DeclareMathSymbol{\varheart}{\mathalpha}{extraup}{86}
\DeclareMathSymbol{\vardiamond}{\mathalpha}{extraup}{87}

\definecolor{labelkey}{rgb}{0,0,1}
\usepackage{bbm}
\usepackage{mathtools} 

\usepackage{tabularx}

\usepackage{listings}
\usepackage[textsize=tiny]{todonotes}
\usepackage{tikz}
\usetikzlibrary{shapes.misc}
\usepackage{etoolbox}
\usepackage{appendix}
\usepackage{subcaption}
\usepackage{wrapfig}
\usepackage{xr}
\usepackage{booktabs}
\usepackage{tikz}
\usetikzlibrary{shapes,shadows.blur,calc,decorations.pathmorphing}

\numberwithin{equation}{section}

\newtheorem{Theorem}{Theorem}[section]
\newtheorem{Proposition}[Theorem]{Proposition}

\newtheorem{Corollary}[Theorem]{Corollary}

\newtheorem{Definition}[Theorem]{Definition}
\newtheorem{Remark}[Theorem]{Remark}

\newtheorem{Method}[Theorem]{Method}

\usepackage[section]{algorithm}      
\usepackage[noend]{algpseudocode}    

\newcommand{\be}{\begin{equation}}
	\newcommand{\ee}{\end{equation}}

\newcommand{\RR}{\mathbb{R}}
\newcommand{\NN}{\mathbb{N}}

\newcommand{\cB}{\mathcal{B}}
\newcommand{\cC}{\mathcal{C}}
\newcommand{\cF}{\mathcal{F}}
\newcommand{\cH}{\mathcal{H}}
\newcommand{\cI}{\mathcal{I}}
\newcommand{\cU}{\mathcal{U}}
\newcommand{\cM}{\mathcal{M}}

\newcommand{\bw}{\mathbf{w}}
\newcommand{\balpha}{\boldsymbol{\alpha}}
\newcommand{\bS}{\boldsymbol{S}}

\newcommand{\EE}{\mathbb{E}}
\newcommand{\Prb}{\mathbb{P}}

\newcommand{\Var}{\text{Var}}

\newcommand{\mps}{\mu}    
\newcommand{\dps}{\pi}    
\newcommand{\sdem}{\widehat{\mu}_{\np, N}} 

\newcommand{\Obs}{\varphi}
\newcommand{\sdest}{\widehat{\Obs}_{N}}  
\newcommand{\rbest}{\widecheck{\Obs}_{N}} 

\newcommand{\mk}{P} 
\newcommand{\vk}{\mathcal{V}} 
\newcommand{\pk}{Q} 
\newcommand{\dpk}{f}
\newcommand{\bpk}{\bar{\pk}} 
\newcommand{\dbpk}{\bar{f}}
\newcommand{\mkmh}{P^\np_{\text{MH}}} 
\newcommand{\mkB}{P^\np_{\text{B}}} 
\newcommand{\mkmhinf}{P^\infty_{\text{MH}}}

\newcommand{\qsp}{X}         
\newcommand{\vsp}{Y}  
\newcommand{\bq}{\mathbf{q}} 
\newcommand{\tbq}{\tilde{\bq}}

\newcommand{\bbq}{\bar{\bq}}
\newcommand{\bu}{\mathbf{u}}
\newcommand{\br}{\mathbf{r}}
\newcommand{\tbr}{\tilde{\mathbf{r}}}
\newcommand{\bv}{\mathbf{v}} 
\newcommand{\bz}{\mathbf{z}}

\newcommand{\bbz}{\bar{\bz}}

\newcommand{\bxi}{\boldsymbol{\xi}}
\newcommand{\tbxi}{\tilde{\boldsymbol{\xi}}}
\newcommand{\bx}{\mathbf{x}}
\newcommand{\ar}{\alpha} 
\newcommand{\art}{\beta} 

\newcommand{\dmu}{\pi} 

\newcommand{\Pot}{\Phi} 
\newcommand{\Ham}{\cH} 
\newcommand{\VPot}{\Psi} 
\newcommand{\hS}{\hat{S}} 
\newcommand{\dt}{\tau} 

\newcommand{\np}{p} 

\newcommand{\dm}{d} 
\newcommand{\indFn}[1]{\mathbbm{1}_{#1}}
\newcommand{\proj}{\text{Pr}} 
\newcommand{\Cn}[1]{|#1|_{\mathcal{C}}}
\newcommand{\Cip}[2]{\langle #1, #2 \rangle_{\mathcal{C}}}

\newcommand{\met}{d}
\newcommand{\lip}{g} 

\title{Mad Props: Parallelism in Markov Chain Monte Carlo 
  Through the Lens of the Infinite Proposal Limit}

\date{\today}
\author{Nathan E.~Glatt-Holtz$^{\varheart}$, Andrew J.~Holbrook$^\vardiamond$,
  Justin A.~Krometis$^{\clubsuit}$, Cecilia F.~Mondaini$^{\spadesuit}$\\
  \scriptsize{\it $^{\varheart}$Department of Statistics and
    Department of Mathematics, Indiana University, Bloomington}\\
  \scriptsize{\it $^\vardiamond$Department of Biostatistics,
    University of California, Los Angeles}\\
  \scriptsize{\it $^{\clubsuit}$National Security Institute
    and Department of Mathematics, Virginia
    Polytechnic Institute and State University}\\
  \scriptsize{\it $^{\spadesuit}$Department of Mathematics,
    Drexel University, Philadelphia, PA}\\
  \scriptsize{emails: negh@iu.edu, aholbroo@g.ucla.edu,
    jkrometi@vt.edu, cf823@drexel.edu}}

\begin{document}

\markboth {Mad Props: Parallelism in MCMC Through the Lens of
  the Infinite Proposal Limit}{Mad Props: Parallelism in MCMC Through the Lens of
  the Infinite Proposal Limit}

\maketitle

\begin{abstract}
  Multiproposal MCMC (MP-MCMC) algorithms use clouds of proposals to
  efficiently traverse state spaces and overcome complex target
  geometries.  While MCMC methods are embarrassingly parallel by
  nature, the non-trivial forms of parallelism provided by the MP-MCMC
  formalism sometimes leads to significant improvements over a naive
  approach. Here, one important tuning parameter is the number of
  proposals $p$ used by a single MP-MCMC iteration. While a number of
  computational strategies have been proposed to efficiently leverage
  large numbers of proposals within the MP-MCMC paradigm, much remains
  unknown about these algorithms, particularly in the large
  $p$-regime.

  In this contribution, we discover surprising results by identifying
  and studying several promising new methods
  (\cref{alg:slingshot:intro}, \cref{alg:MTMpCN},
  \cref{alg:MTMpCN:loc}), ruling out other extant approaches and
  discovering new relationships between different MP-MCMC
  methodologies.  Our analysis is centered on a general state space
  multiproposal involutive theory recently constructed by the authors
  \cite{glatt2024parallel, glatt2024sacred} combined with the
  consideration of the large $p$-limit kernels for MP-MCMC algorithms
  within a variety of different classes of proposal and acceptance
  structures.
\end{abstract}

\begin{small}
\setcounter{tocdepth}{1}
\tableofcontents
\end{small}

\newpage

\section{Introduction}

Generating random samples from complex or high-dimensional probability
distributions remains a central challenge in 21st century science.
Among the most successful strategies is the class of classical
(Metropolis-type) Markov chain Monte Carlo (MCMC) algorithms dating
back to the late 1940's \cite{metropolis1953equation,
  hastings1970monte}.  These methods effectively resolve simple,
low-dimensional target distributions by generating a single proposal
state and then randomly accepting or rejecting this proposal with a
prescribed probability.  Since that time, a wide variety of more
sophisticated MCMC algorithms have been advanced.  This growing
body of Monte Carlo methods, combined with
an ever advancing technological substrate, has allowed for the
resolution of increasingly sophisticated higher-dimensional
statistical models.  Nevertheless, there is always a need to address
ever more challenging models, and the frontier for MCMC algorithm
design and its foundational theory remains open in many directions.

One important class of methods at this frontier are the so called
multiproposal MCMC (MP-MCMC) algorithms \cite{liu2000multiple,
  tjelmeland2004using, frenkel2004speed, delmas2009does, neal2011mcmc,
  calderhead2014general, luo2019multiple, holbrook2023generating, 
  glatt2024parallel,lin2023more,glatt2024sacred, 
  senn2026multiproposal} that use a `cloud'
of proposals at each MCMC iteration to explore a target distribution
more efficiently.  While, in general, MCMC methods are embarrassingly
parallel by nature, the non-trivial forms of parallelism that arise
within the MP-MCMC formalism can lead to significant improvements over
a naive approach.  So far, this MP-MCMC approach is proving especially
beneficial for sampling targets with multimodal and nonlinear
geometries by significantly improving mixing rates while avoiding
costly or infeasible gradient computations, \cite{holbrook2023quantum, 
borggaard2023statistical, glatt2024parallel,lin2023more,
  glatt2024sacred, senn2026multiproposal}.  
On the other hand, applying the MP-MCMC formalism
in practice requires navigating a landscape of seemingly distinct
algorithmic structures (e.g. multiple-try Metropolis, convolutional
proposal structures, local vs global methods, HMC-type approaches...)
and then specifying a number of tuning parameters the exact choices of
which may depend on target geometry and the available computational
resources.  Thus, before a scientist does the work of crafting an
MP-MCMC algorithm and programming GPU code to efficiently scale this
algorithm to use a large number of proposals, they might ask: is it
worth the effort? And, since MP-MCMC comprises a large class of
algorithms, to what extent do the benefits of using large numbers of
proposals depend on algorithmic specifics?

With this backdrop in mind, this contribution provides a unified
analysis of several large classes of MP-MCMC Markov processes.  Our
systematic approach leverages the dual-lenses of the large proposal
$\np \to \infty$ limit, presaged in \cite{liu2000multiple,
  gagnon2023improving}, alongside a general state space involutive
theory for reversible Metropolis-type kernels recently advanced in a
series of works \cite{glatt2020accept, neklyudov2020involutive,
  andrieu2020general, glatt2024parallel, glatt2024sacred}.  This dual
approach clarifies delicate relationships between a number of existing
methods in this complex algorithm landscape while, along the way,
largely ruling out certain extant approaches.  Furthermore we discover
some promising new MP-MCMC methods including the ``slingshot''
\cref{alg:slingshot:intro} as well as two ``Hilbert space'' or
``Preconditioned'' methods in \cref{alg:MTMpCN},
\cref{alg:MTMpCN:loc}.  We refer to the first algorithm as the
``slingshot'' because its acceptance structure includes a bias toward
accepting proposal points far from the current state.  As we unfold
below, this curious kernel has a number of interesting and desirable
properties.  In particular it leaves its target invariant in the large
$\np$-limit despite failing to do so for finite $\np$ while providing
significant degrees of freedom in parameter auto-tuning over other
methods.  Regarding \cref{alg:MTMpCN} and \cref{alg:MTMpCN:loc}, we
provide two new `dimension-free' methods built on a preconditioned
Crank-Nicolson (pCN)-type proposal structure \cite{NealpCNbeforepCN,
  beskos2008mcmc, cotter2013mcmc} in the multiple-try-metropolis (MTM)
vein \cite{liu2000multiple}. These two new methods focus on the global and local
structure of the target measure, respectively.  Finally it is notable
that our analysis provides some interesting technical improvements to
both the large $\np$ and involutive methodologies. For the former we
note \cref{thm:MTM:limit} which leverages the uniform law of Large
Numbers (cf. \cite{talagrand1987glivenko, dudley1991uniform,
  van2014universal, wainwright2019high}) to develop this
$\np \to \infty$ limit for multiple-try Metropolis kernels on general state spaces.  
For the latter, we expand the scope
of the main result \cite[Theorem 2.2]{glatt2024parallel} as
\cref{thm:grand:unifed:inv}, which provides a formulation allowing
us to establish \cref{thm:MTM:unbiased}, again pertinent to MTM type
kernels on general state spaces.

\subsection*{Overview of Main Results}

Let us turn to provide a more systematic overview of our contributions
herein alongside some further commentary on related developments in
the existing literature.  Our aim is to address the MP-MCMC paradigm
over as wide a variety of targets $\mu$ and methods as possible
facilitated by a general state space setting.  As such, it is convenient
to write $\mu$ in the form
\begin{align}\label{eq:stuart:form}
  \mu(d \bq) =  \dmu(\bq) \mu_0(d \bq), 
\end{align}
where $\dmu \in L^1(\mu_0)$ with the base measure $\mu_0$ at least
$\sigma$-finite on a measurable space $(\qsp, \Sigma_\qsp)$. We
sometimes write $\dmu$ in terms of a potential $\Pot$ so that
\begin{align}\label{eq:stuart:form:og}
\pi(\bq) = Z^{-1}\exp( - \Pot(\bq)) \text{  with }
  Z = \int_X \exp( - \Pot(\bbq)) \mu_0(d \bbq) < \infty.
\end{align}

This formulation \eqref{eq:stuart:form} includes the classical
settings where either $X = \RR^\dm$ and $\mu_0$ is the Lebesgue
measure on $X$, or the discrete case when $\mu_0$ is a counting
measure on, e.g., $\mathbb{N}$.  Of course, these two classical cases
by no means exhaust the scope of relevant settings for MP-MCMC
methods.  Indeed we are particularly motivated by the situation
when\footnote{The mean zero condition on $\mu_0$ is imposed only for
  convenience and economy of presentation.  All the forthcoming
  analysis and in particular the formulation of \cref{alg:MTMpCN} ,
  \cref{alg:MTMpCN:loc} can be easily modified to include the general
  case of any non-zero mean.}
\begin{align}\label{eq:s:f:infinite:d}
  X \text{ is a separable Hilbert space and } \mu_0 \text{ is a mean
  zero Gaussian measure with covariance } \cC.
\end{align}
Such targets defined in terms of Gaussian reference measures fall
under the scope of the `preconditioned' or `Hilbert space' algorithms
developed in \cite{NealpCNbeforepCN, beskos2008mcmc,
  murray2010elliptical, cotter2013mcmc, Beskosetal2011,
  glatt2020accept, glatt2024parallel}.  
  These methods are naturally suited for many
high-dimensional problems, a claim which has been set on rigorous
foundations in \cite{hairer2014spectral, eberle2014error,
  glatt2021mixing, bou2021two, CarigiGlattHoltzMondaini2026a}.  
  In regards to application these
methods played a crucial role in the Bayesian approach to PDE inverse
problems \cite{kaipio2006statistical, stuart2010inverse,
  dashti2017bayesian, alexanderian2021optimal, borggaard2020bayesian,
  borggaard2023statistical} and in sampling conditioned diffusions
\cite{beskos2008mcmc, cotter2013mcmc}.  Note that the Hilbert space
approach was recently generalized to our MP-MCMC setting in
\cite{glatt2024parallel}.  We  derive two further new MP-MCMC
methods suited for \eqref{eq:s:f:infinite:d} in this vein herein as
\cref{alg:MTMpCN}, \cref{alg:MTMpCN:loc}.

All of our sampling methods for \eqref{eq:stuart:form} fall under
the concise and often illuminating formulation
\begin{align}  \label{def:mp:ker}
   \mk^{(\vk,\bS, \balpha)}(\bq_0, d\tbq) = \mk(\bq_0, d\tbq) =
  \sum_{j=0}^\np \int_{\vsp} \ar_j (\bq_0, \bv)
  \delta_{\Pi_1 S_j (\bq, \bv)} (d\tbq) \vk(\bq_0, d \bv).
\end{align}
Here, $\mk: \qsp \times \Sigma_\qsp \to [0,1]$ is a Markov transition
kernel defined up to the specification of an auxiliary space $\vsp$,
the `proposal kernel' $\vk: \qsp \times \Sigma_\vsp \to [0,1]$,
forward maps $\bS = (S_0,\ldots, S_\np)$, where
$S_j: \qsp \times \vsp \to \qsp \times \vsp$ and acceptance
probabilities $\balpha= (\ar_0,\ldots, \ar_\np)$ with
$\ar_j: \qsp \times \vsp \to [0,1]$.  The integer $\np$ corresponds to
the number of proposed states at each step, and $\Pi_1(\bq,\bv) = \bq$
is the projection onto the first coordinate.  Thus, algorithmically,
\eqref{def:mp:ker} reads as follows: from the current state
$\bq_0^{(n)}$, (i) draw a state $\bv^{(n+1)} \sim \vk(\bq_0^{(n)}, d \bv)$
(ii) map $(\bq_0^{(n)},\bv^{(n+1)})$ to the `cloud' of proposed new
states $\bbq_j^{(n+1)} := \Pi_1 S_j (\bq_0^{(n)}, \bv^{(n+1)})$,
and then (iii) select the next state $\bq_0^{(n+1)} :=\bbq_j^{(n+1)}$ with
the corresponding probability
$\ar_j^{(n+1)} := \ar_j (\bq_0^{(n)}, \bv^{(n+1)})$.

We emphasize that, without further conditions, \eqref{def:mp:ker} is
only an illuminating way of organizing our analysis and a general
structure which we use to compute the limit as $\np \to \infty$ for a
number of special cases in \cref{sec:large:p:limit},
\cref{sec:HMC:path:sel}.  Indeed, at the given degree of generality,
\eqref{def:mp:ker} has no relation to our target $\mu$.  Here, the
involutive theory of \cite{glatt2024parallel}, recalled and slightly
generalized below as \cref{thm:grand:unifed:inv}, gives a powerful,
concise and broadly applicable criterion for the reversibility of
\eqref{def:mp:ker} with respect to a given $\mu$ as a function of the
elements $\vk, S_j, \alpha_j$.  We make significant further use of
this criteria herein noting in particular an application that broadens
the scope of multiple-try metropolis (MTM) MP-MCMC methods in
\cref{thm:MTM:unbiased} and yielding in particular the new methods
\cref{alg:MTMpCN} and \cref{alg:MTMpCN:loc}.  Also we observe that the
formulation \eqref{def:mp:ker} exposes degeneracies that we use to
essentially rule out `Metropolis-type' multiproposal structures as we
unfold below in \cref{rmk:dengenaries}, \cref{rmk:MH:degenerate} and
\cref{rmk:more:degen}.

Of course we must keep in mind the computational complexity and
implementation details as we proceed. Here we observe that certain
naive unbiased proposal schemes under \cref{thm:grand:unifed:inv}
result in acceptance mechanisms that require burdensome $O(p^2)$
floating-point operations which are not amenable to straightforward
parallelization (see \eqref{def:ar:MH:fd:BB} for example).  On the
other hand, various methods  developed in previous works
\cite{liu2000multiple, tjelmeland2004using, holbrook2023generating,
  holbrook2023quantum, lin2023more, glatt2024parallel} alongside new
methods proposed herein include convenient proposal strategies that
enable simplified acceptance probabilities requiring only $O(p)$
embarrassingly parallel floating-point operations.  In this vein,
\cite{glatt2024parallel} uses a graphics processing unit (GPU) to
parallelize an $O(p)$ MP-MCMC algorithm at $p=100{,}000$ proposals to
successfully target an extremely multimodal example. Similarly,
\cite{holbrook2023quantum} and \cite{lin2023more} develop quantum
computing routines to parallelize simplified acceptance probabilities,
reducing time-complexity from $O(p)$ to $O(\sqrt{p})$ and $O(1)$,
respectively.  Meanwhile in \cite{glatt2024sacred,
  borggaard2023statistical}, $O(p)$ parameter-to-solution solves were
used to resolve some demanding PDE informed inference problems on the
basis of \cite{glatt2024parallel}.  This latter PDE constrained setting
suggests an interesting point of departure for future applications of
\cref{alg:MTMpCN}, \cref{alg:MTMpCN:loc} derived herein.

Although there are a number of MP-MCMC structures we contend with in
this work, perhaps the most natural proposal mechanism is simply to
take
\begin{align}
	\bq_1, \ldots,\bq_\np
	\stackrel{iid}{\sim} Q(\bq_0, \cdot) \text{ conditional on
the current state $\bq_0$},
\label{eq:basic:b:MP-MCMC}
\end{align}	
for a given (single-proposal) kernel $Q$. In this case, $Y = X^\np$,
and
$\vk(\bq_0, d \bq_1, \ldots, d \bq_\np) = \Pi_{k =1}^\np Q(\bq_0, d
\bq_k)$ in the formulation \eqref{def:mp:ker} with the choice of
$\alpha_j$ still to be determined.  To complete the algorithmic
description from \eqref{eq:basic:b:MP-MCMC} one immediate approach,
particularly when $\pk$ is reversible with respect to $\mu_0$, is to
then accept from amongst the $\np$ proposals $\bq_j$ with
probabilities proportional to the target density $\dmu(\bq_j)$.  In
other words we accept proposals most frequently in high-probability---
or perhaps high-likelihood depending on the context---regions.  We dub
this approach the ``bubble bath'' as we lay out in
\cref{subsubsec:bubble:bath}.  Although \cref{subsubsec:bubble:bath}
is biased in general there are several variations to the approach
which result in an unbiased method (cf.  \cref{meth:MTMpCN},
\cref{meth:tj:mPCN}).  Moreover we sometimes obtain an asymptotically
unbiased method if we scale certain algorithmic parameters in $\pk$ as
we send $\np \to \infty$; see \cref{rmk:scale:to:glob}.

Another set of possible acceptance structures from
\eqref{eq:basic:b:MP-MCMC} can be considered which optimizes for
exploring the `local geometry' of the target density around current
state as we explore in \cref{meth:loc:prop}. Indeed, following
previous insights in \cite{zanella2020informed, gagnon2023improving, li2023importance},
we observe that, when $\mu$ has a continuous target density $\dmu$
and $Q(\bq_0,d \bq) = N(\bq_0, \sigma^2 I)(d\bq)$, then we may consider
an acceptance structure with probabilities proportional to
$\sqrt{\dmu(\bq_j)}$.  As we detail in \cref{Prop:alg:scalingatinfty:local} as a
consequence of \cref{cor:large:p:prod}, for large $\np$ and small
$\sigma^2$, we obtain a method which approximates an (overdamped)
Langevin-type dynamic.  Our main novel insight here is that this local
approach can be adapted to the setting of (\ref{eq:s:f:infinite:d}).
Here we consider the case when \eqref{eq:basic:b:MP-MCMC} is a cloud of
pCN type proposals namely where
$Q(\bq_0,d \bq) = N(\rho \bq_0, (1-\rho^2) I)(d\bq)$ for
$\rho \in (0,1)$ and aim at a target measure of the form
\eqref{eq:stuart:form:og}, \eqref{eq:s:f:infinite:d}.  Herein we
observe that when we accept amongst the generated points
proportionally to $\exp(-\rho/(1+\rho) \Pot(\bq_j/\rho))$ we
obtain a Langevin-type method preconditioned on the covariance of
$\mu_0$ in (\ref{eq:s:f:infinite:d}) as in \cite{cotter2013mcmc} for
large $\np$ and $\rho$ close to $1$.  While biased in general this
proposal structure can then be used as within the MTM formalism to
obtain an unbiased method as we unfold below in \cref{meth:loc:pCN},
\cref{alg:MTMpCN:loc}.

A third acceptance structure of interest from
\eqref{eq:basic:b:MP-MCMC} selects proposals with probabilities
proportional to $\pi(\bq_j)/\dpk(\bq_0,\bq_j)$ when
$Q(\bq_0, d \tbq) = \dpk(\bq_0, \tbq) \mu_0(d \tbq)$.  We refer to
this method as the ``slingshot'' as it favors acceptance of proposals
in the tails of $f(\bq_0,\tbq)$, far away from the current point
$\bq_0$.  Algorithmically this approach is summarized as
\begin{algorithm}[H]
	\caption{The Slingshot}
	\begin{algorithmic}[1]\label{alg:slingshot:intro}
          \State Select a number of proposals $\np \geq 1$ 
          per step and an initial state $\bq^{(0)} \in \qsp$.
          \For{$k \geq 0$}
                \State Draw $\bq_1^{(k+1)}, \ldots, \bq_\np^{(k+1)}$
                from $\dpk(\bq^{(k)}, \tbq) \mu_0 (d\tbq)$ independently
                and set $\bq_0^{(k+1)} = \bq^{(k)}$.
                   \State Select $\bq^{(k+1)} := \bq_{j}^{(k+1)}$ from amongst 
                   $\bq_0^{(k+1)}, \bq_1^{(k+1)}, \ldots, \bq_\np^{(k+1)}$
                   with probabilities
                   \begin{align}
                     \alpha_j^{(k+1)} :=
                     \frac{\pi(\bq_j^{(k+1)}) \dpk(\bq^{(k+1)}_0, \bq_j^{(k+1)})^{-1}}
                     {\sum_{l=0}^\np
                         \pi(\bq_l^{(k+1)}) \dpk(\bq^{(k+1)}_0, \bq_l^{(k+1)})^{-1}}.
                   \end{align}
                   \EndFor
                 \end{algorithmic}
\end{algorithm}
\noindent Several observations immediately underline the significance
of \cref{alg:slingshot:intro}. Firstly we notice that its
Rao-Blackwellization is an importance sampler for $\mu$ (see
\eqref{eq:RB:form:SS}).  Secondly we show that the $\np = \infty$
limiting process is simply independent samples from the target $\mu$
(\cref{cor:large:p:prod}, \cref{sec:p:inf:lim:kernels:SS}).  We carry out
empirical studies of \cref{alg:slingshot:intro} in
\cref{sec:The:Slingshot}, providing some initial empirical evidence of
the promise of \cref{alg:slingshot:intro}.  Here a state-dependent,
total variation norm minimizing, proposal standard deviation
$\sigma_f(\bq)$ derived in \cref{sec:bounds:gauss} improves mixing and
the rate of convergence of the associated kernels to the
$\np = \infty$ limit. Interestingly, this strongly state-specific
scaling violates popular adaptive tuning criteria, such as the
diminishing adaptations criterion \cite{roberts2009examples},
established for unbiased kernels, and neither
\cref{alg:slingshot:intro} nor \cref{subsubsec:bubble:bath} yield
generally unbiased Markov kernels that leave the target invariant.

In summary, while each of the above described acceptance structures
from (\ref{eq:basic:b:MP-MCMC}) have their advantages, none of them are
generically unbiased.  Indeed, as we compute below, to have an
unbiased acceptance structure directly from
\eqref{eq:basic:b:MP-MCMC}, we obtain an unwieldy and computationally
expensive acceptance structure (see \eqref{def:ar:MH:fd:BB}), which
requires $O(\np^2)$ floating point operations and which is not easily
amenable to efficient parallelization.  Thus, the first significance
of the large proposal limit (\cref{thm:large:p:lim},
\cref{cor:large:p:prod}) has been made plain as we introduced the
various methods above, not least \cref{alg:slingshot:intro}: it allows
us to identify algorithms that, while biased at the finite $\np$, are
asymptotically unbiased.

Of course it may be computationally infeasible to rely on a
$\np \to \infty$ limit. In previous literature, two approaches have
been proposed to sidestep the inherent bias in
\eqref{eq:basic:b:MP-MCMC} while avoiding the expensive acceptance
structure given by \eqref{def:ar:MH:fd:BB}.  In one direction the
Multiple Try Metropolis (MTM) \cite{liu2000multiple} removes bias
using a two-step acceptance strategy.  Meanwhile the convolutional
proposal approach developed in \cite{tjelmeland2004using,
  holbrook2023generating, glatt2024parallel} does so with a two-step
proposal strategy instead.  Here our involutive perspective allows us
to broaden the scope of both approaches.  Regarding the MTM approach
our general involutive framework allows us to consider this class of
methods in a general state space setting.  Our result here,
\cref{thm:MTM:unbiased}, includes the Hilbert space (or
preconditioned) methods suited for the setting
\eqref{eq:s:f:infinite:d}.  Thus, \cref{thm:MTM:unbiased} provides an
interesting complement/counterpoint to the mpCN methods we developed
recently in \cite{glatt2024parallel, borggaard2023statistical,
  glatt2024sacred}.  We summarize the new MTM-pCN methods that fall
out of \cref{thm:MTM:unbiased} as \cref{meth:MTMpCN},
\cref{alg:MTMpCN} and as \cref{meth:loc:pCN}, \cref{alg:MTMpCN:loc}.
Furthermore, we make clear in \cref{sec:MTM} that the slingshot,
local, and bubble bath methods can all be `corrected for bias' by
introducing an appropriate second accept-reject step (see
\cref{thm:MTM:unbiased}).

Note that while the MTM strategy's use of $p-1$ auxiliary variables
$\bz_1^{(k+1)}, \ldots, \bz_{\np-1}^{(k+1)}$ nearly doubles the amount
of computation required compared to, e.g., the convolutional proposal
approach described in \cref{sec:conv:prop}, the latter approach is not
without issue.  \cref{thm:RB:1Step} shows that, for a single MCMC
iteration, convolutional proposal approaches fail when one tries to
incorporate all proposals within a single estimator, as the variance
of such Rao-Blackwellized estimators does not converge to 0 in the
large $p$-limit. A seeming contradiction arises: convolutional
proposal algorithms leave their target invariant but their
iteration-wise Rao-Blackwellizations' mean squared errors do not
converge to 0 in the large $\np$-limit, while the slingshot
(\cref{alg:slingshot:intro}) fails to leave its target invariant (for
finite $p$) but furnishes kernels with asymptotically exact
Rao-Blackwellizations.  In any event, the relative performance of the
MTM vs convolutional approaches constitutes a delicate open question
which is an interesting topic for future research as we lay out in
\cref{sec:Outlook}.

In addition to shedding light on algorithms that use and avoid
\eqref{eq:basic:b:MP-MCMC}, we discover two degeneracies in the
involutive theory that shut the door on two needlessly computationally
expensive MP-MCMC approaches.  The first degeneracy holds at a high
level of abstraction (see \cref{rmk:dengenaries},
\cref{rmk:other:than:bias}), making plain that various methods with a
Metropolis-type \eqref{def:alphaj:metr:2} rather than a Barker-type
\eqref{def:alphaj:Barker} acceptance mechanism reduce to simple
mixture of single-proposal methods.  From a theoretical perspective,
the latter distinction is particularly pertinent, as the famous
Peskun-optimality \cite{peskun1973, tierney1998note} of the classical
MH kernel does not apply to the multiproposal setting, particularly in
the large $\np$-limit. The practical takeaway is that computationally
intensive MP-MCMC algorithms that use the Metropolis-type acceptance
are exactly as useful as their single-proposal predecessor. Indeed,
this result may accord with intuition. The Metropolis-type acceptance
probabilities of \eqref{def:alphaj:metr:2} have an apparent drawback:
the function $\min (1,\cdot)$ removes the competitive advantage of
proposals with larger target probabilities among those proposals for
which target probabilities exceed unity. Meanwhile, the Barker-type
acceptances feature no such leveling device.

\cref{sec:HMC:path:sel} contains a second degeneracy, which holds for
MP-MCMC algorithms that use intermediate numerical integration points
in an HMC-type scheme as proposals.
\cref{thm:large:p:MH:hmc} establishes that, as long as the numerical
integration technique is consistent in the small step size-limit,
multiple multiproposal HMC kernels degenerate to vanilla HMC with
random integration times.  From a theoretical perspective, the result
is interesting insofar as it holds for both Metropolis- and
Barker-type acceptances, the former being \emph{a fortiori} degenerate
regardless of proposal count, the latter achieving degeneracy only within the
large $p$-limit of the multiproposal HMC context.  The practical
upshot is, again, that certain computationally intensive MP-MCMC
algorithms are exactly as useful as an easy-to-program,
computationally inexpensive, traditional HMC algorithm with small
modification.

Finally let us make a few further remarks on the passage to the
$\np \to \infty$ limit which we carry out in \cref{sec:large:p:limit}
and in \cref{sec:HMC:path:sel}.  Rigorously establishing this limit in
the weak sense is relatively straightforward mathematically for
\eqref{eq:basic:b:MP-MCMC}, its convolutional variant
\eqref{def:vk:Tjelmeland} in \cref{thm:large:p:lim}, and for the
(degenerate) MP-MCMC HMC structures in \cref{thm:large:p:MH:hmc}.  On
the other hand, the MTM $\np \to \infty$ is more challenging due to
non-trivial correlations in the kernels which appear in this case (see
\eqref{eq:MTM:Kernel}).  As described in \cref{thm:MTM:limit}, this limit
yields a traditional Metropolis-Hastings kernel built from a proposal
given by the $\np \to \infty$ limit of the proposal step as defined in
\cref{cor:large:p:prod}.  While analogous results had previously been
considered in \cite{gagnon2023improving} for some continuous targets
on $\RR^d$, we are able to treat the general state space case here.
Our approach uses a delicate compactness argument leveraging the
uniform law of large numbers, \cite{talagrand1987glivenko,
  dudley1991uniform, van2014universal, wainwright2019high} i.e. by
identifying an appropriate Glivenko-Cantelli class in the interaction
between the weighted acceptance structure $\beta$ and the cloud
proposal kernel $\pk$ appearing in \eqref{eq:MTM:Kernel}.

The rest of the manuscript is organized as follows: \cref{sec:MP:MCMC}
sets out the general framework around \eqref{eq:stuart:form},
\eqref{def:mp:ker} which culminates in our extension of the involutive
framework in \cref{thm:grand:unifed:inv}.  Then in
\cref{sec:multi:prop:algs} we comprehensively lay out various methods
including all of our novel algorithmic contributions arising from
\eqref{eq:basic:b:MP-MCMC} and its Multiple Try and Convolution variations.  In
\cref{sec:large:p:limit} we carry out all the $\np \to \infty$ limits
for a broad class of kernels which include all the methods in
\cref{sec:multi:prop:algs} as special cases.  In this section we also
use the derived limit kernels to complete the picture sketched
previously in \cref{sec:multi:prop:algs}.  
\cref{sec:HMC:path:sel} is devoted to establishing the degeneracy of
some HMC-type multiproposal methods.  Meanwhile, \cref{sec:The:Slingshot} gives
some extended case study for \cref{alg:slingshot:intro}. 
Finally \cref{sec:Outlook} provides an outlook and considers
some possible avenues for future research suggested by the results 
herein.  While mathematically
precise statements and sketches of proof are provided in the main
body of the manuscript, rigorous proofs are deferred for
\cref{sec:rig:proofs}.

\section{Preliminaries: The general state space and involutive
  frameworks}
\label{sec:MP:MCMC}

Following our recent works \cite{glatt2024parallel, glatt2024sacred},
we begin by introducing a flexible, general state-space formalism
which provides the setting and notational conventions for all of the
multiproposal algorithms we study in what follows.  In particular, this
framework encompasses a class of `convolutional' multiproposal kernels
considered previously in \cite{tjelmeland2004using, glatt2024parallel,
  glatt2024sacred}, the two stage Multiple Try (MT)
methods from \cite{luo2019multiple, liu2000multiple} as well as some
new algorithms with a similar proposal structure which we introduce
here.  On the other hand, our abstract set-up also admits
multiproposal versions of the HMC algorithm.  In particular this
encompasses the situation where we proceed at each step by selecting
among different points on the (numerically integrated) Hamiltonian
path as we consider below in \cref{sec:HMC:path:sel}.

We proceed as follows: assume $(\qsp, \Sigma_\qsp)$,
$(\vsp, \Sigma_\vsp)$ are measurable spaces, and let
$\vk: \qsp \times \Sigma_\vsp \to [0,1]$ be a Markov kernel. Namely, we
assume $\vk(\bq_0, \cdot)$ is a probability measure for each fixed
$\bq_0 \in X$ and that $\bq_0 \mapsto \vk(\bq_0, A)$ is a measurable
map over all $A \in \Sigma_\vsp$.  Now, for a given number
$\np \geq 1$ of proposals, we consider also a collection of $\np+1$
measurable mappings on $\qsp \times \vsp$, $\bS = (S_0,\ldots, S_\np)$
and we let $\balpha = (\ar_0,\ldots,\ar_\np)$ be a collection of
acceptance probabilities.  Thus, for $j =0, \ldots, p$, we suppose
that $\alpha_j : \qsp \times \vsp \to [0,1]$ is measurable and
$\balpha$ maintains
\begin{align}
  \label{def:mp:ker:prob:cond}
	\sum_{j = 0}^p \int_\vsp \alpha_j(\bq_0,\bv)  \vk(\bq_0,d \bv) = 1,
	\quad \mbox{ for every } \bq_0 \in \qsp.
\end{align}
Then, the triple $(\balpha, \bS, \vk)$ defines a multiproposal
algorithm with a one-step Markov transition kernel
$\mk: \qsp \times \Sigma_\qsp \to [0,1]$ given by \eqref{def:mp:ker}
where $\Pi_1: \qsp \times \vsp \to \qsp$ denotes the projection
operator onto the $\qsp$ component, i.e. $\Pi_1(\bq, \bv) = \bq$ for
all $\bq \in \qsp$, $\bv \in \vsp$, and
$\delta_{\Pi_1 S_j (\bq_0, \bv)}$ is the Dirac measure concentrated at
$\Pi_1 S_j (\bq_0, \bv)$.\footnote{Note that, for many of the rigorous
  results provided below we shall suppose that $\qsp$ is a Polish
  space (this is a separable completely metrizable topological space) with
  $\Sigma_\qsp$ being the usual Borel $\sigma$-algebra generated by the open
  sets.}

Algorithmically,
\eqref{def:mp:ker} leads to MCMC chains obtained
iteratively as follows.  We start from an initial state $\bq^{(0)}$.
Then, for each $n \geq 0$, given the current state $\bq^{(n)}$, we
draw $\bv^{(n)} \sim \vk(\bq^{(n)},d \bv)$ to generate the cloud of
proposals
\begin{align}\label{proposed:states}
    (\bq^{(n+1)}_0, \ldots, \bq^{(n+1)}_\np) :=
	\Pi_1 S_0 (\bq^{(n)}, \bv^{(n)}), \ldots, \Pi_1 S_\np (\bq^{(n)}, \bv^{(n)}).
\end{align}
Finally, the next state $\bq^{(n+1)}$ is selected amongst
$(\Pi_1 S_0 (\bq^{(n)}, \bv^{(n)}), \ldots, \Pi_1 S_\np (\bq^{(n)},
\bv^{(n)}))$ according to probabilities
$(\alpha_0^{(n)}, \ldots, \alpha_\np^{(n)})$ given by
\begin{align*}
  \alpha_j^{(n)} := \alpha_j (\bq^{(n)}, \bv^{(n)}), \quad j = 0,\ldots, \np.
\end{align*}
We can implement this by drawing $u^{(n)} \sim \cU (0,1)$
independently of $\bq^{(n)}, \bv^{(n)}$, and considering the intervals
\begin{align}\label{def:int:I}
  I_0^{(n)} := \left[0, \ar_0^{(n)} \right) \,,\,\,
  I_l^{(n)} :=
  \left[ \sum_{k=0}^{l-1} \ar_k^{(n)}, \sum_{k=0}^l \ar_k^{(n)} \right),
  \quad l=1, \ldots, \np,
\end{align}
and taking $\bq^{(n+1)}$ as
\begin{align}\label{eq:qn:unif:draws:form}
  \bq^{(n+1)} = \sum_{l=0}^\np
  \indFn{I_l^{(n)}} (u^{(n)}) \Pi_1 S_l (\bq^{(n)}, \bv^{(n)}).
\end{align}

\begin{Remark}
  \label{rmk:dengenaries}
  At this level of generality, the kernel $P$ defined by \eqref{def:mp:ker} (and its implementation
  \eqref{eq:qn:unif:draws:form}) is just a convenient
   overarching representation encompassing any
  multiproposal MCMC algorithm.  In fact, as we previously observed in
  \cite{glatt2024parallel}, this kernel includes \emph{any} Markov
  transition semigroup $R$ on a measurable space
  $(\qsp, \Sigma_{\qsp})$ by taking $\vsp= \qsp$, $p=0$,
  $\vk(\bq, d \bv)= R(\bq, d \bv)$, $\alpha_0 = 1$, and finally
  $S_0(\bq, \bv) = (\bv, \bq)$. It is also worth noting that one
  algorithm (or equivalently Markov transition kernel) can have
  multiple (non-trivial) representations of the form
  \eqref{def:mp:ker}.  This is a point we previously explored in
  depth in \cite[Section 3]{glatt2020accept}.

  Here we take the opportunity to point out a particular degeneracy in
  \eqref{def:mp:ker} which we will return to several times below.  To
  explain this degeneracy, given positive constants $a_1,\dots,a_\np$, 
  suppose that we consider any kernel of the Metropolis-type formulation
  \begin{align}\label{def:mp:ker:single degenerate}
    \mk(\bq_0, d\tbq) =&
  \sum_{j=1}^p \int_{\vsp} a_j  (1 \wedge \beta_j (\bq_0, \bv))
                    \delta_{\Pi_1 S_j (\bq_0, \bv)} (d\tbq) \vk(\bq_0, d \bv)
                    \notag\\
                  &+ \int_{\vsp}\left(1 - \sum_{j=1}^p  a_j (1 \wedge
                    \beta_j (\bq_0, \bv))\right)
    \delta_{\Pi_1 S_0 (\bq_0, \bv)} (d\tbq) \vk(\bq_0, d \bv),
    \quad \text{ where } \sum_{j =1}^\np a_j   \leq 1.
  \end{align}
  In this case the expression can be reorganized to show that
  \eqref{def:mp:ker} reduces to the pure mixture:
  \begin{align}
   \label{def:mp:ker:single degenerate:1}
   P(\bq_0, d \tbq) = \sum_{j =1}^p a_j P^j(\bq_0, d \tbq)
   + \bigl(1- \sum_{j =1}^p a_j \bigr) P^0(\bq_0, d \tbq)
  \end{align}
  where $P^0(\bq_0, d \tbq):= \int_{\vsp}
  \delta_{\Pi_1 S_0 (\bq_0, \bv)} (d\tbq) \vk(\bq_0, d \bv)$
  and
  \begin{align}\label{eq:single:prop:degen}
   P^j(\bq, d \tbq) :=
     \int_{\vsp}  (1 \wedge \beta_j (\bq, \bv))
                    \delta_{\Pi_1 S_j (\bq, \bv)} (d\tbq) \vk(\bq, d \bv)
                  + \int_{\vsp} (1 -  (1 \wedge \beta_j) (\bq, \bv))
    \delta_{\Pi_1 S_0 (\bq, \bv)} (d\tbq) \vk(\bq, d \bv),
  \end{align}
  are `single-proposal' Metropolis-type kernels.  Thus, for the
  special case \eqref{def:mp:ker:single degenerate} of
  \eqref{def:mp:ker}, non-trivial parallelism is lost.  Namely, to
  draw from \eqref{def:mp:ker:single degenerate}, one may simply draw
  an index $k \in \{0,1, \ldots, p\}$ with the respective
  probabilities $(1- \sum_{l =1}^\np a_l, a_1, a_2, \ldots, a_\np)$
  and then draw a single proposal from the corresponding $P_k$ as in
  \eqref{eq:single:prop:degen}.   Obviously this later
  formulation in \eqref{def:mp:ker:single degenerate:1},  
  \eqref{eq:single:prop:degen}
  avoids the evaluation of $S_j$ and $\beta_j$ for all $j \not = k$
  in contrast to \eqref{eq:qn:unif:draws:form}.  As our exposition unfolds, we show
  that this degeneracy in \eqref{def:mp:ker:single degenerate} exposes
  the triviality of MCMC methods introduced in previous works; see
  \cref{rmk:MH:degenerate}, \cref{rmk:more:degen} and
  \eqref{def:mk:mh:hmc} below.
\end{Remark}

Given a (measurable) observable $\Obs: \qsp \to \RR$ and number
$N \geq 0$ of algorithmic iterations, the standard empirical estimator
for the average $\int_\qsp \Obs(\bq) \mu(d\bq)$ is given by
\begin{align}\label{eq:phi:p:estimator}
  \sdest := \frac{1}{N} \sum_{j=1}^{N} \Obs(\bq^{(j)})
  = \int_\qsp \Obs(\bq) \sdem (d \bq),
\end{align}
where $\sdem$ is the empirical measure
\begin{align*}
  \sdem(d\bq)
  := \frac{1}{N} \sum_{j=1}^{N} \delta_{\bq^{(j)}} (d \bq)
  = \frac{1}{N} \sum_{j=1}^{N}
  \sum_{l = 0}^p \indFn{I_l^{(j-1)}}(u^{(j-1)})
  \delta_{\Pi_1 S_l (\bq^{(j-1)}, \bv^{(j-1)})} (d \bq),
\end{align*}
with the intervals $I_l^{(j)}$ as defined in \eqref{def:int:I}.

Note that, as in \cite{calderhead2014general, schwedes2021rao,
  yang2018parallelizable}, we can consider the following
Rao-Blackwellization $\rbest$ of $\sdest$.  Taking
the $\sigma$-algebras
\begin{align}
  \cF^{(n)} := \sigma\{ (\bq^{(k)}, \bv^{(k)}) | k =0, \ldots, n\},
  \quad \text{ for } n \geq 0,
\end{align}
and setting $\rbest
 := \frac{1}{N} \sum_{j=1}^{N} \EE(\Obs(\bq^{(j)}) | \cF^{(j-1)})$.  Under this definition
 we compute from \eqref{eq:qn:unif:draws:form} that
\begin{align}\label{eq:RB:form}
  \rbest &
  = \frac{1}{N} \sum_{j=1}^N \sum_{l=0}^\np \EE ( \indFn{I_l^{(j-1)}} (u^{(j-1)})
     \varphi(\Pi_1 S_l (\bq^{(j-1)}, \bv^{(j-1)}))| \cF^{(j-1)} )
           \notag\\
  &=\frac{1}{N} \sum_{j=1}^N \sum_{l=0}^\np \ar_l(\bq^{(j-1)}, \bv^{(j-1)})
  \Obs(\Pi_1 S_l (\bq^{(j-1)}, \bv^{(j-1)}))
  =\frac{1}{N} \sum_{j=1}^N \sum_{l=0}^\np \ar_l(\bq^{(j-1)}, \bv^{(j-1)})
  \Obs(\bq^{(j)}_l).
\end{align}
\begin{Remark}\label{rmk:RB:cald}
  Observe that the final equality in \eqref{eq:RB:form} follows
  from basic properties of the conditional expectation
\begin{align}\label{eq:prop:cond:exp}
    \EE ( \indFn{I_l^{(j-1)}} (u^{(j-1)})
    \varphi(\Pi_1 S_l (\bq^{(j-1)}, \bv^{(j-1)}))| \cF^{(j-1)} )
    &=   \varphi(\Pi_1 S_l (\bq^{(j-1)}, \bv^{(j-1)}))\EE ( \indFn{I_l^{(j-1)}} (u^{(j-1)})
      | \cF^{(j-1)} )
      \notag\\
    &=  \ar_l(\bq^{(j-1)}, \bv^{(j-1)}) \varphi(\Pi_1 S_l (\bq^{(j-1)}, \bv^{(j-1)})),
  \end{align}
  for $j \geq 1$, $l = 0, \ldots, p$.  In particular, for the second equality we are using the
  fact that, given independent random variables $\bz$ and $u$ and any
  integrable $\Psi$, 
  \begin{align}\label{eq:cond:indep:expID}
  \EE (\Psi(\bz, u) | \bz) = \rho(\bz) \text{ where we take }
  \rho(\bbz) = \EE (\Psi(\bbz, u)) \text{ for each fixed } \bbz; 
  \end{align}
  see e.g. \cite[Section
  5.1]{durrett2019probability} for further details.  Note that in
  \cite[Section 3.1]{schwedes2021rao} they instead define
 \begin{align}\label{eq:calderhead:def}
  \rbest' := \EE \left(\frac{1}{N} \sum_{j=1}^{N} \Obs(\bq^{(j)}) |
    \cF^{(N-1)}\right).
  \end{align}
  This is not the same estimator and we should not expect that this
  latter definition \eqref{eq:calderhead:def} will yield an explicit
  structure for the resulting estimator as in \eqref{eq:RB:form}. This
  is because, for $j < N$, $\cF^{(N-1)}$ contains information about
  $u^{(j-1)}$ since it contains the a complete knowledge of $\bq^{(j)}$ and
    $\bv^{(j)}$, namely $u^{(j-1)}$ is not independent of
    $(\bq^{(j)}, \bv^{(j)})$ as this later quantity provides information
    about which interval defined in \eqref{def:int:I} contained $u^{(j-1)}$.
  
 Given a random variable $\bz$ and $\sigma$-algebra $\cF$, recalling 
 the usual definition of conditional variance as
 \begin{align}\label{eq:cond:var:def}
   \Var(\bz | \cF) = \EE(\bz^2| \cF) - (\EE(\bz | \cF))^2
   = \EE( (\bz - \EE(\bz | \cF))^2|\cF)
 \end{align}  
  and noting that 
  \begin{align}\label{eq:RB:var:decomp}
  \Var(\bz) = \EE (\Var(\bz| \cF)) + \Var( \EE( \bz| \cF)) 
  \geq  \Var( \EE( \bz| \cF)) 
  \end{align}
  (see e.g. \cite[Section 5.1]{durrett2019probability}) it is clear
  that, cf. \eqref{eq:phi:p:estimator}, \eqref{eq:calderhead:def},
  $\Var(\rbest') \leq \Var( \sdest)$.  This seems to be 
  basis for the variance reduction claim in \cite[Lemma 3.1]{schwedes2021rao}. 
  However, unfortunately, this
  reasoning does not extend in any obvious way to guarantee variance
  reduction for $\rbest$ relative to $\sdest$ for $N \geq 2$.
\end{Remark}

\subsection{Criteria for Reversibility}

As we observed in \cite{glatt2024sacred} and recalled here in \cref{rmk:RB:cald},
the Markov transition kernels
defined by \eqref{def:mp:ker} are extremely general, covering in fact
\emph{all} Markov kernels as a special case without imposing further
restrictions.  Thus, we need criteria to identify interesting
subclasses in which to consider the $p \to \infty$ limit in
\eqref{def:mp:ker}.  One obvious criterion for determining the elements
$(\vk,\bS, \balpha)$ defining $P$ is a guarantee of unbiasedness of
$\mk$ with respect to the given target measure $\mu$ on
$(\qsp, \Sigma_\qsp)$.  Note at the outset however that unbiasedness
is not an absolute requirement as we discuss extensively below (see
\cref{rmk:other:than:bias}).

In \cite{glatt2024parallel}, we developed a unified `involutive'
theory providing a criteria for reversibility (and thus unbiasedness) of
multiproposal algorithms relative to a given target measure $\mu$ on a
measurable space $(\qsp, \Sigma_\qsp)$.  This work extends earlier
results in \cite{glatt2020accept, andrieu2020general,
  neklyudov2020involutive} from the single step MP-MCMC setting.
We next recall this theory providing a few new refinements here which
will be used in what follows.

In order to proceed we recall a few definitions.  Given measurable
spaces $(Z, \Sigma_Z), (Z', \Sigma_{Z'})$, a probability measure
$\mathcal{N}$ on $(Z, \Sigma_Z)$, and a measurable function
$F: Z \to Z'$, the \emph{pushforward} of $\mathcal{N}$ under $F$,
denoted as $F^*\mathcal{N}$, is the probability measure on
$(Z', \Sigma_{Z'})$ defined as
$F^*\mathcal{N}(A) = \mathcal{N}(F^{-1}(A))$ for any
$A \in \Sigma_{Z'}$.  In other words, if $\bz \sim \mathcal{N}$ then
$F(\bz) \sim F^* \mathcal{N}$.  Next, if
$\mathcal{N}, \tilde{\mathcal{N}}$ are both measures on a measurable
space $(Z, \Sigma_Z)$ we say that $\mathcal{N}$ is absolutely
continuous with respect to $\tilde{\mathcal{N}}$ and write
$\mathcal{N} \ll \tilde{\mathcal{N}}$ if $\mathcal{N}(A) = 0$ for all
$A \in \Sigma_Z$ such that $\tilde{\mathcal{N}}(A) =0$.  When indeed
$\mathcal{N} \ll \tilde{\mathcal{N}}$ and both are $\sigma$-finite
measures then, according to the Radon-Nikodym theorem, there is a
$\tilde{\mathcal{N}}$-unique measurable function, which we denote as
$d \mathcal{N} /d \tilde{\mathcal{N}}$, such that
$\mathcal{N}(d \bq) = d \mathcal{N} /d \tilde{\mathcal{N}}(\bq)
\tilde{\mathcal{N}}(d \bq)$.

We now present the following slight refinement of the main abstract
result in \cite[Theorem 2.2]{glatt2024parallel} as follows:
\begin{Theorem}
  \label{thm:grand:unifed:inv}
  Fix measurable spaces $(\qsp, \Sigma_\qsp)$, $(\vsp, \Sigma_\vsp)$ a
  target probability measure $\mu$ on $(\qsp, \Sigma_\qsp)$ and a
  triple $(\vk,\bS, \balpha)$ producing a kernel $\mk$ of the form
  \eqref{def:mp:ker}, subject to \eqref{def:mp:ker:prob:cond}.
  Denote
  \begin{align}\label{eq:comp:ker}
    \cM(d\bq, d\bv) = \vk(\bq, d \bv) \mu(d\bq).
  \end{align}
  We suppose that $\bS = (S_0,\ldots, S_\np)$ consists of (measurable)
  involutions, namely
  \begin{align}\label{cond:invol:S}
    S_j\circ S_j(\bq,\bv) = (\bq,\bv)
    \text{ for } j = 0, \ldots, p \text{ and any }
    (\bq, \bv) \in \qsp \times \vsp.
  \end{align}
  Furthermore, we assume that\footnote{The formulation in
    \cite[Theorem 2.2]{glatt2024parallel} required that
    $\alpha_j( S_j(\bq, \bv)) S^*_j\cM (d \bq, d \bv) = \alpha_j(\bq, \bv) \cM (d \bq, d \bv)$
    which of course implies \eqref{cond:alpha:rev}.}
  \begin{align}\label{cond:alpha:rev}
    \! \!  \sum_{j =0}^p \int\limits_{\qsp} \! \int\limits_{\vsp} 
    \!  \!\psi(\Pi_1 S_j (\bq, \bv), \bq)
    \alpha_j( S_j(\bq, \bv)) S^*_j\cM (d \bq, d \bv)
   \!  = \! \sum_{j =0}^p \int\limits_{\qsp} \! \int\limits_{\vsp} 
   \!  \! \psi(\Pi_1 S_j (\bq, \bv), \bq )
    \alpha_j(\bq, \bv) \cM (d \bq, d \bv),
  \end{align}
  for any bounded, measurable $\psi: \qsp \times \qsp \to \RR$.  Then,
  under these circumstances,
  \begin{align}\label{eq:rev:gen:def}
    \mk(\bq, d\tbq) \mu(d \bq) = \mk(\tbq, d\bq) \mu(d \tbq),
  \end{align}
  namely $\mk$ generates a reversible algorithm with respect to
  $\mu$.\footnote{In particular \eqref{eq:rev:gen:def} implies
    invariance, $\int_\qsp \mk(\bq, d\tbq) \mu(d \bq) = \mu(d \tbq)$,
    so that this is guarantee of unbiasedness of $P$ with respect to
    $\mu$.}
\end{Theorem}
\noindent For completeness, the proof of this refinement is
provided below in \cref{sec:gen:inv}.

Given the target measure $\mu$ and having decided on a proposal
structure defined by $\vk$ and $\bS$, there are two natural examples
of acceptance probability structures
$\balpha =(\ar_0, \ldots, \ar_\np)$ that satisfy
\eqref{cond:alpha:rev}.  The first `Barker-type' class will guide our
discussions in the rest of the article.  On the other hand a second
class of `Metropolis-Hastings-type' $\balpha$'s are shown to be highly
degenerate in \cref{rmk:other:than:bias}.

For the former setting, we assume that
$\bS$ satisfies the following condition for every $j=0,1,\ldots, \np$
\begin{align}\label{sum:cond}
  \sum_{k=0}^\np (S_j \circ S_k)^*\cM  (d \bq, d \bv)
	= \sum_{k=0}^\np S_k^*\cM  (d \bq, d \bv).
\end{align}
Then the `Barker-type acceptance probabilities' defined as
\begin{align}\label{def:alphaj:Barker}
  \alpha_j(\bq, \bv) =
  \frac{d S_j^*\cM}{d(S_0^*\cM + \cdots + S_p^*\cM)}(\bq, \bv),
  \quad j= 0, \ldots, \np,
\end{align}
maintain \eqref{cond:alpha:rev}.  We refer to \cite[Corollary
2.4]{glatt2024parallel} for a rigorous proof of this.

For the latter setting for $\balpha$ we drop the requirement \eqref{sum:cond} but
assume now that $S_0 = I$ and that $S_j ^* \cM \ll \cM$ for all
$j=1,\ldots, \np$.  In this circumstance we then consider any collection of weights
$a_j \in \RR^+$, $j=1, \ldots, \np$, with $\sum_{j=1}^\np a_j \leq 1$
and formulate the `Metropolis-Hastings-type' acceptance probabilities as
\begin{align}
  \alpha_j(\bq,\bv)
	:= \left[ 1 \wedge \frac{dS_j^*\cM}{d \cM}(\bq,\bv)\right] a_j,
	\quad j = 1,\ldots,p,  \quad \text{ with } \quad
	\alpha_0(\bq,\bv) := 1 - \sum_{j=1}^p \alpha_j(\bq,\bv),
	\label{def:alphaj:metr:2}
\end{align}
Such $\alpha_j$'s also satisfy \eqref{cond:alpha:rev} as recorded in \cite[Corollary
2.6]{glatt2024parallel}.  However, this latter setting leads to a degenerate scheme as follows:

\begin{Remark}\label{rmk:MH:degenerate}
  Selecting the acceptance probabilities \eqref{def:alphaj:metr:2}
  results in a kernel $\mk(\bq, d\tbq)$ of the form
  \eqref{def:mp:ker:single degenerate} in
  \cref{rmk:dengenaries}. Thus, in this situation, $\mk$ is a simple
  mixture of single proposal kernels, namely it has the form
  \eqref{def:mp:ker:single degenerate:1}.  Here the $P^j$ in
  \eqref{eq:single:prop:degen} take the specific form
  \begin{align}
    \label{eq:single:prop:degen:MH}
   P^j(\bq , d\tbq)
   \!=& \! 
      \int_{\vsp} \! \!
      \left( \! 1 \wedge \frac{dS_j^*\cM}{d \cM}(\bq,\bv) \! \right)
  \delta_{\Pi_1 S_j (\bq, \bv)} (d\tbq) \vk(\bq, d \bv)
  \!+ \delta_\bq( d\tbq) \! \!
     \int_{\vsp} \! \!
      \left(\! 1- 1 \wedge \frac{dS_j^*\cM}{d \cM}(\bq,\bv)
      \! \right)\vk(\bq, d \bv).
  \end{align}
  In particular, note that one can readily verify that these $P^j$'s
  in \eqref{eq:single:prop:degen:MH} individually satisfy
  \cref{thm:grand:unifed:inv}, which reduces in the single proposal
  case whose reversibility at this level of generality is addressed in
  earlier results, \cite{glatt2020accept, andrieu2020general}.
\end{Remark}

\begin{Remark}\label{rmk:other:than:bias}
  We should emphasize an unbiasedness requirement is not absolute and
  will not always be imposed in what follows.  For example, while the
  requirement \eqref{sum:cond} is met for classes of (conditionally)
  independent proposal structures, this condition fails to hold in some
  HMC settings we consider; see \eqref{eq:Barker:HMC},
  \eqref{def:mkB:hmc}.  Moreover, the Radon-Nikodym derivative
  $d S_j^* \cM/d \cM$ can have an unwieldy form as in
  \eqref{def:ar:MH:fd:BB}.

  In any event, as we proceed, other motivations for the form of
  $\mk^{(\vk,\bS, \balpha)}$ in \eqref{def:mp:ker} will appear
  suggesting the salience of some biased methods. This includes the
  structure of the associated Rao-Blackwellized estimators
  \eqref{eq:RB:form} and the implied $p \to \infty$ kernel and its
  unbiasedness; see \cref{cor:large:p:prod} and \cref{sec:p:inf:lim:kernels:SS}
  below.  We will also consider situations which up to the present
  merely arise as immediate simplifications starting from an unbiased
  kernel featuring a cheaper to evaluate structure as in
  \cref{subsubsec:bubble:bath}.  In some cases these kernels yield
  desirable empirical properties suggesting rigorously unjustified
  methods worthy of further consideration in the future.
\end{Remark}

\section{A selection of Multiproposal Algorithms}
\label{sec:multi:prop:algs}

In this section, we describe a collection of multiproposal methods
built on a proposal structure based on i.i.d. samples from a fixed
single proposal kernel, conditional on the current state; see
\eqref{eq:basic:b:MP-MCMC}. Thus we are interested here in various
multiproposal generalizations of the random walk Metropolis-Hastings (RWMH). In
\cref{sec:cond:prop}, we consider several different acceptance
mechanisms associated with this proposal structure.  In particular, we
detail a new (biased) method which we refer to as `the Slingshot'
previewed above in the introduction as \cref{alg:slingshot:intro}.  

Since the methods we introduce in \cref{sec:cond:prop} are biased in
general we then describe in \cref{sec:MTM}, \cref{sec:conv:prop} two
different strategies to obtain unbiased methods by modifying slightly
the proposal structure or acceptance mechanisms.  First, in
\cref{sec:MTM}, we consider extensions of the Multiple Try
(MT) approach pioneered in \cite{liu2000multiple}, which is based on
a two stage acceptance structure.  By leveraging
\cref{thm:grand:unifed:inv}, we generalize this Multiple Try approach yielding
an unbiased correction of \cref{alg:slingshot:intro}.  We also
introduce two new Hilbert space methods in the vein of the pCN
algorithm, \cite{beskos2008mcmc, cotter2013mcmc}, as
\cref{alg:MTMpCN}, \cref{alg:MTMpCN:loc}.  In contrast, in
\cref{sec:conv:prop}, we review some `convolutional' methods which add
a preliminary step to the proposal mechanism,
\eqref{def:vk:Tjelmeland}, to provide a different means of obtaining
an unbiased algorithm, including \cite{tjelmeland2004using}, and a
third Multiproposal pCN which we recently derived in
\cite{glatt2024parallel,glatt2024sacred}.

Recall that our main interest herein is to analyze the large proposal
limit of certain relevant subclasses of multiproposal algorithms
falling under the framework of \eqref{def:mp:ker}.  As such, the
particular methods outlined will frame the general setting of
\cref{sec:large:p:limit} below, where our rigorous results will
underline the salience of some methods, most prominently the
slingshot, \cref{alg:slingshot:intro}, while highlighting some
possible disadvantages of others. Note also that, further on in
\cref{sec:HMC:path:sel}, we take up (and then largely rule out the
salience of) some multiproposal versions of the Hamiltonian Monte
Carlo (HMC) approach where our proposals consist of different time steps
along the approximated Hamiltonian dynamic.

\subsection{Independent Proposal Structures}
\label{sec:cond:prop}

Let us begin with some methods around the proposal structures of the
form \eqref{eq:basic:b:MP-MCMC}. To formalize this in our abstract
setup, \eqref{def:mp:ker}, consider a (single proposal) Markov
transition kernel $\pk: \qsp \times \Sigma_\qsp \to [0,1]$ on a
measurable space $(\qsp, \Sigma_\qsp)$ where the target probability
measure $\mu$ of interest is defined.  To complete the formal
description of our proposal structure defined by $(\vk, \bS)$, we take
$\vsp = \qsp^\np$ (with the usual product $\sigma$-algebra) and set
$\bS$ to be the coordinate-flip involutions, namely
\begin{align}\label{def:Sj:coord:flip:0}
    S_j(\bq_0, \ldots, \bq_\np)
  := (\bq_j, \bq_1, \ldots, \bq_{j-1}, \bq_0, \bq_{j+1}, \ldots, \bq_\np),
  \quad (\bq_0, \ldots, \bq_\np) \in \qsp^{\np + 1}, \,\, j=0, \ldots, \np.
\end{align}
We then take
\begin{align}\label{def:vk:before:Tjelmeland}
  \vk (\bq_0, d \bq_1, \ldots, d \bq_\np)
  = \prod_{j=1}^p \pk(\bq_0, d \bq_j),
\end{align}
which indeed yields \eqref{eq:basic:b:MP-MCMC} via
\eqref{proposed:states}.

In what follows, we consider several typical structural assumptions on
the single proposal kernels $\pk$ making up $\vk$.  Often, we can
express $\pk$ in terms of a density with respect to the same base
measure $\mu_0$ as we define our given target $\mu$,
\eqref{eq:stuart:form}, namely
\begin{align}\label{eq:prop:ker:diff:form}
  \pk(\bq_0, d \tbq) = \dpk(\bq_0, \tbq)\mu_0( d\tbq),
\end{align}
for a given measurable `kernel-density'
$\dpk: \qsp \times \qsp \to [0,\infty)$ with
$\int \dpk(\bq, \tbq) \mu_0(d \tbq) = 1$, for all $\bq \in \qsp$.
Here a prototypical case is when $\mu_0$ is the Lebesgue measure.  In
particular, this setting includes the typical Gaussian proposal kernels
\begin{align}\label{eq:Guass:1:prop:classic}
  \pk(\bq_0, d\tbq) = \pk_\sigma(\bq_0, d\tbq) := N(\bq_0, \sigma^2 \cC)(d\tbq),
\end{align}
where $\sigma^2$ is the proposal variance parameter and $\cC$ is
symmetric positive definite.  See also the adaptive-covariance
variation on \eqref{eq:Guass:1:prop:classic} in \eqref{eq:sphere:g:ker}
below.

In the infinite dimensional setting, \eqref{eq:s:f:infinite:d}, where
$\mu_0$ is a Gaussian measure, we can consider instead the pCN type
kernel
\begin{align}\label{eq:Guass:1:prop:pCN}
  \pk(\bq_0, d\tbq) = \pk_\rho(\bq_0, d\tbq)
  := N(\rho \bq_0, (1- \rho^2) \cC)(d\tbq),
\end{align}
where $\cC$ is symmetric, positive-definite and trace class, and
$\rho \in [0,1]$ is an algorithmic parameter.\footnote{As described in
  \cite{cotter2013mcmc}, this method can be derived from a
  Crank-Nicolson approximation of a preconditioned Langevin dynamic
  for $\mu_0$ as in \eqref{eq:s:f:infinite:d}.  This corresponds to
  \eqref{eq:precond:LA:dynam}, \eqref{eq:pCN:Mala:disc} below where we
  drop the terms involving $\Pot$.}  While both
\eqref{eq:Guass:1:prop:classic}, \eqref{eq:Guass:1:prop:pCN} are well
defined even when $X$ is e.g. a separable Hilbert space, only the
latter case \eqref{eq:Guass:1:prop:pCN} should be considered in the
infinite dimensional setting \eqref{eq:s:f:infinite:d} in view of the
Feldman-Hayek theorem, \cite{DPZ2014}.  In this `Hilbert space'
setting, the formulation \eqref{eq:prop:ker:diff:form} is unavailable
for \eqref{eq:Guass:1:prop:pCN}.  See e.g. \cite{cotter2013mcmc,
  glatt2020accept} for further context on this latter class of
proposal kernels.

To complete the picture and to obtain an unbiased algorithm from
\eqref{def:Sj:coord:flip:0}, \eqref{def:vk:before:Tjelmeland}, we note
that, according to \cite[Theorem 2.10]{glatt2024parallel},
\eqref{sum:cond} is satisfied under \eqref{def:Sj:coord:flip:0},
\eqref{def:vk:before:Tjelmeland}.  However, this does not typically
lead to a desirable structure for $\alpha$ according to
\eqref{def:alphaj:Barker}.  For example, when $\pk$ is of the form in
\eqref{eq:prop:ker:diff:form} and recalling the notation in \eqref{eq:stuart:form}, we have
\begin{align}\label{def:ar:MH:fd:BB}
    \frac{dS_j^*\cM}{d(S_0^*\cM+\cdots + S_\np^*\cM)}(\bq_0, \ldots, \bq_p)
  = \frac{\dmu(\bq_j) \prod_{\stackrel{i=0}{i\neq j}}^\np
  \dpk(\bq_j, \bq_i)}
  {\sum_{l =0}^\np\dmu(\bq_l)
  \prod_{\stackrel{i=0}{i\neq l}}^\np  \dpk(\bq_l, \bq_i)},
  \text{ for } j = 0,\ldots, \np.
\end{align}
This identity follows along the lines of \cite[Section 4.1]{glatt2024parallel},
where we use that
\begin{align}\label{eq:change:of:Ms:1}
  \cM(d\bq_0, \ldots, d \bq_\np) :=
  \dmu(\bq_0) \prod_{i=1}^\np \dpk(\bq_0, \bq_i) \cM_0(d\bq_0, \ldots,
  d \bq_\np)
  \quad \text{ with }
  \cM_0(d\bq_0, \ldots, d \bq_\np) = \prod_{l =0}^\np \mu_0(d\bq_l),
\end{align}
so that
\begin{align}\label{eq:change:of:Ms:2}
  S_j^*\cM(d\bq_0, \ldots, d \bq_p) :=
  \dmu(\bq_j) \prod_{\stackrel{i=0}{i\neq j}}^\np
  \dpk(\bq_j, \bq_i) \cM_0(d\bq_0, \ldots, d \bq_\np),
\end{align}
for $j = 0, \ldots, p$.

Clearly \eqref{def:ar:MH:fd:BB} has an unwieldy structure, one which
may be computationally onerous to evaluate. Below in \cref{sec:MTM},
\cref{sec:conv:prop}, we describe two different competing ways to
obtain an unbiased multiproposal method with a less costly acceptance
mechanism. However, before turning to these unbiased modifications, we
first consider three classes of algorithms falling under
\eqref{def:Sj:coord:flip:0}, \eqref{def:vk:before:Tjelmeland} which do
not satisfy \eqref{cond:alpha:rev} but which maintain their salience
for other reasons.
\begin{Method}[The `Bubble Bath' Algorithm]
  \label{subsubsec:bubble:bath}
  One natural idea is to simply accept amongst the current state
  $\bq_0$ and the proposals $(\bq_1, \ldots, \bq_\np)$ from
  \eqref{def:vk:before:Tjelmeland} proportionally to $\dmu$ in
  \eqref{eq:stuart:form}.  Hence, for targets of the form
  \eqref{eq:stuart:form}, we now complete our definition of $\mk$ in
  \eqref{def:mp:ker} from \eqref{def:Sj:coord:flip:0},
  \eqref{def:vk:before:Tjelmeland}, by taking
  \begin{align}\label{eq:BB:algo}
  \ar_j(\bq_0, \ldots, \bq_\np)
    = \frac{\dmu(\bq_j)}{\sum_{l=0}^\np \dmu(\bq_l)},
    \quad j = 0,\ldots, \np,
  \end{align}
  under the arbitrary convention that $0/0 = 1$.  Under a target measure
  $\mu$ of the type \eqref{eq:stuart:form:og}, these $\alpha_j$'s take
  the form
  \begin{align}\label{eq:BB:algo:Stuart:form}
  \ar_j(\bq_0, \ldots, \bq_\np)
    = \frac{\exp( - \Pot(\bq_j))}{\sum_{l=0}^\np  \exp( - \Pot(\bq_l))},
    \quad j = 0,\ldots, \np,
  \end{align}
  whose unbiased corrections appear below as \cref{alg:MTMpCN} and as
  \cite[Algorithm 4.2]{glatt2024parallel}.  On the other hand, in the
  case when $\pk$ in \eqref{def:vk:before:Tjelmeland} takes the form
  \eqref{eq:prop:ker:diff:form}, this formulation in
  \eqref{eq:BB:algo} amounts to blithely removing all interaction
  terms, namely deleting $\dpk(\bq_j, \bq_i)$ for $i\neq j$ from
  \eqref{def:ar:MH:fd:BB}.  See \cref{tbl:beta:form} and
  \cref{alg:barker:mMCMC} for an explicit algorithmic expression of
  the method arising from \eqref{eq:BB:algo}.
  
  The method defined by \eqref{eq:BB:algo} is biased in
  general.\footnote{Based on the single proposal theory, as developed
    in \cite[Section 5]{dashti2017bayesian}, \cite{glatt2020accept},
    one might presume that the `prior reversibility condition' on
    $\pk$ and $\mu_0$ given in \eqref{eq:rev:spec} would be sufficient
    for unbiasedness here.  In fact \eqref{eq:rev:spec} generally
    yields unbiasedness only in the single proposal case or under
    careful modifications to \eqref{eq:BB:algo} or to
    \eqref{def:vk:before:Tjelmeland} given below in \cref{meth:MTMpCN}
    or \cref{meth:tj:mPCN} respectively.} However, there are several
  situations where \eqref{eq:BB:algo} is approximately unbiased.  For
  one, suppose that $\pk$ is of the form \eqref{eq:prop:ker:diff:form}
  whose proposal density $\dpk$ has a constant diagonal, namely where
  $\dpk(\bq,\bq)$ is independent of $\bq$, and assume
  \begin{align}\label{eq:ss:unbiased:cond}
    \prod_{i=0}^\np  \dpk(\bq_l, \bq_i) \approx
    \prod_{i=0}^\np  \dpk(\bq_m,\bq_i)
    \quad \text{ for every } l, m = 0, \ldots \np.
  \end{align}
  Then the acceptance structure \eqref{eq:BB:algo} is indeed a close
  approximation of the unbiased method with \eqref{def:ar:MH:fd:BB}.  In
  particular, proposal kernels $\vk$ involving independent draws from
  $\mu_0$ (namely, when $f \equiv 1$) yield an unbiased method under
  \eqref{eq:BB:algo}.
  
  Further approximately unbiased situations appear in various large
  $\np$-$\pk$ proposal parameter asymptotics.  For example, we may
  consider $\np$ large and $\sigma^2$ large in \eqref{eq:Guass:1:prop:classic} 
  or $\np$ large and $0 \leq \rho \ll 1$
  in \eqref{eq:Guass:1:prop:pCN}.  We begin to flesh out and rigorously
  justify these parameter asymptotic statements in
  \cref{prop:Glob:param:Limit} and \cref{rmk:scale:to:glob} once we
  derive the $\np = \infty$ kernels corresponding to
  \eqref{eq:BB:algo} in \cref{cor:large:p:prod} and
  \eqref{inf:ker:bb}.
\end{Method}

\begin{Method}[`Local' proposals]
  \label{meth:loc:prop}
  When generating a proposal cloud via
  \eqref{def:vk:before:Tjelmeland}, an `aggressive' (e.g. high
  variance) $\pk$ can be understood as a `global' search for high
  probability regions of the target measure.  By contrast, a `conservative'
  $\pk$ may be interpreted as exploring the local structure of the
  target measure around the current state.  In this latter conservative setting,
  different acceptance probabilities can be used to achieve an
  (over-damped) Langevin type method in the large $\np$ and small
  single-proposal variance $\sigma^2$ limit.

  To make this explicit, consider the case of a continuous target,
  $\mu( d \tbq) = \dmu(\tbq) d \tbq$, and a Gaussian single proposal
  kernel of the form
  $\pk_\sigma(\bq_0, d\tbq) = N(\bq_0, \sigma^2 I)(d \tbq)$. That is,
  $\pk$ in \eqref{def:vk:before:Tjelmeland} is of the form
  \eqref{eq:Guass:1:prop:classic} with $\cC = I$.  Then, following
  \cite{gagnon2023improving}, we take
  \begin{align}\label{eq:Zan:Loc:algo}
  \ar_j(\bq_0, \ldots, \bq_\np)
    = \frac{\sqrt{\dmu(\bq_j)}}{\sum_{l=0}^\np \sqrt{\dmu(\bq_l)}},
    \quad j = 0,\ldots, \np.
  \end{align}
  Below in \cref{sec:alg:scalingatinfty:local} and
  \cref{Prop:alg:scalingatinfty:local}, we flesh out the relationship
  between the method implied by \eqref{eq:Zan:Loc:algo} and an
  over-damped Langevin diffusion, \eqref{eq:trad:LA:dynam}, as
  well as the associated Metropolis adjusted (MALA) method,
  \cite{besag1994comments, roberts1996exponential}.
\end{Method}

\begin{Method}[Local proposals for preconditioned methods]
    \label{meth:loc:prop:precond}
    The idea of a local multiproposal method can be adapted to the
    case of when $\mu(d \tbq) \propto \exp(- \Pot(\tbq)) \mu_0(d\tbq)$
    takes the form
    \eqref{eq:stuart:form:og}--\eqref{eq:s:f:infinite:d}, that is to
    the situation when we are interested to resolve an infinite
    dimensional target $\mu$ defined relative to a Gaussian measure
    $\mu_0$.  Here, we may consider a `preconditioned approach' by
    selecting proposals with the probabilites
  \begin{align}\label{eq:Loc:pcN:algo}
  \ar_j(\bq_0, \ldots, \bq_\np)
    = \frac{\exp( -
    \frac{\rho}{1+\rho}\Pot(\bq_j/\rho))}
    {\sum_{l=0}^\np \exp( -
 \frac{\rho}{1+\rho}\Pot(\bq_l/\rho))}
    \quad j = 0,\ldots, \np.
  \end{align}
  In this case, our method is considered relative to a pCN single proposal
  kernel $\pk = \pk_\rho$ of the form \eqref{eq:Guass:1:prop:pCN} with
  $\rho \in (0,1)$, so that the resulting method makes sense even in a
  function space setting.  A complete algorithmic expression arising
  from \eqref{eq:Loc:pcN:algo} and $\pk_\rho$ is found below in
  \cref{tbl:beta:form}, \cref{alg:barker:mMCMC}.

  The form of \eqref{eq:Loc:pcN:algo} is based on the preconditioned
  over-damped Langevin equation considered in \cite{cotter2013mcmc}.
  We detail this relationship in \cref{sec:alg:scalingatinfty:local}
  and \cref{Prop:alg:scalingatinfty:local} below.  Note that while
  this is again a biased method, we establish an unbiased correction
  below within the Multiple Try paradigm as \cref{alg:MTMpCN}.
\end{Method}  

\begin{Method}[The Slingshot Algorithm]
  \label{subsubsec:slingshot}

  Another new class of algorithms arising out of the proposal structure
  \eqref{def:Sj:coord:flip:0}, \eqref{def:vk:before:Tjelmeland} has a
  general tendency to accept proposals that are far from the current
  point, leading to our moniker `the Slingshot'.  Although biased in
  general, this (class of) methods has many interesting and
  desirable properties.  In addition to the theoretical results in
  \cref{sec:large:p:limit} (see in particular \cref{sec:p:inf:lim:kernels:SS}), we
  carry out some numerical case studies of this method below in
  \cref{sec:The:Slingshot}.

  To formulate this method, we suppose that
  \begin{align}\label{eq:ss:abs:cont:cond}
    \mps \ll \pk(\bq_0,\cdot) \quad \text{ for every }
    \bq_0 \in \qsp.
  \end{align}
  Typically, we realize \eqref{eq:ss:abs:cont:cond} by considering the
  setting \eqref{eq:stuart:form}, \eqref{eq:prop:ker:diff:form} and
  supposing e.g. that
  \begin{align}\label{eq:ss:ker:strict:pos}
    \dpk(\bq, \tbq) > 0,  \quad \bq, \tbq \in  \qsp.
  \end{align}  
  The slingshot algorithm is then defined starting from
  \eqref{def:Sj:coord:flip:0}, \eqref{def:vk:before:Tjelmeland} by
  taking
  \begin{align}\label{eq:ar:SS:gen}
    \ar_j(\bq_0,\ldots, \bq_\np)
    = \frac{\frac{d\mu}{d \pk(\bq_0,\cdot)}(\bq_j)}
    {\sum_{l=0}^\np \frac{d\mu}{d \pk(\bq_0,\cdot)}(\bq_l)}
    = \frac{ \dmu(\bq_j) \dpk(\bq_0,\bq_j)^{-1}}
    {\sum_{l=0}^\np \dmu(\bq_l) \dpk(\bq_0,\bq_l)^{-1}} ,
  \end{align}
  for $j=0,\ldots, \np$.  We already previewed the resulting method
  above as \cref{alg:slingshot:intro}.

  There are a number of reasons to consider this particular acceptance
  structure \eqref{eq:ar:SS:gen}.  First, let us note several
  situations where \eqref{eq:ar:SS:gen} exactly or approximately
  corresponds to the unbiased acceptance structure
  \eqref{def:ar:MH:fd:BB}.  One case of interest where slingshot is
  unbiased appears when $\{\bq_j\}_{j=1}^\np$ are
  i.i.d. samples from a fixed $\bq_0$-independent measure. Namely, we
  take $\pk(\bq_0, d \tbq) =: \nu(d \tbq)$ in \eqref{def:vk:before:Tjelmeland} and assume $\mu \ll \nu$ so
  that \eqref{eq:ss:abs:cont:cond} holds.  Hence, in the notation of
  \eqref{def:vk:before:Tjelmeland}, \eqref{eq:comp:ker},
  \begin{align*}
  \vk(\bq_0, d \bq_1,\ldots , d \bq_\np) = \prod_{j=1}^\np
  \nu(d\bq_j), \quad \text{ and } \quad
  \cM(d\bq_0, d \bq_1, \ldots, d \bq_\np) = \mu(d\bq_0) \prod_{j=1}^\np
  \nu(d\bq_j).
  \end{align*}
  Denoting $\widetilde{\cM}(d\bq_0, d \bq_1, \ldots, d\bq_\np) = 
  \prod_{j=0}^\np \nu(d\bq_j)$, it follows that since $\mu \ll \nu$,
  then $\cM \ll \widetilde{\cM}$ and, consequently,
  $S_j^* \cM \ll S_j^* \widetilde{\cM}$. But
  $S_j^* \widetilde{\cM} = \widetilde{\cM}$, thus
  $S_j^* \cM \ll \widetilde{\cM}$ and moreover, 
  \begin{align*}
    \frac{d S_j^* \cM}{d\widetilde{\cM}} (\bq_0, \bq_1, \ldots, \bq_\np)
  = \frac{d\mu}{d\nu}(\Pi_1 S_j (\bq_0, \bq_1, \ldots, \bq_\np)) =
    \frac{d\mu}{d\nu}(\bq_j) \quad \mbox{for all } j=0,\ldots, \np,
  \end{align*}    
    so that
    \begin{align*}
      \frac{d (S_j^* \cM)}{d \left(
      \sum_{k=0}^\np S_k^* \cM \right)} (\bq_0, \bq_1, \ldots, \bq_\np)
  = \frac{\frac{d (S_j^* \cM)}{d \widetilde{\cM}} (\bq_0, \bq_1, \ldots,
    \bq_\np)}{\sum_{k=0}^\np \frac{d S_k^* \cM}{d \widetilde{\cM}}
    (\bq_0, \bq_1, \ldots, \bq_\np)} = \frac{\frac{d\mu}{d \nu}
      (\bq_j)}{\sum_{k=0}^\np \frac{d\mu}{d \nu}(\bq_k)}.
  \end{align*}    
  Thus the unbiasedness of \cref{alg:slingshot:intro} follows from
  \cref{thm:grand:unifed:inv} via \eqref{def:alphaj:Barker}
  by noting that the condition \eqref{sum:cond} is satisfied
  according to \cite[Theorem 2.10]{glatt2024parallel}.

  A second reason for considering \eqref{eq:ar:SS:gen} is that the
  corresponding Rao-Blackwellization, \eqref{eq:RB:form}, is the so
  called importance sampling method at each MCMC step.  Indeed, under
  \eqref{eq:ar:SS:gen}, we have,
  \begin{align}\label{eq:RB:form:SS}
  \rbest 
    =\frac{1}{N} \sum_{j=1}^N 
    \frac{\sum_{l=0}^\np \frac{d\mu}{d \pk(\bq_0^{(j)},\cdot)}(\bq_l^{(j)})
    \Obs(\bq^{(j)}_l)}
    {\sum_{l=0}^\np \frac{d\mu}{d \pk(\bq_0^{(j)},\cdot)}(\bq_l^{(j)})}.
  \end{align}
  where $(\bq_0^{(j)}, \ldots, \bq_p^{(j)})$ is the proposal cloud
  at step $j-1$ around the element $\bq_0^{(j)} :=\bq^{(j-1)}$ previously selected
  at step $j-2$.

  Finally, we observe that, while the slingshot is in general biased
  for fixed $p$, its limit as $p \to \infty$ is asymptotically
  unbiased.  Indeed, this limit kernel is just $\mu$ itself, so that
  we are simply considering Monte Carlo samples from $\mu$ in this
  limit.  See \cref{cor:large:p:prod} and \cref{sec:p:inf:lim:kernels:SS} below.
\end{Method}

Here and in the rest of the manuscript, it will be useful to consider
\cref{subsubsec:bubble:bath}, \cref{meth:loc:prop},
\cref{meth:loc:prop:precond} and \cref{subsubsec:slingshot} as special
cases of a broader class of methods.  To this end, we consider the
Barker-like acceptance probability structure of the form
\begin{align}\label{def:ar:simple}
  \ar_j (\bq_0, \ldots, \bq_\np)
  = \frac{\art(\bq_j, \bq_0)}{\sum_{l=0}^\np \art(\bq_l, \bq_0)},
  \quad j = 0, \ldots, p.
\end{align}
In other words, we accept $\bq_j$ with a probability proportional to
$\art(\bq_j, \bq_0)$, up to the selection of proportionality factor
$\art: \qsp  \times \qsp  \to (0,\infty)$.  The resulting class of proposal
kernels thus takes the form
\begin{align}\label{eq:gen:bk:ker:single:lay}
    \mk^\np(\bq_0, d \tbq)  = \sum_{j=0}^\np \int_{\qsp^\np}
    \delta_{\bq_j}(d \tbq) \frac{\art(\bq_j, \bq_0)}
    {\sum_{k=0}^\np \art(\bq_k, \bq_0)}
    \prod_{l=1}^\np \pk(\bq_0, d \bq_l),
\end{align}
which is itself a particular case of the general formulation
\eqref{def:mp:ker} found by selecting $\bS$ and $\vk$
according to \eqref{def:Sj:coord:flip:0} and
\eqref{def:vk:before:Tjelmeland}, respectively.

This formulation, \eqref{def:ar:simple}, includes \eqref{eq:BB:algo},
\eqref{eq:BB:algo:Stuart:form}, \eqref{eq:Zan:Loc:algo},
\eqref{eq:Loc:pcN:algo} and \eqref{eq:ar:SS:gen} for an appropriate
choice of $\art$.  For convenience and later reference, we summarize
our menu of possible choices for $\art$ as \cref{tbl:beta:form}.  An
algorithmic expression of \eqref{eq:gen:bk:ker:single:lay} is provided
as \cref{alg:barker:mMCMC}.

\begin{table}[htbp]
  \centering
  \begin{tabular}{cccc}
 \toprule
 $\art(\tbq, \bq_0)$ & 
   Primary Scope of $\pk(\bq_0, d\tbq)$ & 
    Biased Methods & Unbiased Corrections \\
    \midrule
    $\dmu(\tbq)$ & $N(\bq_0,\sigma^2I)$&
                                                                \ref{subsubsec:bubble:bath}
                                                                                  &  \ref{meth:MTMpCN}
    \\
        $\exp(- \Pot(\tbq))$ &  $N(\rho \bq_0,(1 - \rho^2) \cC )$ &
                                                                \ref{subsubsec:bubble:bath}
                                                                                  &  \cref{alg:MTMpCN}
    \\

    $\sqrt{\dmu(\tbq)}/\sqrt{\dmu(\bq_0)}$
                                                     & $N(\bq_0,\sigma^2I)$ 
                                                     &  \ref{meth:loc:prop} &
                                               \ref{meth:loc:MTM} \\
        $\exp\left( -
    \frac{\rho}{1+\rho}(\Pot(\frac{\tbq}{\rho})- \Pot(\bq_0))\right)$
      & $N(\rho \bq_0,(1 - \rho^2) \cC )$ 
      &
                                                             \ref{meth:loc:prop:precond} &
                                                            \cref{alg:MTMpCN:loc}\\
    $\dmu(\tbq)\dpk(\bq_0, \tbq)^{-1}$
                                                     &$\dpk(\bq_0, \tbq)\mu_0( d\tbq)$&
                                                                                     \cref{alg:slingshot:intro} &
                                                                                     \cref{meth:MTMslingshot}\\
 \bottomrule
  \end{tabular}
  \caption{Form of $\art$ and $\pk$ as appears in each of the methods
    introduced in this section under the general scope of
    \eqref{eq:gen:bk:ker} or \eqref{eq:MTM:Kernel}.}
  \label{tbl:beta:form}
\end{table}

\begin{algorithm}[H]
	\caption{Barker Type MP-MCMC (General Form)}
	\begin{algorithmic}[1]\label{alg:barker:mMCMC}
          \State Select $\art: \qsp \times \qsp \to (0,\infty)$ (see
          \cref{tbl:beta:form})  and a single proposal
          kernel $\pk: \qsp \times \Sigma_\qsp \to [0,1]$.
          \State Fix an initial point $\bq^{(0)} \in \qsp$.
          \For{$k \geq 0$}
          \State Draw $\bq_1^{(k+1)}, \ldots, \bq_\np^{(k+1)}$ from
          $\pk(\bq^{(k)}, d \tbq)$ independently and set
          $\bq^{(k+1)}_0 := \bq^{(k)}$.
          \State Determine  $\bq^{(k+1)}$ by drawing from the collection
          $\{\bq_0^{(k+1)}, \bq_1^{(k+1)}, \ldots, \bq_\np^{(k+1)}\}$ with
          the probabilities
          \begin{align}
            \alpha_j^{(k+1)}
            :=  \frac{\art (\bq_j^{(k+1)}, \bq^{(k+1)}_0)}
                   {\sum_{l=0}^\np \art(\bq_l^{(k+1)},
            \bq^{(k+1)}_0)},
            \quad  j = 0, \ldots, \np.
          \end{align}
          \EndFor
        \end{algorithmic}
\end{algorithm}

\begin{Remark}
\label{rmk:more:degen}
Two further variations on
\eqref{def:ar:simple}--\eqref{eq:gen:bk:ker:single:lay} are
notable. Firstly we may force the algorithm to choose a proposal point
by setting $\ar_0 (\bq_0, \ldots, \bq_\np) \equiv 0$ and taking
\begin{align}\label{def:ar:simple:no:0}
  \ar_j (\bq_0, \ldots, \bq_\np)
  = \frac{\art(\bq_j, \bq_0)}{\sum_{l=1}^\np
  \art(\bq_l, \bq_0)},
  \quad j = 1, \ldots, p,
  \quad \ar_0 (\bq_0, \ldots, \bq_\np) \equiv 0.
\end{align}
Although of course \eqref{def:ar:simple:no:0} is distinct from
\eqref{def:ar:simple}, our result in  \cref{thm:large:p:lim} notes that
their $p \to \infty$ limits are the same.

A second structure is to consider a Metropolis-Hastings type
alternative:
\begin{align}
  \ar_j (\bq_0, \ldots, \bq_\np) = 
   a_j
  \left[ 1 \wedge \frac{\art(\bq_j, \bq_0)}{\art(\bq_0,
  \bq_j)}\right],
  \quad j=1,\ldots, \np, 
  \quad
  \ar_0(\bq_0, \ldots, \bq_\np) =
  1 - \sum_{j=1}^\np \ar_j(\bq_0, \ldots, \bq_\np),
        \label{def:ar:MH:2}
\end{align}
where $\sum_{j =1}^\np a_j =1$.  Here notice that we are falling under
the scope of \cref{rmk:dengenaries} (cf. \cref{rmk:MH:degenerate}).
In this special case, each $P^j$ in \eqref{eq:single:prop:degen}
reduces to the same single proposal kernel $P^j = \bar{P}$ where
\begin{align}
\label{eq:ar:MH:ind:prop}
     \bar{P}(\bq_0, d \tbq)=
  \int_{\qsp}  1 \wedge \frac{\art(\bbq, \bq_0)}
                             {\art(\bq_0,  \bbq)}
                    \delta_{\bbq} (d\tbq) \pk(\bq_0, d \bbq)
  + \delta_{\bq_0} (d\tbq)
  \int_{\qsp} \left(1 -  1 \wedge \frac{\art(\bbq, \bq_0)}
                             {\art(\bq_0,  \bbq)}\right)
    \pk(\bq_0, d \bbq).
\end{align}
Thus, \eqref{def:Sj:coord:flip:0}, \eqref{def:vk:before:Tjelmeland}
supplemented with \eqref{def:ar:MH:2} turns out to be nothing more
than a convoluted formulation of the single proposal kernel
$P = \bar{P}$ .
\end{Remark}

\subsection{The Multiple-Try (MT) Approach}
\label{sec:MTM}

We turn next to provide our general state space formulation of the
Multiple Try  (MT) approach.  As noted in the introduction,
this family of (unbiased) algorithms has been previously introduced in
\cite{liu2000multiple} and studied in a number of subsequent works
\cite{Craiu2007,Bedard2012,chang2022rapidly,Yang2023,
  luo2019multiple}.  Expanding on this paradigm, we now show that each
of the (biased) algorithms laid out in \cref{subsubsec:bubble:bath},
\cref{meth:loc:prop}, \cref{meth:loc:prop:precond} and
\cref{subsubsec:slingshot} (essentially) provide different preliminary
proposal steps for different (unbiased) Multiple Try algorithms.

Given a target measure $\mu$ on a measurable space
$(\qsp, \Sigma_\qsp)$, we consider to this end any (measurable)
$\art: \qsp \times \qsp \to (0,\infty)$, Markov transition kernel
$\pk: \qsp \times \Sigma_\qsp \to [0,1]$, and number of proposals
$\np$.  Given the single proposal kernel $\pk$ as in
\eqref{def:vk:before:Tjelmeland} we take
\begin{align}\label{def:eta:MTM}
  \eta (d \bq, d \tbq) = \pk (\bq, d \tbq) \mu(d \bq), \quad
  \mbox{and } \,\, \eta^\perp(d \bq, d \tbq) = \eta (d \tbq, d \bq)
  \quad \text{ and assume that } \eta \ll \eta^\perp,
\end{align}
so that these two measures are mutually absolutely continuous.  Then,
relative to this target measure $\mu$ and the choice of the elements
$\art, \pk, \np$, we define the associated Multiple Try type kernel
$P(\bq_0, d \tbq)$ as
\begin{align}\label{eq:MTM:Kernel}
  \sum_{j = 1}^p \int\limits_{\qsp^{2 \np -1}} \!  \! \! 
  \frac{\art (\bq_j, \bq_0)}{\sum_{l=1}^\np \art(\bq_l, \bq_0)}
     \biggl(& \!   \bar{\alpha}_j( \bq_0, \ldots, \bq_\np, \bz_1, \ldots, \bz_{\np-1})
              \delta_{\bq_j}(d \tbq)
  \notag\\
     &+ (1 -\bar{\alpha}_j( \bq_0, \ldots, \bq_\np, \bz_1, \ldots, \bz_{\np-1}))
     \delta_{\bq_0}(d \tbq)  \!  \biggr)  \! \! 
     \prod_{k =1}^{p-1} \! \pk(\bq_j, d \bz_k)  \! \!  \prod_{m =1}^p \!\pk(\bq_0, d \bq_m),
\end{align}
where
\begin{align}\label{eq:MTM:ar}
  \bar{\alpha}_j(\bq_0, \ldots, \bq_\np, \bz_1, \ldots, \bz_{\np-1})
  :=      1 \wedge
  \left(\frac{d \eta^\perp}{d \eta}(\bq_0, \bq_j)
  \frac{\art(\bq_0, \bq_j)}{\art(\bq_j, \bq_0)} \frac{\sum_{l =1}^p\art(\bq_l, \bq_0)}
     {\art(\bq_0, \bq_j)+ \sum_{l =1}^{p-1}\art(\bz_l, \bq_j)} \right).
\end{align}
We summarize the algorithmic expression of \eqref{eq:MTM:Kernel} as
follows:
\begin{algorithm}[H]
  \caption{Multiple Try-Metropolis (General Form)}
  \begin{algorithmic}[1]\label{alg:MTM}
    \State Select the algorithm parameters:
    $\art: \qsp \times \qsp \to (0,\infty)$,
    $\pk: \qsp \times \Sigma_\qsp \to [0,1]$, and $\np \geq 1$.
    \State Set $\eta, \eta^\perp$ as in \eqref{def:eta:MTM}, where $\mu$ is
    the target measure.
    \State Select a $\bq^{(0)} \in \qsp$.
    \For{$k \geq 0$}
        \State Draw $\bq_1^{(k+1)}, \ldots, \bq_\np^{(k+1)}$ from
        $\pk(\bq^{(k)}, d \tbq)$ independently.
        \State Select $\bbq^{(k+1)} := \bq_{j}^{(k+1)}$ amongst 
        $\bq_1^{(k+1)}, \ldots, \bq_\np^{(k+1)}$
        with the probabilities
        \begin{align}\label{eq:MTM:prelim:sel:alg}
          \alpha_j^{(k+1)} :=
          \frac{\art (\bq_j^{(k+1)},\bq^{(k)})}{\sum_{l=1}^\np \art(\bq_l^{(k+1)},
          \bq^{(k)})},
          \quad j= 1, \ldots, \np.
        \end{align}
        \State Draw $\bz_1^{(k+1)}, \ldots, \bz_{\np-1}^{(k+1)}$ from
        $\pk(\bbq^{(k+1)}, d \tbq)$
        independently.
        \State Set $\bq^{(k+1)} = \bbq^{(k+1)}$ with probability 
        \begin{align}\label{eq:MTM:ar:alg}
          \bar{\alpha}^{(k+1)} := 
          1 \wedge
          \left( \frac{d \eta^\perp}{d \eta}(\bq^{(k)},\bbq^{(k+1)})
          \frac{\art(\bq^{(k)}, \bbq^{(k+1)})}{\art(\bbq^{(k+1)}, \bq^{(k)})}
          \frac{\sum_{l=1}^\np \art(\bq_l^{(k+1)}, \bq^{(k)})}
          { \art(\bq^{(k)}, \bbq^{(k+1)})  + \sum_{l =1}^{\np-1} \art(\bz_l^{(k+1)}, \bbq^{(k+1)})} \right),
        \end{align}
        \Statex \hspace{\algorithmicindent} and otherwise take $\bq^{(k+1)} = \bq^{(k)}$.
        \EndFor
      \end{algorithmic}
\end{algorithm}     

\begin{Remark}
\label{eq:full:abs:form:MTM}
The careful reader will notice that the formulation
\eqref{eq:MTM:Kernel} is not, strictly speaking, written in the form
\eqref{def:mp:ker:prob:cond}.  We provide such formulation below in
\cref{sec:MTM:unbiased} as
\eqref{eq:MTM:full:caff:0}--\eqref{eq:MTM:full:caff:4}.
\end{Remark}

\begin{Remark}
\label{eq:red:alpha:MTM}
In the special case when $\art$ satisfies the balance condition
\begin{align}
    \label{eq:art:cond}
    \art (\tbq, \bq_0) \pk(\bq_0, d \tbq)\mu(d \bq_0)  
     = \art (\bq_0, \tbq) \pk(\tbq, d \bq_0)\mu(d \tbq),  
\end{align}
we have, cf. \eqref{def:eta:MTM}, that
\begin{align}
  \label{eq:cond}
	\frac{d \eta^\perp}{d \eta}(\bq_0, \br) \frac{\art(\bq_0,
  \br)}{\art(\br, \bq_0)} = 1,
\end{align}
for $\eta$-almost every $(\bq_0, \br)$.  Hence the acceptance
probability of keeping the selected item in the proposal cloud,
\eqref{eq:MTM:ar} (cf. \eqref{eq:MTM:ar:alg}), reduces to
\begin{align}\label{eq:MTM:ar:red}
  \bar{\alpha}_j(\bq_0, \ldots, \bq_\np, \bz_1, \ldots, \bz_{\np-1})
  :=      1 \wedge
    \left(\frac{\sum_{l =1}^p\art(\bq_l, \bq_0)}
     {\art(\bq_0, \bq_j)+ \sum_{l =1}^{p-1}\art(\bz_l, \bq_j)} \right).
\end{align}

This condition \eqref{eq:art:cond} holds for many (but not all) of the
specific methods of interest here.  For example, we often assume our
target measure $\mu$ has the form \eqref{eq:stuart:form} and impose
the `prior reversibility' or balance condition
\begin{align}\label {eq:rev:spec}
    \pk(\bq, d \tbq) \mu_0 (d \bq)  = \pk(\tbq, d \bq)\mu_0(d \tbq),
\end{align}
relative to the `base measure' $\mu_0$ appearing in
\eqref{eq:stuart:form}.  Then, in this situation,
\begin{align}
  \label{eq:eta:etaper:form}
  \frac{d \eta^\perp}{d \eta}(\bq_0, \br) = \frac{\dmu(\br)}{\dmu(\bq_0)},
\end{align}
and hence, when $\beta(\bq, \bq_0) = \dmu(\bq)$, we obtain
\eqref{eq:cond}.

Note that the prior reversibility condition \eqref{eq:rev:spec} is
easily achieved under \eqref{eq:prop:ker:diff:form} when $\dpk$ is
symmetric, namely when
\begin{align}\label{eq:prop:ker:diff:form:sym}
  \dpk(\tbq, \bq)= \dpk(\bq, \tbq) \quad
  \text{ for $\mu_0 \otimes \mu_0$ almost every } \tbq, \bq
  \in \qsp,
\end{align}
which holds, for example, under \eqref{eq:Guass:1:prop:classic}.
We also achieve \eqref{eq:rev:spec} when
$Q(\bq_0, d\tbq) = N(\rho\bq_0, (1 - \rho^2) \cC)$ is the pCN type
proposal kernel as in \eqref{eq:Guass:1:prop:pCN} and
$\mu_0 = N(0,\cC)$ as in \eqref{eq:s:f:infinite:d}.  Here
\eqref{eq:rev:spec} is established by noting that
$\pk(\bq, d \tbq) \mu_0 (d \bq)$ is Gaussian and computing its
characteristic function; see  \cite{cotter2013mcmc,
  glatt2020accept}.
\end{Remark}

The following result establishing the unbiasedness of
\eqref{eq:MTM:Kernel}, namely \cref{alg:MTM}, includes
\cite{liu2000multiple} as a special case.

\begin{Theorem}
  \label{thm:MTM:unbiased}
  Fix a target probability measure $\mu$ on $(\qsp,
  \Sigma_\qsp)$. Consider a measurable function
  $\art: \qsp \times \qsp \to (0,\infty)$, a Markov transition kernel
  $\pk: \qsp \times \Sigma_\qsp \to [0,1]$, and number of proposals
  $\np \geq 1$. Suppose that $\eta$ and $\eta^\perp$ defined in
  \eqref{def:eta:MTM} satisfy $\eta^\perp \ll \eta$.  Then $P$ defined
  according to \eqref{eq:MTM:Kernel} is reversible with respect to
  $\mu$.
\end{Theorem}
\noindent The proof of \cref{thm:MTM:unbiased} is based on
\cref{thm:grand:unifed:inv}, and is provided below in
\cref{sec:MTM:unbiased}.

We now turn to provide the Multiple Try correction to each of the methods
defined previously in \cref{sec:cond:prop}.
\begin{Method}[Multiple Try Corrected Bubble Bath]
  \label{meth:MTMpCN}
  Notice that with a target measure of the form \eqref{eq:stuart:form}
  we can take
  \begin{align}
    \label{eq:BB:adj}
    \art (\tbq, \bq_0) = \dmu(\tbq)
  \end{align}
  and obtain that the corresponding Multiple Try kernel, \eqref{eq:MTM:Kernel},
  is an unbiased adjustment of \cref{subsubsec:bubble:bath}.  Under
  the `prior reversibility' assumption, \eqref{eq:rev:spec}, the condition
  \eqref{eq:art:cond} holds so that \eqref{eq:MTM:ar:alg} reduces to
  \eqref{eq:MTM:ar:red}. In particular, when $\pk$ is the pCN kernel
  \eqref{eq:Guass:1:prop:pCN} with our target measure of the form
  \eqref{eq:stuart:form:og} we obtain a `global multiple-try' version
  of the pCN algorithm, with
  \begin{align}\label{def:art:MTpCN}
  	\art(\tbq, \bq_0) = e^{-\Pot (\tbq)}.
  \end{align}  
  This is counterpoint to the mpCN method
  derived recently in \cite{glatt2024parallel}.  As such, we record this
  new method as \cref{alg:MTMpCN}.
\end{Method}

\begin{algorithm}[H]
  \caption{Multiple-Try Preconditioned Crank-Nicolson (MTpCN)}
  \begin{algorithmic}[1]\label{alg:MTMpCN}
     \State Select the algorithm parameter $\rho \in [0,1]$ and
          set $\pk(\bq_0, d \tbq) = N(\rho \bq_0, (1 - \rho^2)\cC)$.
          \State Select a $\bq^{(0)} \in \qsp$.
          \For{$k \geq 0$}
                   \State Draw $\bq_1^{(k+1)}, \ldots,
                   \bq_\np^{(k+1)}$
                   from $\pk(\bq^{(k)}, d \tbq)$ independently.
                   \State Select $\bbq^{(k+1)} := \bq_{j}^{(k+1)}$
                   amongst 
                   $\bq_1^{(k+1)}, \ldots, \bq_\np^{(k+1)}$
                   with the corresponding probabilities
                   \begin{align}
                   	\alpha_j^{(k+1)} :=  \frac{\exp(-\Phi(\bq_j^{(k+1)}))}
                     {\sum_{l=1}^\np \exp(-\Phi(\bq_l^{(k+1)}))}, \quad j=1, \ldots, \np.
                   \end{align}
                   \State Draw $\bz_1^{(k+1)}, \ldots,
                   \bz_{\np-1}^{(k+1)}$
                   from $\pk(\bbq^{(k+1)}, d \tbq)$ independently.
                   \State Set $\bq^{(k+1)} := \bbq^{(k+1)}$ with probability 
                    \begin{align}
                   	\bar{\alpha}^{(k+1)} := 1 \wedge
	                \frac{ \sum_{l=1}^\np \exp(-\Phi(\bq_l^{(k+1)}))}
                      { \exp(- \Phi(\bq^{(k)}))
                      +\sum_{l=1}^{\np-1} \exp( - \Phi(\bz_l^{(k+1)}) )}.
                   \end{align}
                   \State Otherwise take $\bq^{(k+1)} := \bq^{(k)}$.
                   \EndFor
                 \end{algorithmic}
\end{algorithm}

\begin{Method}[Local Multiple Try]
  \label{meth:loc:MTM}
  We can also correct for bias in the local approach introduced in
  \cref{meth:loc:prop}.  Here we are in the case where we aim to
  resolve a continuously distributed $\mu(d\bq) = \dmu(\bq) d \bq$
  using a random walk type single proposal kernel
  $\pk$ of the form \eqref{eq:Guass:1:prop:classic}.  Given our
  formulation \eqref{eq:Zan:Loc:algo}, we take
  \begin{align}
    \label{eq:zanella:adj}
    \art(\tbq, \bq_0) = \sqrt{\frac{\dmu(\tbq)}{\dmu(\bq_0)}},
  \end{align}
  as we already previewed in \cref{tbl:beta:form}.  The resulting Multiple Try
  kernel, \eqref{eq:MTM:Kernel}, corresponds to the local algorithm
  introduced in \cite{gagnon2023improving}.  Here note that
    $\art (\tbq, \bq_0) \pk(\bq_0, d \tbq)\mu(d \bq_0)  \propto
    \sqrt{\dmu(\tbq)\dmu(\bq_0)} \exp( -
    (2 \sigma^2)^{-1}|\bq_0-\tbq|^2)
    d \bq_0 d\tbq$,
    so that the condition \eqref{eq:art:cond} is maintained
    and the simplified acceptance structure \eqref{eq:MTM:ar:red}
    also applies in this case.
\end{Method}

\begin{Method}[Local MTpCN]
   \label{meth:loc:pCN}
  On the other hand, when we consider the Hilbert space setting where
  our target is of the type \eqref{eq:stuart:form:og} and thus use a
  pCN type single proposal kernel of the form \eqref{eq:Guass:1:prop:pCN},
  then the resulting local acceptance structure \eqref{eq:Loc:pcN:algo}
  corresponds to
  \begin{align}\label{eq:loc:pCN:beta}
    \beta(\tbq, \bq_0) := \exp\left( -
    \frac{\rho}{1+\rho}(\Pot(\tbq/\rho)- \Pot(\bq_0))\right),
  \end{align}
  as our choice for  $\beta$.  In this situation, we have, cf. \eqref{eq:prop:ker:diff:form:sym},
  \begin{align}\label{eq:RN:eta:etaperp}
    \frac{d \eta^\perp}{d \eta}(\bq_0, \tbq) = \exp( - \Pot(\tbq) + \Pot(\bq_0))
  \end{align}
  so that \eqref{eq:MTM:ar} is in this case
  \begin{align}
    \bar{\alpha}_j( &\bq_0, \ldots, \bq_\np, \bz_1, \ldots,
                    \bz_{\np-1})
    \notag\\
    &= 1 \wedge \frac{\exp( - \Pot(\bq_j)
                    - \frac{\rho}{1+\rho}
                    \Pot(\bq_0/\rho))}
                    {\exp(- \Pot(\bq_0) -
                    \frac{\rho}{1+\rho}
                    \Pot(\bq_j/\rho)) }
    \frac{ \sum_{l =1}^\np \exp( -\frac{\rho}{1+\rho}\Pot(\bq_l/\rho))}
      { \exp( -\frac{\rho}{1+\rho}\Pot(\bq_0/\rho))  +
                    \sum_{l=1}^{p-1} \exp(
                        -\frac{\rho}{1+\rho}\Pot(\bz_l/\rho))}.
  \end{align}
  We therefore have a second new (local) Multiple Try multiproposal pCN method which we
  record here as \cref{alg:MTMpCN:loc}.
\end{Method}

\begin{algorithm}[H]
  \caption{Local Multiple Try Preconditioned Crank-Nicolson (lMTpCN)}
  \begin{algorithmic}[1]\label{alg:MTMpCN:loc}
     \State Select the algorithm parameter $\rho \in (0,1]$ and
          set $\pk(\bq_0, d \tbq) = N(\rho \bq_0, (1 - \rho^2)\cC)$.
          \State Select a $\bq^{(0)} \in \qsp$.
          \For{$k \geq 0$}
                   \State Draw $\bq_1^{(k+1)}, \ldots,
                   \bq_\np^{(k+1)}$
                   from $\pk(\bq^{(k)}, d \tbq)$ independently.
                   \State Select $\bbq^{(k+1)} := \bq_{j}^{(k+1)}$
                   amongst 
                   $\bq_1^{(k+1)}, \ldots, \bq_\np^{(k+1)}$
                   with the corresponding probabilities
                   \begin{align*}
                     \alpha_j^{(k+1)} :=
                     \frac{\exp(-\frac{\rho}{1+\rho}\Phi(\bq_j^{(k+1)}/\rho))}
                     {\sum_{l=1}^\np \exp(-\frac{\rho}{1+\rho}
                     \Phi(\bq_l^{(k+1)}/\rho))}.
                   \end{align*}
                   \State Draw $\bz_1^{(k+1)}, \ldots,
                   \bz_{\np-1}^{(k+1)}$
                   from $\pk(\bbq^{(k+1)}, d \tbq)$ independently.
                   \State Set $\bq^{(k+1)} := \bbq^{(k+1)}$ with probability 
                    \begin{align*}
                      \! \! \! \!\!\bar{\alpha}^{(k+1)} :=1\wedge 
                      \frac{\exp( - \Pot(\bbq^{(k+1)})
                    - \frac{\rho}{1+\rho}\Pot(\bq^{(k)}/\rho))}
                    {\exp(- \Pot(\bq^{(k)}) -
                    \frac{\rho}{1+\rho}
                                                \Pot(\bbq^{(k+1)}/\rho))}
	                \frac{ \sum_{l=1}^\np \exp(-\frac{\rho}{1+\rho}\Phi(\bq_l^{(k+1)}/\rho))}
                      { \exp(- \frac{\rho}{1+\rho}\Phi(\bq^{(k)}/\rho))
                      +\sum_{l=1}^{\np-1} \exp( - \frac{\rho}{1+\rho}\Phi(\bz_l^{(k+1)}/\rho) )}.
                   \end{align*}
                   \State Otherwise take $\bq^{(k+1)} := \bq^{(k)}$.
                   \EndFor
                 \end{algorithmic}
\end{algorithm}

\begin{Method}[Multiple Try Slingshot]
  \label{meth:MTMslingshot}
  Finally, we record the Multiple Try correction to the Slingshot algorithm we
  introduced in \cref{subsubsec:slingshot} resulting in
  \cref{alg:slingshot:intro}.  Here, cf. \eqref{eq:ar:SS:gen}, we take
  \begin{align}
    \label{eq:SS:adj}
    \art (\tbq, \bq_0) = \dmu(\tbq)/ \dpk(\bq_0,\tbq)
  \end{align}
  where we are under our strict positivity assumption,
  \eqref{eq:ss:ker:strict:pos} on the proposal kernel
  $\pk(\bq_0, d \tbq) = \dpk(\bq_0,\tbq) \mu_0(d \tbq)$, for a target
  $\mu$ given as in \eqref{eq:stuart:form}.  Regarding the acceptance
  probabilities $\bar{\alpha}_j$ in the resulting kernel
  \eqref{eq:MTM:Kernel}, we note once again that \eqref{eq:art:cond}
  holds.
\end{Method}

\subsection{Convolutional Proposal Structures}
\label{sec:conv:prop}

A different proposal strategy which avoids the biased approaches outlined in
\cref{subsubsec:bubble:bath} or \cref{subsubsec:slingshot} is to draw
a preliminary point around the current state and only then to draw the
proposal cloud from around this preliminary point.  This leads to
proposal structures of the form
\begin{align}\label{def:vk:Tjelmeland}
  \vk (\bq_0, d \bq_1, \ldots, d \bq_\np)
  = \int_\qsp \prod_{j=1}^p \pk(\tbq, d \bq_j)
  \bpk (\bq_0, d\tbq)
\end{align}
for any given Markov kernels
$\pk, \bpk: \qsp \times \cB(\qsp) \to [0,1]$. This structure for $\vk$
in \eqref{def:vk:Tjelmeland} was also considered in our previous work
\cite{glatt2024parallel}, and is inspired by an algorithmic
construction given in \cite{tjelmeland2004using}.  In this case, given
$n \geq 1$ and a current state $\bq^{(n)} = \bq_0^{(n+1)}$, a cloud
of proposals $(\bq_1^{(n+1)}, \ldots, \bq_\np^{(n+1)})$ is obtained by
first drawing
\begin{align}\label{eq:tj:prelim:draw}
	\bbq_0^{(n+1)} \sim \bpk (\bq^{(n)}, d \tbq)
\end{align}
and then independently drawing 
\begin{align}\label{eq:tj:cloud:draw}
  \bq_j^{(n+1)} \sim \pk(\bbq_0^{(n+1)}, d\tbq),
\end{align}
for $j=1, \ldots, \np$.

Note that \eqref{def:vk:before:Tjelmeland} is a special case of
\eqref{def:vk:Tjelmeland} when we take
$\bpk (\bq_0, d\tbq) = \delta_{\bq_0}(d \tbq)$.  Furthermore,
\cite[Theorem 2.10]{glatt2024parallel} applies in this greater
generality, namely \eqref{sum:cond} holds for any choice of
$\pk, \bpk$ in \eqref{def:vk:Tjelmeland}, and under the choice of
$\bS$ given by the coordinate-flip involutions as in
\eqref{def:Sj:coord:flip:0}. However, in contrast to
{\eqref{def:vk:before:Tjelmeland}, if we consider a target of the form
  \eqref{eq:stuart:form} and take the assumption that
\begin{align}\label{eq:rev:spec:tj}
    \pk(\bq, d \tbq) \mu_0 (d \bq)  = \bpk(\tbq, d \bq)\mu_0(d \tbq),
\end{align}
then, in this case, the simplified acceptance structure appearing in
\eqref{eq:BB:algo} which we imposed in \cref{subsubsec:bubble:bath}
leads to a provably unbiased scheme.  Indeed, according to
\cite[Theorem 2.11]{glatt2024parallel}, if $\vk$ is of the form
\eqref{def:vk:Tjelmeland} with $\pk,\bpk$ satisfying \eqref{eq:rev:spec:tj}, $\bS$ is as in \eqref{def:Sj:coord:flip:0}, and we take $\mu$ as in \eqref{eq:stuart:form} with $\dmu(\bq) > 0$ $\mu_0$-a.e., 
then $S_j^* \cM \ll \cM$ for every $j = 0,\ldots, \np$,
where $\cM$ is as
in \eqref{eq:comp:ker}, and
\begin{align}\label{eq:simp:TJ:RNDform}
  \frac{d S_j^* \cM}{d \cM}(\bq_0, \ldots, \bq_p) 
  = \frac{\dmu (\bq_j)}{\dmu(\bq_0)}, \quad j = 0, \ldots, \np.
\end{align}

\begin{Method}[Tjelmeland, Multiproposal pCN]
  \label{meth:tj:mPCN}
  Specializing mostly to the case $\pk = \bpk$ in
  \eqref{def:vk:Tjelmeland}, where \eqref{eq:rev:spec:tj} reduces to
  \eqref{eq:rev:spec}, and maintaining $\bS$ as in
  \eqref{def:Sj:coord:flip:0}, we obtain several concrete algorithms
  which have been previously considered.  Indeed, under these
  circumstances, when $\qsp = \RR^\dm$ and we are aiming to resolve a
  continuously distributed target $\mu$, that is where $\mu_0$ is the
  Lebesgue measure in \eqref{eq:stuart:form}, then \eqref{eq:rev:spec}
  holds for any symmetric kernel as in
  \eqref{eq:prop:ker:diff:form:sym}.  This is the method discovered in
  \cite{tjelmeland2004using}.  Note that the more flexible condition
  \eqref{eq:rev:spec:tj} allows for the consideration of some
  non-symmetric proposal structures; see \cite[Algorithm
  4]{glatt2024parallel}.

  In the infinite-dimensional setting \eqref{eq:s:f:infinite:d},
  taking $\pk = \bpk$ as in \eqref{eq:Guass:1:prop:pCN} leads to a
  multiproposal version of the pCN algorithm from
  \cite{NealpCNbeforepCN, beskos2008mcmc, cotter2013mcmc}, which we
  derived in \cite{glatt2024parallel}, \cite{glatt2024sacred}.  Here
  $\dmu$ is often written as in \eqref{eq:stuart:form:og}, with
  $\Pot(\bq)$ being the log-likelihood in a Bayesian context, so that
  the acceptance probabilities are written as
  \begin{align}\label{eq:mpCN:ar}
    \alpha_j(\bq_0, \cdots, \bq_\np) :=
    \frac{\exp(-\Pot(\bq_j))}
    {\sum_{l =0}^p \exp(-\Pot(\bq_l))}.
  \end{align} 
\end{Method}

\begin{Remark}
  When \eqref{eq:rev:spec:tj} fails to hold, the resulting acceptance
  probabilities are even more cumbersome.  Here, in the setting where
  $\pk(\bq_0, d \tbq) =\dpk(\bq_0, \tbq) \mu_0(d \tbq)$ and
  $\bpk(\bq_0, d \tbq) = \dbpk(\bq_0, \tbq) \mu_0(d \tbq)$, an argument
  analogous to \eqref{eq:change:of:Ms:1}, \eqref{eq:change:of:Ms:2},
  yields
  \begin{align}\label{def:ar:Barker:fd}
    \frac{d S_j^* \cM}{ d (S_0^* \cM + \cdots + S_{\np}^*\cM)}{}
    (\bq_0, \ldots, \bq_p)
    = \frac{\dmu(\bq_j) \int
    \prod_{\stackrel{i=0}{i\neq j}}^\np \dpk(\bq, \bq_i)
    \dbpk(\bq_j, \bq) \mu_0(d \bq) }{\sum_{l=0}^\np \dmu(\bq_l)
    \int \prod_{\stackrel{i=0}{i\neq l}}^\np \dpk(\bq,\bq_i)
    \dbpk(\bq_l, \bq) \mu_0(d \bq) }, \quad j = 0,\ldots, \np,
\end{align}
\end{Remark}

To summarize, we can express the convolutional proposal approach
\eqref{def:vk:Tjelmeland}, which is implemented by \cref{meth:tj:mPCN},
as the kernel
\begin{align}\label{eq:gen:bk:ker}
    \mk^\np(\bq_0, d \tbq)  = \sum_{j=0}^\np \int_{\qsp^\np}
    \delta_{\bq_j}(d \tbq) \frac{\art(\bq_j, \bq_0)}
    {\sum_{k=0}^\np \art(\bq_k, \bq_0)}
    \int_{\qsp}\prod_{l=1}^\np \pk(\bbq, d \bq_l) \bpk(\bq_0, d \bbq),
\end{align}
which is a special case of \eqref{def:mp:ker}.  

We record the algorithm expression of \eqref{eq:gen:bk:ker}
as
\begin{algorithm}[H]
	\caption{Convolutional MP-MCMC (General Form)}
	\begin{algorithmic}[1]\label{alg:cov:mMCMC}
          \State Determine a relative selection
          probability function $\art: \qsp \times \qsp \to (0,\infty)$
          (cf. \cref{tbl:beta:form}) and single proposal
          kernels $\bpk, \pk: \qsp \times \Sigma_\qsp \to [0,1]$.
          \State  Pick an initial state $\bq^{(0)} \in \qsp$.
          \For{$k \geq 0$}
          \State Draw $\bbq^{(k+1)}$ from $\bpk(\bq^{(k)}, d \bbq)$.
          \State Draw $\bq_1^{(k+1)}, \ldots, \bq_\np^{(k+1)}$ from
          $\pk(\bbq^{(k+1)}, d \tbq)$ independently and set $\bq_0^{(k+1)} = \bq^{(k)}$.
          \State Select
          $\bq^{(k+1)} := \bq_{j}^{(k+1)}$ amongst
          $\{\bq_0^{(k+1)}, \bq_1^{(k+1)}, \ldots, \bq_\np^{(k+1)}\}$ with
          the probabilities
          \begin{align}
            \alpha_j^{(k+1)}
            :=  \frac{\art (\bq_j^{(k+1)}, \bq^{(k+1)}_0)}
                   {\sum_{l=0}^\np \art(\bq_l^{(k+1)}, \bq^{(k+1)}_0)}.
          \end{align}
          \EndFor
        \end{algorithmic}
\end{algorithm}     

Note that the formulation
in \eqref{eq:gen:bk:ker} and its \cref{alg:cov:mMCMC} contains
\eqref{eq:gen:bk:ker:single:lay} (and \cref{alg:barker:mMCMC}) as a
special case by taking $\bpk(\bq_0, d \tbq) = \delta_{\bq_0}(d \tbq)$.

\begin{Remark}\label{rmk:MH:limit}
  The Metropolis-Hastings type acceptance probabilities
  \eqref{def:ar:MH:2}, with $\vk$ as in \eqref{def:vk:Tjelmeland},
  result in a multiproposal kernel $\mk$ from \eqref{def:mp:ker} which
  exhibits the degeneracy described in \cref{rmk:MH:degenerate}.
  Here, similarly to \cref{rmk:more:degen}, this is simply a single
  proposal Metropolis-Hastings kernel as in \eqref{eq:ar:MH:ind:prop}
  with $\pk$ replaced by
\begin{align}\label{eq:lim:ker:large:p}
  \hat{\pk}(\bq_0, d \tbq)
  = \int_\qsp \pk(\bbq, d \tbq) \bpk(\bq_0, d \bbq),
  \quad \bq_0 \in \qsp,
\end{align} 
in the expression for $\bar{P}$.  In the case of a continuous target,
the analog of \cref{subsubsec:slingshot} reduces to a random walk
metropolis method.  This reduction also holds in the special case of a
symmetric proposal kernel for \cref{subsubsec:bubble:bath} and for
\cref{meth:tj:mPCN}.  Moreover, in the Gaussian case,
\eqref{eq:Guass:1:prop:classic}, the kernel in \eqref{eq:lim:ker:large:p} simply
doubles the variance of the proposal step.  Similarly, in the pCN proposal setting
\eqref{eq:Guass:1:prop:pCN}, the analogues of
\cref{subsubsec:bubble:bath} and \cref{meth:tj:mPCN} simply reduce to (single proposal) pCN,
where in the latter case
$\hat{\pk}(\bq_0, d \tbq) = N(\rho^2 \bq_0, (1- \rho^4) \cC)(d\tbq)$.
\end{Remark}

\begin{Remark}
  \label{rmk:resamp}
  Note that additional `resampling steps' can be added to
  \cref{alg:cov:mMCMC} as follows. After generating the cloud of
  points
  $\mathfrak{P} = (\bq_0^{(k+1)}, \bq_1^{(k+1)}, \ldots,
  \bq_\np^{(k+1)})$ and computing the associated acceptance
  probabilities $\alpha_j^{(k+1)}$, draw multiple points (with
  replacement) from this collection $\mathfrak{P}$ with these
  $\alpha_j^{(k+1)}$ probabilities.  To continue, we use the last
  resampled draw to generate the next cloud in a subsequent iteration.
  Note that as regards unbiasedness \cref{thm:grand:unifed:inv} was
  adapted to the resampled setting in \cite[Section
  3]{glatt2024parallel}.

  In \cite{schwedes2021rao} this resampling strategy is advocated as
  an approximation to the Rao-Blackwellization as in \eqref{eq:RB:form} for
  various multiproposal methods.  Such a resampling approach may thus
  provide some of the advantages of Rao-Blackwellization but with less
  arduous memory requirements.  As we discuss in \cref{sec:Outlook},
  this is an interesting direction for future research around the
  various methods introduced in this section.
\end{Remark}

\section{Large proposal limit for conditionally independent proposals}
\label{sec:large:p:limit}

In this section, we analyze the large proposal limit for a general
class of conditionally independent Barker-type multiproposal
algorithms motivated by and encompassing each of the different methods
introduced across \cref{sec:multi:prop:algs}.  Specifically, we
consider this $\np \to \infty$ limit for kernels of the form
\eqref{eq:gen:bk:ker} or \eqref{eq:MTM:Kernel}, where we recall
\cref{tbl:beta:form}, which summarizes how we recover each of the
concrete methods in \cref{sec:multi:prop:algs} with an appropriate
choice of $\art: \qsp \times \qsp \to (0,\infty)$ and the single proposal
kernels $\pk$, $\bpk$.

Our analysis is divided into five subsections.
\cref{subsec:large:p:Barker} deals with the weak limit as
$p \to \infty$ for both \eqref{eq:gen:bk:ker} and
\eqref{eq:MTM:Kernel}, given in \cref{thm:large:p:lim} and
\cref{thm:MTM:limit}, respectively.  \cref{thm:MTM:limit} lays out a
novel approach to the $\np \to \infty$ limit for the Multiple Try methods
introduced in \cref{sec:MTM}.  Next, in \cref{sec:p:inf:lim:kernels}, we
explain how \cref{thm:large:p:lim} and \cref{thm:MTM:limit} apply to
each of the methods summarized in \cref{tbl:beta:form}.  In particular,
we justify the asymptotic unbiasedness of \cref{alg:slingshot:intro} on the
basis of \cref{cor:large:p:prod}
(\cref{sec:p:inf:lim:kernels:SS}, \cref{prop:Slingshot:p:lim}). Here, we also
carefully address the infinite dimensional scope of \cref{alg:MTMpCN} and
\cref{alg:MTMpCN:loc} (\cref{prop:MTM:glob:p:inf:lim},
\cref{prop:local:MTM:pinf:lim}).  In
\cref{sec:alg:scalingatinfty}, we then consider various algorithmic
parameter limits at $\np = \infty$ from the kernels derived in
\cref{subsec:large:p:Barker}.  This analysis further motivates and
classifies the `global' (\cref{subsubsec:bubble:bath}) and `local'
(\cref{meth:loc:prop}, \cref{meth:loc:prop:precond}) approaches as
well as their unbiased counterparts provided in \cref{sec:MTM},
\cref{sec:conv:prop}.  \cref{sec:bounds} provides quantitative
convergence results under more restrictive conditions on $\art$ in
total variation distance for \eqref{eq:gen:bk:ker}.  In particular, the
main result in this subsection (\cref{thm:tv:bound}) plays a central
role in our tuning analysis for the Slingshot algorithm below in
\cref{sec:The:Slingshot}.  Finally, in \cref{sec:RB::1step:Limit}, we
estimate the variance of the one step Rao-Blackwellization associated
to \eqref{eq:gen:bk:ker}.  Here one notable consequence is that this
variance vanishes in the case where there is no preliminary proposal
as in \cref{subsubsec:bubble:bath} and \cref{subsubsec:slingshot}.
This suggests an advantage of these methods over
\cref{meth:tj:mPCN}.

\subsection{Weak Convergence Results}
\label{subsec:large:p:Barker}

Our first result concerns the weak limit of multiproposal kernels
with a particular `Barker-type' acceptance structure. 

\begin{Theorem}\label{thm:large:p:lim}
  Let $\qsp$ be a Polish space.  For each $\np \geq 1$, consider the
  Markov kernel $\mk$ as in \eqref{def:mp:ker}, with $\vk$ as in
  \eqref{def:vk:Tjelmeland} and $(\ar_0, \ldots, \ar_\np)$ as in
  \eqref{def:ar:simple}, so that $P = P^\np$ takes the form
  \eqref{eq:gen:bk:ker}.  Suppose that, for every
  $\bq, \bbq \in \qsp$, the mapping
  $\art(\cdot, \bq): \qsp \to (0,\infty)$ is
  $\pk(\bbq, \cdot)$-integrable. Then, for any probability measure
  $\nu \in \Pr(\qsp)$, the limit
  \begin{align}
  \label{def:mk:infty}
    \int \mk^\infty(\bq_0, d \tbq) \nu( d \bq_0)
    = \lim_{\np \to \infty}  \int \mk^\np(\bq_0, d \tbq) \nu ( d \bq_0) 
  \end{align}
  exists in the sense of weak convergence\footnote{This means, by definition that
    $\int \int \phi(\tbq) \mk^\infty(\bq_0, d \tbq) \nu( d \bq_0)= \lim_{\np \to \infty}
   \int  \int \phi(\tbq) \mk^\np(\bq_0, d \tbq)\nu( d \bq_0)$, for every
    $\phi: \qsp \to \RR$ which is continuous and bounded.  Note that 
    this of course implies this limit exists pointwise, namely that
    $\mk^\infty(\bq_0, d \tbq) = \lim_{\np \to \infty}
    \mk^\np(\bq_0, d \tbq)$ weakly for every $\bq_0 \in \qsp$.}, and we
  have the following explicit form for the limit kernel
  \begin{align}\label{eq:p:inf:w:lim:conv:barker}
    \mk^\infty(\bq_0, d \tbq)
    = \int_\qsp \frac{\art(\tbq, \bq_0) \pk(\bbq, d \tbq)}
          {\int_\qsp \art(\bq', \bq_0) \pk(\bbq, d \bq')} \bpk(\bq_0, d \bbq).
  \end{align}
\end{Theorem}

To understand the structure of the limit kernel $\mk^\infty$ in
\eqref{eq:p:inf:w:lim:conv:barker}, we can proceed from a
representation of the form \eqref{eq:qn:unif:draws:form}.
Specializing to our present situation, \eqref{def:vk:Tjelmeland},
\eqref{def:ar:simple}, typically we can draw
$\bq \sim P^\np(\bq_0, d\tbq)$ as
\begin{align}\label{eq:exp:express:ptoinf}
  \bq  := \sum_{j=0}^\np
  \indFn{I_j} (u) \bq_j.
\end{align}
Here in \eqref{eq:exp:express:ptoinf} we draw $\bbq_0$ as
\eqref{eq:tj:prelim:draw} and then obtain $\bq_j$ for
$j = 1,\ldots, \np$ according to \eqref{eq:tj:cloud:draw}.  We then
independently draw $u \sim \cU (0,1)$ and define the subintervals of
$[0,1]$ as
\begin{align*}
  I_0 :=
  \left[0, \frac{\art(\bq_0, \bq_0)}{\sum_{k=0}^\np \art(\bq_k,
  \bq_0)}\right),
  \quad
  I_j :=
  \left[ \frac{ \sum_{m=0}^{j-1}\art(\bq_m, \bq_0)}{\sum_{k=0}^\np
  \art(\bq_k, \bq_0)},
  \frac{ \sum_{m=0}^{j}\art(\bq_m, \bq_0)}{\sum_{k=0}^\np
  \art(\bq_k, \bq_0)} \right),
  \quad j=1, \ldots, \np.
\end{align*}
Now, given any bounded and continuous observable $\phi: X \to \RR$,
and taking
\begin{align}\label{eq:barker:quick:proof:sigma:alg}
  \mathcal{F}_\np = \sigma\{ \bq_0, \bq_1, \ldots, \bq_\np\},
  \quad
  \mathcal{G} = \sigma\{ \bbq_0 \},
\end{align}
we have, cf. \cref{rmk:RB:cald},
\begin{align}
  \EE \phi(\bq) &=
  \sum_{j =0}^\np \EE (\indFn{I_j} (u)\phi(\bq_j))
  = \sum_{j =0}^\np \EE (\EE (\indFn{I_j} (u)\phi(\bq_j))|
  \mathcal{F}_\np ))
  = \sum_{j =0}^\np \EE ( \phi(\bq_j) \EE (\indFn{I_j} (u))|
                  \mathcal{F}_\np ))
  \notag\\
  &= \sum_{j =0}^\np \EE \left( 
    \frac{ \phi(\bq_j) \art(\bq_j, \bq_0)}
    {\sum_{k=0}^\np \art(\bq_k,  \bq_0)}\right)
    = \EE \left( 
    \frac{ \tfrac{1}{\np}\sum_{j =0}^\np \phi(\bq_j) \art(\bq_j,
    \bq_0)}
    {\tfrac{1}{\np}\sum_{k=0}^\np \art(\bq_k, \bq_0)}\right).
    \label{eq:int:cond:arg:lrg:p1}
\end{align}
Then, with \eqref{eq:cond:indep:expID}, the Law of Large Numbers and
the Dominated Convergence theorem, we find that
\begin{align}
   \lim_{\np \to \infty} \int \phi(\tbq) P^p( \bq_0, d\tbq)
     &= \EE \left(\lim_{\np \to \infty} \EE\left(  
     \frac{ \tfrac{1}{\np}\sum_{j =0}^\np \phi(\bq_j) \art(\bq_j,
     \bq_0)}
   {\tfrac{1}{\np}\sum_{k=0}^\np \art(\bq_k, \bq_0)} |\mathcal{G} \right)
   \right) 
  = \EE \left( \frac{\int_X \phi(\tbq) \art(\tbq, \bq_0) \pk(\bbq_0, d \tbq)}
       {\int_\qsp \art(\bq', \bq_0) \pk(\bbq_0, d \bq')} \right)
       \notag \\
  &= \int_\qsp  \frac{\int_X \phi(\tbq) \art(\tbq, \bq_0) \pk(\bbq, d \tbq)}
          {\int_\qsp \art(\bq', \bq_0) \pk(\bbq, d \bq') } \bpk(\bq_0, d \bbq).
           \label{eq:int:cond:arg:lrg:p1:2}
\end{align}
We provide a detailed measure theoretic proof of
\cref{thm:large:p:lim} applicable at the stated level of generality
below in \cref{sec:Large:p:w:l}.

As we already noted above, for the particular case when
$\bpk(\bq_0, d \bbq)$ is the Dirac measure $\delta_{\bq_0}(d \bbq)$, the
proposal kernel $\vk^\np$ from \eqref{def:vk:Tjelmeland} reduces to
\eqref{def:vk:before:Tjelmeland}.  For this special case, we record the
following immediate corollary of \cref{thm:large:p:lim} as follows:

\begin{Corollary}\label{cor:large:p:prod}
 Let $\qsp$ be a Polish space.  For each $\np \geq 1$,
  consider the Markov kernel $\mk^\np$ as in \eqref{def:mp:ker}, with
  $\vk^\np$ as in \eqref{def:vk:before:Tjelmeland} and
  $(\ar_0, \ldots, \ar_\np)$ as in \eqref{def:ar:simple}. Suppose
  that, for every $\bq_0 \in \qsp$, the mapping
  $\art(\cdot, \bq_0): \qsp \to (0,\infty)$ is $\pk(\bq_0,
  \cdot)$-integrable. Then, for every $\bq_0 \in \qsp$, the limiting
  kernel $\mk^\infty$ in \eqref{def:mk:infty} exists in the sense of
  weak convergence, and we have
  \begin{align}\label{eq:p:inf:w:lim:conv:barker:bb}
		\mk^\infty(\bq_0, d \tbq) = \frac{\art(\tbq, \bq_0)
          \pk(\bq_0, d \tbq)}
          {\int_\qsp \art(\bq', \bq_0) \pk(\bq_0, d \bq')}.
  \end{align}
\end{Corollary}
\noindent The desirable limiting behavior of the Slingshot algorithm,
\cref{subsubsec:slingshot}, is a particularly notable consequence of
\cref{cor:large:p:prod} as we lay out below in
\cref{sec:p:inf:lim:kernels:SS}.

 We turn now to address the $\np \to \infty$ limit for
 Multiple Try (MT) type kernels introduced in
 \cref{sec:MTM} (cf.  \cref{alg:MTM}).  Although the following result
 imposes a number of technical conditions on the Multiple Try kernel
 \eqref{eq:MTM:Kernel} in comparison to \cref{thm:large:p:lim} for
 \eqref{eq:gen:bk:ker}, this result nevertheless has a broad scope of
 applicability that includes
 \cref{meth:MTMpCN}--\cref{meth:MTMslingshot}.  We spell this out in
 detail immediately below in \cref{sec:p:inf:lim:kernels}.
 \begin{Theorem}
  \label{thm:MTM:limit}
  Let $(\qsp, \met)$ be a Polish space endowed with the usual
  Borel $\sigma$-algebra and let $\mu$ be a target probability
  measure on $\qsp$. Consider any Multiple Try type kernel $P^\np$ of the form
  \eqref{eq:MTM:Kernel} corresponding to the choices of measurable
  function $\art: \qsp \times \qsp \to (0,\infty)$, Markov kernel
  $\pk: \qsp \times \Sigma_\qsp \to [0,1]$, and number of proposals
  $\np \geq 1$. Assume that:
  \begin{enumerate}[label={(H\arabic*)}]
  \item\label{A3} For every compact subset $A \subset \qsp$, it holds
    that
    \begin{align}\label{inf:int:beta:nu}
      0 < \inf_{\br \in A} \int_\qsp \art(\tbq, \br) \pk(\br, d \tbq) < \infty,
    \end{align}
    and, for all $\bq_0 \in \qsp$,
    \begin{align}\label{cond:eta:art}
      \sup_{\br \in A} \left| \frac{d \eta^\perp}{d \eta}(\bq_0, \br)
      \frac{\art(\bq_0, \br)}{\art(\br, \bq_0)} \right| < \infty,
      \quad \sup_{\br \in A} |\art(\bq_0, \br)| < \infty,
    \end{align}
    with $\eta, \eta^\perp$ as in \eqref{def:eta:MTM}.
  \item\label{A1} There exists a probability measure $\nu$ on $\qsp$
    and a measurable function $F: \qsp \times \qsp \to \qsp$ such that
    $\pk(\bq_0, \cdot) = F(\bq_0, \cdot)^* \nu$ for all
    $\bq_0 \in \qsp$.
  \item\label{A4} For every $\varepsilon > 0$ and compact set
    $A \subset \qsp$, one can find a
    $\delta = \delta_{\varepsilon, A} > 0$ and measurable function
    $g = g_{\varepsilon, A}: \qsp \to \RR^+$ such that
  	\begin{align}\label{cont:art}
           |\art(F(\br, \bu), \br) - \art(F(\tilde{\br},\bu),
          \tilde{\br})| < g(\bu),  \quad
          \text{ whenever } d(\br, \tbr) < \delta \text{ and } \br, \tbr
          \in A,
  	\end{align}
        and such that
        \begin{align}
          \label{eq:g:loc:MTM:cond}
          \int_{\qsp} g(\bu) \nu(d\bu) < \epsilon,
        \end{align}
        where $\nu$ is the measure in \ref{A1}.
         \item\label{A2} For every $\bq_0 \in \qsp$ and $\varepsilon > 0$,
    there exists a compact subset $A \subset \qsp$ such that
    \begin{align}
          \int_\qsp \art(\tbq, \bq_0) \indFn{A^c} (\tbq)
          \pk(\bq_0, d\tbq)< \varepsilon.
		\label{eq:comp:exterior:demand:MTM}
	\end{align}
  \end{enumerate}
  Then, under \ref{A3}-\ref{A2}, we have, for any $\nu \in \Pr(\qsp)$, that
  \begin{align}
  \label{eq:weak:conv:MTM}
   \lim_{\np \to \infty} \int P^\np(\bq_0, d \tbq) \nu(d\bq_0) = \int P^\infty(\bq_0, d \tbq) \nu(d \bq_0),
  \end{align}
  in the sense of weak convergence.  Here,
  \begin{align}\label{eq:MTM:inf:ker}
    P^\infty(\bq_0, d \tbq) =
    \bar{\alpha}^\infty( \tbq, \bq_0) Q^\infty(\bq_0, d\tbq)
    + \delta_{\bq_0}(d\tbq) \int_\qsp (1 - \bar{\alpha}^\infty( \bbq, \bq_0))
    Q^\infty(\bq_0, d\bbq),
  \end{align}
  where
  \begin{align}\label{eq:MTM:inf:ker:prop}
    Q^\infty(\bq_0,d \tbq) = \frac{ \art(\tbq, \bq_0) Q(\bq_0, d \tbq)}
    {\int_\qsp \art(\bbq,\bq_0) Q(\bq_0, d \bbq) } 
  \end{align}
  and, recalling again the notation  \eqref{def:eta:MTM}, 
  \begin{align}\label{eq:MTM:inf:ker:ar}
    \bar{\alpha}^\infty( \tbq, \bq_0) :=
    1 \wedge \left( \frac{d \eta^\perp}{d \eta}(\bq_0, \tbq)
  \frac{\art(\bq_0, \tbq)}{\art(\tbq, \bq_0)}
    \frac{\int_\qsp \art(\bq',\bq_0) Q(\bq_0, d \bq') }
    { \int_\qsp \art(\bq',\tbq) Q(\tbq, d \bq')} \right). 
  \end{align}
\end{Theorem}

Analogously to the proof of \cref{thm:large:p:lim},
\eqref{eq:int:cond:arg:lrg:p1}-\eqref{eq:int:cond:arg:lrg:p1:2}, the
kernel \eqref{eq:MTM:Kernel} can be decomposed in such a way to expose
\eqref{eq:MTM:inf:ker} as a Law of Large Numbers (LLN) limit.  Indeed,
for any bounded continuous function $\phi: X \to \RR$,
\begin{align}
  \int  \phi(\tbq)& P^\np( \bq_0, d \tbq)
         \notag\\
  =& \EE
  \left( \frac{\np^{-1}\sum_{j = 1}^\np \phi(\bq_j) \art(\bq_j, \bq_0)
  \bar{\alpha}^\infty(\bq_j, \bq_0)}
     {\np^{-1}\sum_{j = 1}^\np \art(\bq_j, \bq_0)} \right)
  + \phi(\bq_0) \EE
  \left( \frac{\np^{-1}\sum_{j = 1}^\np \art(\bq_j, \bq_0)(1 -
  \bar{\alpha}^\infty(\bq_j, \bq_0))}
     {\np^{-1}\sum_{j = 1}^\np \art(\bq_j, \bq_0)} \right)
     \notag\\
     &+  \EE \left( \frac{\np^{-1}\sum_{j = 1}^\np (\phi(\bq_j)  - \phi(\bq_0)) \art(\bq_j, \bq_0)
  (\bar{\alpha}_j(\bq_0, \bq_1, \ldots, \bq_\np, \bz_1^j, \ldots,\bz_{\np-1}^j)-
    \bar{\alpha}^\infty(\bq_j, \bq_0))}
       {\np^{-1}\sum_{j = 1}^\np \art(\bq_j, \bq_0)} \right),
      \label{eq:overview:MTM:lim}
\end{align}
where $\bq_j$ are drawn independently from $\pk(\bq_0, d\tbq)$ and
$\bz_k^j := F(\bq_j, \bu_k)$ with $\bu_k$ taken independently from
$\nu$ in $\ref{A1}$.  Clearly, the first two terms in
\eqref{eq:overview:MTM:lim} limit to
$\int \phi(\tbq) P^\infty( \bq_0, d \tbq)$ with $P^\infty$
as in \eqref{eq:MTM:inf:ker}.  However, the final `error term' is nontrivial
due to the delicate structure of \eqref{eq:MTM:inf:ker}.

To handle this final term in \eqref{eq:overview:MTM:lim} we develop a
uniform version of the Law of Large Numbers based on a compactness
argument following the formulation in \cite[Proposition
5.1]{borggaard2020consistency} (see also \cite{talagrand1987glivenko,
  dudley1991uniform, van2014universal, wainwright2019high}).  To see
how this comes about, observe that, for any compact $A \subset \qsp$,
the last error term in \eqref{eq:overview:MTM:lim} is estimated by
\begin{align}
  \label{eq:MTM:ULLN:est::pinf:lim}
  \|\phi\|_\infty \left( \EE  \left(\sup_{\br \in A}\epsilon_\np(\br) \right)
  + 4 \EE \left(  \frac{\np^{-1}\sum_{j = 1}^\np 
  	\art(\bq_j, \bq_0) \indFn{\bq_j \in A^c}}
  {\np^{-1}\sum_{j = 1}^\np \art(\bq_j, \bq_0)} \right) \right)
\end{align}
where
\begin{align*}
   \epsilon_\np(\br)
  := \biggl|1 \wedge&
   \left(\frac{d \eta^\perp}{d \eta}(\bq_0, \br)
   \frac{\art(\bq_0, \br)}{\art(\br, \bq_0)} \frac{\np^{-1}\sum_{l =1}^p\art(\bq_l, \bq_0)}
      {\np^{-1}\art(\bq_0, \br)+ \np^{-1}\sum_{l =1}^{\np -1}\art(F(\br, \bu_l), \br)}
      \right)
       \notag\\
                    &\qquad \qquad \qquad
                      - 1 \wedge \left( \frac{d \eta^\perp}{d \eta}(\bq_0, \br)
   \frac{\art(\bq_0, \br)}{\art(\br, \bq_0)}
     \frac{\int_\qsp \art(\bq',\bq_0) Q(\bq_0, d \bq') }
     { \int_\qsp \art(\bq',\br) Q(\br, d \bq')} \right)
     \biggr| .
\end{align*}
Thus, our imposed conditions \ref{A3}--\ref{A4} allow us to establish
that the LLN holds uniformly over $\br$ in any compact $A$, so that the first
term in \eqref{eq:MTM:ULLN:est::pinf:lim} vanishes as
$\np \to \infty$.  Meanwhile, the last assumption \ref{A2} clearly
addresses the second term in \eqref{eq:MTM:ULLN:est::pinf:lim}. The
complete, rigorous proof of \cref{thm:MTM:limit} is provided below in
\cref{app:mtm:limit}.

\subsection{Infinite $\np$ Kernels for Some Concrete Methods}
\label{sec:p:inf:lim:kernels}

Leveraging \cref{thm:large:p:lim} and \cref{thm:MTM:limit}, we turn now
to describe the $\np = \infty$ kernels for each of the methods
introduced in \cref{sec:multi:prop:algs}.  Our presentation is
organized in terms of the Global, Local and Slingshot type approaches
successively.

\subsubsection{Global Methods}
\label{sec:p:inf:lim:kernels:GM}
Start with the `Bubble Bath' approach defined by
\cref{subsubsec:bubble:bath} alongside other associated `global
approaches' that correct for bias in \cref{subsubsec:bubble:bath} as
\cref{meth:MTMpCN}, \cref{meth:tj:mPCN}.  \cref{thm:large:p:lim}
covers the setting of \cref{subsubsec:bubble:bath} (via
\cref{cor:large:p:prod}) as well as \cref{meth:tj:mPCN}, which includes
the multiproposal method in \cite{tjelmeland2004using}, and the mpCN
algorithm in \cite[Algorithm 4]{glatt2024parallel}.  Here, for a
target of the form \eqref{eq:stuart:form}, we take
\begin{align}
  \label{eq:trgt:beta:frm}
    \art (\tbq, \bq) = \dmu(\tbq);
\end{align}
see \cref{tbl:beta:form}.  Regarding \cref{subsubsec:bubble:bath},
with \cref{cor:large:p:prod} and this choice of $\art$ we obtain the
$\np = \infty$ limit as
\begin{align}
  \label{inf:ker:bb}
  \mk^\infty_{BB}(\bq_0, d \tbq) :=
  \frac{\dmu(\tbq) \pk(\bq_0, d \tbq)}
  {\int_\qsp \dmu(\bq') \pk(\bq_0, d \bq')}.
\end{align}
On the other hand, for \cref{meth:tj:mPCN}, we take $\pk = \bpk$ in
\eqref{eq:p:inf:w:lim:conv:barker} and therefore find from
\cref{thm:large:p:lim} that the limit kernel $\mk^\infty$ is given by
\begin{align}
  \label{inf:ker:cond:ind}
  \mk^\infty_{TJ}(\bq_0, d \tbq) =
  \int_\qsp \frac{\dmu(\tbq) \pk(\bbq, d \tbq)}
  {\int_\qsp \dmu(\bq') \pk(\bbq, d \bq')} \pk(\bq_0, d \bbq).
\end{align}

The scope of $\dmu$ and $\pk$ where we can rigorously justify these
limits, \eqref{inf:ker:bb}, \eqref{inf:ker:cond:ind} is very broad
and can be deduced as a direct corollary of \cref{thm:large:p:lim} as
follows:
\begin{Corollary}
  The kernel \eqref{inf:ker:bb} is the $\np \to \infty$ weak limit of
  \eqref{eq:gen:bk:ker:single:lay} with
  $\art (\tbq, \bq) = \dmu(\tbq)$ at least so long as
  $\dmu \in L^1(Q(\bq_0, \cdot))$ for every $\bq_0 \in \qsp$.
  Similarly, \eqref{inf:ker:cond:ind} is the weak limit for the kernel
  \eqref{eq:gen:bk:ker} with the same choice of $\art$ and
  $\pk = \bar{\pk}$ and under the same integrability condition on
  $\dmu$ with respect to $\pk$.
\end{Corollary}

We turn next to \cref{meth:MTMpCN}, in particular \cref{alg:MTMpCN},
whose $\np = \infty$ kernel we derive under \cref{thm:MTM:limit}.
From \eqref{eq:MTM:inf:ker:ar}, we compute
\begin{align}
  \label{eq:MTM:ar:p:inf}
  \bar{\alpha}^\infty_{gMT}( \tbq, \bq_0) :=
    1 \wedge \left( \frac{d \eta^\perp_0}{d \eta_0}(\bq_0, \tbq)
    \frac{\int_\qsp \dmu(\bq')  Q(\bq_0, d \bq') }
  { \int_\qsp \dmu(\bq') Q(\tbq, d \bq')} \right)
\end{align}
where
\begin{align*}
  \eta_0(d \bq, d \tbq) = Q(\bq, d \tbq)\mu_0(d \bq),
  \quad \eta_0^\perp(d \bq, d \tbq) = Q(\tbq, d \bq)\mu_0(d \tbq).
\end{align*}
Under the prior invariance condition \eqref{eq:rev:spec}, these two
measures $\eta_0$ and $\eta_0^\perp$ are the same, so that
$\frac{d \eta^\perp_0}{d \eta_0}(\bq_0, \tbq) \equiv 1$.  Thus, the
$\np = \infty$ kernel of \cref{meth:MTMpCN} takes the form
\begin{align}
    P^\infty_{gMT}(\bq_0, d \tbq) =&
     \left[1 \wedge \left(
    \frac{\int_\qsp \dmu(\bq')  Q(\bq_0, d \bq') }
                                     { \int_\qsp \dmu(\bq') Q(\tbq, d \bq')} \right) \right]
                                      \frac{\dmu(\tbq) \pk(\bq_0, d \tbq)}
  {\int_\qsp \dmu(\bq') \pk(\bq_0, d \bq')}
                         \notag\\
  &+ \delta_{\bq_0}(d\tbq) \int_\qsp
    \left(1 -      1 \wedge \left(
    \frac{\int_\qsp \dmu(\bq')  \pk(\bq_0, d \bq') }
    { \int_\qsp \dmu(\bq') \pk(\bbq, d \bq')} \right)
    \right)
                                          \frac{\dmu(\bbq) \pk(\bq_0, d \bbq)}
  {\int_\qsp \dmu(\bq') \pk(\bq_0, d \bq')}.
    \label{eq:MTM:inf:ker:BB}
\end{align}

While the application of \cref{thm:large:p:lim} is straightforward for
relevant cases in \eqref{inf:ker:bb} and \eqref{inf:ker:cond:ind},
determining the scope of the conditions \ref{A3}--\ref{A2} in
\cref{thm:MTM:limit} is a more delicate task.  Here, for simplicity, we
restrict our attention to two relevant Gaussian single proposal
kernels $\pk$, namely \eqref{eq:Guass:1:prop:classic} or
\eqref{eq:Guass:1:prop:pCN} in \eqref{eq:MTM:inf:ker:BB}.
\begin{Proposition}
  \label{prop:MTM:glob:p:inf:lim}
  Consider the Multiple Try kernel \eqref{eq:MTM:Kernel}, with $\art$ as in
  \eqref{eq:BB:adj}, namely $\art(\tbq, \bq_0) = \dmu(\tbq)$.  Then
  \eqref{eq:MTM:Kernel} converges weakly to \eqref{eq:MTM:inf:ker:BB}
  as $\np \to \infty$ under either of the following two circumstances:
  \begin{itemize}
  \item[(i)] We take $\qsp = \RR^d$ and suppose that we are given a
    continuously distributed target measure of the form
    $\mu(d\tbq) = \dmu (\tbq) d \tbq$ and a Gaussian single proposal
    kernel
    $\pk(\bq_0, d\tbq) = \pk_\sigma(\bq_0, d\tbq) = N(\bq_0, \sigma^2
    \cC)(d \tbq)$ as in \eqref{eq:Guass:1:prop:classic}.  We assume
    that $\pi$ is everywhere positive and that, for every
    $\epsilon > 0$ and $R > 0$, there exist a
    $\delta = \delta_{\epsilon, R} > 0$ and a measurable function
    $g = g_{\epsilon, R}: \RR^d \to \RR^+$ with
    \begin{align}
            \label{eq:fd:dmu:MTM:2}
      \int g( \bu) \nu(d\bu) < \epsilon
      \quad
      \text{ where } \nu = N(0, \cC)
    \end{align}
    and
    \begin{align}
            \label{eq:fd:dmu:MTM:3}
      | \dmu( \sigma \bu + \br) - \dmu( \sigma \bu + \tbr)|
      \leq g(\bu),
      \text{ for all } \bu \in \RR^d
      \text{ so long as }
      |\br - \tbr| < \delta
      \text{ and }
      \max\{|\br|, |\tbr|\} < R.
    \end{align}
  \item[(ii)] Let $\qsp$ be a separable Hilbert space and consider a
    target measure $\mu$ on $\qsp$ of the form
    \eqref{eq:stuart:form:og}, where we assume as in
    \eqref{eq:s:f:infinite:d} $\mu_0 = N(0,\cC)$ for some symmetric,
    positive-definite and trace class covariance operator $\cC$ on
    $\qsp$.   Fix $\rho \in [0,1]$ and consider a proposal kernel
    $\pk_\rho(\bq_0, d\tbq) = N( \rho \bq_0, (1 - \rho^2) \cC)$ as
    formulated in \eqref{eq:Guass:1:prop:pCN}.  We suppose that
    $\Pot: \qsp \to \RR$, the potential function in
    \eqref{eq:stuart:form:og}, is continuous and bounded from below,
    namely
    \begin{align}
            \label{cond:Pot:F:lip:1}
       M := \inf_{ \bq \in \qsp} \Pot(\bq) > - \infty.
    \end{align}
    Additionally, we assume, for any given $\varepsilon > 0$ and
    $A \subset \qsp$ compact, that there exists a corresponding
    $\delta = \delta_{\varepsilon, A}$ as well as a measurable
    function $\lip: \qsp \to \RR^+$ with
    \begin{align}
                  \label{cond:Pot:F:lip:2}
      \int_\qsp \lip (\bu) \mu_0(d \bu) < \varepsilon
    \end{align}
    and such that
    \begin{align}
      \label{cond:Pot:F:lip}
    |\Pot(\rho \br + \sqrt{1 - \rho^2} \bu)
    - \Pot (\rho \tbr + \sqrt{1 - \rho^2} \bu)| \leq
    \lip(\bu) 
     \text{ for all } \bu \in \qsp \text{ so long as } | \br -
      \tbr| < \delta, \br,\tbr \in A.
    \end{align}
\end{itemize}
\end{Proposition}

The boundedness condition in
\eqref{cond:Pot:F:lip:1} is easy to understand and to verify in many
situations of interest.  On the other hand, \eqref{eq:fd:dmu:MTM:2},
\eqref{eq:fd:dmu:MTM:3} or \eqref{cond:Pot:F:lip:2},
\eqref{cond:Pot:F:lip} can be seen to hold when $\dmu$ or $\Pot$,
respectively, are continuously differentiable with their derivatives
growing at no more than a small quadratic exponential rate at
`spatial infinity'; see \eqref{eq:C1:fd:special:case}.  In particular,
the conditions imposed in the second case in
\cref{prop:MTM:glob:p:inf:lim} include in its scope some interesting
targets arising in the Bayesian approach to PDE inverse problems.  We
make these observations precise in \cref{rmk:MTM:glob:p:inf:lim:scope}
immediately below.

Having surmounted the challenge of identifying a set of tractable
conditions on the target, the task of verifying the conditions in
\cref{thm:MTM:limit} for the proof of \cref{prop:MTM:glob:p:inf:lim}
is otherwise mostly routine.  The one subtle point is establishing
\ref{A2} in case (ii), which involves the use of compactness in the
infinite dimensional Gaussian measure theory setting.  Complete
details are provided below in \cref{app:MTM:glob:p:inf:lim}.

\begin{Remark}\label{rmk:MTM:glob:p:inf:lim:scope}
  Let us explain how the conditions in \cref{prop:MTM:glob:p:inf:lim} are
  fulfilled in specific, concrete circumstances.  Regarding
  the assumptions in the first case, (i), we observe that the conditions
  \eqref{eq:fd:dmu:MTM:2}--\eqref{eq:fd:dmu:MTM:3} hold when
    \begin{align}
      \label{eq:C1:fd:special:case}
      \dmu \in C^{1}(\RR^d)  \text{ and }
      \kappa := \sup_{\bx \in \RR^d} \left(
      \exp( - \gamma  |\cC^{-1/2} \bx|^2) |\nabla
      \dmu(\bx)| \right)< \infty,
    \end{align}
    for any $0< \gamma < (4 \sigma^2)^{-1}$. To justify this claim, we 
    observe that, with the Mean Value theorem,
    we have under \eqref{eq:C1:fd:special:case}
    \begin{align}
      | \dmu( \sigma \bu + \br) - \dmu( \sigma \bu + \tbr)|
      \leq& \sup_{t \in [0,1]} | \nabla \dmu( \sigma \bu + t\br +
            (1-t) \tbr)| |\br - \tbr|
            \notag\\
      \leq& \kappa |\br - \tbr|  \sup_{t \in [0,1]}
            \exp\left( \gamma |\cC^{-1/2}(\sigma \bu + t\br +(1-t) \tbr)|^2\right)
            \notag\\
      \leq&   \kappa |\br - \tbr|
            \exp\left( 2\gamma \max\{ |\cC^{-1/2} \br|^2,  |\cC^{-1/2} \tbr|^2\}\right)
            \exp\left( 2\gamma\sigma^{2}  |\cC^{-1/2}\bu|^2\right).
    \end{align}
    Hence, for any given $\epsilon, R> 0$, we can take e.g.
    \begin{align}\label{def:delta:g:rmk}
      \delta =\epsilon
      \kappa^{-1}\frac{
      \exp\left(- 2\gamma \|\cC^{-1/2}\|^2 R^2
      \right)}
      { \int \exp\left( 2\gamma \sigma^{2}
      |\cC^{-1/2}\bu|^2\right)\nu(d \bu) },
      \quad g(\bu) = \epsilon
      \frac{
      \exp\left( 2\gamma\sigma^{2}  |\cC^{-1/2}\bu|^2\right)}
      { \int \exp\left( 2\gamma \sigma^{2}  |\cC^{-1/2}\bu|^2\right)\nu(d \bu) },
    \end{align}
    to obtain \eqref{eq:fd:dmu:MTM:2}, \eqref{eq:fd:dmu:MTM:3}, where
    $\| \cC^{-1/2}\|$ denotes the operator norm of $\cC^{-1/2}$ and we
    recall that $\nu = N(0,\cC)$. Note that the condition
    $0< \gamma < (4 \sigma^2)^{-1}$ ensures that the integral in the
    denominators from \eqref{def:delta:g:rmk} is finite. This
    completes the proof in the first case.

    The scope of applicability of (ii) in
    \cref{prop:MTM:glob:p:inf:lim} includes a variety of Bayesian PDE
    inverse problems of the type developed in e.g.
    \cite{stuart2010inverse, borggaard2020bayesian}.  Here in
    principle we could formulate general conditions of the type
    \eqref{eq:C1:fd:special:case}, but the functional-analytic
    considerations are much more delicate and vary greatly in
    specifics from problem to problem.  Thus, it is more useful and
    illuminating to illustrate how things play out for specific model
    problems.  Here, we treat a problem concerning the Bayesian
    estimation of a velocity field from the concentration of a solute
    sitting in this flow, as we developed previously in
    \cite{borggaard2020bayesian, borggaard2020consistency,
      glatt2021mixing, glatt2024parallel}.

    For this model problem, our target is of the type
    \eqref{eq:stuart:form}--\eqref{eq:s:f:infinite:d} where the
    potential takes the form
    \begin{align}
      \label{eq:AD:pot:available}
      \Pot(\bq) = | \Gamma^{-1/2} ( \mathcal{Y} - \mathcal{O}(
      \theta(\bq)))|^2.
    \end{align}
    Here, $\mathcal{Y} \in \RR^m$ represents the observed concentrations
    of a solute, and $\Gamma$ is a symmetric positive definite matrix
    coming from our model of data acquisition, namely,  
    \begin{align}
      \mathcal{Y} = \mathcal{O}(
      \theta(\bq)) + \epsilon,
    \end{align}
    where our observations are polluted by additive, unbiased Gaussian
    error $\epsilon \sim N(0,\Gamma)$.  The `solute concentration'
    $\theta = \theta(\bq)$ is the solution of
    \begin{align}
      \label{eq:AD:problem}
      \partial_t \theta + \bq \cdot \nabla \theta = \kappa \Delta
      \theta, \quad \theta(0) = \theta_0,
    \end{align}  
    evolving on a two dimensional periodic box $\mathbb{T}^2$ with the
    unknown quantity $\bq: \mathbb{T}^2 \to \RR^2$ which is a
    divergence free vector field, taken here to be time-independent in
    our toy model.  The diffusivity $\kappa$ is a positive
    constant.  For convenience, since the mean $m(t) =\int \theta(t,\bx)
    d\bx$ is preserved by \eqref{eq:AD:problem}, we suppose that
    $\int \theta_0(\bx) d \bx = 0$.

    For all sufficiently regular $\bq$ (and any given initial
    condition $\theta_0$), these solutions $\theta(\bq)$ take values in
    $Z_s := C([0,T]; H^s(\mathbb{T}^2))$, that is $\theta(\bq)$
    evolves continuously in time with values in the Sobolev space of
    mean free functions $H^s(\mathbb{T}^2)$, for any $s \geq 0$.  Note
    furthermore that this map $\bq \mapsto \theta(\bq)$ is continuously
    differentiable from $H^s(\mathbb{T}^2)$ into this $Z_s$, for any
    such $s \geq 0$; see \cite[Proposition 8.1, 8.2]{glatt2021mixing}
    for further details.  The observation operator
    $\mathcal{O}: Z_s \to \RR^m$ maps these solutions of
    \eqref{eq:AD:problem} to a collection of $m$ observations.
    Regarding this observation procedure $\mathcal{O}$, typically we
    are interested in space-time point observations as
    \begin{align}
      \label{eq:pt:obs:AD}
      \mathcal{O}(\theta(\bq))
      = (\theta(t_1, \bx_1; \bq), \ldots, \theta(t_m, \bx_m; \bq)),
    \end{align}  
    or in spatial-temporal averages taking the form
    \begin{align}
      \label{eq:spt:obs:AD}
       \mathcal{O}(\theta(\bq)) = \left( \int_0^T \int_{\mathbb{T}^2}
      \theta(t, \bx; \bq) g_1(t, \bx) d\bx dt, \cdots, \int_0^T \int_{\mathbb{T}^2}
      \theta(t, \bx; \bq) g_m(t, \bx) d\bx dt
      \right),
    \end{align}  
    where $g_1, \ldots, g_m$ are suitable basis functions.

    The results in \cite[Section 8]{glatt2021mixing}, \cite[Appendix
    A]{borggaard2020consistency} can be adapted to show that target
    measures of the form
    \eqref{eq:stuart:form}--\eqref{eq:s:f:infinite:d} with potentials
    given as \eqref{eq:AD:pot:available} fall under the scope
    of \cref{prop:MTM:glob:p:inf:lim}, (ii).  For this, we need to specify
    the conditions on the class of observation operators $\mathcal{O}$ and
    covariance $\cC$ defining the prior $\mu_0$ such that the required
    conditions \eqref{cond:Pot:F:lip:1}--\eqref{cond:Pot:F:lip} are
    met.  Regarding $\mathcal{O}$, we suppose that it is bounded and
    linear on $Z_s$, for $s > 1$. This means that
    \begin{align}
      \label{eq:Obs:regular:AD}
      | \mathcal{O} (\psi) | \leq c_0 \sup_{t \in [0,T]} \|\psi(t)\|_{H^s}.
    \end{align}
    Thus, in view of the Sobolev embedding of $H^s(\mathbb{T}^2)$ into
    $C(\mathbb{T}^2)$ for $s > 1$, we are covering the two important
    concrete observational regimes \eqref{eq:pt:obs:AD},
    \eqref{eq:spt:obs:AD} under this general assumption.  With this
    regularity \eqref{eq:Obs:regular:AD} in mind, we suppose that
    $\qsp = H^s(\mathbb{T}^2)$ with $s > 1$ and assume that our prior
    $\mu_0$ is defined by a covariance operator $\cC$ maintaining
    \eqref{eq:s:f:infinite:d} as can for example be achieved by
    assuming that $\cC$ is a solution operator of
    $(-\Delta)^\gamma \mathbf{w} = \bq$, for $\gamma > 1$.  See also
    \cite{borggaard2020bayesian} for different variations on this
    theme motivated by two-dimensional turbulence.

    With these specifications for $\qsp$ and $\cC$, the regularity
    requirements on $\Phi$ are given by \cite[Proposition 8.1,
    8.2]{glatt2021mixing}, as we already mentioned.  Here,
    \eqref{cond:Pot:F:lip:1} is just a consequence of $\Phi$  in \eqref{eq:AD:pot:available} being
    non-negative.  Now, regarding
    \eqref{cond:Pot:F:lip:2}--\eqref{cond:Pot:F:lip}, observe that
    \begin{align}
      \label{eq:pot:der:AD}
      \nabla \Pot( \bbq)\bw
      = - 2\langle \Gamma^{-1/2}
      (\mathcal{Y} -\mathcal{O}(\theta(\bbq))),
      \Gamma^{-1/2} \mathcal{O}(\psi(\bbq, \bw))
      \rangle.
    \end{align}
    where $\psi := \psi(\bbq, \bw)$ obeys
    \begin{align}
      \label{eq:theta:der:AD}
      \partial_t  \psi + \bbq \cdot \nabla \psi =
      \kappa \Delta \psi - \bw \cdot \nabla \theta(\bbq),
      \quad \psi(0) = 0.
    \end{align}
    Thus, from our assumption \eqref{eq:Obs:regular:AD} and the Mean
    Value theorem, we have, for any $\br, \tbr, \bu \in H^s(\mathbb{T}^2) =
    \qsp$ and $\rho \in [0,1]$,
    \begin{align}
      |\Pot(\rho \br + \sqrt{1 - \rho^2}& \bu)
      - \Pot (\rho \tbr + \sqrt{1 - \rho^2} \bu)|
                  = \rho |\nabla \Pot(\bbq_\tau) (\br - \tbr)|
             \notag\\     
      \leq& c \left(1 + \sup_{t \in [0,T]} \| \theta(t;\bbq_\tau)\|_{H^s}\right)
            \left( \sup_{t \in [0,T]}\|\psi(t;\bbq_\tau, \br -
            \tbr)\|_{H^s}\right)
            \label{eq:pot:int:lip:bnd:AD}
    \end{align}
    where
    $\bbq_\tau := \rho (\tau\br + (1-\tau)\tbr) + \sqrt{1 - \rho^2}
    \bu$, for some appropriate $\tau \in [0,1]$ and the constant $c$
    depends only on $\mathcal{Y}$, $\Gamma$ and $\rho$.  From here, we
    claim that
    \begin{align}
      \label{eq:theta:Hs:bnd:1}
      \sup_{t \in [0,T]} \| \theta(t;\bbq)\|_{H^s}
      \leq  \exp(c (\| \bbq\|_{H^s}^a+1))
    \end{align}
    and that
    \begin{align}
      \label{eq:theta:Hs:bnd:2}
      \sup_{t \in [0,T]} \| \psi(t;\bbq, \bw)\|_{H^s}
      \leq  \| \bw\|_{H^s} \exp(c (\| \bbq\|_{H^s}^a+1)).
    \end{align}
    where, critically, the constants
    $c = c(\kappa, s,T, \|\theta_0\|_{H^s})$ and
    $a = a(s) < 2$ are independent of $\bbq, \bw$.
    Hence, combining \eqref{eq:pot:int:lip:bnd:AD} with
    \eqref{eq:theta:Hs:bnd:1}, \eqref{eq:theta:Hs:bnd:2}, we obtain
    \begin{align}
           |\Pot(\rho \br + \sqrt{1 - \rho^2} \bu)
      - \Pot (\rho \tbr + \sqrt{1 - \rho^2} \bu)|
                \leq    \| \br - \tbr\|_{H^s}
                  \exp(c( \| \bu\|^a +  \|\br \|^a + \|
      \tbr\|^a + 1)),
                  \label{eq:pot:int:lip:bnd:AD:1}
    \end{align}
    where again
    $c = c(\kappa, s, T, \|\theta_0\|_{H^s}, \mathcal{Y}, \Gamma,
    \rho)$ is independent of $\br, \tbr, \bu$.  Thus, for any compact
    subset $A$ of $H^s(\mathbb{T}^2)$ and $\epsilon >0$, we simply take
    $g(\bu) = \delta e^{c( \| \bu\|^a+ 2 \sup_{\br \in A} \|\br
      \|^a+ 1)}$ for an appropriately small
    $\delta = \delta(\epsilon, A)$ to obtain \eqref{cond:Pot:F:lip:2},
    \eqref{cond:Pot:F:lip}.

    It remains to justify \eqref{eq:theta:Hs:bnd:1},
    \eqref{eq:theta:Hs:bnd:2}.  These bounds are essentially contained
    in \cite[Proposition 8.1, 8.2]{glatt2021mixing} with one crucial
    tweak to the estimates found therein.  Introducing the notation
    $\Lambda^s = (-\Delta)^{s/2}$ and remembering that $\bbq$
    is divergence free, we have
    \begin{align}
      \label{eq:energy:AD}
      \frac{1}{2}\frac{d}{dt} \|\theta \|^2_{H^s} + \kappa\|\theta
      \|^2_{H^{s+1}} =
      \int \Lambda^s(\bbq\cdot \nabla \theta) \Lambda^s \theta d \bx
      = \int
      (\Lambda^s(\bbq\cdot \nabla \theta)
      -\bbq\cdot \nabla \Lambda^s\theta  )\Lambda^s \theta d \bx
    \end{align}
    for $\theta = \theta(\bbq)$.  We now recall the Kenig-Ponce-Vega
    commutator estimate that 
    \begin{align}
      \label{eq:KPV}
      \| \Lambda^s ( \bv \cdot \nabla \psi) - \bv\cdot \nabla \Lambda^s
      \psi\|_{L^q}
      \leq c (\|\nabla \bv\|_{L^{q_1}} \|\Lambda^s \psi\|_{L^{q_2}}
      + \| \Lambda^s \bv\|_{L^{q_3}} \|\nabla \psi\|_{L^{q_4}}),
    \end{align}  
    valid for any $\bv$, $\psi$ sufficiently smooth and mean free, and
    any $q, q_i \in (1,\infty)$ with
    $q^{-1} = q_1^{-1} + q_2^{-1} = q_3^{-1} + q_4^{-1}$ where the
    constant $c = c(q,q_i, s)$ is independent of $\bv$, $\psi$; see
    \cite[Appendix A] {KenigPonceVega1993_gKdV} or the reformulation
    in e.g. \cite[Appendix B]{constantin2014unique}.  Therefore, with
    \eqref{eq:KPV}, the Sobolev embedding bound
    $ \|\psi\|_{L^r} \leq c \|\Lambda^{1 - 2/r} \psi\|_{L^2}$ valid
    for $r \in [2,\infty)$, the interpolation inequality
    $\|\Lambda^r \psi \|_{L^2} \leq \|\Lambda^{\gamma_l} \psi \|_{L^2}^{\frac{\gamma_u
        -r}{\gamma_u-\gamma_l} }
    \|\Lambda^{\gamma_u}\psi\|_{L^2}^{\frac{r-\gamma_l }{\gamma_u-\gamma_l}
    }$ which holds whenever $0 \leq \gamma_l < r < \gamma_u$, we
    obtain, under a suitable choice of $q \in (2,\infty)$ that, 
    \begin{align}
      \left|\int
      (\Lambda^s(\bbq\cdot \nabla \theta)  -\bbq\cdot \nabla
      \Lambda^s\theta  )\Lambda^s \theta d \bx\right|
      \leq&
          c(  \|\nabla \bbq\|_{L^{2q/(q-2)}}\|\Lambda^s\theta\|_{L^2}
                +  \|\Lambda^s
            \bbq\|_{L^2}\|\nabla\theta\|_{L^{2q/(q-2)}}) \|\Lambda^s\theta\|_{L^q}
                        \notag\\
            \leq&
           c \|\bbq\|_{H^s}\|\theta\|_{H^s}^{2-\gamma}
                  \|\theta\|_{H^{s+1}}^{\gamma}
                  \leq       c \|\bbq\|_{H^s}^{\frac{2}{2-\gamma}}\|\theta\|_{H^s}^2
                  +\frac{\kappa}{2} \|\theta\|_{H^{s+1}}^2,
                  \label{eq:KPV:pcked} 
    \end{align}
    for some $\gamma = \gamma(s) \in (0,1)$ and $c = c(s,\kappa)$.
    Thus, \eqref{eq:KPV:pcked} with \eqref{eq:energy:AD} yields
    \eqref{eq:theta:Hs:bnd:1} as in \cite[(8.10)]{glatt2021mixing}
    with $a < 2$.  For the second bound \eqref{eq:theta:Hs:bnd:2} we
    can reason precisely as in \cite[Proposition 8.2]{glatt2021mixing}
    on the basis of the first bound.
\end{Remark}

\subsubsection{Local Methods}
\label{sec:p:inf:lim:kernels:LM}

Next let us derive the $\np = \infty$ kernels for each of the `local
algorithms' in \cref{sec:multi:prop:algs} -- \cref{meth:loc:prop},
\cref{meth:loc:prop:precond}, \cref{meth:loc:MTM},
\cref{meth:loc:pCN} -- on the basis of
\eqref{eq:p:inf:w:lim:conv:barker:bb} and \eqref{eq:MTM:inf:ker} in
the biased and Multiple Try corrected cases, respectively. Regarding
\cref{meth:loc:prop}, referring back to \cref{tbl:beta:form} we take
$\beta(\tbq, \bq_0) = \sqrt{\dmu(\tbq)/\dmu(\bq_0)}$ and consider the
single proposal kernel $\pk_\sigma$ defined according to
\eqref{eq:Guass:1:prop:classic}.  Thus, the $p = \infty$ kernel for
\cref{meth:loc:prop} takes the form
\begin{align}
  \label{eq:prop:kern:loc:classic}
  \mk^\infty_\sigma(\bq_0, d \tbq)
  =\frac{\sqrt{\dmu(\tbq)/ \dmu(\bq_0)}\pk_\sigma(\bq_0, d\tbq)}
  {\int\sqrt{\dmu(\bq')/\dmu(\bq_0)}\pk_\sigma(\bq_0, d \bq')}
  =\frac{\sqrt{\dmu(\tbq)}\pk_\sigma(\bq_0, d\tbq)}
  {\int\sqrt{\dmu(\bq')}\pk_\sigma(\bq_0, d \bq')}.
\end{align}
Turning next to \cref{meth:loc:MTM}, where again
$\beta(\tbq, \bq_0) = \sqrt{\dmu(\tbq)/\dmu(\bq_0)}$, we have from
\eqref{eq:MTM:inf:ker:prop} that its $\np = \infty$ proposal kernel
coincides with \eqref{eq:prop:kern:loc:classic}
(cf. \eqref{eq:p:inf:w:lim:conv:barker:bb} and
\eqref{eq:MTM:inf:ker:prop}), namely
\begin{align}\label{def:pkinfty:localMTM}
	\pk^\infty_\sigma(\bq_0, d \tbq) =\frac{\sqrt{\dmu(\tbq)}\pk_\sigma(\bq_0, d\tbq)}
	{\int\sqrt{\dmu(\bq')}\pk_\sigma(\bq_0, d \bq')},
\end{align}
whereas its $p = \infty$ accept-reject mechanism is obtained from
\eqref{eq:MTM:inf:ker:ar}. Here note that, due to the symmetry of
$\pk_\sigma$, we obtain that
$\frac{d \eta^\perp}{d \eta}(\bq_0, \tbq) =
\frac{\dmu(\tbq)}{\dmu(\bq_0)}$.  Thus, from
\eqref{eq:MTM:inf:ker:ar}, we conclude that
\begin{align}
  \label{eq:prop:AR:loc:classic}
  \bar{\alpha}^\infty_{\sigma}( \tbq, \bq_0)
  &=    1 \wedge \left(
  \frac{\int   \sqrt{\dmu(\bq')/ \dmu(\bq_0)}  Q_\sigma(\bq_0, d \bq') }
  { \int \sqrt{\dmu(\bq') / \dmu(\tbq)}  Q_\sigma(\tbq, d \bq')}
  \right).
\end{align}
Hence, the limit kernel for \cref{meth:loc:MTM} is given as
\begin{align}
  \label{eq:loc:MTM:infty:ker}
  \mk_{\sigma,lMT}^\infty ( \bq_0, d\tbq)
  =   \bar{\alpha}^\infty_{\sigma}( \tbq, \bq_0)
  \pk^\infty_\sigma(\bq_0, d\tbq)
  + \delta_{\bq_0}( d \tbq) \int
  (1 - \bar{\alpha}^\infty_{\sigma}( \bq', \bq_0))
  \pk^\infty_\sigma(\bq_0, d \bq').
\end{align}
We provide a rigorous basis for this $\np \to \infty$ limit
\eqref{eq:loc:MTM:infty:ker} as follows.
\begin{Proposition}
  \label{prop:local:MTM:pinf:lim:0}
  Let $\qsp = \RR^d$ and suppose we are given a
    continuously distributed target measure of the form
    $\mu(d\tbq) = \dmu (\tbq) d \tbq$, and Gaussian single proposal
    kernel
    $\pk(\bq_0, d\tbq) = \pk_\sigma(\bq_0, d\tbq) = N(\bq_0, \sigma^2
    \cC)(d \tbq)$ as in \eqref{eq:Guass:1:prop:classic}.  Assume that
    \begin{align}
      \label{eq:MTM:loc:strong:dmu:cond}
      \dmu \in L^\infty(\RR^d) \cap C(\RR^d)
      \text{ and is strictly positive}.
    \end{align}
    Additionally, we assume, for any given $\varepsilon > 0$ and
    $A \subset \qsp$ compact, that there exists a corresponding
    $\delta = \delta_{\varepsilon, A}$ as well as a measurable
    function $\lip = \lip_{\varepsilon,A}: \RR^d \to \RR^+$ with
     \begin{align}
    	\label{cond:Pot:F:lip:2:loc}
    	\int g( \bu) \nu(d\bu) < \epsilon,
    	\quad
    	\text{ where } \nu = N(0, \cC),
    \end{align}
    and
    \begin{align}
    	\label{cond:Pot:F:lip:loc}
    	| \dmu( \sigma \bu + \br) - \dmu( \sigma \bu + \tbr)|
    	\leq g(\bu),
    	\text{ for all } \bu \in \RR^d
    	\text{ so long as }
    	|\br - \tbr| < \delta, \br,\tbr \in A.
    \end{align} 
    Then, the Multiple Try kernel \eqref{eq:MTM:Kernel} with this choice of
    $\pk$ and with $\art(\tbq, \bq_0) = \sqrt{\dmu(\tbq)/\dmu(\bq_0)}$
    converges weakly to \eqref{eq:loc:MTM:infty:ker} as
    $\np \to \infty$.
\end{Proposition}
\noindent See \cref{app:local:local:MTM:pinf:lim:0} for the proof.

Let us now address the preconditioned cases introduced in
\cref{meth:loc:prop:precond} and \cref{meth:loc:pCN}.  We compute the
limiting kernel for \cref{meth:loc:prop:precond} from
\eqref{eq:p:inf:w:lim:conv:barker:bb} referring back to
\cref{tbl:beta:form} to find, for $\rho \in (0,1)$, that
\begin{align}
  \label{eq:pcn:loc:uncor:pinf:ker}
  \mk^\infty_\rho(\bq_0,d\tbq) =
 \frac{\exp\left( -
    \frac{\rho}{1+\rho}(\Pot(\tbq/\rho)-
  \Pot(\bq_0))\right) \pk_\rho(\bq_0, d \tbq)}
  {\int\exp\left( -
    \frac{\rho}{1+\rho}(\Pot(\bq'/\rho)-
  \Pot(\bq_0))\right) \pk_\rho(\bq_0, d \bq')}.
\end{align}
Here recall that, as in \eqref{eq:Guass:1:prop:pCN},
$\pk_\rho(\bq_0, d \bq') = N( \rho \bq_0, (1 - \rho^2) \cC)$.  As
regards \cref{meth:loc:pCN} and the associated \cref{alg:MTMpCN:loc},
first note that, according to \eqref{eq:MTM:inf:ker:prop}, its
$p = \infty$ proposal kernel coincides with
\eqref{eq:pcn:loc:uncor:pinf:ker}, namely
\begin{align}\label{def:Qrhoinf}
	\pk^\infty_\rho(\bq_0,d\tbq) =
	\frac{\exp\left( -
		\frac{\rho}{1+\rho}(\Pot(\tbq/\rho)-
		\Pot(\bq_0))\right) \pk_\rho(\bq_0, d \tbq)}
	{\int\exp\left( -
		\frac{\rho}{1+\rho}(\Pot(\bq'/\rho)-
		\Pot(\bq_0))\right) \pk_\rho(\bq_0, d \bq')}.
\end{align}
Next, referring back to \eqref{eq:RN:eta:etaperp}, we then compute its
$p=\infty$ accept-reject mechanism from \eqref{eq:MTM:inf:ker:ar} as
\begin{align}
  \label{eq:pcn:loc:AR:pinf}
  \bar{\alpha}^\infty_{\rho}( \tbq, \bq_0)
  =1 \wedge \frac{\exp\left(  -\Pot(\tbq) -
    \frac{\rho}{1+\rho}\Pot(\bq_0/\rho)\right)}
  {\exp\left( -\Pot(\bq_0) -
    \frac{\rho}{1+\rho}\Pot(\tbq/\rho)\right) }
  \frac{\int\exp\left(  -   \frac{\rho}{1+\rho}\Pot(\bq'/\rho)
     \right) \pk_\rho(\bq_0, d \bq')}
  {\int \exp\left( - \frac{\rho}{1+\rho}\Pot(\bq'/\rho)
     \right)\pk_\rho(\tbq, d \bq')}
\end{align}
and then define the associated kernel $\mk^\infty_{\rho,lMT}$ in direct
analogy to \eqref{eq:loc:MTM:infty:ker}.

\begin{Proposition}
  \label{prop:local:MTM:pinf:lim}
  Let $\qsp$ be a separable Hilbert space upon which is defined a
  target probability measure $\mu$ of the form
  \eqref{eq:stuart:form:og}, where $\mu_0 = N(0,\cC)$ for some
  symmetric, positive-definite and trace class covariance operator
  $\cC$ on $\qsp$ as in \eqref{eq:s:f:infinite:d}.  Regarding the
  potential $\Pot: \qsp \to \RR$ in the definition of $\mu$, we posit
  continuity and assume the global lower boundedness condition given
  in \eqref{cond:Pot:F:lip:1}.  Furthermore, we posit that, for any
  $\epsilon > 0$ and any compact set $A \subset \qsp$, there is a
  corresponding $\delta> 0$ and measurable function $g: \qsp \to \RR^+$, with
  $\int g(\bu) \nu (d \bu)< \epsilon$ and
  \begin{align}
    \label{eq:some:local:pot}
     \left|\Pot\biggl(\br +
         \frac{\sqrt{1 - \rho^2}}{\rho} \bu\biggr)
    -
    \Pot\biggl(\tbr +
    \frac{\sqrt{1 - \rho^2}}{\rho} \bu\biggr)
    \right|
    \leq g(\bu),
  \end{align}
  for any $|\br - \tbr| < \delta, \br, \tbr \in A$.  Then, under these
  circumstances, the Multiple Try kernel $P^\np$ in \eqref{eq:MTM:Kernel}
  specified by \cref{alg:MTMpCN:loc} converges weakly to the kernel
  $P^\infty$ of the general form \eqref{eq:MTM:inf:ker} specified
  by \eqref{eq:pcn:loc:uncor:pinf:ker}, \eqref{eq:pcn:loc:AR:pinf}.
\end{Proposition}
\noindent See \cref{app:local:local:MTM:pinf:lim} for the detailed proof.

\begin{Remark}
  The observations in \cref{rmk:MTM:glob:p:inf:lim:scope} concerning
  the scope of applicability of \cref{prop:MTM:glob:p:inf:lim} also
  apply equally well to \cref{prop:local:MTM:pinf:lim:0},
  \cref{prop:local:MTM:pinf:lim} with minor adjustments.
\end{Remark}

\subsubsection{The large proposal asymptotic for the Slingshot:
  unbiased samples from the target.}
\label{sec:p:inf:lim:kernels:SS}

The $\np \to \infty$ limiting behavior of the Slingshot,
\cref{alg:slingshot:intro}, \cref{subsubsec:slingshot}, is another
notable consequence of \cref{cor:large:p:prod}.  Indeed, recall that
in this Slingshot case, cf.  \cref{tbl:beta:form}, we fix a Markov
kernel $\pk$ as in \eqref{eq:prop:ker:diff:form} and take $\art$ as in
\eqref{eq:SS:adj}, namely
\begin{align} 
\label{eq:ss:gen:beta}
  \art(\tbq, \bq_0) = \frac{\dmu(\tbq)}{\dpk(\bq_0, \tbq)}
  = \frac{d\mu}{d \pk(\bq_0, \cdot)}(\tbq),
\end{align}
so that, according to \eqref{eq:p:inf:w:lim:conv:barker:bb}, the
limiting kernel of $\mk^\np$ given by \eqref{eq:gen:bk:ker:single:lay}
as $\np \to \infty$ is
\begin{align}\label{eq:sling:shot}
  \mk^\infty(\bq_0, d \tbq)
  = \frac{\frac{d\mu}{d \pk(\bq_0, \cdot)}(\tbq) \pk(\bq_0, d \tbq)}
  {\int_\qsp \frac{d\mu}{d \pk(\bq_0, \cdot)}(\bq') \pk(\bq_0, d \bq')}
  = \frac{\mu(d \tbq)}{\int_\qsp \mu(d \bq')} = \mu(d \tbq).
\end{align}
In other words, the large $\np$ limit of \cref{subsubsec:slingshot}
coincides with the target $\mu$ so that, algorithmically, we are simply
taking i.i.d.  samples from this given $\mu$.

Let us furthermore observe that the Multiple Try adjusted
\cref{meth:MTMslingshot} also reduces to an independence sampler on the
target $\mu$ in the $\np \to \infty$ limit.  Thus, comparing with our
observations in the previous paragraph, the Multiple Try correction for
\cref{alg:slingshot:intro} is superfluous for $\np$ sufficiently
large.  To justify this claim about \cref{meth:MTMslingshot}, recall
that we are again assuming $\pk$ takes the form
\eqref{eq:prop:ker:diff:form} and $\beta$ is defined as in
\eqref{eq:SS:adj}.  Comparing these formulations with
\eqref{eq:MTM:inf:ker:prop}, we see that indeed
$Q^\infty(\bq_0,d \tbq) = \mu(d\tbq)$.  Furthermore, recalling
\eqref{eq:MTM:inf:ker:ar} and noting that in this
case, cf. \eqref{def:eta:MTM},
\begin{align}
  \label{eq:etaperp:eta:SS}
  \frac{d \eta^\perp}{d \eta}(\bq_0, \tbq)
  = \frac{ \dpk(\tbq, \bq_0) \pi(\tbq)}{\dpk(\bq_0, \tbq) \pi(\bq_0)},
\end{align}
it follows with \eqref{eq:prop:ker:diff:form}, \eqref{eq:SS:adj}
that $\bar{\alpha}^\infty( \tbq, \bq_0) \equiv 1$.

Let us record some situations where we can rigorously justify
the $\np \to \infty$ limits for \cref{subsubsec:slingshot} and
\cref{meth:MTMslingshot}.  The scope and proof for the first
uncorrected case is obvious from \cref{cor:large:p:prod}, whereas the
details for the second case are provided below in
\cref{app:proof:of:slingshot:lim}.
\begin{Proposition}
\label{prop:Slingshot:p:lim}
\mbox{}
\begin{itemize}
\item[(i)] Consider a target probability measure $\mu$ and a Markov
  kernel $\pk$ on a Polish space $\qsp$, for which the absolute
  continuity condition \eqref{eq:ss:abs:cont:cond} is
  maintained. Then, the kernel associated with
  \cref{alg:slingshot:intro}, namely \eqref{eq:gen:bk:ker:single:lay},
  with $\art$ given in \eqref{eq:ss:gen:beta} converges weakly to
  $\mu$ as $\np \to \infty$.
\item[(ii)] Let us now suppose that $\mu(d \tbq) = \dmu(\tbq) d \tbq$
  is continuously distributed on $\RR^d$ with $\dmu \in C(\RR^d)$.
  Furthermore, for any $M > 0$ we suppose there is a
  $\gamma = \gamma_M: \RR^d \to \RR^+$ with
  \begin{align}
    \label{eq:uniform:den:bnd}
    \sup_{|\bq| \leq M} \dmu( \bq + \bu) \leq \gamma_M(\bu)\dmu(\bu)
    \text{ and } \int \gamma_M(\bu)\dmu(\bu) d \bu < \infty.
  \end{align}
  Regarding the form of the single proposal kernel, we assume that
  $\pk(\bq_0, d \tbq) = N(\bq_0, \sigma^2 \cC)$ for some $\sigma >0$ and
  $\cC$ symmetric positive definite as in
  (\ref{eq:Guass:1:prop:classic}).  Then, under these conditions, the
  Multiple Try corrected kernel \eqref{eq:MTM:Kernel} associated with
  \cref{alg:slingshot:intro} (where the acceptance probabilities
  $\bar{\alpha}_j$ are as in \eqref{eq:MTM:ar:red} and $\art$ is as in
  \eqref{eq:ss:gen:beta}) converges weakly to $\mu$ as
  $\np \to \infty$.
\end{itemize}
\end{Proposition}

\subsection{Algorithmic Parameter Limits at $\np = \infty$: Local vs Global}
\label{sec:alg:scalingatinfty}

Having now computed the $\np \to \infty$ limits for some concrete
methods in \cref{sec:p:inf:lim:kernels} based on
\cref{thm:large:p:lim} and \cref{thm:MTM:limit}, we now turn to some
algorithmic parameter limits at $\np = \infty$. Specifically, we
consider certain limits for the parameters $\sigma$ in
\eqref{eq:Guass:1:prop:classic} and $\rho$ in
\eqref{eq:Guass:1:prop:pCN}, treating the `global' and `local' methods
from \cref{sec:multi:prop:algs} successively. In the former `global'
situation, we establish the limiting cases $\sigma \to \infty$ or
$\rho \to 0$ in the proposal mechanism, toward i.i.d. Monte Carlo
draws from the target measure in \cref{prop:Glob:param:Limit}.
Meanwhile, \cref{Prop:alg:scalingatinfty:local} shows that the `local'
methods converge towards appropriate Langevin dynamics in the opposite
extreme, that is either $\sigma \to 0$ or $\rho \to 1$ in the proposal
structure.

\subsubsection{`Global' Proposal Structures --
  Scaling to directly sample the Target Measure}
\label{sec:alg:scalingatinfty:global}

Two limits in the algorithmic parameters in
\cref{subsubsec:bubble:bath} and \cref{meth:MTMpCN},
\cref{meth:tj:mPCN} at the limit $\np = \infty$ are apparent from
\eqref{inf:ker:bb}, and from \eqref{inf:ker:cond:ind},
\eqref{eq:MTM:inf:ker:BB}.  First, let us consider the case when the
target measure $\mu$ is defined with respect to a Gaussian base
measure as in \eqref{eq:stuart:form}--\eqref{eq:s:f:infinite:d}, and
$\pk(\bq_0, d \tbq)$ is a pCN type proposal kernel as in
\eqref{eq:Guass:1:prop:pCN}, namely
$\pk(\bq_0, d \tbq) := \pk_\rho(\bq_0, d \tbq) = N(\rho\bq_0, (1-
\rho^2) \cC)(\tbq)$ for $\rho \in [0,1]$.  Here we observe that
$\pk_0(\bq, d \tbq) = \mu_0(d\tbq)$ so that, for this choice
$\rho = 0$, and noting that in this case
$\dmu(\tbq) = \exp( - \Pot(\tbq))$ we see, from direct inspection of
\eqref{inf:ker:bb}--\eqref{eq:MTM:inf:ker:BB} that
\begin{align}\label{eq:rho:to:zero:Pinf}
  \mk^\infty_{BB, 0}(\bq, d \tbq)
  =  \mk^\infty_{TJ, 0}(\bq, d\tbq)
  = \mk^\infty_{MT, 0}(\bq, d \tbq) = \mu(d\tbq).
\end{align}
Here we have adopted the notation that, for $\rho \in [0,1]$,
$\mk^\infty_{BB, \rho}$, $\mk^\infty_{TJ, \rho}$,
$\mk^\infty_{MT, \rho}$ denote the kernels in
\eqref{inf:ker:bb}, \eqref{inf:ker:cond:ind},
\eqref{eq:MTM:inf:ker:BB}, respectively, with $\pk := \pk_\rho$ according to
\eqref{eq:Guass:1:prop:pCN}.

Secondly, we consider the case when the target
$\mu(d \tbq) = \dmu(\tbq) d \tbq$ is continuously distributed on
$\RR^d$, that is where $\mu_0$ is the Lebesgue measure in
\eqref{eq:stuart:form}. In this setting, the individual proposals are
taken from the Gaussian kernel
$\pk(\bq_0, d \tbq) := \pk_\sigma(\bq_0, d \tbq) = N(\bq_0, \sigma^2
\cC)(d\tbq)$ as in \eqref{eq:Guass:1:prop:classic}. Analogously to the
previous situation, we fix the notation $\mk^\infty_{BB,\sigma}$,
$\mk^\infty_{TJ, \sigma}$, and $\mk^\infty_{MT, \sigma}$ to represent
the kernels in \eqref{inf:ker:bb}, \eqref{inf:ker:cond:ind} and
\eqref{eq:MTM:inf:ker:BB}, respectively, in the case where
$\pk = \pk_\sigma$ as in \eqref{eq:Guass:1:prop:classic} for
$\sigma^2 > 0$.

Regarding in particular the kernel $\mk^\infty_{BB,\sigma}$, we fix
the notation $\Cn{\tbq} = |\cC^{-1/2} \tbq|$, $\tbq \in \RR^d$, and
note that for any (bounded and continuous) test function
$\phi: \RR^d \to \RR$
\begin{align}
	&\left| \int \phi(\tbq) \mk^\infty_{BB,\sigma}(\bq_0, d \tbq) - \int
	\phi(\tbq)  \dmu( \tbq)  d \tbq\right|
	=\left|
	\frac{\int \phi(\tbq)
		\dmu( \tbq) \exp( - \frac{1}{2 \sigma^2} \Cn{\tbq -
		\bq_0}^2) d \tbq}
	{\int \dmu( \tbq) \exp( - \frac{1}{2 \sigma^2} \Cn{\tbq -
		\bq_0}^2) d \tbq} - \int
	\phi(\tbq)  \dmu( \tbq)  d \tbq\right|
	\notag\\
	&\qquad \quad =
	\left| \frac{ \int \phi(\tbq)
		\dmu( \tbq) ( \exp( - \frac{1}{2 \sigma^2} \Cn{\tbq -
		\bq_0}^2) - 1) d \tbq + (\int
		\phi(\tbq)  \dmu( \tbq)  d \tbq ) \int 
		\dmu( \tbq) ( 1 - \exp( - \frac{1}{2 \sigma^2} \Cn{\tbq -
		\bq_0}^2) ) d \tbq}{ \int \dmu( \tbq) \exp( - \frac{1}{2 \sigma^2} \Cn{\tbq -
		\bq_0}^2) d \tbq }
	\right|
	\notag\\
	&\qquad \quad \leq 
	2 \| \phi \|_\infty \frac{ \int \dmu(\tbq) ( 1 - \exp( - \frac{1}{2 \sigma^2} \Cn{ \tbq -
		\bq_0}^2) )  d \tbq}{  \int \dmu( \tbq) \exp( - \frac{1}{2 \sigma^2} \Cn{\tbq -
		\bq_0}^2) d \tbq }.
	\label{eq:weak:conv:iid}
\end{align}
Thus, this
estimate, \eqref{eq:weak:conv:iid}, suggests the weak convergence for
the $\np = \infty$ kernels defined by
\cref{subsubsec:bubble:bath}, \cref{meth:MTMpCN} and
\cref{meth:tj:mPCN} to the kernel $\mu(d \tbq)$ for i.i.d. samples
from the target measure in the limit as $\sigma \to \infty$.

We now summarize our convergence results at $\np = \infty$ in the
global case as follows:
\begin{Proposition}
  \label{prop:Glob:param:Limit}
  \begin{itemize}
  \item[(i)] Let $\mu$ be a target measure of the form
    \eqref{eq:stuart:form}--\eqref{eq:s:f:infinite:d} defined on a
    separable Hilbert space $\qsp$.  Assume that
    $\exp( - \Pot(\bq)) \in L^1(\mu_0) \cap C(\qsp)$ with $\Phi$
    bounded from below.  Consider $\mk^\infty_{BB,\rho}$,
    $\mk^\infty_{TJ, \rho}$ and $\mk^\infty_{MT, \rho}$ as in
    \eqref{inf:ker:bb}, \eqref{inf:ker:cond:ind} and
    \eqref{eq:MTM:inf:ker:BB}, respectively, defined relative to
    $\pk := \pk_\rho$ as in \eqref{eq:Guass:1:prop:pCN}. Then
  \begin{align}\label{eq:rho:to:zero:Pinf:prop}
  \lim_{\rho \to 0} \mk^\infty_{BB,\rho}(\bq_0, d \tbq)
  =   \lim_{\rho \to 0} \mk^\infty_{TJ, \rho}(\bq_0, d\tbq)
  =  \lim_{\rho \to 0}  \mk^\infty_{MT, \rho}(\bq_0, d \tbq) = \mu(d\tbq),
  \end{align}
  weakly.
\item[(ii)] On the other hand, in the case when the target $\mu$ is
  continuously distributed, with $\mu(d \tbq) = \dmu(\tbq) d \tbq$,
  and we instead take $\mk^\infty_{BB,\sigma}$,
  $\mk^\infty_{TJ, \sigma}$, $\mk^\infty_{MT, \sigma}$ in
  \eqref{inf:ker:bb}, \eqref{inf:ker:cond:ind} and
  \eqref{eq:MTM:inf:ker:BB} with respect to the proposal kernel
  $\pk_\sigma$ defined in \eqref{eq:Guass:1:prop:classic}, it holds
  that
  \begin{align}\label{eq:sigma:to:zero:Pinf}
    \lim_{\sigma \to \infty} \mk^\infty_{BB,\sigma}(\bq_0, d \tbq)
      =   \lim_{\sigma \to \infty} \mk^\infty_{TJ, \sigma}(\bq_0, d\tbq)
  =  \lim_{\sigma \to \infty}  \mk^\infty_{MT, \sigma}(\bq_0, d \tbq)
   = \mu(d\tbq),
  \end{align}
  again in the sense of weak convergence.
  \end{itemize}
\end{Proposition}
\noindent Given our discussions above around
\eqref{eq:rho:to:zero:Pinf} and \eqref{eq:weak:conv:iid}, the proofs of
\eqref{eq:rho:to:zero:Pinf:prop} and \eqref{eq:sigma:to:zero:Pinf} are
more or less routine.  We provide the complete details below in
\cref{app:proof:of:large:sig}.

\begin{Remark}[Large $\np$ Algorithmic Parameter Scaling: `Global'
  Proposal Structures]
  \label{rmk:scale:to:glob}
  Our claims about $\rho$ and $\sigma$ in \eqref{eq:rho:to:zero:Pinf},
  \eqref{eq:sigma:to:zero:Pinf} are made here at $\np = \infty$.  The
  optimal relationship between $\rho$ and $\np$, or between $\sigma$
  and $\np$ is a very interesting question for future work as we
  discuss in \cref{sec:Outlook}.
\end{Remark}

\begin{Remark}
  Although \cref{subsubsec:bubble:bath} is biased in general, the
  limits in \eqref{eq:rho:to:zero:Pinf:prop},
  \eqref{eq:sigma:to:zero:Pinf} suggest that this bias is small for
  large $p$ and $\sigma$ large or $0 \leq \rho \ll 1$ as appropriate
  for the structure of the target measure and proposal kernel.
\end{Remark}

\subsubsection{`Local' Proposal Structures -- Over-Damped Langevin Scaling}
\label{sec:alg:scalingatinfty:local}

At the other extreme are the acceptance structures appearing under
\cref{meth:loc:prop}, \cref{meth:loc:prop:precond} and the associated
unbiased Multiple Try counterparts: \cref{meth:loc:MTM}, \cref{meth:loc:pCN}.
Each of these methods are asymptotically comparable to an appropriate
over-damped Langevin dynamic as we shall explain next.

Regarding \cref{meth:loc:prop} and its Multiple Try adjustment,
\cref{meth:loc:MTM}, we consider the continuous in time dynamic defined as
\begin{align}\label{eq:trad:LA:dynam}
	d \bq = \frac{1}{2} \nabla  \log \dmu(\bq) dt + dW,
\end{align}
which is easily seen to be reversible (and therefore invariant) with
respect to $\mu(d \tbq) = \dmu(\tbq) d\tbq$.  Recall that
\eqref{eq:trad:LA:dynam} has the infinitesimal generator
\begin{align}
	\label{eq:inf:gen:trad:LA:dynam}
	\mathcal{L} \psi(\bq_0)
	=  \frac{1}{2} \nabla \log \dmu(\bq_0) \cdot \nabla \psi(\bq_0)
	+ \frac{1}{2}\Delta \psi(\bq_0).
\end{align}
In practice, in order to computationally generate proposals inspired by
\eqref{eq:trad:LA:dynam} in an MCMC method targeting
$\mu(d \tbq) = \dmu(\tbq) d \tbq$, we consider an Euler-Maruyama
numerical approximation of \eqref{eq:trad:LA:dynam} as
\begin{align*}
	\bq := \bq_0 + \frac{\sigma^2}{2} \nabla \log \dmu(\bq_0) + \sigma
	\xi,
\end{align*}
for a time step of size $\sigma^2$ and where $\xi \sim N(0, I)$.  That is,
we have a proposal kernel of the form
\begin{align}
	\bar{Q}_\sigma(\bq_0, d \tbq)
	&= N(\bq_0 + \frac{\sigma^2}{2}\nabla \log \dmu(\bq_0), \sigma^2
	I)= 
	\frac{1}{(2\pi \sigma^2)^{d/2}} \exp\left( - \frac{1}{2 \sigma^2}|\tbq -
	\bq_0 - \frac{\sigma^2}{2}\nabla \log \dmu(\bq_0)|^2\right) d \tbq
	\notag\\
	&=
	\exp\left( \frac{1}{2} \langle \nabla \log \dmu(\bq_0),
	\tbq - \bq_0 \rangle
	- \frac{\sigma^2}{8} | \nabla \log \dmu(\bq_0)|^2
	\right)
	\pk_\sigma(\bq_0, d \tbq).
	\label{eq:MALA:Prop:Kernel}
\end{align}
where $\pk_\sigma(\bq_0, d \tbq)= N( \bq_0, \sigma^2 I)(d \tbq)$ as in
\eqref{eq:Guass:1:prop:classic}. 

Of course, the proposal based on \eqref{eq:MALA:Prop:Kernel} is
typically biased with respect to $\mu(d \tbq) = \dmu(\tbq) d \tbq$. As
observed going back at least to
\cite{besag1994comments,roberts1996exponential}, we may correct for
this with a Metropolis-Hastings step defined by an acceptance
probability of the form
\begin{align}
	\label{eq:MALA:AR}
	\alpha  (\tbq, \bq_0) =
	1 \wedge \left(
	\frac{\dmu(\tbq)
		\exp( - \tfrac{1}{2 \sigma^2}|\bq_0 - \tbq -
		\tfrac{\sigma^2}{2}\nabla \log \dmu(\tbq)|^2)}
	{\dmu(\bq_0) \exp( - \tfrac{1}{2 \sigma^2}|\tbq -
		\bq_0 - \frac{\sigma^2}{2}\nabla \log \dmu(\bq_0)|^2)} \right).
\end{align}
With this $\alpha$, we then obtain an unbiased method defined by the `MALA
kernel'
\begin{align*}
	R_\sigma(\bq_0, d \tbq) =
	\alpha(\tbq, \bq_0)\bar{Q}_\sigma(\bq_0, d \tbq)
	+ \delta_{\bq_0}(d \tbq)
	\int (1 -   \alpha(\bq', \bq_0))\bar{Q}_\sigma(\bq_0, d \bq').
\end{align*}

Let us now formally compare the Langevin proposal kernel
$\bar{Q}_\sigma$ and its unbiased MALA correction $R_\sigma$ with the
$\np = \infty$ limits $\mk^\infty_\sigma$ for \cref{meth:loc:prop}
given in \eqref{eq:prop:kern:loc:classic} and $P^\infty_{\sigma,lMT}$ for
\cref{meth:loc:MTM} in \eqref{eq:loc:MTM:infty:ker}, respectively.  To
this end, observe that we have with Taylor's theorem up to first order
that
\begin{align}
	\label{eq:taylor:relation}
	\beta(\bq', \bbq)
	= \frac{ \sqrt{\dmu(\bq')} }{\sqrt{\dmu(\bbq)}}
	= \exp\left(  \frac{1}{2}(\log \dmu(\bq')
	-  \log \dmu(\bbq)) \right)
	= \exp\left( \frac{1}{2}\langle \nabla \log \dmu( \bbq), \bq' - \bbq
	\rangle + o(|\bq' - \bbq|)\right)
\end{align}
for any $\bq', \bbq$.  Regarding $\pk^\infty_\sigma$, we apply
\eqref{eq:taylor:relation} to \eqref{eq:prop:kern:loc:classic},
ignoring the error terms.  Note that under this use of the Taylor
approximation, the normalizing constant in
\eqref{eq:prop:kern:loc:classic} acts as the moment generating
function of $N(0, \sigma^2I)$ evaluated at $\frac{1}{2}\nabla \log \dmu(\bq_0)$,
namely
\begin{align}
	\label{eq:MGM:Fn:obs}
	\int \exp\left(
	\frac{1}{2}\langle\nabla \log \dmu(\bq_0),
	\tbq -\bq_0\rangle
	\right) \pk_\sigma(\bq_0, d \tbq)
	= \exp\left( \frac{\sigma^2}{8} | \nabla \log \dmu(\bq_0)|^2\right).
\end{align}  
Thus, we formally recover $\bar{Q}_\sigma$ in \eqref{eq:MALA:Prop:Kernel}
from $\mk^\infty_\sigma$ in \eqref{eq:prop:kern:loc:classic} when
$\sigma^2 \ll 1$.

In order to see how the acceptance term in MALA, \eqref{eq:MALA:AR},
matches the one from the Multiple Try limit, \eqref{eq:prop:AR:loc:classic}, notice that,
again blithely dropping error terms in \eqref{eq:taylor:relation}
and using \eqref{eq:MGM:Fn:obs}, we find
\begin{align*}
	\frac{\int   \sqrt{\dmu(\bq')/\dmu( \bq_0) }  Q_\sigma(\bq_0, d \bq') }
	{ \int \sqrt{\dmu(\bq')/\dmu( \tbq)}  Q_\sigma(\tbq, d \bq')}
	\approx
	\frac{\int
		e^{\frac{1}{2}\langle \nabla \log \dmu( \bq_0), \bq' - \bq_0\rangle}
		Q_\sigma(\bq_0, d \bq')}
	{\int
		e^{\frac{1}{2}\langle \nabla \log \dmu( \tbq), \bq' - \tbq\rangle}
		Q_\sigma(\tbq, d \bq')}
	= \frac{\exp\left( \frac{\sigma^2}{8} | \nabla \log
		\dmu(\bq_0)|^2\right)}
	{\exp\left( \frac{\sigma^2}{8} | \nabla \log \dmu(\tbq)|^2\right)}.
\end{align*}
On the other hand, working from \eqref{eq:MALA:AR},
and then using \eqref{eq:taylor:relation}, ignoring
error terms and reading right to left, we conclude
\begin{align*}
	\frac{\dmu(\tbq)
		e^{- \tfrac{1}{2 \sigma^2}|\bq_0 - \tbq -  \tfrac{\sigma^2}{2}\nabla \log \dmu(\tbq)|^2}}
	{\dmu(\bq_0)
		e^{ - \tfrac{1}{2 \sigma^2}|\tbq -  \bq_0 - \frac{\sigma^2}{2}\nabla \log \dmu(\bq_0)|^2}}
	=&
	\frac{e^{ \frac{\sigma^2}{8} | \nabla \log  \dmu(\bq_0)|^2}}
	{e^{ \frac{\sigma^2}{8} | \nabla \log \dmu(\tbq)|^2}}
	\frac{\dmu(\tbq)
		\exp( \tfrac{1}{2}\langle
		\nabla \log \dmu(\tbq) ,\bq_0 - \tbq \rangle)}
	{\dmu(\bq_0) \exp( \tfrac{1}{2}\langle
		\nabla \log \dmu(\bq_0) ,\tbq - \bq_0 \rangle) }
	=  \frac{e^{ \frac{\sigma^2}{8} | \nabla \log  \dmu(\bq_0)|^2}}
	{e^{ \frac{\sigma^2}{8} | \nabla \log \dmu(\tbq)|^2}}.
\end{align*}  
Thus, with these additional observations on the acceptance
probabilities, we formally conclude that
$P^\infty_\sigma \approx R_\sigma$ when $\sigma \ll
1$.%

We make these formal arguments linking $\bar{Q}_\sigma$, $R_\sigma$ to
$\mk^\infty_\sigma$, $\mk^\infty_{\sigma,lMT}$ for small $\sigma$ rigorous at
the level of infinitesimal generators immediately below in
\cref{Prop:alg:scalingatinfty:local}. Specifically, we show that,
appropriately scaled, the $\np=\infty$ generators for
\cref{meth:loc:prop} and \cref{meth:loc:MTM} converge to
\eqref{eq:inf:gen:trad:LA:dynam} as $\sigma^2 \to 0$.  In the first
case from \eqref{eq:prop:kern:loc:classic}, noting that the size of
the time step in the numerical scheme is $\sigma^2$, the associated
infinitesimal generator is defined as:
\begin{align}
	\label{eq:prop:kern:loc:classic:IG}
	\mathcal{L}_\sigma \psi( \bq_0) =
	\frac{\int (\psi( \tbq)- \psi( \bq_0)) \pk^\infty_\sigma (\bq_0, d\tbq) }{\sigma^2}
	= \frac{\int (\psi(\tbq) - \psi(\bq_0)) \sqrt{\dmu(\tbq)}\pk_\sigma(\bq_0, d\tbq)}
	{\sigma^2\int\sqrt{\dmu(\bq')}\pk_\sigma(\bq_0, d \bq')}.
\end{align}
Similarly, for the Multiple Try corrected scheme \eqref{eq:loc:MTM:infty:ker},
the associated infinitesimal generator is
\begin{align}
	\mathcal{L}_{\sigma,lMT} \psi( \bq_0) =&
	\frac{\int (\psi( \tbq)- \psi( \bq_0)) P^\infty_\sigma (\bq_0,
		d\tbq) }{\sigma^2}
	\notag\\
	=& \frac{\int (\psi(\tbq) - \psi(\bq_0))
		\left( 1 \wedge \left(
		\frac{  \sqrt{\dmu( \tbq)} }
		{ \sqrt{\dmu( \bq_0) } }
		\frac{\int   \sqrt{\dmu(\bq')}  Q_\sigma(\bq_0, d \bq') }
		{ \int \sqrt{\dmu(\bq')}  Q_\sigma(\tbq, d \bq')}
		\right) \right)
		\sqrt{\dmu(\tbq)}\pk_\sigma(\bq_0, d\tbq)}
	{\sigma^2\int\sqrt{\dmu(\bq')}\pk_\sigma(\bq_0, d \bq')}.
	\label{eq:loc:MTM:infty:IG}
\end{align}

The preconditioned case defined by \eqref{eq:Loc:pcN:algo} is
justified by a similar logic.  Here, following \cite{cotter2013mcmc},
we consider the preconditioned dynamic
\begin{align}\label{eq:precond:LA:dynam}
	d \bq + \frac{1}{2} (\bq + \cC\nabla \Pot(\bq)) dt =\sqrt{\cC} dW.
\end{align}
This dynamic has an infinitesimal generator of the form
\begin{align}\label{eq:inf:gen:pLangevin}
	\mathcal{L}\psi(\bq) =
	- \frac{1}{2} (\bq + \cC\nabla \Pot(\bq)) \cdot \nabla \psi
	+ \frac{1}{2} \mbox{Tr} (\cC D^2 \psi(\bq)).
\end{align}  

In order to discretize \eqref{eq:precond:LA:dynam} in time, we follow
\cite{cotter2013mcmc} and take a Crank-Nicolson approximation in
the linear part of the drift term and leave $\cC\nabla \Pot$ fully
explicit to obtain
\begin{align}
	\label{eq:pCN:Mala:disc}
	\bq := \bq_0 - \frac{\delta^2}{4}( \bq + \bq_0) -
  \frac{\delta^2}{2}  \cC\nabla \Pot(\bq_0) + \delta \sqrt{\cC} \xi
\end{align}
for a time step of size $\delta^2 > 0$. Thus, rearranging in
\eqref{eq:pCN:Mala:disc}
we obtain the $\infty$-MALA proposal kernel
\begin{align}    \label{eq:pCN:Mala:prop}
	\bar{Q}_\rho(\bq_0, d \tbq) =
	N(\rho \bq_0 - (1 - \rho) \cC\nabla \Pot(\bq_0), (1 - \rho^2)\cC),
	\quad
	\text{ where } \rho(\delta) := \frac{1 - \tfrac{\delta^2}{4}}{1 + \tfrac{\delta^2}{4}},
\end{align}
i.e. $\delta(\rho) = 2 \sqrt{\frac{1-\rho}{1+\rho}}$.
Hence, we find that\footnote{Formally,
	\begin{align*}
		\bar{Q}_\rho(\bq_0, d \tbq)
		&= Z^{-1} \exp\left( - \frac{1}{2 (1- \rho^2)} | \cC^{-1/2}(\tbq
		-  \rho \bq_0 + (1 - \rho) \cC\nabla \Pot(\bq_0))|^2  \right) d
		\tbq\\
		&= Z(\bq_0)^{-1}
		\exp\left( -\frac{1 - \rho}{1- \rho^2}
		\langle \cC^{1/2}\nabla \Pot(\bq_0),
		\cC^{-1/2}(\tbq-  \rho \bq_0) \rangle
		\right)
		\exp\left( - \frac{1}{2 (1- \rho^2)} | \cC^{-1/2}(\tbq
		-  \rho \bq_0)|^2 \right) d \tbq.
	\end{align*}
	However, if we want to consider the kernel
	\eqref{eq:pCN:Mala:prop} as an object defined on an infinite
	dimensional space, the identity \eqref{eq:inf:MALA:ABS:Cont} can be
	derived as a consequence of the Cameron-Martin Theorem
	\cite{DPZ2014}.}
\begin{align}\label{eq:inf:MALA:ABS:Cont}
	\bar{Q}_\rho(\bq_0, d \tbq)
	&= 
	\exp\left(- \frac{1}{1+ \rho}\left(
	\langle \nabla \Pot(\bq_0),
	(\tbq-  \rho \bq_0) \rangle
	+ \frac{1-\rho}{2} | \cC^{1/2} \nabla \Pot(\bq_0)|^2 \right)
	\right) \pk_\rho(\bq_0, d \tbq)
	\notag\\
	&= Z(\bq_0)^{-1}
	\exp\left(- \frac{1}{1 +\rho}
	\langle \nabla \Pot(\bq_0),
	(\tbq-  \rho \bq_0) \rangle 
	\right) \pk_\rho(\bq_0, d \tbq),
\end{align}  
where $\pk_\rho(\bq_0, d \tbq)$ is the pCN proposal kernel as in
\eqref{eq:Guass:1:prop:pCN}, and $Z(\bq_0)$ is a normalization
constant, namely
$$Z(\bq_0) = \int \exp\left(- \frac{1}{1 +\rho} \langle \nabla
  \Pot(\bq_0), (\tbq- \rho \bq_0) \rangle \right) \pk_\rho(\bq_0, d
\tbq).$$ Observe also that $Z(\bq_0)$ is the moment generating
function of $N(0, (1 - \rho^2) \cC)$ evaluated at
$- (1+ \rho)^{-1} \nabla \Pot (\bq_0),$ so that
\begin{align}\label{eq:Z:barQrho}
  Z(\bq_0) =
  \exp \left( \frac{1-\rho}{2 (1+\rho)} | \cC^{1/2} \nabla \Pot(\bq_0)|^2 \right).
\end{align}

As in the previous case, we can consider $\bar{Q}_\rho$ as a proposal
step in an unbiased algorithm as was advocated for in
\cite{cotter2013mcmc} as the $\infty$-MALA algorithm.  To compute
the associated accept-reject mechanism as found in
\cite{cotter2013mcmc, glatt2020accept}, we can apply
\cref{thm:grand:unifed:inv} with
\begin{align*}
	\mathcal{M}( d \bq_0, d \tbq)
	=  \bar{Q}_\rho(\bq_0, d \tbq)\mu(d \bq_0),
	\quad \text{ and } \quad S(\bq_0, \tbq) = ( \tbq, \bq_0).
\end{align*}
Using that $\frac{ dS^* \mathcal{M} }{d\mathcal{M}}$ maintains
\eqref{cond:alpha:rev} and that $\mu_0$ is reversible with respect
to $\pk_\rho$, we therefore obtain
\begin{align}\label{def:alpha:inf:MALA}
	\alpha(\tbq, \bq_0) = 1 \wedge \frac{ dS^* \mathcal{M} }{d\mathcal{M}}(\tbq, \bq_0) =
	1 \wedge \frac{\exp\left(- \Pot(\tbq) - \frac{1}{1+ \rho}\left(
		\langle \nabla \Pot(\tbq),
		\bq_0-  \rho \tbq \rangle  + \frac{1-\rho}{2} | \cC^{1/2} \nabla \Pot(\tbq)|^2 \right)\right)}
	{\exp\left(- \Pot(\bq_0) - \frac{1}{1+ \rho}\left(
		\langle \nabla \Pot(\bq_0),
		\tbq-  \rho \bq_0 \rangle  + \frac{1-\rho}{2} | \cC^{1/2} \nabla \Pot(\bq_0)|^2 \right)\right)}.
\end{align}

For $\alpha$ as in \eqref{def:alpha:inf:MALA}, we thus have the
$\infty$-MALA transition kernel, given by
\begin{align}
	R_\rho (\bq_0, d \tbq) =
	\alpha(\tbq, \bq_0)\bar{Q}_\rho(\bq_0, d \tbq)
	+ \delta_{\bq_0}(d \tbq)
	\int (1 -   \alpha(\bq', \bq_0))\bar{Q}_\rho(\bq_0, d \bq').
\end{align}

Now we compare the $\infty$-MALA proposal kernel $\bar{Q}_\rho$ and
the corresponding transition kernel $R_\rho$ with the $p=\infty$
kernels $\mk^\infty_\rho$ from \cref{meth:loc:prop:precond} in
\eqref{eq:pcn:loc:uncor:pinf:ker} and $\mk^\infty_{\rho,lMT}$ from
\cref{meth:loc:pCN} as following from \eqref{eq:MTM:inf:ker} with
\eqref{def:Qrhoinf} and \eqref{eq:pcn:loc:AR:pinf}, respectively.

Recall that for both \cref{meth:loc:prop:precond} and
\cref{meth:loc:pCN} (cf. \eqref{eq:Loc:pcN:algo},
\cref{tbl:beta:form}), we take $\beta$ as in \eqref{eq:loc:pCN:beta}
so that a first order Taylor expansion yields
\begin{align}\label{eq:beta:pcond:MTM:TE}
	\beta(\bq', \bbq) = \exp\left( - \frac{\rho}{1 + \rho} (\Pot(
	\bq'/\rho) - \Pot( \bbq))\right) =
	\exp\left(- \frac{1}{1+ \rho}
	\langle \nabla \Pot(\bbq),
	\bq'-  \rho \bbq \rangle + o(|\bq' - \rho\bbq|)
	\right).
\end{align}
Thus, when $\rho$ is close to $1$ (or equivalently when
$\delta^2$ is small), we obtain from \eqref{eq:beta:pcond:MTM:TE} and \eqref{eq:Z:barQrho} that
\begin{align}
	\int \beta(\bq', \bbq) \pk_\rho(\bbq, d \bq') 
	&\approx  
	\int \exp\left(- \frac{1}{1+ \rho}
	\langle \nabla \Pot(\bbq),
	\bq'-  \rho \bbq \rangle \right) \pk_\rho(\bbq, d \bq') 
	\notag\\
	&= \exp \left( \frac{1-\rho}{2 (1+\rho)} | \cC^{1/2} \nabla \Pot(\bbq)|^2 \right).
	\label{int:beta:Qrho}
\end{align}
Now observe that by ignoring the error term in
\eqref{eq:beta:pcond:MTM:TE} and plugging the resulting approximation
for $\beta$ together with \eqref{int:beta:Qrho} into the expressions
defining $\bar{\alpha}^\infty_{\rho}$ and $\pk^\infty_\rho$ in
\eqref{eq:pcn:loc:AR:pinf} and \eqref{def:Qrhoinf}, respectively, we
deduce under this formal argument that
$\bar{Q}_\rho \approx \mk^\infty_\rho$ and
$R_\rho \approx \mk^\infty_{\rho,lMT}$ for $\rho$ sufficiently close
to $1$.

 Regarding the infinitesimal generators for our preconditioned local
methods, the scaling is different from \eqref{eq:prop:kern:loc:classic:IG}
and \eqref{eq:loc:MTM:infty:IG}.  Here, from
\eqref{eq:pCN:Mala:disc}--\eqref{eq:pCN:Mala:prop} we have that
$\rho \in (0,1)$ corresponds to a time step
\begin{align}
  \label{eq:CN:ts}
\delta^2(\rho) = 4 \frac{1-\rho}{1+\rho}
\end{align}
in a (Crank-Nicolson)
discretization of \eqref{eq:precond:LA:dynam}.  Thus, we take
\begin{align}
  \label{eq:inf:gen:precond}
  \mathcal{L}_\rho \psi(\bq_0)
  = \frac{1+\rho}{4(1-\rho)} \int (\psi(\tbq) - \psi(\bq_0))
  \pk^\infty_\rho(\bq_0, d\tbq),
\end{align}  
and
\begin{align}
  \mathcal{L}_{\rho,lMT} \psi(\bq_0) &=
   \frac{1+\rho}{4(1-\rho)}\int (\psi(\tbq) - \psi(\bq_0))
  \mk^\infty_\rho(\bq_0, d\tbq)
  \notag\\
  &=  \frac{1+\rho}{4(1-\rho)} \int (\psi(\tbq) - \psi(\bq_0))
   \bar{\alpha}^\infty_{\rho}( \tbq, \bq_0)
  \pk^\infty_\rho(\bq_0, d\tbq)
  \notag\\
  &=\frac{1+\rho}{4(1-\rho)} \int (\psi(\tbq) - \psi(\bq_0))
  ( \bar{\alpha}^\infty_{\rho}( \tbq, \bq_0) -1)
  \pk^\infty_\rho(\bq_0, d\tbq) +   \mathcal{L}_\rho \psi(\bq_0),
     \label{eq:inf:gen:precond:MTM}
\end{align}  
as the infinitesimal generators at $\np = \infty$ corresponding to
\cref{meth:loc:prop:precond} and \cref{meth:loc:pCN}, respectively.

We can make the above formal discussions relating the MALA and
$\infty$-MALA algorithms rigorous in terms of their associated
infinitesimal generators as follows. In the statement below, we
consider the notation $C_b^2(\RR^d)$ (respectively $C_b^2(X)$) to
denote all real-valued functions on $\RR^d$ (respectively $X$) that
are twice continuously Fr\'echet differentiable and bounded with all
derivatives up to second order also bounded.
  
  \begin{Proposition}
    \label{Prop:alg:scalingatinfty:local}
   \begin{itemize}
   \item[(i)] Let $\mathcal{L}_\sigma$ and $\mathcal{L}_{\sigma, lMT}$
     be the infinitesimal generators corresponding to
     \cref{meth:loc:prop} and to \cref{meth:loc:MTM} as defined in
     \eqref{eq:prop:kern:loc:classic:IG} and
     \eqref{eq:loc:MTM:infty:IG}, respectively. We assume that the
     target density $\dmu$ is such that $\ln \dmu \in \cC_b^2(\RR^d)$.
     Then, each of these generators converge to the generator
     $\mathcal{L}$ for \eqref{eq:trad:LA:dynam} as $\sigma \to 0$, in
     the sense that
     \begin{align}
       \label{eq:inf:gen:zanella:conv}
       \lim_{\sigma \to 0}\mathcal{L}_\sigma \psi(\bq_0)
       = \lim_{\sigma \to 0} \mathcal{L}_{\sigma, lMT}\psi(\bq_0)
       = \frac{1}{2} \nabla \log \dmu(\bq_0) \cdot \nabla \psi(\bq_0)
            + \frac{1}{2}\Delta \psi(\bq_0),
     \end{align}
     for all $\bq_0 \in \RR^d$ and $\psi \in C^2_b( \RR^d)$.
   \item[(ii)] On the other hand, let $\mathcal{L}_\rho$ and
     $\mathcal{L}_{\rho, lMT}$ be the infinitesimal generators
     arising from \cref{meth:loc:prop:precond} and from
     \cref{meth:loc:pCN} (which corresponds to \cref{alg:MTMpCN:loc})
     as given in \eqref{eq:inf:gen:precond}
     and \eqref{eq:inf:gen:precond:MTM}. Assume that the potential function 
     $\Pot$ belongs to $C^2_b(X)$. In this case,
     $\mathcal{L}_\rho$ and  $\mathcal{L}_{\rho, lMT}$ converge
     to the infinitesimal generator 
     $\mathcal{L}$ for  \eqref{eq:precond:LA:dynam} as $\rho \to 1$,
     in the sense that
     \begin{align}
              \label{eq:inf:gen:prcond:conv}
            \lim_{\rho \to 1}\mathcal{L}_\rho \psi(\bq_0)
       = \lim_{\rho \to 1} \mathcal{L}_{\rho, lMT}\psi(\bq_0)
       =  - \frac{1}{2} (\bq_0 + \cC\nabla \Pot(\bq_0))\cdot \nabla \psi(\bq_0)
  + \frac{1}{2} \mbox{Tr} (\cC D^2\psi(\bq_0)),
     \end{align}
    for all $\bq_0 \in X$ and over all test functions $\psi \in C_b^2(X)$.
  \end{itemize}
\end{Proposition}

In \cref{app:proof:of:local:scaling}, we provide the complete details
on the proof of the preconditioned case, namely item (ii) from
\cref{Prop:alg:scalingatinfty:local}. The proof of the
finite-dimensional case in item (i) follows with analogous arguments
and is thus omitted.

\subsection{Convergence rates}
\label{sec:bounds}

We turn next to more qualitative results concerning the convergence
\eqref{def:mk:infty}.  Here we show that, under more restrictive
conditions on $\art$, we obtain a limit to $P^\infty$,
\eqref{eq:p:inf:w:lim:conv:barker}, with a rate $O(\sqrt{p})$.
Notably, we establish this convergence in the stronger sense of the
total variation distance.

Let us recall that, given two probability measures $\mu$ and $\nu$ on
a measurable space $(\qsp, \Sigma_\qsp)$, the total variation distance
between $\mu$ and $\nu$ is defined as
\begin{align}\label{eq:TV:dist:def}
	\| \mu - \nu\|_{TV}  =
  \frac{1}{2}\sup_{ \phi: \qsp \to [-1, 1]}
  \left|\int \phi(\bbq) \mu( d \bbq)  - \int \phi(\bbq) \nu(d \bbq) \right|
	= \sup_{A \in \Sigma_\qsp} | \mu(A) - \nu(A)|,
\end{align}
where the first supremum ranges over all measurable functions
$\phi: \qsp \to [-1,1]$.

\begin{Theorem}\label{thm:tv:bound}
  Under the conditions of \cref{thm:large:p:lim} consider the kernels
  $P = P^\np$ as in \eqref{eq:gen:bk:ker} and its weak
  $\np \to \infty$ limit $P = P^\infty$ given in
  \eqref{eq:gen:bk:ker}.  In addition to the assumptions of
  \cref{thm:large:p:lim} we furthermore suppose that, for a given
  $\bq_0 \in \qsp$,
  \begin{align}\label{eq:bart:inf:cond}
    \bar{\art}(\bq_0)
    = \inf_{\bbq \in X} \int_{\qsp} \art(\bq', \bq_0) \pk(\bbq, d \bq')
    > 0,
    \quad\text{ with }\quad \sup_{\bbq \in X} \int_{\qsp} \art(\bq',
    \bq_0) \pk(\bbq, d \bq') < \infty
  \end{align}
  and that
  \begin{align}\label{eq:bart:2nd:mom:cond}
    \hat{\art}(\bq_0)
    = \int_{\qsp}  \left(\int_{\qsp} 
    \left(\art(\bq_1, \bq_0) -\int_{\qsp} \art(\bq', \bq_0) \pk(\bbq, d \bq')\right)^2
    \pk(\bbq, d \bq_1)\right)^{1/2}
    \bpk(\bq_0, d \bbq) < \infty.
  \end{align}
  Then, for any such $\bq_0$,
  \begin{align}\label{eq:ss:tv:bound}
		\| \mk^\np(\bq_0, d \tbq) - P^\infty(\bq_0, d \tbq)\|_{TV} \leq 
		\frac{ \art(\bq_0,\bq_0)}{\art(\bq_0,\bq_0) + p \bar{\art}(\bq_0)}
		+
		\frac{\hat{\art}(\bq_0)}{ 2\sqrt{p} \bar{\art}(\bq_0)},
  \end{align}
  and moreover, for any probability measure $\nu$,
  \begin{align}\label{eq:ss:tv:bound:msr}
   \biggl\| \int_{\qsp}  P^\np(\bq, d \tbq) \nu(d\bq)
    - \int_{\qsp} P^\infty(\bq,d \tbq)  \nu(d\bq) \biggr\|_{TV}\leq
    \int_{\qsp}\left(\frac{ \art(\bq,\bq)}
    {\art(\bq,\bq)+ p \bar{\art}(\bq)}
    +\frac{\hat{\art}(\bq)}{ 2\sqrt{p}\bar{\art}(\bq)}\right)
                                              \nu(d \bq).
  \end{align}
\end{Theorem}

\begin{Remark}\label{rmk:cond:TV:conv:ind:prop}
  In the particular case when the outer kernel $\bpk(\bq_0, \cdot)$ is
  the Dirac measure $\delta_{\bq_0}$, as it occurs for the algorithms
  discussed in \cref{sec:cond:prop}, the conditions in
  \eqref{eq:bart:inf:cond} simplify to
	\begin{align}\label{eq:bart:inf:cond:ip}
		 0 < \bar{\art}(\bq_0)
		= \int_{\qsp} \art(\bq', \bq_0) \pk(\bq_0, d \bq')
		< \infty,
	\end{align}
whereas \eqref{eq:bart:2nd:mom:cond} becomes
\begin{align}\label{eq:bart:2nd:mom:cond:ip}
	\hat{\art}(\bq_0)
	= \left(\int_{\qsp} 
  \left(\art(\bq_1, \bq_0) -\int_{\qsp}
  \art(\bq', \bq_0) \pk(\bbq, d \bq')\right)^2
	\pk(\bq_0, d \bq_1)\right)^{1/2} < \infty.
\end{align}
\end{Remark}

To prove \cref{thm:tv:bound} we demonstrate how to control the
interaction terms in the denominator of the acceptance terms in
\eqref{eq:gen:bk:ker}.  This leads to an expression that more
explicitly exhibits the formation of $\mk^\infty$, for large $\np$.
To address these denominator terms we note that, for large $p$, and
any fixed $\bbq \in \qsp$,
\begin{align}
  \label{eq:LLN:sep:heurst}
  \sum_{k=0}^\np \art(\bq_k, \bq_0)
  \approx \art(\bq_0, \bq_0) + p \int_{\qsp} \art(\tbq, \bq_0)
  \pk(\bbq, d \tbq)\;
  \text{ when } \{\bq_k\}_{k=1}^d \text{ are i.i.d. samples from }
  \pk(\bbq, d \tbq),
\end{align}
by the Law of Large numbers.   Replacing $\sum_{k=0}^\np \art(\bq_k,
\bq_0)$ by this mean we find after some computations carried out
below, \eqref{int:mk:np:varphi:1}, that
\begin{align}
  \mk^\np&(\bq_0, d \tbq) - P^\infty(\bq_0, d \tbq)
  \notag\\
  =& - \int_{\qsp} \frac{\art(\bq_0, \bq_0)}
    {\art(\bq_0, \bq_0) + p\int_{\qsp} \art(\bq', \bq_0) \pk(\bbq, d \bq')} 
     \frac{\art(\tbq, \bq_0)}{\int_{\qsp} \art(\bq',
       \bq_0) \pk(\bbq, d \bq')}
       \pk(\bbq, d \tbq) \bpk(\bq_0, d \bbq)
       \notag\\
    &+\delta_{\bq_0}(d \tbq)  \int_{\qsp} \frac{\art(\bq_0, \bq_0)}
      {\art(\bq_0, \bq_0)
      + p \int_{\qsp} \art(\bq', \bq_0) \pk(\bbq, d \bq')}\bpk(\bq_0, d \bbq) 
        \label{eq:diff:Ppinf}\\
         &- \sum_{j=0}^\np \int_{\qsp}
           \int_{\qsp^\np}   \frac{\delta_{\bq_j}(d \tbq)\art(\bq_j, \bq_0)}
    {\sum_{k=0}^\np \art(\bq_k, \bq_0)} \frac{\tfrac{1}{p}
    \sum_{k=1}^\np (\art(\bq_k, \bq_0) -\int_{\qsp} \art(\bq', \bq_0) \pk(\bbq, d \bq'))}
      {p^{-1}\art(\bq_0, \bq_0) + \int_{\qsp} \art(\bq', \bq_0)
      \pk(\bbq, d \bq')} \prod_{l=1}^\np \pk(\bbq, d \bq_l) \bpk(\bq_0, d \bbq).
	\notag
\end{align}
One may directly compare the first term on the right in
\eqref{eq:diff:Ppinf} (or in \eqref{int:mk:np:varph:0}) to
\eqref{def:mk:infty}.  In any event, the first two terms in this
expression \eqref{eq:diff:Ppinf} yield the $O(p^{-1})$ first term in
\eqref{eq:ss:tv:bound}.   Here, an interesting and illustrative special 
case appears when we consider \eqref{eq:gen:bk:ker} with $\art \equiv 1$.
In this case, we find
\begin{align}
 \mk^\np(\bq_0, d \tbq) 
  = \frac{p}{1+p} \int_{\qsp} 
       \pk(\bbq, d \tbq) \bpk(\bq_0, d \bbq)
    + \frac{\delta_{\bq_0}(d \tbq)}{p+1}, 
    \quad
     \mk^\infty(\bq_0, d \tbq)  =  \int_{\qsp} 
       \pk(\bbq, d \tbq) \bpk(\bq_0, d \bbq).
\end{align}
Note that this situation arises for example in
\cref{subsubsec:bubble:bath} and in \cref{meth:tj:mPCN} with the
target measure $\mu = \mu_0$ in \eqref{eq:stuart:form}.  The last term
in \eqref{eq:diff:Ppinf} is responsible for the $O(p^{-1/2})$ rate
in \eqref{eq:ss:tv:bound}. This again seems to be an optimal rate in
view of the mean convergence rate in the Law of Large Numbers.

The complete proof of \cref{thm:tv:bound} is reserved for
\cref{app:bounds:proofs}.  We investigate the implications of this
bound, \eqref{eq:ss:tv:bound:msr}, for the particular case of the
Slingshot Algorithm, \cref{subsubsec:slingshot}, below in
\cref{sec:The:Slingshot}.

\subsection{Rao-Blackwellization}
\label{sec:RB::1step:Limit}

We turn next to establish some large $p$ results for the
Rao-Blackwellization, \eqref{eq:RB:form}, corresponding to the
Barker-type kernel \eqref{eq:gen:bk:ker}.  Here we show that variance
of one step of this estimator decays as $O(p^{-1})$ under quite
general conditions on $\art$.

\begin{Theorem}\label{thm:RB:1Step}
  For any measurable, bounded observable $\phi: \qsp \to \RR$ consider
  the one step ($N =1$) Rao-Blackwell type estimator,
  \eqref{eq:RB:form} defined relative to the kernel
  \eqref{eq:gen:bk:ker} so that, for any $\bq_0 \in \qsp$,
  \begin{align}\label{eq:RB:est:gen:bark}
    \check{\phi}_p(\bq_0) = \sum_{j = 0}^p
    \frac{\art(\bq_j, \bq_0)}{\sum_{l=0}^\np \art(\bq_l, \bq_0)}
    \phi(\bq_j),
    \text{ with } (\bq_1, \ldots, \bq_\np) \sim \vk(\bq_0, d \bq_1,
    \ldots, d \bq_\np),
  \end{align}
  with $(\bq_1, \ldots, \bq_\np)$ drawn from $\vk$ as in
  \eqref{def:vk:Tjelmeland} so that $\bbq_0 \sim \bpk (\bq_0, d \tbq)$
  and
  $(\bq_1, \ldots, \bq_\np) \sim \prod_{j=1}^p \pk(\bbq_0, d \bq_j)
  \sim \vk(\bq_0, d \bq_1, \ldots, d \bq_\np)$.  Assume that, 
  \begin{align*}
    \EE[\art(\bq_1, \bq_0)^2]< \infty \quad \text{ and } \quad
     \kappa_{\art} := \EE\left(\frac{\Var (\art(\bq_1, \bq_0)| \bbq_0)}
  {\EE (\art(\bq_1, \bq_0)|\bbq_0)^2} \right) < \infty,
  \end{align*}
  for any $\bq_0 \in \qsp$.  Then, in this situation,
  \begin{align}
    \frac{1}{2}&\Var\left( \frac{\EE  (  \art(\bq_1, \bq_0) \phi(\bq_1)|\bbq_0)
        +p^{-1}\art(\bq_0,\bq_0) \phi(\bq_0) }
        {\EE  (  \art(\bq_1, \bq_0) |\bbq_0)+ p^{-1}\art(\bq_0,\bq_0)} \right)-
                 \frac{\| \phi\|_\infty^2}{\np}   \kappa_{\art}
                   \leq  \Var(\check{\phi}_p(\bq_0))
                 \notag\\
               &\leq \frac{4\| \phi\|_\infty^2}{\np}
                 (\kappa_{\art}+1)
                             + 2\Var\left( \frac{\EE  (  \art(\bq_1, \bq_0) \phi(\bq_1)|\bbq_0)
        +p^{-1}\art(\bq_0,\bq_0) \phi(\bq_0) }
        {\EE  (  \art(\bq_1, \bq_0) |\bbq_0)+ p^{-1}\art(\bq_0,\bq_0)} \right),
       \label{eq:rb:upper:lower}
  \end{align}
  so that
  \begin{align*}
        \frac{1}{2} \Var\!\left( \frac{\EE  (  \art(\bq_1, \bq_0) \phi(\bq_1)|\bbq_0)}
   {\EE  (  \art(\bq_1, \bq_0) |\bbq_0)} \right)
\leq
    \liminf_{p \to \infty}\Var(\check{\phi}_p(\bq_0))\leq
    \limsup_{p \to \infty} \Var(\check{\phi}_p(\bq_0))\leq
    2\Var \!\left( \frac{\EE  (  \art(\bq_1, \bq_0) \phi(\bq_1)|\bbq_0)}
   {\EE  (  \art(\bq_1, \bq_0) |\bbq_0)} \right).
  \end{align*}
  In particular, if $\bpk (\bq_0, d \tbq) = \delta_{\bq_0}(d \tbq)$ then
  \eqref{eq:rb:upper:lower} reduces to 
  \begin{align}\label{eq:no:convolution:RB:est}
\Var(\check{\phi}_p(\bq_0))
    \leq \frac{4\| \phi\|_\infty^2}{\np}
    \left(\frac{\Var (\art(\bq_1, \bq_0))}
  {[\EE (\art(\bq_1, \bq_0))]^2} +1\right),
  \end{align}
  so that in this case $\lim_{p \to \infty}\Var(\check{\phi}_p(\bq_0)) = 0$.
\end{Theorem}

The upper and lower bound \eqref{eq:rb:upper:lower} make essential use
of the conditional variance identity \eqref{eq:cond:var:def}.  Indeed,
to understand the origin of the upper and lower bounds, observe that,
with the \eqref{eq:RB:var:decomp} identity and the conditionally
independent structure of $\vk$, \eqref{def:vk:Tjelmeland},
\begin{align}
  \Var\left( \sum_{j = 0}^p \art(\bq_j,
        \bq_0)\phi(\bq_j)\right)
  &= 
  \EE\left( \Var\left(\sum_{j = 0}^p \art(\bq_j, \bq_0)\phi(\bq_j) |  \bbq_0\right)\right)
        + \Var\left(\EE\left(\sum_{j = 0}^p \art(\bq_j,
    \bq_0)\phi(\bq_j)|  \bbq_0\right)\right)
    \notag\\
   &=p
  \EE\left( \Var\left(\art(\bq_1, \bq_0)\phi(\bq_1) |  \bbq_0\right)\right)
     +p^2 
     \Var\left(\EE\left( \art(\bq_1, \bq_0)\phi(\bq_1)|  \bbq_0\right)\right).
\end{align}
Handling the normalization term in \eqref{eq:RB:est:gen:bark}
analogously to \eqref{eq:LLN:sep:heurst}, \eqref{eq:diff:Ppinf} and
employing further repeated conditioning arguments we obtain
\eqref{eq:rb:upper:lower}.  The detailed proof is provided below in
\cref{eq:RB:Var:Estimate}.

\begin{Remark}\label{rmk:RB:ss:over:tj}
  The bounds \eqref{eq:rb:upper:lower},
  \eqref{eq:no:convolution:RB:est} suggest an advantage of
  \cref{subsubsec:bubble:bath} and \cref{subsubsec:slingshot} over
  \cref{meth:tj:mPCN}.  Indeed, in the case of the first two
  approaches, $\vk$ has the form \eqref{def:vk:before:Tjelmeland} so
  that there is no preliminary step in the proposal structure and the
  variance of the Rao-Blackwellized estimator is $O(p^{-1})$. On the
  other hand regarding the unbiased algorithms \cref{meth:tj:mPCN} we see
  that the Variance of such one step estimators do not generically
  decay.
\end{Remark}

\section{Degeneracy in the case of HMC sample path selection}
\label{sec:HMC:path:sel}

As we said, the methods introduced in the previous two sections,
\cref{sec:multi:prop:algs}, \cref{sec:large:p:limit}, can be viewed as
multiproposal versions of the RWM algorithm and its many variants and
generalizations.  On the other hand, it is also interesting to
consider multiproposal implementations of the Hamiltonian Monte Carlo
(HMC) approach to sampling, \cite{duane1987hybrid, neal2011mcmc}.
Actually there are a number of non-trivial and non-equivalent ways to
introduce multiproposal structure into the HMC paradigm.  Here we
focus on the case where the proposal cloud represent the position
coordinates of different steps along the (numerical) solution path of
the Hamiltonian dynamic associated to the target measure of interest.
This approach has been previously considered in
\cite{calderhead2014general, glatt2024parallel} and here,
unfortunately, our main take away is that such schemes should be
avoided in practice. We mention some other possible approaches to
multiproposal HMC sampling not covered in our analysis herein as
an outlook for future work below in \cref{sec:Outlook}.

\subsection{General State Space Formulation}
We start by recalling some elements of the HMC approach within the
general state space formalism, \cref{sec:MP:MCMC}, for our purposes
here as follows; see e.g. \cite{neal2011mcmc, bou2018geometric,
  glatt2020accept, glatt2024parallel, glatt2024sacred} for further
general background.  Note that, for simplicity, we restrict the
presentation here to the `finite-dimensional case' where our target
measure $\mu$ has a $C^1$ density supported on $\RR^n$.  Nevertheless
many of the results presented here and below in \cref{sec:large:p:HMC}
can be easily adapted to other situations; particularly the
Hilbert-space (or preconditioned dimension-free) setting introduced in
\cite{Beskosetal2011} and considered in e.g. \cite{glatt2020accept,
  glatt2021mixing}.

To proceed, we set $\qsp = \vsp = \RR^\dm$, for some $\dm \in \NN$, and
consider a target measure of the form
$\mu(d \bq) = Z^{-1} e^{-\Pot(\bq)} d \bq$, with
$Z = \int_{\RR^\dm} e^{-\Pot(\bq)} d \bq$, for some potential function
$\Pot: \RR^\dm \to \RR$ with $\Pot \in C^1(\RR^\dm)$ and
$\bq \mapsto e^{-\Pot(\bq)} \in L^1(\RR^\dm)$. To sample from $\mu$
via an HMC-type algorithm, the idea consists in selecting a suitable
Hamiltonian function $\Ham: \RR^{2 \dm} \to \RR$ for which the
associated Gibbs measure
\begin{align}\label{def:cM:hmc}
  \cM(d \bq, d \bv) = \frac{1}{Z_\Ham}
  e^{-\Ham (\bq, \bv)}  d\bq \, d \bv,
  \quad Z_\Ham = \int_{\RR^{2\dm}} e^{-\Ham(\bq, \bv)}
  d \bq\, d \bv,
\end{align}
has as its marginal with respect to the ``position'' variable $\bq$
the target measure $\mu$. As such, the Hamiltonian function may be
written in the following generic form
\begin{align}\label{def:ham:hmc}
	\Ham(\bq, \bv) = \Pot(\bq) + \VPot(\bq, \bv) 
\end{align}
for some measurable function $\VPot: \RR^{2\dm} \to \RR$ where we
assume additionally that $\VPot \in C^1$ and the map
\begin{align}
  \label{eq:mom:marg:condition}
  \bq \mapsto \int_{\RR^{\dm}} e^{-\VPot(\bq, \bv)} d \bv
  \text{ is constant},
\end{align}
so that $\cM(d \bq, d \bv)$ indeed maintains $\mu$ as its momentum
marginal. We then consider the associated Hamiltonian dynamic in the
``position'' and ``momentum'' variables $\bq$ and $\bv$, respectively,
as
\begin{align}\label{hmc:dyn}
  \frac{d }{dt} (\bq, \bv)
  = J^{-1} \nabla \Ham (\bq, \bv),
  \quad (\bq(0), \bv(0)) = (\bq_0, \bv_0), 
\end{align}
where $J$ is a $2 \dm \times 2 \dm$ real matrix that is invertible and
antisymmetric and we assume that $\Pot$ and $\VPot$, and hence $\Ham$,
are sufficiently regular so that \eqref{hmc:dyn} admits a unique and
globally defined classical solution for any initial datum
$(\bq_0, \bv_0)$.  In addition to the above we typically impose
that $\VPot$ is symmetric in $\bv$, i.e.
\begin{align}\label{VPot:sym:v}
  \VPot(\bq, \bv) = \VPot(\bq, - \bv) \quad
  \mbox{for all } (\bq, \bv) \in \RR^{2\dm}.
\end{align}
and that
\begin{align}
  R J^{-1} R = J^{-1} \text{ where }
†  R =
  \begin{pmatrix}
    I & 0\\
    0 & -I
  \end{pmatrix}.
\end{align}  

Under these various conditions on $\Pot$ and $\VPot$, the dynamics
defined by \eqref{hmc:dyn} preserves the measure $\cM$ in
\eqref{def:cM:hmc} as an invariant, which implies that its marginal in
$\bq$ is also invariant under the projection of this dynamic in $\bq$.
To see this and as prelude to the formulation of various HMC schemes,
observe that for each $t \in \RR$, we may define the solution map
\begin{align}\label{def:sol:hmc}
  \hS_t: \RR^{2\dm} \to \RR^{2\dm},
  \quad \hS_t(\bq_0, \bv_0) = (\bq(t), \bv(t)),
\end{align}
where $(\bq(t), \bv(t))$ obeys \eqref{hmc:dyn} at each time
$t \in \RR$ and emanates from the initial pair $(\bq_0, \bv_0)$.  By
direct computation, we see that $\{\hS_t \}_{t \in \RR}$ preserves
$\Ham$ as an invariant, i.e.
\begin{align}\label{Ham:inv:hSt}
  \Ham(\hS_t(\bq,\bv)) = \Ham(\bq, \bv)
  \quad \mbox{for all } (\bq, \bv) \in \RR^{2\dm}, \,\, t \in \RR.
\end{align}
Furthermore, $\hS_t$ is symplectic, namely we have that
\begin{align}\label{Ham:simpl}
  \nabla\hS_t(\bq,\bv)^* J  \nabla\hS_t(\bq,\bv)= J
  \quad \mbox{ over each } (\bq, \bv) \in \RR^{2\dm}, \,\, t \in \RR,
\end{align}
a condition which immediately implies that
\begin{align}\label{Ham:vol}
  |\det \nabla \hS_t(\bq,\bv)| = 1, \quad \text{ for }
  (\bq, \bv) \in \RR^{2\dm}, \,\, t \in \RR.
\end{align}

Hence, with these two properties
\eqref{Ham:inv:hSt}, \eqref{Ham:vol}, we infer
\begin{align}
  \hS_t^* \cM(d\bq, d \bv) = \frac{1}{Z_\Ham}
  e^{-\Ham (\hS_t^{-1}(\bq, \bv))}  |\det \nabla \hS_t^{-1}(\bq, \bv)| d\bq d \bv
  = \frac{1}{Z_\Ham}
  e^{-\Ham (\hS_{-t}(\bq, \bv))}  d\bq d \bv
  = \cM(d\bq, d \bv).
\end{align}  
Finally we note that $\hS_t$ maintains the involutive structure
\begin{align}\label{eq:Invol:map:cont}
  R\circ\hS_t \circ   R\circ\hS_t(\bq,\bv) = (\bq,\bv)
  \quad \mbox{ for every } (\bq, \bv) \in \RR^{2\dm}, t \in \RR.
\end{align}  
This `reversibility' in the dynamics \eqref{hmc:dyn} corresponds to
the required involutive structure for our theory,
\cref{thm:grand:unifed:inv}.

From the above idealization, we derive the following practical strategy
for sampling: from the current state $\bq_0$ draw $\bv_0$ from the
momentum marginal
\begin{align}\label{def:cM:vk:hmc}
   \bv_0 \sim \vk(\bq, d \bv) := e^{-\VPot(\bq, \bv)} d \bv,
\end{align}
of $\cM$ \eqref{def:cM:hmc} and then propose a new state $\bq$ by
numerically integrating \eqref{hmc:dyn} from the starting point
$(\bq_0, \bv_0)$ up to a time $T > 0$ yielding $(\bq(T), \bv(T))$ and
taking $\bq := \bq(T)$. Here, to fix notations for our numerical
integration of \eqref{hmc:dyn} we take, for a given time step size
$\dt > 0$, a generic numerical integrator map
$\Xi_\dt: \RR^{2 \dm} \to \RR^{2\dm}$ and let $\Xi_\dt^k$ denote the
$k$-fold composition of $\Xi_\dt$ with itself.  Thus, for a time of
integration $T > 0$ and number of integration steps $\np$, we set
\begin{align}
  \hS^\np_T (\bq_0, \bv_0) :=\Xi_{T/\np}^\np (\bq_0, \bv_0),
  \label{eq:numer:int:HAM}
\end{align}  
as our approximation of $\hS_T$.  Thus, with these notations the
proposals corresponding to a numerical scheme $\Xi$ integrated
up to time $T >0$ in $\np$ steps is given as
\begin{align}\label{prop:HMC}
  \Pi_1 \hS^\np_T (\bq_0, \bv_0)  = \Pi_1 R \hS^\np_T (\bq_0, \bv_0)
  \quad \text{ where } \bv_0 \sim \vk(\bq, d \bv)
\end{align}
where  we recall that $\Pi_1: \RR^{2\dm} \to \RR^\dm$ is the projection
map onto the $\bq$-variable, namely $\Pi_1(\bq, \bv) = \bq$.

We notice that, if we select $\Xi$ such that, cf. \eqref{Ham:vol},
\eqref{eq:Invol:map:cont}
\begin{align}\label{eq:vol:HMC:prev:step}
     |\det \nabla \Xi_\dt (\bq,\bv)| = 1, \quad \text{ for }
  (\bq, \bv) \in \RR^{2\dm},
\end{align}  
and 
\begin{align}\label{eq:Inv:HMC:prev:step}
    R\circ \Xi_\dt \circ   R\circ \Xi_\dt (\bq,\bv) = (\bq,\bv)
  \quad \mbox{ for every } (\bq, \bv) \in \RR^{2\dm},
\end{align}
for every $\dt > 0$, then $S = R \hS^\np_T$ defines an involution
and we have
\begin{align}\label{eq:AR:HMC:classic}
  \frac{d S^*\cM}{d\cM}(\bq_0,\bv_0)
  = \exp( - \Ham(S(\bq_0, \bv_0)) + \Ham(\bq_0, \bv_0))
  = \exp( - \Ham(\hS^\np_T (\bq_0, \bv_0)) + \Ham(\bq_0, \bv_0)).
\end{align}  
Hence, in this situation, according to \cref{thm:grand:unifed:inv},
\eqref{prop:HMC} with an acceptance probability defined in terms of
\eqref{eq:AR:HMC:classic} yields a $\mu$ unbiased scheme.  In other
words the kernel
\begin{align}\label{def:mk:gen:hmc}
  \mk_T(\bq, d \tbq)
  = &\int_{\RR^\dm}
    \left[ 1 \wedge e^{ \Ham(\bq, \bv) - \Ham(R 
    \hS_T^p(\bq, \bv))}  \right]
  \delta_{\Pi_1 \hS^\np_{T}(\bq, \bv)} (d \tbq)
  e^{- \VPot(\bq,
      \bv)} d \bv
  \notag\\
    &+ \delta_{\bq}(d \tbq) \int_{\RR^\dm}
      \left[1- 1 \wedge e^{ \Ham(\bq, \bv) - \Ham(R 
    \hS_T^p(\bq, \bv))}  \right]
    e^{- \VPot(\bq, \bv)} d \bv,
\end{align}
is reversible with respect to $\mu$.

Note that typically $\Ham$ in \eqref{def:ham:hmc} has a `separable'
form where $\VPot(\bq, \bv)$ is $\bq$ independent.  Here, in this
situation we can typically select $\Xi_\dt$ according to the leap-frog
splitting scheme
\begin{align}
  \Xi_\dt = \Xi_{\dt/2}^{(P)} \circ \Xi_\dt^{(M)}\circ \Xi_{\dt/2}^{(P)}
  \; \text{ where } \; \Xi_{\dt}^{(P)}(\bq,\bv) := (\bq + \dt \nabla
  \VPot(\bv), \bv) \quad
  \Xi_{\dt}^{(M)}(\bq,\bv) := (\bq, \bv - \dt \nabla \Pot(\bq)).
\end{align}
This choice yields an integrator which maintains
\eqref{eq:vol:HMC:prev:step} and \eqref{eq:Inv:HMC:prev:step}, as can
easily be verified by direct computation.  Moreover if $\VPot$ is
quadratic in $\bv$ the resulting $\vk$ is Gaussian and therefore is easy
and cheap to draw from.  Nevertheless, our results below,
\cref{thm:large:p:MH:hmc} do not make any unbiasedness requirements as
in \eqref{Ham:vol}, \eqref{eq:Invol:map:cont}.  Nor do we impose any
separability requirements so that we include the case of a position
dependent kinetic energy which is considered in more advanced HMC
methods; see \cite{girolami2011riemann}.

\subsection{Multiproposal Formulations}
We turn now to define the type of multiproposal HMC algorithms
previously considered in \cite{calderhead2014general,
  glatt2024parallel}.  As in the single proposal case, we fix total
integration time $T > 0$.  Our multiproposal parameter $\np \geq 1$ is
the desired number of (uniform) time steps so that consider a time
step size $\dt = T/\np$.  We then set
\begin{align}\label{def:Sj:hmc}
  S_0 = I, \quad
  S_j = R \circ \hS^\np_{jT/\np} = R \circ \Xi_{T/\np}^j,
  \quad j = 1, \ldots, \np,
\end{align}
where, as above, $R(\bq, \bv) = (\bq, - \bv)$.  Then, given a current
state $\bq_0 \in \qsp$ and $\bv_0 \sim \vk(\bq_0, \cdot)$, $\np$
proposals are generated by `integrating along the steps of the
solution path' as
$\bq_j = \Pi_1 \hS^\np_{jT/\np}(\bq_0, \bv_0) = \Pi_1 \Xi_{T/\np}^j
(\bq_0, \bv_0)$, $j=1, \ldots, \np$.

With the proposal structure in hand, \eqref{def:Sj:hmc}, we can then
consider two different distinct approaches which correspond to our
acceptance probabilities which take a Metropolis-type,
\eqref{def:alphaj:metr:2} and Barker-type, \eqref{def:alphaj:Barker}
form respectively.
Regarding the acceptance probabilities $\ar_0, \ldots, \ar_\np$, we
first consider the following Metropolis-Hastings-type definition:
\begin{gather}
  \ar_j(\bq, \bv) =
  a_j
  \left[ 1 \wedge \exp \left( \Ham(\bq, \bv)
      - \Ham(S_j(\bq, \bv)) \right) \right],
  \quad j = 1, \ldots, \np, \label{def:ar:mh:hmc:1}\\
  \ar_0(\bq, \bv) = 1 - \sum_{j=1}^\np \ar_j(\bq,
  \bv), \label{def:ar:mh:hmc:2}
\end{gather}
with $\Ham$ as defined in \eqref{def:ham:hmc} and $a_j \geq 0$ such
that $\sum_{j =1}^\np a_j \leq 1$.  This situation falls under the
scope of \cref{rmk:MH:degenerate}.  Thus, under the choices of $\vk$,
$(S_0, \ldots, S_\np)$, and $(\ar_0, \ldots,\ar_\np)$ given by
\eqref{def:cM:vk:hmc}, \eqref{def:Sj:hmc}, and
\eqref{def:ar:mh:hmc:1}-\eqref{def:ar:mh:hmc:2}, respectively, we fix
the following notation for the resulting multiproposal kernel
$\mk^{(\vk,\bS, \balpha)}$ from \eqref{def:mp:ker} is actually
just the mixture kernel
\begin{align}
  \label{def:mk:mh:hmc}
  \mkmh(\bq, d \tbq) = \sum_{j =1}^\np  a_j \mk_{Tj/\np}(\bq, d \tbq) +
  (1 - \sum_{j=1}^\np  a_j )  \delta_{\bq}(d \tbq)
\end{align}
where $\mk_{Tj/\np}$ are the single proposal HMC kernel integrated out
to a time $Tj/\np$, \eqref{def:mk:gen:hmc}.  Thus we see that this is
just the random integration time HMC approach advocated in
e.g. \cite{mackenze1989improved,neal2011mcmc,
  betancourt2015fundamental, bourabee2017randomized} where the
distribution of random integration times are dictated by the weights
$a_j$.

Next, we consider the Barker-type acceptance probabilities 
\begin{align}\label{eq:Barker:HMC}
  \ar_j(\bq, \bv )
  = \frac{\exp(- \Ham(S_j(\bq, \bv)))}
  {\sum_{k=0}^\np \exp(-\Ham(S_k(\bq,\bv)))},
  \quad j = 0, \ldots, \np,
\end{align}
where again $\Ham$ and $S_j$ are as defined in \eqref{def:ham:hmc} and
\eqref{def:Sj:hmc}, respectively.  By complementing these choices with
the kernel $\vk$ again as in \eqref{def:cM:vk:hmc}, we then obtain the
following multiproposal kernel as a special case of
\eqref{def:mp:ker}:
\begin{align}\label{def:mkB:hmc}
	\mkB(\bq, d \tbq) = 
  \sum_{j=0}^\np \int_{\RR^\dm}
  \frac{\exp(- \Ham(S_j(\bq, \bv)))}{\sum_{k=0}^\np \exp(-\Ham(S_k(\bq,\bv)))}
  \delta_{\Pi_1 \hS^\np_{jT/\np}(\bq, \bv)} (d \tbq) e^{- \VPot(\bq, \bv)} d \bv.
\end{align}
Notice that \eqref{eq:Barker:HMC} entails accepting along proposal
path with probabilities proportional to the lowest relative
`Hamiltonian energy'.  Here, in contrast to
\cref{subsubsec:bubble:bath} or \cref{meth:tj:mPCN}, it is not clear
that this acceptance \eqref{eq:Barker:HMC} has a particular bias
towards high probability regions of the target measure $\mu$; due to
the influence of the kinetic energy term $\Psi$ in $\Ham$, $\ar_j$ is
not proportional to $\exp(-\Pot(\Pi_1
S_j(\bq,\bv)))$. 

It is also worth noting that, in contrast to \eqref{def:mk:mh:hmc}
there is no reason to believe that \eqref{def:mkB:hmc} corresponds to
an unbiased method for finite $\np$, even if
\eqref{eq:vol:HMC:prev:step}, \eqref{eq:Inv:HMC:prev:step} so that
\eqref{eq:AR:HMC:classic}.  Indeed, in this situation \eqref{sum:cond}
amounts to requiring that, cf. \eqref{eq:numer:int:HAM},
\eqref{eq:AR:HMC:classic},
\begin{align}
  \sum_{k =1}^\np
  \exp( - \Ham(\Xi_{T/\np}^{j+k}(\bq_0, \bv_0)))
  =   \sum_{k =1}^\np
  \exp( - \Ham(\Xi_{T/\np}^{k} (\bq_0, \bv_0)))
  \quad 
  \text{ for every } j =1, \ldots, k.
\end{align}
This is a very restrictive condition and should not be expected to
hold outside of very special situations.

\subsection{Large proposal limit for solution path multiproposal HMC
  kernels}
\label{sec:large:p:HMC}

We finally turn to consider the $\np \to \infty$ asymptotic for the
different multiproposal HMC kernels $\mkmh$, $\mkB$ defined in
\eqref{def:mk:mh:hmc}, \eqref{def:mkB:hmc} respectively.  Although the
former is just a simple mixture and the latter is expected to be biased,
the fact that they both approach the same limit (when $a_j = 1/p$ in
\eqref{def:ar:mh:hmc:1}) as $\np \to \infty$ fills out the picture.
See also \cref{rmk:HMC:ass:unbiased}, \cref{rmk:HMC:ass:weight} after
the statement of \cref{thm:large:p:MH:hmc}.

To proceed, we first set down an appropriate notion of numerical
consistency.
\begin{Definition}\label{def:num:int:hmc}
  Fix $T > 0$ and $\np \geq 1$. Then, we define for any $t \in [0,T]$
  the map $\hS_t^\np: \RR^{2\dm} \to \RR^{2\dm}$ such that
  \begin{align}\label{def:num:int:hmc:nodes}
    \hS^\np_t(\bq_0, \bv_0) = \Xi_{T/\np}^j (\bq_0, \bv_0)
    \quad \mbox{for } t = jT/\np, \quad j = 1, \ldots, \np,
  \end{align}
  and, for all other values of $t \in [0,T]$,
  $\hS_t^\np(\bq_0, \bv_0)$ is any interpolation of the values in
  \eqref{def:num:int:hmc:nodes} such that $\hS_t^\np$ is a measurable
  function.
\end{Definition}

With this in hand our large $\np$ regime result is precisely
formulated as follows
\begin{Theorem}\label{thm:large:p:MH:hmc}
  Fix $T > 0$, and let $\Ham$ be as in \eqref{def:ham:hmc}, with
  $\Pot, \VPot$ satisfying the following conditions:
  $\Pot, \VPot \in C^1$,
  $\bq \mapsto e^{-\Pot(\bq)} \in L^1(\RR^\dm)$,
  $\int_{\RR^\dm} e^{-\VPot(\bq, \bv)} d \bv$ does not depend on
  $\bq \in \RR^\dm$, and \eqref{VPot:sym:v}. Assume additionally that
  $\Pot$ and $\VPot$ are sufficiently regular so that \eqref{hmc:dyn}
  admits a unique and globally defined classical solution, for each
  $(\bq_0, \bv_0) \in \RR^{2\dm}$, so that the solution map $\hS_t$ in
  \eqref{def:sol:hmc} is well-defined, for all $t \in [0,T]$.
  Moreover, for each $\np \geq 1$ and $t \in [0,T]$, let $\hS_t^\np$
  be as in \cref{def:num:int:hmc}. Suppose that
  \begin{align}\label{ass:hSt:hSnpt}
          \lim_{\np \to \infty} \sup_{t \in [0,T]}
          \| \hS_t (\bq, \bv) - \hS^\np_t (\bq, \bv)\| = 0
          \quad \mbox{for any } \,\, (\bq, \bv) \in \RR^{2 \dm},
  \end{align}
  where $\| \cdot\|$ denotes the standard Euclidean norm in
  $\RR^{2\dm}$.  Then, for every $\bq \in \RR^\dm$, the limits
  \begin{align}\label{lim:mkmh:mkB}
    \mkmhinf (\bq, d \tbq) = \lim_{\np \to \infty} \mkmh
    (\bq, d \tbq) = \lim_{\np \to \infty} \mkB (\bq, d \tbq)
  \end{align}
  exist in the sense of weak convergence, and we have
  \begin{align}\label{def:mkmhinf}
          \mkmhinf (\bq, d \tbq)
          = \frac{1}{T} \int_{\RR^\dm}
          \int_0^T \delta_{\Pi_1 \hS_t (\bq, \bv)} (d \tbq) e^{-\VPot(\bq, \bv)} dt \, d \bv.
  \end{align}
  Here $\mkmh$ is as in \eqref{def:mk:mh:hmc} where we take
  $a_j = 1/\np$ for all $j$, and $\mkB$ is taken according to
  \eqref{def:mkB:hmc}.
\end{Theorem}
\noindent The proof of this result is given in
\cref{app:proof:large:p:MH:hmc}.

\begin{Remark}\label{rmk:HMC:ass:unbiased}
  Note that our result, \eqref{thm:large:p:MH:hmc} does not require
  \eqref{eq:vol:HMC:prev:step}, \eqref{eq:Inv:HMC:prev:step}.  Thus,
  the numerical consistency condition, \eqref{ass:hSt:hSnpt} is all
  that is required for $\mkmh$ and $\mkB$ to be asymptotically
  unbiased.
\end{Remark}

\begin{Remark}\label{rmk:HMC:ass:weight}
  As in  \eqref{def:ar:mh:hmc:1} different weights $a_j^\np$ can be chosen
  to define $\mkmh$ at each finite $\np$.    In this more general
  situation suppose that
  \begin{align}
     \nu^{\np}(dt) = \sum_{j =1}^p a_j \delta_{T (j/\np)}(dt)
  \end{align}
  and we have that $\nu^{\np} \to \nu$ as $\np \to \infty$, in the sense of weak
  convergence. Then, taking $\mkmh$ to be the corresponding kernel
  defined as \eqref{def:mk:mh:hmc}, 
  \begin{align}
    \lim_{\np \to \infty} \mkmh  (\bq, d \tbq)
       = \int_{\RR^\dm}
          \int_0^T \delta_{\Pi_1 \hS_t (\bq, \bv)} (d \tbq)
    e^{-\VPot(\bq, \bv)}  \nu(dt) \, d \bv.
  \end{align}  
\end{Remark}  

\section{The Slingshot Algorithm}
\label{sec:The:Slingshot}

We turn in this final section to some further study of the Slingshot
algorithm introduced as \cref{alg:slingshot:intro},
\cref{subsubsec:slingshot} above.  \cref{sec:numexp:kernel} considers
a Gaussian proposal kernel with a state dependent covariance to
optimize the rate of convergence to the large proposal limit in some
basic examples.  We then take up the efficacy of the Slingshot against
other MCMC algorithms in \cref{sec:numexp:mcmc}.  Here we provide some
promising initial empirical results, particularly when we employ an
adaptive proposal variance. Code and data are available at
\url{https://github.com/andrewjholbrook/infiniteProposals}.

\subsection{Optimizing Convergence Rates in the Large Proposal
  Limit}
\label{sec:numexp:kernel}

One notable property of this class of methods is that we can introduce
a state dependence to the proposal variance without introducing bias
at the $\np = \infty$ limit.  As we will see, appropriately optimizing
over possible variance structures leads to a faster convergence of the
one step proposal kernel to the target measure in some representative
examples.

For this aim, we revisit the total variation upper bound from 
\cref{thm:tv:bound} in our particular context here. In this situation,
according to \eqref{eq:ar:SS:gen}, we have
$\art(\tbq, \bq_0) = \dmu(\tbq) \dpk(\bq_0,\tbq)^{-1}$. Thus, in view
of \cref{rmk:cond:TV:conv:ind:prop}, we note that the expressions
\eqref{eq:bart:inf:cond:ip} and \eqref{eq:bart:2nd:mom:cond:ip} reduce
to
\begin{align}
  \label{eq:tv:consta:trans}
  \bar{\art}(\bq_0) \equiv 1,
  \quad
  \hat{\art}(\bq_0) = \sqrt{\int \dmu(\tbq)^2 \dpk(\bq_0,\tbq)^{-1} \mu_0(d \tbq) - 1 },
\end{align}  
so that \eqref{eq:ss:tv:bound} reduces to
\begin{align}\label{eq:ss:tv:bound:red}
  \| \mk^\np(\bq_0, d \tbq) - P^\infty(\bq_0, d \tbq)\|_{TV} \leq
  \frac{\dmu(\bq_0) }{\dmu(\bq_0)  + p \dpk(\bq_0,\bq_0)}
  + \frac{1}{2} \sqrt{ \frac{\int \dmu(\tbq)^2 \dpk(\bq_0, \tbq)^{-1} \mu_0(d \tbq) - 1 }{p}}.
\end{align}

\subsubsection{Scaling Results in the Gaussian Case}
\label{sec:bounds:gauss}
We first apply \eqref{eq:ss:tv:bound:red} to the case where the target
distribution and the proposal distribution are both Gaussians in $d$
dimensions. In particular, taking $\mu_0$ as the Lebesgue measure in
$\qsp = \RR^d$, we consider spherical Gaussian proposals
\begin{align}\label{eq:sphere:g:ker}
  \dpk (\bq_0, \tbq)
  = (2\pi \sigma_\dpk^2(\bq_0))^{-d/2}
  \exp\left(  -\frac{1}{2 \sigma_\dpk^2(\bq_0)} |\bq_0 - \tbq|^2\right),
\end{align}
and a centered, spherical Gaussian target with density
\begin{align*}
	\dps(\bq)
  = (2\pi \sigma_\dps^2)^{-d/2}
  \exp\left( - \frac{1}{2 \sigma_\dps^2} |\bq|^2\right).
\end{align*}

In view of \eqref{eq:bart:2nd:mom:cond:ip} and
\eqref{eq:tv:consta:trans}, we compute
\begin{align*}
  \int &\dps(\tbq)^2 \dpk(\bq_0, \tbq)^{-1} d \tbq  
	= 
   \frac{(2\pi \sigma_\dpk^2(\bq_0))^{d/2}}{(2\pi \sigma_\dps^2)^{d}}
   \int \exp\left(- \frac{1}{\sigma_\dps^2} |\tbq|^2
   + \frac{1}{2 \sigma_\dpk^2(\bq_0)} |\bq_0 - \tbq|^2\right) d \tbq
   \nonumber \\
	&= \begin{cases}
		\exp\left( |\bq_0|^2 / (2\sigma_\dpk^2(\bq_0) - \sigma_\pi^2)\right)
		\sigma_\dpk(\bq_0)^{2d}\sigma_\dps^{-d} (2\sigma_\dpk^2(\bq_0) - \sigma_\pi^2)^{-d/2}
                &\text{ when }  2\sigma_\dpk^2(\bq_0) - \sigma_\pi^2 > 0, \\
		\infty &\text{ otherwise.}
              \end{cases}
\end{align*}
Thus, satisfying \eqref{eq:bart:2nd:mom:cond:ip} requires that
\begin{align}\label{eq:ss:conv:cond:2:gauss}
  \sigma_\dpk(\bq_0) > \frac{1}{\sqrt{2}}\sigma_\dps, \quad
  \text{ for every } \bq_0 \in \RR^d.
\end{align}
When this condition is met, applying \cref{thm:tv:bound} via
\eqref{eq:ss:tv:bound:red}, we find
\begin{align}\label{eq:ss:tv:bound:gauss}
	\| &P^\np(\bq_0, d \tbq) - \pi(\tbq) d \tbq\|_{TV} \\
	& \quad\leq 
	\frac{1}{\np \sigma_\dps^d \sigma_\dpk^{-d}(\bq_0)\exp\left( |\bq_0|^2 / 2 \sigma_\dps^2 \right) + 1}
	+
	\frac{1}{\np^{1/2}} \left[ \exp\left( |\bq_0|^2 / (2\sigma_\dpk^2(\bq_0) - \sigma_\pi^2) \right)
   \frac{\sigma_\dpk(\bq_0)^{2d}}
   {\sigma_\dps^{d} (2\sigma_\dpk^2(\bq_0) - \sigma_\pi^2)^{d/2}} -1 \right]^{1/2}.
   \notag
\end{align}

Note that by minimizing the second (order $\mathcal{O}(\np^{-1/2})$)
term in \eqref{eq:ss:tv:bound:gauss} with respect to the proposal
standard deviation $\sigma_\dpk$, for fixed $\bq_0$, we can derive the
corresponding $\sigma_\dpk(\bq_0)$ that yields the kernel closest to
the target density:
\begin{align}\label{eq:ss:gauss:propstd}
	{\sigma_\dpk^\star}^2 (\bq_0)
  = \frac{3}{4}\sigma_\dps^2
  + \frac{1}{2}|\bq_0|^2 d^{-1}
  +\frac{1}{4}\sqrt{ \sigma_\dps^4
  +12\sigma_\dps^2|\bq_0|^2 d^{-1}
  +4|\bq_0|^4 d^{-2}}.
\end{align}
Note that $\sigma_\dpk^\star \approx \sigma_\dps$ when
$\sigma_\dps \gg |\bq_0|/\sqrt{d}$ -- i.e., when the current sample is
close to the center of the target distribution, the proposal variance
should be similar to the target variance -- and that the relationship
is exact when $|\bq_0|=0$, the center of the target. On the other hand,
$\sigma_\dpk^\star \approx |\bq_0|/\sqrt{d}$ when
$|\bq_0|/\sqrt{d} \gg \sigma_\dps$; that is, when the current sample is
far from the center of the target distribution, the proposal
distribution should be aggressive enough to return to the bulk of the
target. In \cref{sec:numexp:kernel}, we further explore how
$\sigma_\dpk$ can be tuned to more efficiently approximate the
target. 

We explore the empirical convergence of the slingshot kernel to the
target distribution as the number of proposals $\np$ grows
large. \cref{fig:tune:kernel:oneDexample} shows a one-dimensional
Gaussian target and Gaussian proposals, as examined in
\cref{sec:bounds:gauss}. The target distribution is taken to be the
standard normal distribution, while the starting sample is 4, far in
the tail of the target. The left plot shows a case where the proposal
kernel is too narrow, so many proposals are required to converge to
the target distribution. The right plot shows the case where the
proposal standard deviation is taken to be the optimal value from
\eqref{eq:ss:gauss:propstd}; as a result, convergence is achieved with
orders of magnitude fewer proposals.

\begin{figure}[!t]
	\centering
	\includegraphics[width=0.9\textwidth]{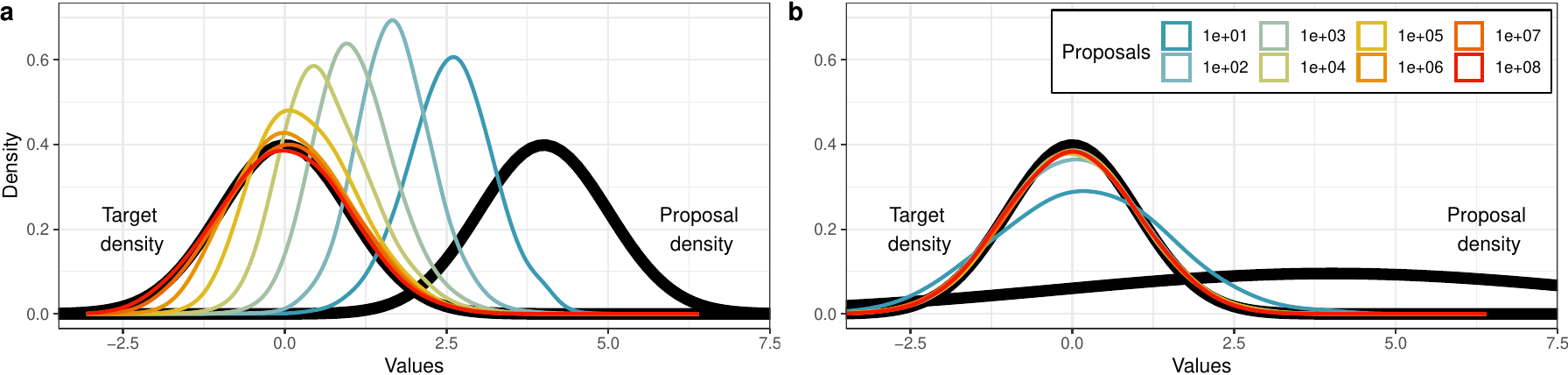}
	\caption{As the number of proposals $p$ grows large, a single
          iteration of the slingshot kernel converges to the target
          distribution. For each proposal count setting, we generate
          10,000 independent samples from the slingshot kernel with
          Gaussian proposal distributions (\textbf{a}) N$(4,1^2)$ and
          (\textbf{b}) N$(4, {\sigma_\dpk^\star}^2 )$ for a standard
          normal target. For faster convergence in $p$, we use
          \eqref{eq:ss:gauss:propstd} to obtain the optimal proposal
          standard deviation ${\sigma_\dpk^\star}=4.18$.  }
	\label{fig:tune:kernel:oneDexample}
\end{figure}

\subsubsection{Non-Gaussian Targets}
\label{sec:ban:target:tune}

To investigate a more complicated example, we consider the ``tilted
banana'' distribution given by
\begin{align}\label{eq:tiltedbanana}
  \dps(\bq) = \frac{1}{Z} \exp\left[-a q_1^2 - c \bq_1
  - \frac{1}{2}(q_2 + B q_1^2 - b B)^2 \right],
\end{align}
where $a = 0.005, b=100, c=0.05, B=0.1$. The density of this
distribution is shown in the left panel of \cref{fig:tune:tiltbanana:app} below. We measure
convergence of the slingshot kernel $P^{p, \sigma_f}$, with $p$ proposals and
Gaussian proposal kernel with variance $\sigma_f$, to the target via 
a loss function defined to be the sum of
the relative errors in the first six moments of this target
distribution $\pi$ as
\begin{align}\label{eq:tiltedbanana:loss}
  \mathcal{L}(\sigma_f,p)
  = \sum_{k=1}^6
  \frac{|\int |\tbq|^k P^{p, \sigma_f}
  (\bq_0, d \tbq)  - \int |\tbq|^k \pi(\tbq) d \tbq|}
  {\int |\tbq|^k \pi(\tbq) d \tbq}.
\end{align}
The ``true'' moments of $\pi$ were approximated numerically via increasingly
refining a grid until the values converged; the normalization constant
$Z$ was also estimated in this fashion.

The middle and right panels of \cref{fig:tune:tiltbanana:app} show the
convergence of the slingshot kernel originating at the origin to the
target distribution, as measured by the loss
\eqref{eq:tiltedbanana:loss}, by number of proposals $\np$ and
proposal standard deviation $\sigma_\dpk$. In the middle plot, we see
that additional proposals provide very little value when $\sigma_\dpk$
is poorly tuned; the convergence is much faster for appropriately
tuned $\sigma_\dpk$. The plot on the right shows the same data with
$\sigma_\dpk$ along the $x$-axis; here we see that that the window for
the optimal $\sigma_\dpk$ seems to grow wider as $\np$ increases, but
also that the optimal $\sigma_\dpk$ depends on $\np$. This aligns with
intuition that some of the benefit of additional proposals should be
used to make them more aggressive; however, we note that the
derivation of \eqref{eq:ss:gauss:propstd} for the tuning proposal
standard deviation for Gaussian distributions does not depend on
$\np$, indicating that there may be room for improving the argument to
select a more optimal value.

\noindent
\textbf{A Tuning Heuristic}

\begin{figure}[htbp]
	\centering
	\includegraphics[width=\textwidth]{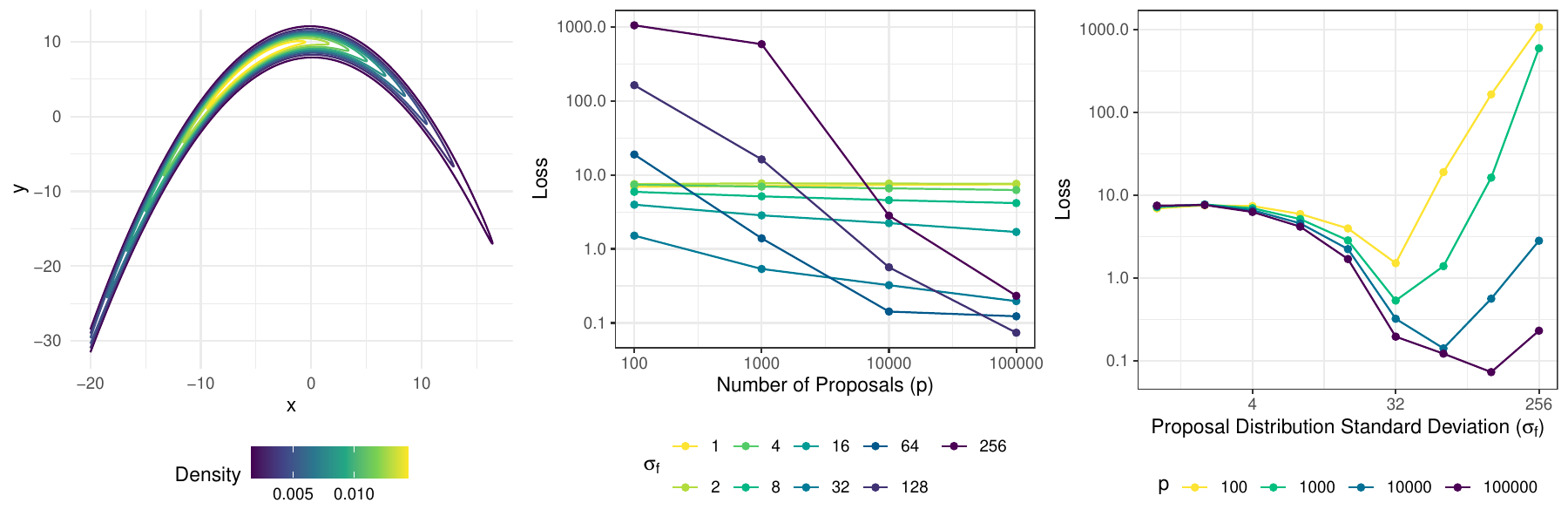}
	\caption{Left: Density of the tilted banana distribution given
          in \eqref{eq:tiltedbanana}. Middle and Right: Tuning
          $\sigma_\dpk$ for the tilted banana distribution. Each point
          represents a sample of 100,000 redraws from the slingshot
          kernel centered at the origin. Left: Loss by number of
          proposals $\np$. Right: Loss by proposal standard deviation
          $\sigma_\dpk$. Loss decreases uniformly with $\np$, though
          benefits are limited unless $\sigma_\dpk$ is tuned
          appropriately.}
	\label{fig:tune:tiltbanana:app}
\end{figure}

As described in the proof of \cref{thm:tv:bound} given in
\cref{app:bounds:proofs}, a key term in the error in total variation
norm between the slingshot kernel and the target is bounded by the
difference between the mean of the proposal weights
\begin{align}\label{eq:alpha:bar}
  \bar{\ar}_\np
  := \frac{1}{\np} \sum_{l=1}^\np\frac{\pi(\bq_l)}{\dpk(\bq_l,\bq_0)}
\end{align}
and its expected value (1 for normalized posteriors).  See the third
term in \eqref{int:mk:np:varphi:1} noting that
$\bar{Q}(\bq_0, d \tbq) = \delta_{\bq_0}(d \tbq)$.  Here in view of
\eqref{eq:t3:rate:TV:bnd} we expect that $\bar{\ar}_\np \to 1$ as
$\np \to \infty$; choosing a $\sigma_\dpk$ that yields a faster rate of convergence as $\np \to \infty$ in
this term might improve the overall rate of convergence to the
straight Monte Carlo sampling of our target $P^\infty(\bq_0, \tbq)
=\mu(d\tbq)$.

\begin{figure}[htbp]
	\centering
	\includegraphics[width=\textwidth]{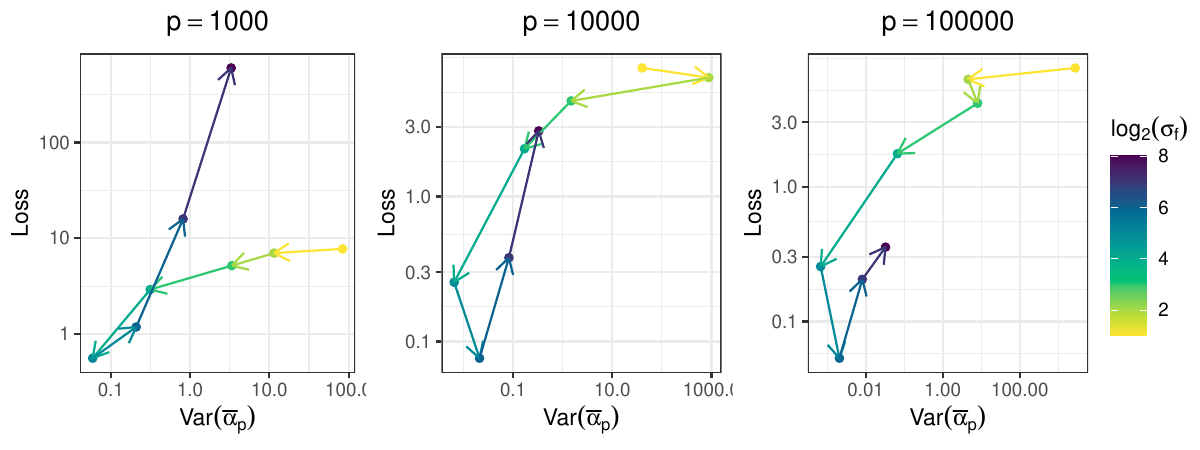}
	\caption{Approximation of the tilted banana distribution by
          the slingshot kernel, by number of proposals $\np$. Each
          point represents a sample of 100,000 redraws from the
          slingshot kernel centered at the origin. The $x$-axis shows
          the variance of $\bar{\ar}_\np$ from \eqref{eq:alpha:bar}
          and the $y$-axis shows the loss, while arrows show
          increasing $\sigma_\dpk$. All plots show that accurate
          approximation of the target is associated with low
          $\Var(\bar{\ar}_\np)$.  }
	\label{fig:tune:kernel:tiltbanana:var:approx}
\end{figure}

To test this hypothesis, we could tune $\sigma_\dpk$ until, for
example, $\left| \bar{\ar}_\np - Z \right|^2$ is small. However, for
unnormalized targets, $Z$ will typically be unknown. A more general
proxy would be to measure the variance of $\bar{\ar}_\np$ across
multiple draws of $\np$ proposals -- a metric that, while potentially
computationally expensive, is generally computable across a broad
class of proposals. \cref{fig:tune:kernel:tiltbanana:var:approx}
compares the variance of $\bar{\ar}_\np$ across redraws from the
slingshot kernel against the resulting loss, for different numbers of
proposals $\np$ and different proposal standard deviations
$\sigma_\dpk$. We see that low loss values -- that is, strong
agreement between the slingshot kernel and the target distribution --
correspond to low variance of $\bar{\ar}_\np$, i.e., a consistent
clustering around its mean value. This therefore provides a
potentially powerful tuning heuristic for tuning $\sigma_\dpk$ on
general target distributions, at the expense of having to evaluate
$\bar{\ar}_\np$ for multiple sets of proposals. It also provides
empirical evidence that error in the approximation
$\bar{\ar}_\np \approx \EE \beta = Z$ is a key bottleneck for
convergence of the Slingshot kernel to the target distribution as
$\np$ grows large.

\subsection{MCMC Case Studies}
\label{sec:numexp:mcmc}

We next explore some of the properties of the slingshot sampler \cref{alg:slingshot:intro}
in three separate case studies.

\subsubsection{The Effect of the Proposal Kernel on MCMC Convergence}
\label{sec:numexp:mcmc:proposal}

In our first study, we investigate the tuning of the proposal
distribution in \cref{alg:slingshot:intro} to achieve improved MCMC
convergence. In particular, we will increase or decrease the proposal
standard deviation adaptively to target a given acceptance rate,
defined for multiproposal MCMC as selecting one of the $\np$ new
proposals rather than keeping the current sample for another iteration
\cite{holbrook2023generating}. We made these proposal distribution
adjustments by multiplying the precision matrix by the ratio between
the target and measured acceptance rates, clamped to
$[\frac{1}{2},2]$; these adjustments were made at intervals of
exponentially increasing length. As an example target distribution, we
again use the tilted banana distribution given in
\eqref{eq:tiltedbanana}. \cref{fig:tune:tiltbanana} shows the results
of a series of 200,000-sample slingshot MCMC chains for varying
numbers of proposals and target acceptance rates. The left plot shows
the proposal standard deviation $\sigma$ adaptively selected for each
target acceptance rate; smaller, less aggressive values are of course
chosen as the target acceptance rate goes up, while larger $\np$
allows more aggressive proposals across a range of target acceptance
rates. The middle plot shows that the target acceptance rates were
largely achieved by the given choices of $\sigma$; the exceptions were
when $\np$ was very small, in which case the Barker acceptance
formulation places a limit on how high the acceptance rate can go.
Finally, the right plot shows the loss by acceptance rate; the loss
decreases as $\np$ increases and, for sufficiently large $\np$, the
convergence is roughly uniform across a wide range of target
acceptance rates. \cref{fig:tune:triMix} in the appendix shows a
similar set of results for a target distribution made up of a mixture
of Gaussians.

\subsubsection{Off-the-Shelf Parallelization}
\label{sec:numexp:mcmc:conv}

Exploitation of MP-MCMC structure via parallelization is explored in
\cite{holbrook2023quantum,glatt2024parallel,lin2023more}. Here, we
focus on easy, off-the-shelf parallelization of the slingshot
algorithm with the likes of \textsc{TensorFlow}
\cite{abadi2016tensorflow}. \cref{fig:parallel} demonstrates the
natural parallelization of the slingshot algorithm and the ability to
efficiently scale the algorithm to massive proposal counts with
relatively accessible GPUs.  First, with $p=100{,}000$ Gaussian
proposals scaled according to \eqref{eq:ss:gauss:propstd}, we use a
single GPU to speed up proposals and target evaluations, and compare
to CPU implementations for multivariate Gaussian targets of increasing
dimensions.  We see that our simple GPU implementations are hundreds
of times faster than the CPU implementations, with relative speedups
increasing with target dimension.  Next, we demonstrate how additional
simultaneous parallelization across the $M$ components of a uniform
mixture distribution
$\frac{1}{M}\sum_{m=0}^{M-1}$N$_2(10m\boldsymbol{1},\boldsymbol{I})$
facilitates even more proposals when within a fixed timing budget. A
GPU-powered chain is able to use 100,000 proposals, while a
CPU-powered chain only uses 100. In this experiment, both GPU and CPU
setups adapt proposal standard deviations targeting a 50\% acceptance
rate. Both take the same amount of time per iteration but achieve
different effective sample sizes (ESS) after 10,000 iterations.  We
see that GPU-based parallelization becomes necessary for as few as
$M=50$ mixture components and barely suffers for the massively
multimodal target with $M=1{,}000$ modes.

\subsubsection{Empirical Convergence}

While theoretical results indicate that the slingshot algorithm
generates samples from the target as $\np\rightarrow \infty$,
\cref{fig:fourPlots} describes empirical experiments that demonstrate
accuracy for moderate proposal counts ($\np \leq 512$), albeit for
relatively simple target distributions: the 2-dimensional normal
N$_2(\boldsymbol{0},\boldsymbol{I})$, the uniform mixture of
2-dimensional normals
$\frac{1}{2}$N$_2(\boldsymbol{0},\boldsymbol{I})+\frac{1}{2}$N$_2(\boldsymbol{5},\boldsymbol{I})$,
the 2-dimensional banana distribution \eqref{eq:tiltedbanana}, with
$a=1/2,b=1,c=0,B=1$, and the 4-dimensional normal distribution
N$_4(\boldsymbol{0},\boldsymbol{I})$. To isolate the effect of
proposal count, we use the same proposal variance regardless of
proposal count for each target. All ESSs are more than $1{,}000$
regardless of proposal count, so minimal Monte Carlo error biases
results.  In general, convergence as a function of proposal count $p$
appears slowest for the highly-nonlinear banana distribution, followed
by the mixture distribution, the 4-dimensional normal and the
2-dimensional normal, in order. That said, all distributions
demonstrate the same positive trend as proposals increase.

\subsubsection{Select Comparisons}

Finally, we are interested in the slingshot algorithm's empirical
performance in comparison to a select collection of
non-gradient-informed algorithms, including random walk Metropolis and
a number of unbiased MP-MCMC algorithms.  \cref{tab:results} describes
an experiment that compares these algorithms' mixing speeds and
estimator accuracies while increasing target dimensionality and
keeping slingshot's proposal count fixed at $1{,}000$.  Despite this
moderate proposal count, the biased slingshot's 1st and 2nd moment
estimators surprisingly outperform those of its unbiased competitors.
That said, increasing target dimensionality with $\np$ fixed leads to
a breakdown in estimator accuracy, pointing to the need for a future
theoretical investigation of the $d,p\rightarrow \infty$ joint limit.
In addition to providing accurate estimators, the slingshot
outperforms its competitors in terms of mixing speed as measured by
both ESS per iteration and ESS per second.

\begin{figure}[!t]
	\centering
	\includegraphics[width=\textwidth]{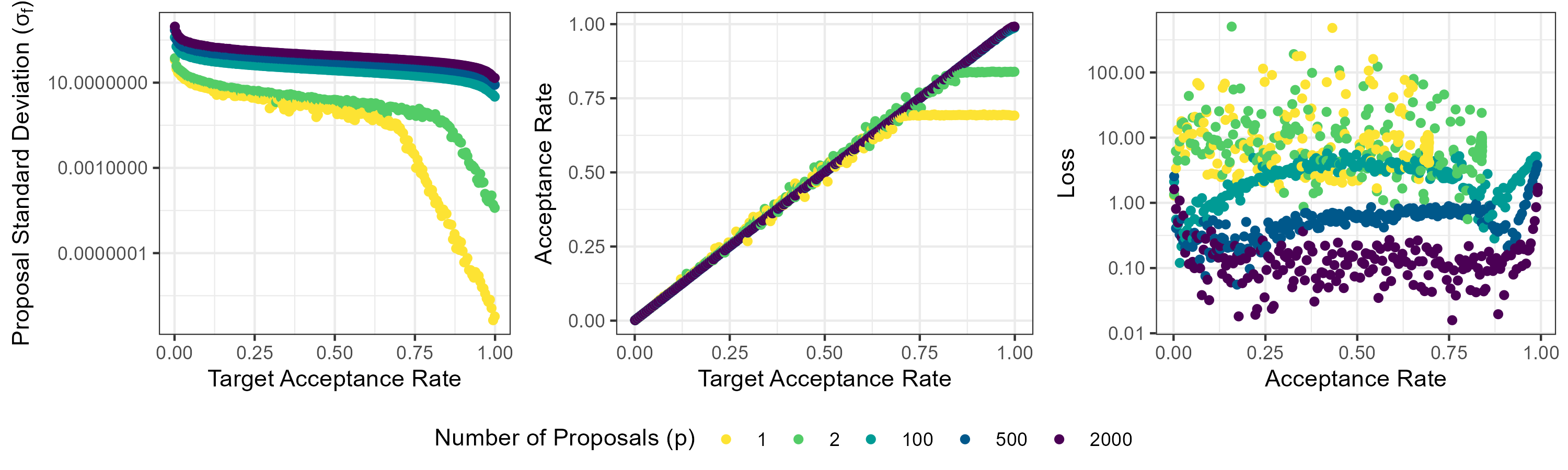}
	\caption{Tuning $\sigma_\dpk$ for the tilted banana
          distribution, by number of proposals $\np$. Each point
          represents a chain of 200,000 MCMC samples with
          $\sigma_\dpk$ adaptively tuned to a given target acceptance
          rate. Left: The choice of $\sigma_\dpk$ by target acceptance
          rate. Middle: The acceptance rate by target acceptance rate,
          showing that the target acceptance rate was achieved up to
          the limits of the Barker formulation. Right: Loss by
          acceptance rate, showing accurate approximation across a
          broad range of acceptance rates.}
	\label{fig:tune:tiltbanana}
\end{figure}

\begin{figure}[htbp]
	\centering
	\includegraphics[width=0.7\textwidth]{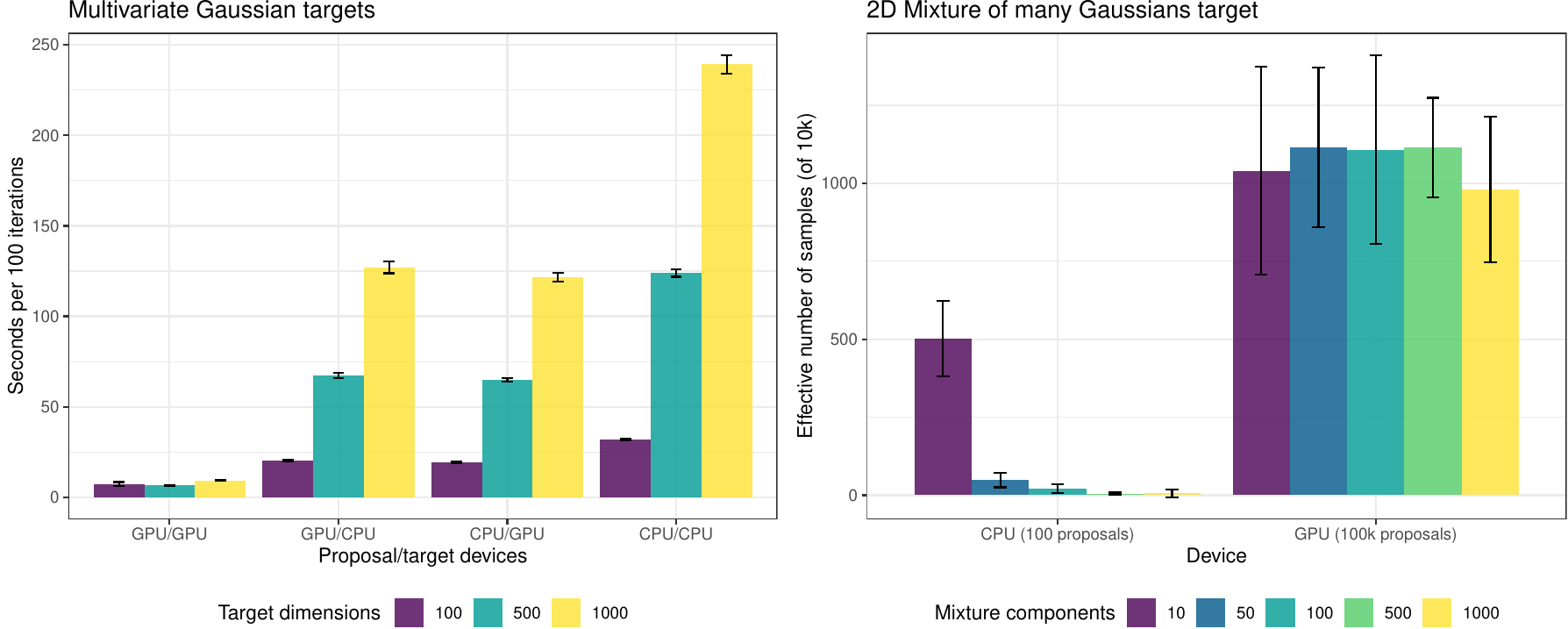}
	\caption{Thanks to its simplified proposal and acceptance
          structures, the slingshot allows for easy off-the-shelf
          parallelization with, e.g., \textsc{TensorFlow}
          \cite{abadi2016tensorflow} and an Nvidia T4 GPU, enabling
          the use of larger numbers of proposals.  Left: with
          $p=100{,}000$ Gaussian proposals scaled according to
          \eqref{eq:ss:gauss:propstd}, we use a single GPU to speed up
          proposal and/or target evaluation steps and compare to CPU
          implementation for standard multivariate Gaussian targets
          with varying dimensions. Right: parallelization across the
          $M$ components of the uniform mixture distribution
          $\frac{1}{M}\sum_{m=0}^{M-1}$N$_2(10m\boldsymbol{1},\boldsymbol{I})$
          facilitates more proposals when within a fixed timing
          budget. A GPU-powered chain uses 100,000 proposals, a
          CPU-powered chain uses 100, and both setups adapt proposal
          standard deviations targeting a 50\% acceptance rate. Both
          take the same amount of time per iteration but achieve
          different effective sample sizes after 10,000 iterations. }
	\label{fig:parallel}
\end{figure}


\begin{figure}[!t]
	\centering
	\includegraphics[width=0.9\textwidth]{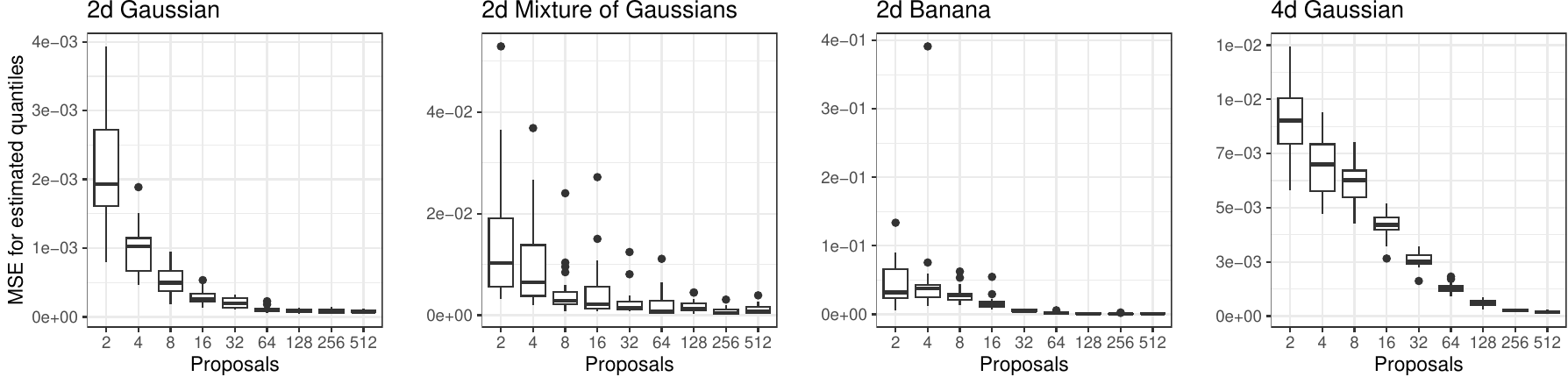}
	\caption{As the number of proposals grows moderately large,
          MCMC samples obtained using the slingshot kernel converge to
          target distributions: N$_2(\boldsymbol{0},\boldsymbol{I})$;
          uniform mixture of normals
          $\frac{1}{2}$N$_2(\boldsymbol{0},\boldsymbol{I})+\frac{1}{2}$N$_2(\boldsymbol{5},\boldsymbol{I})$;
          2-dimensional banana distribution \eqref{eq:tiltedbanana}
          with $a=1/2,b=1,c=0,B=1$; and
          N$_4(\boldsymbol{0},\boldsymbol{I})$. Boxes describe first,
          second and third quartiles, and whiskers span remaining
          points with length no greater than 1.5$\times$ the
          interquartile length for 20 independent MCMC simulations
          $i=1,\dots,20$. For each proposal count setting, we generate
          100,000 MCMC samples and remove the first 50,000 as
          burn-in. Averaging over target dimensions, we then obtain a
          mean-squared error between slingshot-sampled marginal order
          statistics $q_{i(j)}$ and those sampled directly $y_{i(j)}$
          from target marginal distributions:
          MSE$_i=\sum_{j=1}^{50{,}000}
          (q_{i(j)}-y_{i(j)})^2/50{,}000$.  For each target, we use
          the same proposal variance regardless of proposal
          count. Long chains guarantee large ($>1{,}000$) effective
          sample sizes regardless of proposal count, thereby
          eliminating Monte Carlo error as a biasing factor. }
	\label{fig:fourPlots}
\end{figure}

\begin{table}[!t]
	\centering
	\resizebox{0.6\linewidth}{!}{\begin{tabular}{@{}lllllllll@{}}
			\toprule
			&          &  \multicolumn{2}{l}{Mean}&\multicolumn{2}{l}{Minimum}& \multicolumn{2}{l}{Mean squared error} \\\cmidrule(l){3-4} \cmidrule(l){5-6} \cmidrule(l){7-8}
			Dimension $(d)$  & Algorithm &  ESS &  ESS/s &  ESS &  ESS/s & 1st Moment  & 2nd Moment \\ 
			\midrule
			$2$   & Slingshot & \cellcolor{blue!15}9999.9 & 1925.0 & \cellcolor{blue!15}9934.2 & 1912.4 & \cellcolor{blue!15}1.1e-04 & \cellcolor{blue!15}1.6e-04 \\ 
			& Tjelmeland & 4356.9 & \cellcolor{blue!15}2173.5 & 4223.6 & \cellcolor{blue!15}2107.0 & 3.3e-04 & 4.3e-04 \\ 
			& Naive MP-MCMC$^*$ & 1229.4 & 0.9 & 1187.7 & 0.8 & 4.1e-04 & 1.3e-03 \\ 
			& Indep MP-MCMC & 4352.1 & 604.5 & 4238.4 & 588.7 & 4.0e-04 & 5.5e-04 \\ 
			& RWM & 1346.7 & 467.2 & 1303.1 & 452.1 & 6.4e-04 & 1.3e-03 \\ 
			& Simple & 1162.1 & 185.2 & 1130.4 & 180.1 & 7.9e-04 & 1.0e-03 \\ \cmidrule(l){3-8}
			$4$    & Slingshot & \cellcolor{blue!15}9984.0 & 1317.2 & \cellcolor{blue!15}9738.5 & 1284.8 & \cellcolor{blue!15}8.6e-05 & \cellcolor{blue!15}2.3e-04 \\ 
			& Tjelmeland & 3669.4 & \cellcolor{blue!15}1365.3 & 3457.8 & \cellcolor{blue!15}1286.6 & 3.5e-04 & 8.5e-04 \\ 
			& Indep MP-MCMC & 4012.3 & 335.7 & 3791.6 & 317.3 & 3.0e-04 & 3.6e-04 \\ 
			& RWM & 746.6 & 212.6 & 708.6 & 201.8 & 1.1e-03 & 1.8e-03 \\ 
			& Simple & 1397.9 & 175.9 & 1334.7 & 168.0 & 5.9e-04 & 7.8e-04 \\ \cmidrule(l){3-8}
			$8$ & Slingshot &  \cellcolor{blue!15}9676.4 &  \cellcolor{blue!15}508.5 &  \cellcolor{blue!15}9244.3 &  \cellcolor{blue!15}485.8 &  \cellcolor{blue!15}9.5e-05 & \cellcolor{blue!15}5.4e-04 \\ 
			& Tjelmeland & 2614.0 & 304.2 & 2375.3 & 276.5 & 3.8e-04 & 3.4e-03 \\ 
			& Indep MP-MCMC & 3142.5 & 159.3 & 2760.6 & 139.9 & 3.7e-04 & 7.8e-04 \\ 
			& RWM & 395.1 & 108.3 & 357.0 & 97.8 & 2.5e-03 & 3.0e-03 \\ 
			& Simple & 1106.3 & 128.1 & 1032.3 & 119.5 & 8.7e-04 & 1.1e-03 \\ \cmidrule(l){3-8}
			$16$   & Slingshot & \cellcolor{blue!15}7552.1 & \cellcolor{blue!15}172.6 & \cellcolor{blue!15}7056.9 & \cellcolor{blue!15}161.3 & \cellcolor{blue!15}1.5e-04 & 4.8e-02 \\ 
			& Tjelmeland & 1649.7 & 133.7 & 1440.9 & 116.8 & 6.1e-04 & 8.0e-03 \\ 
			& Indep MP-MCMC & 2277.2 & 65.5 & 1746.0 & 50.2 & 5.8e-04 & \cellcolor{blue!15}8.3e-04 \\ 
			& RWM & 204.3 & 53.2 & 174.1 & 45.4 & 5.1e-03 & 6.3e-03 \\ 
			& Simple & 759.8 & 70.1 & 694.2 & 64.0 & 1.3e-03 & 1.5e-03 \\
			\bottomrule
	\end{tabular}}
      \caption{Comparing the slingshot algorithm with other
        multiproposal MCMC (MP-MCMC) algorithms and random walk
        Metropolis (RWM) all targeting products of $d$ standard
        normals. Across 20 independent chains of length $10{,}000$
        (15,000 total $-$ 5,000 burnin), we measure performance using
        effective sample size (ESS), ESS per second (ESS/s) and
        empirical mean squared error estimating first and second
        moments. We consider mean and minimum ESS across target
        dimensions and use 1,000 proposals for slingshot with Gaussian
        proposals, MP-MCMC using the Tjelmeland correction with
        Gaussian proposals, naive MP-MCMC using iid multivariate
        Gaussian proposals conditioned on the current state and
        MP-MCMC using a multivariate Gaussian independent (indep) of the
        current state. RWM and the simplicial sampler (Simple) use 1
        and $d$ proposals, respectively. We do not show results for
        the naive algorithm past $d=2$ as its $O(p^2)$ complexity
        grinds it to a halt. RWM uses optimal scaling
        $2.4/\sqrt{d}$, and slingshot uses our location-adaptive optimal scaling \eqref{eq:ss:gauss:propstd}; other
        algorithms auto-adapt to a 50\% acceptance rate. Surprisingly,
        the slingshot performs well despite using a moderate proposal
        count $p$. Using this proposal count, the slingshot begins to
        perform poorly at $d=16$.}\label{tab:results}
\end{table}

\section{Summary and Outlook}
\label{sec:Outlook}

The large proposal $\np$-limit and general state space frameworks set
out above provide a new and useful lens to study different forms of
non-trivial parallelism within the MP-MCMC paradigm. Following this
paradigm we identify a collection of promising new methods
(\cref{alg:slingshot:intro}, \cref{alg:MTMpCN} and
\cref{alg:MTMpCN:loc} and see also \cref{sec:The:Slingshot},
\cref{thm:MTM:unbiased}, \cref{meth:MTMpCN}) while largely ruling out
the use of other forms of parallelism that have been suggested in
previous works (\cref{rmk:more:degen}, \cref{sec:large:p:HMC},
\cref{thm:large:p:MH:hmc}).  Moreover, in our analysis of the large
$\np$-properties of the one-step Rao-Blackwellization of a class of
convolutional methods, (\cref{thm:RB:1Step}) we encounter some
possible disadvantages of existing methods.  Finally, our quantitative
results (\cref{thm:tv:bound}) suggest a means of deriving optimal
scaling results as we begin to explore in \cref{sec:numexp:kernel}.

The implications of the results herein and in particular our
asymptotic $\np \to \infty$ approach suggest a number of very
interesting and fruitful directions of future research.
\begin{enumerate}[leftmargin=.5cm]
\item \underline{The $P^\infty$ kernel as an intrinsic object to
    study:} The chains derived in \cref{thm:large:p:lim} and
  \cref{thm:MTM:limit} provide a means of assessing the ultimate
  limits of multiproposal parallelization for a number of algorithms
  considered herein.  In particular, we have in mind the $P^\infty$
  formulation of \cite{tjelmeland2004using} and the multiproposal pCN
  methods we recently derived in \cite{glatt2024parallel,
    glatt2024sacred} alongside its Multiple Try variation given here as
  \cref{alg:MTM}, \cref{alg:MTMpCN} and \cref{alg:MTMpCN:loc}.  In
  both cases, the study of mixing rates for both $\infty$ kernels are
  of interest in comparative perspective to their single proposal
  analogues; cf. \cite{hairer2014spectral, glatt2021mixing,
    andrieu2024explicit}.  Some preliminary steps in this
  direction have been carried out in soon to be completed work
  \cite{CarigiGlattHoltzMondaini2026a}.
\item \underline{Large $\np$-properties under parameter scaling:} The
  limits established in \cref{thm:large:p:lim}, \cref{thm:MTM:limit}
  and \cref{thm:tv:bound} are only a starting point.  Here, we note
  that the general kernels \eqref{eq:MTM:Kernel} and
  \eqref{eq:gen:bk:ker} are typically defined up to particular
  algorithmic parameters, call them $\rho$, which can be varied
  with $\np$ as $p \to \infty$.  In a number of these methods, we may
  hope to derive an `optimal' choice $\rho^*(\np)$.

  As a preliminary step in this direction we have carried out some
  analysis of algorithmic parameter limits directly at $\np = \infty$
  in \cref{sec:alg:scalingatinfty}.  For
  \cref{subsubsec:bubble:bath}, \cref{meth:MTMpCN},
  \cref{meth:tj:mPCN} we identify in \cref{prop:Glob:param:Limit} some
  algorithmic parameter limits leading to Monte Carlo sampling
  directly from the target measure.  This limiting result includes,
  for example \cref{alg:MTMpCN} and its convolutional counterpart
  derived in \cite{glatt2024parallel} with $\np = \infty$ and
  $\rho = 0$.  On the other hand, for \cref{alg:MTMpCN:loc}, we
  identify limiting behavior to a preconditioned overdamped Langevin
  dynamic derived in \cite{cotter2013mcmc} in the `small time step
  regimen'.  See \cref{Prop:alg:scalingatinfty:local} and related
  earlier results from \cite{gagnon2023improving}.

\item \underline{Dimensional scaling in the large $p$-limit:} While we
  have shown previously \cite{glatt2024parallel, glatt2024sacred} that
  MP-MCMC methods are particularly effective for resolving measures with
  multimodal or complex geometric structure, it is not clear if the
  `non-preconditioned' MP-MCMC forms of parallelism originally
  proposed in \cite{tjelmeland2004using, liu2000multiple} or our
   slingshot method (\cref{alg:slingshot:intro})
  are more effective for high-dimensional problems.

  We have several approaches in immediate view that can be used to get
  a better handle on this question.  Regarding the method
  \cite{tjelmeland2004using} recalled here as \cref{meth:tj:mPCN}, one
  may aim to carry out a diffusion limit analysis in the vein of
  \cite{gelman1997weak, roberts1998optimal, mattingly2012diffusion,
    pillai2012optimal, beskos2013optimal} for the $p =\infty$
  processes defined as special cases of the kernel
  \eqref{eq:p:inf:w:lim:conv:barker} in \cref{thm:large:p:lim}. One
  interesting question is how the number of samples required to
  accurately resolve the target measure under an optimal scaling of
  algorithmic parameters in this $\infty$-limit method compares to
  benchmarks for RWM, MALA and HMC ($O(d), O(d^{1/3}), O(d^{1/4})$) in
  the aforementioned works.  Regarding our slingshot method, we
  observe above in \cref{sec:p:inf:lim:kernels:SS}, the implied $p = \infty$-limit
  is simply taking i.i.d.~samples from the target measure, an obvious
  dimension-free method.  However, here it is not clear how fast the
  convergence as $\np \to \infty$ degenerates as a function of the
  dimension.
\item \underline{Rigorous bounds and diagnostics for bias:} As we lay
  out above, the most natural MP-MCMC methods described in
  \cref{sec:cond:prop} can be corrected by using the Multiple Try
  approach \cref{sec:MTM} or by introducing a convolutional mechanism
  \cref{sec:conv:prop}.  However, as hinted by \cref{thm:RB:1Step},
  these unbiased corrections may have disadvantages over their biased
  counterpart.  Moreover, as we emphasize in
  \cref{sec:p:inf:lim:kernels:SS} the slingshot algorithm of
  \cref{subsubsec:slingshot} and \cref{sec:The:Slingshot} is unbiased
  in the $p \to \infty$ asymptotic.  Note also
  \cref{rmk:scale:to:glob} in regards to \cref{subsubsec:bubble:bath}.
  Thus, further systematic case studies supplemented by the
  development of rigorous frameworks for bias would be desirable to
  complete the picture and determine the viability of the Slingshot
  and other `uncorrected' MP-MCMC methods as general purpose tools.
\item \underline{Other forms of parallelism and their large
    $p$-limits:} The multiproposal formulations we lay out in
  \cref{sec:multi:prop:algs} are not the only possibilities under the
  general multiproposal formalism.  In particular, while we largely
  rule out the use of multiproposal sample-path HMC approaches in
  \cref{sec:large:p:HMC}, other different multiproposal HMC
  formulations remain to be explored.  For example, we can consider
  HMC with multiple momentum draws at each step and, in this context, consider
  the associated large $\np$-limit.  On the other hand, we note the
  new multi-proposal generalization of the elliptic slice sampler,
  MESS, developed in \cite{senn2026multiproposal}.  While MESS can be
  viewed as a (degenerate) surrogate HMC trajectory method its
  structure is quite different from the asymptotically accurate HMC setup
  we described in \cref{sec:large:p:HMC}.
  
\item \underline{Rao-Blackwellization and resampling:} Other questions
  remain in regards to the Rao-Blackwellization \eqref{eq:RB:form}
  associated to multiproposal methods.  Here, for example, our
  analysis in \cref{sec:RB::1step:Limit} does not touch on the
  Multiple Try or HMC methods.  Moreover, one of the aspects of the
  resampling methods, briefly mentioned above in \cref{rmk:resamp}, is
  that this approach may give us some of the advantages of
  Rao-Blackwellizations while mitigating some of the associated costs
  in terms of memory storage.  Thus, we may ask how many resampling
  steps are required to obtain comparable estimator efficiency and
  ascertain whether conclusions hold in the methods' respective large
  $p$-limits.
\end{enumerate}

\section*{Acknowledgements}

Our efforts are supported under the grants NSF-DMS-2108790,
NSF-DMS-2510856 (NEGH), NSF-DMS-2108791 (JAK), NIH-K25-AI153816,
NSF-DMS-2152774 and NSF-DMS-2236854 (AJH) and NSF-DMS-2239325 (CFM).
We would like to thank Giulia Carigi, Claudius Zlhot Gypetee, Faming
Liang, Guillermina Senn, H{\aa}kon Tjelmeland and Andrew Warren for
inspiring discussions and helpful feedback on this work.


\begin{footnotesize}
	\addcontentsline{toc}{section}{References}
	\bibliographystyle{alpha}
	\bibliography{refs}
\end{footnotesize}

\newpage
\begin{multicols}{2}
\noindent
Nathan E. Glatt-Holtz\\ {\footnotesize
Department of Statistics and Department of Mathematics\\
Indiana University, Bloomington\\
Web: \url{https://negh.pages.iu.edu}\\
Email: \url{negh@iu.edu}} \\[.2cm]

\noindent
Andrew J. Holbrook\\  {\footnotesize
Department of Biostatistics\\
University of California, Los Angeles\\
Web: \url{https://andrewjholbrook.github.io}\\
Email: \url{aholbroo@g.ucla.edu}}\\[.2cm]

\columnbreak

\noindent Justin Krometis\\
{\footnotesize
National Security Institute and Department of Mathematics\\
Virginia Polytechnic Institute and State University\\
Web: \url{https://krometis.github.io/}\\
Email: \href{mailto:jkrometis@vt.edu}{\nolinkurl{jkrometis@vt.edu}}} \\[.2cm]

\noindent Cecilia F. Mondaini \\
{\footnotesize
  Department of Mathematics\\
  Drexel University\\
  Web: \url{https://www.math.drexel.edu/~cf823}\\
  Email: \url{cf823@drexel.edu}}\\[.2cm]
 \end{multicols}

\newpage

\appendix

\renewcommand{\thefigure}{A\arabic{figure}}
\setcounter{figure}{0}

\section{Rigorous Proofs}
\label{sec:rig:proofs}

This appendix collects all of the complete and rigorous proofs of the
main theorems above.

\subsection[Proof of Theorem~\ref{thm:grand:unifed:inv}]
    {Proof of \cref{thm:grand:unifed:inv} }
\label{sec:gen:inv}

In order to establish \eqref{eq:rev:gen:def} we fix any bounded
measurable test function $\psi: \qsp \times \qsp \to \RR$ and observe
that
\begin{align*}
  \int_{\qsp}  \int_{\qsp}&\psi(\bq, \tbq)  \mk(\bq, d\tbq) \mu(d \bq)
  = \sum_{j =0}^p \int_{\qsp} \int_{\vsp} \psi(\bq , \Pi_1 S_j (\bq, \bv)) 
     \alpha_j(\bq, \bv) \cM (d \bq, d \bv)\\
  =& \sum_{j =0}^p \int_{\qsp} \int_{\vsp}
     \psi(\Pi_1 (S_j (S_j(  \bq, \bv))) , \Pi_1 S_j (\bq, \bv))
      \alpha_j(S_j( S_j(\bq, \bv))) \cM (d \bq, d \bv)\\
   =&    \sum_{j =0}^p \int_{\qsp} \int_{\vsp} \psi(\Pi_1 S_j (\bq,
      \bv), \bq)
      \alpha_j(S_j(\bq, \bv)) S_j^*\cM (d \bq, d \bv)\\
  =&    \sum_{j =0}^p \int_{\qsp} \int_{\vsp} \psi(\Pi_1 S_j (\bq,
      \bv), \bq)
     \alpha_j(\bq, \bv)\cM (d \bq, d \bv)
   =   \int_{\qsp}  \int_{\qsp}\psi(\tbq, \bq)  \mk(\bq, d\tbq) \mu(d \bq)
\end{align*}  
Notice that the second equality in the above computation uses the
involutive condition \eqref{cond:invol:S} whereas the penultimate
equality is simply \eqref{cond:alpha:rev}.  Since the test function
$\psi$ in this computation was arbitrary, the proof is complete.

\subsection[Proof of Theorem~\ref{thm:MTM:unbiased}]
{Proof of \cref{thm:MTM:unbiased}}
\label{sec:MTM:unbiased}

We begin by rewriting \eqref{eq:MTM:Kernel} in the formulation
\eqref{def:mp:ker} in order to apply
\cref{thm:grand:unifed:inv}.  To this end, take $\vsp = \qsp^{\np^2}$
and set the notation:
\begin{align}
  \label{eq:MTM:full:caff:0}
  \bv := (\bq_1, \ldots, \bq_\np, \bz_1^{1}, \ldots, \bz_{\np
  -1}^{1}, \ldots,
  \bz_1^{\np}, \ldots, \bz_{\np-1}^{\np})
  := (\bv^{(\bq)}, \bv^{(\bz)}).
\end{align}
On this space $\vsp$, we extend the definition of the involutions
\eqref{def:Sj:coord:flip:0} to act on the first set of
`$\bq$-coordinates', namely
\begin{align}
  S_j&(\bq_0, \bq_1, \ldots, \bq_{j-1}, \bq_j, \bq_{j+1},
       \ldots, \bq_\np, \bz_1^{1}, \ldots, \bz_{\np -1}^{1}, \ldots,
       \bz_1^{p}, \ldots, \bz_{\np-1}^{p})
  \notag\\
     &= (\bq_j, \bq_1, \ldots, \bq_{j-1}, \bq_0, \bq_{j+1},
       \ldots,\bq_\np, \bz_1^{1}, \ldots, \bz_{\np  -1}^{1}, \ldots,
       \bz_1^{p}, \ldots, \bz_{\np-1}^{p}),
         \label{eq:MTM:full:caff:1}
\end{align}
for $j = 0, \ldots, p$.  To complete the formulation, we set
\begin{align}
  \alpha_j (\bq_0, \bv)
  &:= \frac{\art (\bq_j, \bq_0)}{\sum_{l=1}^\np \art(\bq_l, \bq_0)}
  \left[
  1 \wedge
    \left(\frac{d \eta^\perp}{d \eta}(\bq_0, \bq_j)
    \frac{\art(\bq_0, \bq_j)}{\art(\bq_j, \bq_0)}
    \frac{\sum_{l =1}^p\art(\bq_l, \bq_0)}
  {\art(\bq_0, \bq_j)+ \sum_{l =1}^{p-1} \art(\bz_l^j, \bq_j)} \right) \right]
  \notag\\
  &= 
  \frac{ \art (\bq_j, \bq_0)}{\sum_{l=1}^\np \art(\bq_l, \bq_0)} 
  \wedge
  \left(
  \frac{d \eta^\perp}{d \eta}(\bq_0, \bq_j)  \frac{\art(\bq_0, \bq_j)}
  {\art(\bq_0, \bq_j)+ \sum_{l =1}^{p-1}\art(\bz_l^j, \bq_j)} 
  \right)
  \label{eq:MTM:full:caff:2}
\end{align}  
for $j = 1, \ldots,\np$, and
\begin{align}
  \alpha_0(\bq_0, \bv) &:= \sum_{j =1}^\np
  \frac{\art (\bq_j, \bq_0)}{\sum_{l=1}^\np \art(\bq_l, \bq_0)}
   \left\{ 1 - \left[
   1 \wedge
            \left(\frac{d \eta^\perp}{d \eta}(\bq_0, \bq_j)
                         \frac{\art(\bq_0, \bq_j)}{\art(\bq_j, \bq_0)}
                         \frac{\sum_{l =1}^p\art(\bq_l, \bq_0)}
   {\art(\bq_0, \bq_j)+ \sum_{l =1}^{p-1}\art(\bz_l^j, \bq_j)} \right) \right]
   \right\}
  \notag\\
  &= 1 - \sum_{j = 1}^p \alpha_j(\bq_0, \bv),
  \label{eq:MTM:full:caff:3}
\end{align}
and then take
\begin{align}
  \vk(\bq_0, d \bv)
  &= \vk(\bq_0, d \bq_1, \ldots, d \bq_\np,d \bz_1^{(1)}, \ldots, d\bz_{\np
  -1}^{(1)}, \ldots,
    d\bz_1^{(p)}, \ldots, d\bz_{\np-1}^{(p)})
  \notag\\
  &=  \prod_{l=1}^p\prod_{k =1}^{p-1} \prod_{m =1}^p
    \pk(\bq_l, d \bz_k^{l}) \pk(\bq_0, d \bq_m).
    \label{eq:MTM:full:caff:4}
\end{align}

Now, since $S_0$ is the identity and in view of \eqref{cond:alpha:rev}
to establish the desired result it is sufficient to show that, under
\eqref{eq:MTM:full:caff:0}-\eqref{eq:MTM:full:caff:4}
\begin{align}
  \int_{\qsp}\int_{\qsp^{\np^2}} \psi(\bq_0, \bq_j)
    \alpha_j( \bq_0, \bv) \cM (d \bq_0, d \bv)
     =  \int_{\qsp} \int_{\qsp^{\np^2}}   \psi(\bq_j, \bq_0 )
  \alpha_j(\bq_0, \bv) \cM (d \bq_0, d \bv),
  \label{eq:j:wise:cond:alpha:rev:ap}
\end{align}
for each $j = 1, \ldots, p$ and any measurable bounded
$\psi: \qsp \times \qsp \to \RR$.  Now observe that $(\np -1)^2$ of the
$\bz_k^l$ coordinates integrate out on each
side of \eqref{eq:j:wise:cond:alpha:rev:ap} so that our goal is
finally to show:
\begin{align*}
  &\int\limits_{\qsp^{2\np}}
    \! \psi(\bq_0, \bq_j) \\
    &\qquad
    \cdot \left[
    \frac{ \art (\bq_j, \bq_0)}{\sum_{l=1}^\np \art(\bq_l, \bq_0)} 
    \wedge
    \left(
    \frac{d \eta^\perp}{d \eta}(\bq_0, \bq_j)  \frac{\art(\bq_0, \bq_j)}
    {\art(\bq_0, \bq_j)+ \sum_{l =1}^{p-1}\art(\bz_l^j, \bq_j)} 
    \right)
    \right]
      \prod_{k =1}^{p-1} \prod_{m =1}^p\!
    \pk(\bq_j, d \bz_k^{j}) \pk(\bq_0, d \bq_m) \mu(d \bq_0)                            
  \\
  &\! \!= \! \!  \int\limits_{\qsp^{2\np}} \! \! \psi(\bq_j, \bq_0 ) \\
  &\qquad
  \cdot \left[
  \frac{ \art (\bq_j, \bq_0)}{\sum_{l=1}^\np \art(\bq_l, \bq_0)} 
  \wedge
  \left(
  \frac{d \eta^\perp}{d \eta}(\bq_0, \bq_j)  \frac{\art(\bq_0, \bq_j)}
  {\art(\bq_0, \bq_j)+ \sum_{l =1}^{p-1} \art(\bz_l^j, \bq_j)} 
  \right)
  \right]
  \prod_{k =1}^{p-1} \prod_{m =1}^p\!
  \pk(\bq_j, d \bz_k^{j}) \pk(\bq_0, d \bq_m) \mu(d \bq_0)  
\end{align*}
Working from the left hand side of this desired identity, we define
\begin{align*}
  &H_j(\bq_0, \bq_j) := \\
  &\quad \int\limits_{\qsp^{2\np-2}}
  \left[
   \frac{ \art (\bq_j, \bq_0)}{\sum_{l=1}^\np \art(\bq_l, \bq_0)} 
   \wedge
   \left(
   \frac{d \eta^\perp}{d \eta}(\bq_0, \bq_j)  \frac{\art(\bq_0, \bq_j)}
   {\art(\bq_0, \bq_j)+ \sum_{l =1}^{p-1} \art(\bz_l^j, \bq_j)} 
   \right)
   \right]
      \prod_{k =1}^{p-1} \prod_{m =1, m \not= j}^p\!
                          \pk(\bq_j, d \bz_k^{j}) \pk(\bq_0, d \bq_m).
\end{align*}

Noting that the assumption $\eta^\perp \ll \eta$ implies that
$\eta \ll \eta^\perp$ and
\begin{align*}
  \frac{d \eta}{d \eta^\perp}(\bq, \tbq)
  =  \frac{d \eta^\perp}{d \eta}(\tbq, \bq),
  \quad (\bq, \tbq) \in \qsp \times \qsp,
\end{align*}
we thus obtain that
\begin{align*}
  H_j(\bq_0, \bq_j) \frac{d \eta}{d \eta^\perp} (\bq_0, \bq_j)
  = H_j(\bq_j, \bq_0). 
\end{align*}

Hence, with Fubini's theorem, we have 
\begin{align*}
	LHS 
	&= \int\limits_{\qsp^2}
	\psi(\bq_0, \bq_j)  H_j(\bq_0, \bq_j)  \pk(\bq_0, d \bq_j)  \mu(d \bq_0)
	= \int\limits_{\qsp^2}	\psi(\bq_0, \bq_j)  H_j(\bq_0, \bq_j) \eta (d \bq_0, d \bq_j) \\
	&= \int\limits_{\qsp^2} \psi(\bq_0, \bq_j)  H_j(\bq_0, \bq_j) \frac{d \eta}{d \eta^\perp} (\bq_0, \bq_j) \eta^\perp (d \bq_0, d \bq_j) \\
	&= \int\limits_{\qsp^2} \psi(\bq_0, \bq_j)  H_j(\bq_j, \bq_0) \eta^\perp(d \bq_0, d \bq_j)
	= \int\limits_{\qsp^2} \psi(\bq_j, \bq_0)  H_j(\bq_0, \bq_j)  \eta(d \bq_0, d \bq_j) = RHS.
\end{align*}
The proof is complete.

\subsection[Proof of Theorem~\ref{thm:large:p:lim}]
{Proof of \cref{thm:large:p:lim}}
\label{sec:Large:p:w:l}

We address \eqref{def:mk:infty} for any Dirac measure
$\delta_{\bq_0}$; the general case then follows by the Dominated
Convergence Theorem.  Let $\varphi: \qsp \to \RR$ be a continuous and
bounded function. Fix $\bq \in \qsp$ and note that, from the
definitions of $\mk^\np$, $\vk^\np$, and $(\ar_0, \ldots, \ar_\np)$ in
\eqref{def:mp:ker}, \eqref{def:vk:Tjelmeland}, and
\eqref{def:ar:simple}, respectively, we have
\begin{align}
	\int_\qsp \varphi(\tbq) \mk^\np(\bq, d \tbq) 
  &= \sum_{j=0}^\np \int_\qsp \int_{\qsp^\np}
    \ar_j (\bq, \bq_1, \ldots, \bq_\np) \varphi(\bq_j)
    \prod_{l=1}^\np \pk(\bbq, d \bq_l) \bpk(\bq, d \bbq) \nonumber \\
  &= \int_\qsp \int_{\qsp^\np} \frac{\art(\bq, \bq) \varphi(\bq)
    + \sum_{j=1}^\np \art(\bq_j, \bq) \varphi(\bq_j)}{\art(\bq, \bq)
    + \sum_{k=1}^\np \art(\bq_k, \bq)} \prod_{l=1}^\np \pk(\bbq, d \bq_l) \bpk(\bq, d \bbq).
	\label{int:mk:np:varphi}
\end{align}

Fix $\bq, \bbq \in \qsp$. Let $\qsp^\NN$ denote the space of sequences
of elements in $\qsp$. For each finite subset of indices
$\cI \subset \NN$, let $|\cI|$ be the number of elements in $\cI$, and
let $\proj_\cI : \qsp^\NN \to \qsp^{|\cI|}$ denote the projection
mapping $\proj_{\cI}((\bq_n)_{n \in \NN}) = (\bq_n)_{n \in
  \cI}$. Moreover, for each finite subset $\cI$, we consider the
probability measure $\pk(\bbq, \cdot)^{\otimes |\cI|}$. Note that,
since $\qsp^{|\cI|}$ is a Polish space with the product topology, then
$\pk(\bbq, \cdot)^{\otimes |\cI|}$ is a tight measure on $\qsp^{\cI}$,
see e.g. \cite[Theorem 12.7]{Aliprantis2013}. It thus follows from the
Kolmogorov extension theorem (see \cite[Theorem 2.4.3]{tao2011}) that
there exists a probability measure $\nu_{\bbq}$ on $\qsp^\NN$ such
that $\pk(\bbq, \cdot)^{|\cI|} = \proj_{\cI}^* \nu_{\bbq}$ for any
finite subset $\cI \subset \NN$. In particular,
$\pk(\bbq, \cdot)^{\otimes p} = \proj_{\{1,\ldots, \np\}}^*
\nu_{\bbq}$ for any $\np \geq 1$. We may thus write
\eqref{int:mk:np:varphi} as
\begin{align}\label{int:mk:np:varphi:b}
	\int_\qsp \varphi(\tbq) \mk^\np(\bq, d \tbq) 
	=  \int_\qsp \int_{\qsp^\NN} \frac{\art(\bq, \bq) \varphi(\bq)
  + \sum_{j=1}^\np \art(\bq_j, \bq) \varphi(\bq_j)}{\art(\bq, \bq)
  + \sum_{k=1}^\np \art(\bq_k, \bq)}
  \nu_{\bbq} (d (\bq_n)_{n \in \NN}) \bpk(\bq, d \bbq).
\end{align}

Now observe that, for each $p \geq 1$, the projection maps
$\proj_1, \ldots, \proj_\np$ form a collection of independent and
identically distributed random variables on the probability space
$(\qsp^\NN, \nu_{\bbq})$.  Indeed, $\proj_j$ is distributed as
$\proj_j^* \nu_{\bbq} = \pk(\bbq, \cdot)$ for all $j=1,\ldots, \np$,
and their joint distribution
$(\proj_1, \ldots, \proj_\np)^* \nu_{\bbq} =
\proj_{\{1,\ldots,\np\}}^* \nu_{\bbq}$ is the product measure
$\pk(\bbq, \cdot)^{\otimes \np}$. Consequently,
$\art(\proj_j(\cdot),\bq): \qsp^\NN \to \RR$, $j=1, \ldots, \np$, are
independent and identically distributed real-valued random variables
on $(\qsp^\NN, \nu_{\bbq})$. Moreover, from the integrability
assumption on $\art$, it follows that
\begin{align*}
  \EE \art(\proj_j(\cdot), \bq)
  = \int_{\qsp^\NN} \art(\proj_j((\bq_n)_n), \bq)
    \nu_{\bbq}(d (\bq_n)_n)
	= \int_\qsp \art(\tbq, \bq) \proj_j^* \nu_{\bbq} (d \tbq)
	= \int_\qsp \art(\tbq, \bq) \pk(\bbq, d \tbq) < \infty
\end{align*}
for all $j=1, \ldots, \np$. Thus, by the strong Law of Large Numbers,
\begin{align}\label{LLN:art}
	\lim_{p\to \infty} \frac{1}{\np} \sum_{k=1}^\np \art(\proj_k((\bq_n)_n),\bq) 
	= \int_\qsp \art(\tbq, \bq) \pk(\bbq, d \tbq)
\end{align}
for $\nu_{\bbq}$-a.e. $(\bq_n)_{n \in \NN} \in \qsp^\NN$.

Since $\varphi$ is bounded, it follows with similar arguments that
\begin{align}\label{LLN:art:varphi}
  \lim_{p\to \infty} \frac{1}{\np}
  \sum_{j=1}^\np \art(\proj_j((\bq_n)_n),\bq)
  \varphi(\proj_j((\bq_n)_n))
	= \int_\qsp \art(\tbq, \bq) \varphi(\tbq) \pk(\bbq, d \tbq)
\end{align}
for $\nu_{\bbq}$-a.e. $(\bq_n)_{n \in \NN} \in \qsp^\NN$.

Note that 
\begin{align}
  &\int_{\qsp^\NN} \frac{\art(\bq, \bq) \varphi(\bq)
    + \sum_{j=1}^\np \art(\bq_j, \bq) \varphi(\bq_j)}{\art(\bq, \bq)
    + \sum_{k=1}^\np \art(\bq_k, \bq)} \nu_{\bbq} (d (\bq_n)_{n \in \NN}) 
   \nonumber \\
  &\qquad=  \int_{\qsp^\NN} \frac{ \frac{1}{\np}\art(\bq, \bq) \varphi(\bq)
    + \frac{1}{\np} \sum_{j=1}^\np \art(\proj_j((\bq_n)_n), \bq) \varphi(\proj_j((\bq_n)_n))}{ \frac{1}{\np} \art(\bq, \bq)
    + \frac{1}{\np} \sum_{k=1}^\np \art(\proj_k((\bq_n)_n), \bq)} \nu_{\bbq} (d (\bq_n)_{n \in \NN}) .
	\label{int:LLN}
\end{align}

Since the integrand in \eqref{int:LLN} is bounded above by
$\|\varphi\|_\infty = \sup_{\bq' \in \qsp} |\varphi(\bq')|$, then,
together with \eqref{LLN:art} and \eqref{LLN:art:varphi}, it follows
by the Dominated Convergence theorem that
\begin{align*}
	\lim_{\np \to \infty} \int_{\qsp^\NN} \frac{\art(\bq, \bq)
  \varphi(\bq)
  + \sum_{j=1}^\np \art(\bq_j, \bq) \varphi(\bq_j)}{\art(\bq, \bq)
  + \sum_{k=1}^\np \art(\bq_k, \bq)} \nu_{\bbq} (d (\bq_n)_{n \in \NN}) 
  &= \int_{\qsp^\NN} \frac{\int_\qsp \art(\tbq, \bq)
    \varphi(\tbq) \pk(\bbq, d \tbq)}{ \int_\qsp \art(\bq', \bq)
    \pk(\bbq, d \bq')}
   \nu_{\bbq} (d (\bq_n)_{n \in \NN})  \\
	&= \frac{\int_\qsp \art(\tbq, \bq) \varphi(\tbq) \pk(\bbq, d
   \tbq)}
   { \int_\qsp \art(\bq', \bq) \pk(\bbq, d \bq')}
\end{align*}
for all $\bbq \in \qsp$. By a second application of the Dominated
Convergence theorem, we thus obtain from \eqref{int:mk:np:varphi:b} that
\begin{align*}
	\lim_{\np \to \infty} \int_\qsp \varphi(\tbq) \mk^\np(\bq, d \tbq) 
	= \int_\qsp \frac{\int_\qsp \art(\tbq, \bq) \varphi(\tbq)
  \pk(\bbq, d \tbq)}
  { \int_\qsp \art(\bq', \bq) \pk(\bbq, d \bq')} \bpk(\bq, d \bbq), 
\end{align*}
as desired.

\subsection[Proof of Theorem~\ref{thm:MTM:limit}]
{Proof of \cref{thm:MTM:limit}}
\label{app:mtm:limit}

As in the proof of \cref{thm:large:p:lim} we establish the desired
convergence, \eqref{eq:weak:conv:MTM}, in the special case of Dirac
measures and then bootstrap to the general case with the Dominated
Convergence Theorem. Fix any continuous bounded $\phi: \qsp \to \RR$
and observe that
\begin{align}\label{int:phi:Pp}
  \int & \phi(\tbq) P^\np( \bq_0, d \tbq)
         \notag\\
  =& \EE
  \left( \frac{\sum_{j = 1}^\np \phi(\bq_j) \art(\bq_j, \bq_0)
  \bar{\alpha}_j(\bq_0, \bq_1 \ldots, \bq_\np, \bz_1^j, \ldots, \bz_{\np-1}^j)}
     {\sum_{j = 1}^\np \art(\bq_j, \bq_0)} \right)
     \notag\\
  &+ \phi(\bq_0) \EE
  \left( \frac{\sum_{j = 1}^\np \art(\bq_j, \bq_0)(1 -
  \bar{\alpha}_j(\bq_0, \bq_1 \ldots, \bq_\np, \bz_1^j, \ldots, \bz_{\np-1}^j))}
    {\sum_{j = 1}^\np \art(\bq_j, \bq_0)} \right)
    \notag\\
  =& \EE
  \left( \frac{\np^{-1}\sum_{j = 1}^\np \phi(\bq_j) \art(\bq_j, \bq_0)
  \bar{\alpha}^\infty(\bq_j, \bq_0)}
     {\np^{-1}\sum_{j = 1}^\np \art(\bq_j, \bq_0)} \right)
  + \phi(\bq_0) \EE
  \left( \frac{\np^{-1}\sum_{j = 1}^\np \art(\bq_j, \bq_0)(1 -
  \bar{\alpha}^\infty(\bq_j, \bq_0))}
     {\np^{-1}\sum_{j = 1}^\np \art(\bq_j, \bq_0)} \right)
     \notag\\
     &+ \EE \mathfrak{E}_p(\bq_0, \bq_1 \ldots, \bq_\np, \bz_1^1,
       \ldots, \bz_{\np-1}^1, \ldots,  \bz_1^\np, \ldots, \bz_{\np-1}^\np),
\end{align}
where\footnote{Here, $\bq_j$, $\bz_k^j$, $k=1,\ldots,\np-1$, $j=1,\ldots,\np$, are
taken as random variables defined on a suitable ($\np$-independent)
probability space $\Omega$, which is obtained by arguing via the
Kolmogorov extension theorem, similarly as in the proof of
\cref{thm:large:p:lim} given in \cref{sec:Large:p:w:l}. As such, the
expected value in \eqref{int:phi:Pp} represents an integral with
respect to a suitable probability measure $\nu_{\bq_0}$ on $\Omega$.}
\begin{align}\label{qj:zkj:iid}
  \{\bq_j\}_{j=1}^\np \text{ i.i.d. from } \pk(\bq_0, d\tbq),
  \quad
  \bz_k^j := F(\bq_j, \bu_k)
  \text{ with } \{\bu_k\}_{k=1}^{\np -1} \text{ i.i.d. from } \nu \text{ in \ref{A1}},
\end{align}  
and
\begin{align}
 \mathfrak{E}_\np= \mathfrak{E}_\np(&\bq_0, \bq_1, \ldots, \bq_\np, \bz_1^1,
  \ldots, \bz_{\np-1}^1, \ldots,
    \bz_1^\np, \ldots, \bz_{\np-1}^\np, \phi)
                   \notag\\
  =&  \frac{\np^{-1}\sum_{j = 1}^\np (\phi(\bq_j)  - \phi(\bq_0)) \art(\bq_j, \bq_0)
  (\bar{\alpha}_j(\bq_0, \bq_1 \ldots, \bq_\np, \bz_1^j, \ldots,\bz_{\np-1}^j)-
    \bar{\alpha}^\infty(\bq_j, \bq_0))}
     {\np^{-1}\sum_{j = 1}^\np \art(\bq_j, \bq_0)} . 
\end{align}
Thus, to establish the desired result, it suffices to show that
$\lim_{\np \to \infty} \EE \mathfrak{E}_\np = 0$.

To this end, recalling our definitions of $\bar{\alpha}_j$,
$\bar{\alpha}^\infty$, in \eqref{eq:MTM:ar},  and
\eqref{eq:MTM:inf:ker:ar} respectively,  let
\begin{align}
  \epsilon_\np(\br)
  :=& \biggl|1 \wedge
   \left(\frac{d \eta^\perp}{d \eta}(\bq_0, \br)
   \frac{\art(\bq_0, \br)}{\art(\br, \bq_0)} \frac{\np^{-1}\sum_{l =1}^p\art(\bq_l, \bq_0)}
      {\np^{-1}\art(\bq_0, \br)+ \np^{-1}\sum_{l =1}^{\np -1}\art(F(\br, \bu_l), \br)}
      \right)
      \notag\\
  & \qquad - 1 \wedge \left( \frac{d \eta^\perp}{d \eta}(\bq_0, \br)
  \frac{\art(\bq_0, \br)}{\art(\br, \bq_0)}
    \frac{\int_\qsp \art(\bq',\bq_0) Q(\bq_0, d \bq') }
    { \int_\qsp \art(\bq',\br) Q(\br, d \bq')} \right)
    \biggr| .
\end{align}
We observe that, for any compact set $A \subset \qsp$
\begin{align}
   |\EE \mathfrak{E}_p|
  &\leq \EE \!
         \left( \sup_{\br \in A}\epsilon_\np(\br)
         \frac{\np^{-1}\sum_{j = 1}^\np |\phi(\bq_j)  - \phi(\bq_0)| \art(\bq_j, \bq_0)}
         {\np^{-1}\sum_{j = 1}^\np \art(\bq_j, \bq_0)}\right)
   \!+  2 \EE \!\left(
    \frac{\np^{-1}\sum_{j = 1}^\np |\phi(\bq_j)  - \phi(\bq_0)|
    \art(\bq_j, \bq_0) \indFn{\bq_j \in A^c}}
  {\np^{-1}\sum_{j = 1}^\np \art(\bq_j, \bq_0)}
  \right) \notag \\
  &\leq 2 \|\phi\|_\infty \EE  \left( \sup_{\br \in A}\epsilon_\np(\br) \right)
  + 4 \|\phi\|_\infty \EE \left(  \frac{\np^{-1}\sum_{j = 1}^\np 
  	\art(\bq_j, \bq_0) \indFn{\bq_j \in A^c}}
    {\np^{-1}\sum_{j = 1}^\np \art(\bq_j, \bq_0)} \right),
    \label{ineq:E:fEp}
\end{align}
where $\|\phi\|_\infty := \sup_{\bq \in \qsp} |\phi(\bq)| < \infty$.
Hence, due to \eqref{qj:zkj:iid} and assumption \eqref{inf:int:beta:nu} in
\ref{A3}, we may invoke the Law of Large Numbers to obtain that
\begin{align}
  \limsup_{\np \to \infty} |\EE \mathfrak{E}_p| \leq 2 \|\phi\|_\infty \left(
  \limsup_{\np\to \infty} \EE \left(  \sup_{\br \in A}\epsilon_\np(\br) \right)
   +  2   \frac{\int_\qsp \art(\tbq, \bq_0) \indFn{A^c} (\tbq)
		\pk(\bq_0, d\tbq)}{\int_\qsp \art(\tbq, \bq_0)
  \pk(\bq_0, d \tbq)} \right)
  \label{lim:T2p}
\end{align}
which holds for any compact $A \subset \qsp$.

In order to estimate the first term in the upper bound
\eqref{lim:T2p}, we observe that, since $x \mapsto 1 \wedge x$ is
a Lipschitz function with Lipschitz constant $1$, it follows that
\begin{align}
	\epsilon_\np(\br) 
	&\leq
	\left| \frac{d \eta^\perp}{d \eta} (\bq_0, \br) \frac{\art(\bq_0, \br)}{\art(\br, \bq_0)} \right| 
	\left| \frac{\np^{-1}\sum_{l =1}^p\art(\bq_l, \bq_0)}
	{\np^{-1}\art(\bq_0, \br)+ \np^{-1}\sum_{l =1}^{\np - 1} \art(F(\br, \bu_l), \br)}
	- \frac{\int_\qsp \art(\bq',\bq_0) Q(\bq_0, d \bq') }
   { \int_\qsp \art(\bq',\br) Q(\br, d \bq')} 	\right|
   \notag \\
	&\leq 
		\left| \frac{d \eta^\perp}{d \eta} (\bq_0, \br) \frac{\art(\bq_0, \br)}{\art(\br, \bq_0)} \right| 
		\left[ 
		\frac{\left|  \np^{-1}\sum_{l =1}^p\art(\bq_l, \bq_0)
   - \int_\qsp \art(\bq',\bq_0) Q(\bq_0, d \bq')\right| }
   {\np^{-1}\art(\bq_0, \br)+ \np^{-1}\sum_{l =1}^{\np - 1}
   \art(F(\br, \bu_l), \br)} \right.
   \label{ineq:epsilonp}\\
	&\qquad\qquad
		+ \left.  \frac{\left| \np^{-1}\art(\bq_0, \br)+
                      \np^{-1}\sum_{l =1}^{\np - 1}
                        \art(F(\br, \bu_l), \br) -  \int_\qsp
   \art(\bq',\br) Q(\br, d \bq')\right| }
   {\np^{-1}\art(\bq_0, \br)+ \np^{-1}\sum_{l =1}^{\np - 1} \art(F(\br, \bu_l), \br)}
		\cdot 
		\frac{\int_\qsp \art(\bq',\bq_0) Q(\bq_0, d \bq') }
		{ \int_\qsp \art(\bq',\br) Q(\br, d \bq')} 
   \right].
   \notag
\end{align}
We now claim that assumptions \ref{A1}, \ref{A4}, and
\eqref{inf:int:beta:nu} in \ref{A3} imply that
\begin{align}\label{lim:p:ulln}
  \lim_{\np \to \infty} \sup_{\br \in A}
  \left| \frac{1}{\np -1} \sum_{l=1}^{\np - 1} \art(F(\br, \bu_l),
  \br)
  - \int_{\qsp} \art(F(\br, \bu), \br) \nu(d \bu) \right| = 0 \quad \mbox{a.s.},
\end{align}
for any compact set $A$.  Stipulating \eqref{lim:p:ulln}, for the
moment from \eqref{ineq:epsilonp}, it thus follows with assumption
\ref{A3} that
$\lim_{\np \to \infty} \sup_{\br \in A} \epsilon_\np(\br) = 0$. Since
also $\epsilon_\np(\br) \leq 2$ for all $\br$, then by the Dominated
Convergence theorem, we deduce that
\begin{align}\label{lim:T1p}
  \lim_{\np \to \infty}
  \EE \sup_{\br \in A} \epsilon_\np(\br)  = 0 \quad
  \text{ for any compact } A \subset \qsp.
\end{align}
Hence from \eqref{lim:T2p}, \eqref{lim:T1p}, and \ref{A2},
\eqref{eq:comp:exterior:demand:MTM} we infer that, for any
$\epsilon > 0$,
\begin{align*}
  \limsup_{\np \to \infty} \EE \mathfrak{E}_\np
  \leq \frac{4 \| \phi \|_\infty \epsilon}
  {\int_\qsp \art(\bq',\bq_0) Q(\bq_0, d \bq')}. 
\end{align*}
Thus, with \ref{A3}, \eqref{inf:int:beta:nu} we conclude that
$\lim_{\np \to \infty} \EE \mathfrak{E}_\np = 0$, as desired.

Thus, to complete the proof it remains to justify our claim
\eqref{lim:p:ulln}.  To this end define
\begin{align*}
    h(\bu, \br) := \art(F(\br, \bu), \br) - \int_\qsp \art(F(\br,
  \bu'),
  \br) \nu(d \bu'), \quad
  \text{ for }  \br,\bu \in \qsp.
\end{align*}
Observe that, under our assumption \ref{A4}, \eqref{cont:art}, for any
$\varepsilon > 0$, and compact $A \subset \qsp$ there is a
$\delta = \delta(\varepsilon, A)$ and a $g: \qsp \to \RR^+$
so that $\int g(\bu) \nu(d \bu) < \varepsilon$ and
\begin{align}
  \label{eq:MTM:lip:cond}
  |h(\bu, \br) - h(\bu, \tbr)|  \leq  g(\bu) + \varepsilon,
  \quad
  \text{ for every $\bu \in \qsp$ so long
  as $d( \tbr, \br) < \delta$ and $\br, \tbr \in A$}.
\end{align}

With \eqref{eq:MTM:lip:cond} in hand we fix an arbitrary
$\varepsilon > 0$ and any compact set $A \subset \qsp$.  For each
$\br \in A$ we consider a ball
$B_{\delta(\varepsilon, A)}(\br) = \{ \tilde{\br} \in X \,:\,
\met(\br, \tilde{\br}) < \delta(\varepsilon, A) \}$ where
$\delta(\varepsilon, A) > 0$ maintains the relationship in
\eqref{eq:MTM:lip:cond} for the given $\varepsilon, A$.  In this
fashion we form an open cover
$A \subset \bigcup_{\br \in A} B_{\delta(\varepsilon,A)} (\br)$. Now,
since $A$ is compact, we then choose a finite collection of points
$\br_1, \ldots, \br_n \in A$ such that
$A \subset \bigcup_{i=1}^n B_{\delta (\varepsilon,A)}
(\br_i)$. Now, given any $\br \in A$ select $j \in \{1, \ldots, n\}$
such that $\br \in B_{\delta(\varepsilon, A)}(\br_j)$.  For this $\br$
we observe
\begin{align*}
  \left| \frac{1}{\np-1} \sum_{l=1}^{\np-1} h(\bu_l, \br) \right|
  &\leq
    \left| \frac{1}{\np-1} \sum_{l=1}^{\np-1} ( h (\bu_l, \br)
                        - h (\bu_l, \br_j)) \right|
   + \left| \frac{1}{\np-1} \sum_{l=1}^{\np-1} h (\bu_l, \br_j) \right| \\
  &\leq     \frac{1}{\np-1} \sum_{l=1}^{\np-1} \left|  h (\bu_l, \br)
                        - h (\bu_l, \br_j)  \right|
    + \max_{m=1,\ldots,n}
    \left| \frac{1}{\np-1} \sum_{l=1}^{\np-1}   h(\bu_l, \br_m) \right|\\
  &\leq     \frac{1}{\np-1} \sum_{l=1}^{\np-1} (g(\bu_l) + \varepsilon)
    + \max_{m=1,\ldots,n} \left| \frac{1}{\np-1} \sum_{l=1}^{\np-1} h(\bu_l, \br_m) \right|.
\end{align*}

Taking the supremum over $\br \in A$ and then the $\limsup$ as
$\np \to \infty$, it follows again from \eqref{qj:zkj:iid} and the Law
of Large Numbers that
\begin{align}
  \limsup_{\np \to \infty} \sup_{\br \in A} \left| \frac{1}{\np-1} \sum_{l=1}^{\np-1} h(\bu_l, \br) \right|
  \leq \int g(\bu) \nu(d\bu) + \varepsilon +
  \limsup_{\np \to \infty} \max_{j=1,\ldots,n} \left| \frac{1}{\np-1}
  \sum_{l=1}^{\np-1} h(\bu_l, \br_j) \right|
  \leq 2\varepsilon,
  \label{eq:MTM:final:LLN}
\end{align}
almost surely.  Note that for the second inequality we also used the
fact that $\int_\qsp h(\bu, \br) \nu(d\bu) = 0$ for any $\br \in A$.

Hence, from \eqref{eq:MTM:final:LLN}, the set
\begin{align*}
  \Omega_\varepsilon := \left\{ \limsup_{\np \to \infty}
  \sup_{\br \in A} \left| \frac{1}{\np-1}
  \sum_{l=1}^{\np-1} h (\bu_l, \br) \right| < \varepsilon \right\},
\end{align*}
is such that $\Prb(\Omega_\varepsilon) = 1$ for all $\varepsilon >
0$. Therefore, defining
$\Omega_0 := \bigcap_{k=1}^\infty \Omega_{\frac{1}{k}}$, it follows by
continuity of measures that
\begin{align*}
  \Prb(\Omega_0)
  = \Prb \left( \lim_{\np \to \infty}
  \sup_{\br \in A} \left| \frac{1}{\np-1} \sum_{l=1}^{\np-1} h (\bu_l,
  \br) \right| = 0 \right)
	= \lim_{K \to \infty} \Prb(\bigcap_{k=1}^K \Omega_{\frac{1}{k}})
  = 1,
\end{align*}
which shows \eqref{lim:p:ulln}.  With this in hand the proof of
\cref{thm:MTM:limit} is now complete.

\subsection[Proof of Proposition~\ref{prop:MTM:glob:p:inf:lim}]
{Proof of \cref{prop:MTM:glob:p:inf:lim}}
\label{app:MTM:glob:p:inf:lim}

The proof consists in showing that the conditions \ref{A3}--\ref{A2}
of \cref{thm:MTM:limit} hold for each of the given cases (i) and (ii).
Start with (i), beginning with the verification of \ref{A3}.  Here,
cf. \eqref{eq:trgt:beta:frm}, we have
\begin{align}
  \label{prop:MTM:glob:pf:1}
  \int \beta(\tbq, \br) \pk(\br, d \tbq)
  = Z^{-1} \int \dmu(\tbq)
  \exp\left( - \frac{1}{2 \sigma^2}
  (|\cC^{-1/2} \tbq|^2  - 2\langle \cC^{-1} \tbq, \br \rangle + |\cC^{-1/2} \br|^2)  
  \right) d \tbq
\end{align}
where $Z = \sqrt{(2 \pi )^{d} \det(\sigma^2\cC)}$.  This implies the
first bound \eqref{inf:int:beta:nu}.  Regarding \eqref{cond:eta:art},
we observe that the first bound holds trivially since \eqref{eq:cond}
occurs here, as observed in \cref{eq:red:alpha:MTM}, given our assumed
form of the target $\mu$ and single proposal kernel $\pk$.  Meanwhile,
the second bound in \eqref{cond:eta:art} is immediate since, due to
the integrability of $\dmu$ in $\RR^d$, implied by $\mu$ being a
probability measure, we can, if necessary, redefine $\dmu$ on a set of
measure zero in $\RR^d$ so that it is finite everywhere.

Regarding \ref{A1}, observe that we
may take
\begin{align}
  \label{eq:push:it:gauss:1}
  F(\bq_0, \bu) = \bq_0 + \sigma \bu \text{ and } \nu = N(0, \cC).
\end{align}
As far as \ref{A4} is concerned, notice our assumptions in (i),
\eqref{eq:fd:dmu:MTM:2} and \eqref{eq:fd:dmu:MTM:3}, immediately imply
\eqref{cont:art}, \eqref{eq:g:loc:MTM:cond}.  Finally, as concerns the
final condition \ref{A2}, observe that for any $\bq_0$ and compact
$A \subset \qsp$
\begin{align*}
  \int \beta(\tbq, \bq_0)  \indFn{A^c}(\tbq)\pk(\bq_0, d\tbq)
  =   \int \dmu(\tbq)  \indFn{A^c}(\tbq)\pk(\bq_0, d\tbq)
  \leq Z^{-1} \int \dmu(\tbq)  \indFn{A^c}(\tbq) d \tbq,
\end{align*}
where $Z^{-1}$ is the normalizing constant for $\pk$ as in
\eqref{prop:MTM:glob:pf:1}. Since any compact set in $\RR^d$ is
bounded, the final condition \eqref{eq:comp:exterior:demand:MTM} now
follows from the fact that $\dmu$ is a probability density, and hence
integrable. This completes the proof in the first case.

We turn to address the second case (ii).  For \ref{A3}, recalling that
$\art$ is given as in \eqref{def:art:MTpCN} and then employing our
assumption \eqref{cond:Pot:F:lip:1}, the required upper bound in
\eqref{inf:int:beta:nu} follows immediately. Regarding the lower bound
in \eqref{inf:int:beta:nu}, we notice that
\begin{align}
  \int \art(\tbq, \br) \pk(\br, d \tbq)
  = \int \exp(- \Pot( \rho \br + \sqrt{1 - \rho^2} \bxi))
  \mu_0(d \bxi)
  \label{eq:a3:pushforward:rep:pcond}
\end{align}
and then argue by contradiction as follows.  Suppose, to the contrary,
that we can find a compact set $A$ such that
$\inf_{\br \in A}\int \art(\tbq, \br) \pk(\br, d \tbq) =0$.  This in
turn would imply the existence of a sequence of
$\{\br_k\}_{k \geq 1} \subset A$ with
$\int \art(\tbq, \br) \pk(\br_k, d \tbq) \leq 1/k$.  Due to the
compactness of $A$, there would exist a convergent subsequence
$\{\br_{k_j}\}_{j \geq 1}$ whose limit $\bar{\br}$ lies in $A$.  But
again with the lower bound condition \eqref{cond:Pot:F:lip:1}, the
representation \eqref{eq:a3:pushforward:rep:pcond} and the Dominated
Convergence theorem, we would in turn find that
$\int \exp(- \Pot( \rho \bar{\br} + \sqrt{1 - \rho^2} \bxi)) \mu_0(d
\bxi)= 0$.  Since, under our assumptions $\exp(- \Pot(\bz)) > 0$ for
every $\bz$, we have reached a contradiction.  For
\eqref{cond:eta:art}, as shown in \cref{eq:red:alpha:MTM}, we recall
that under the assumed form of the target measure
\eqref{eq:stuart:form:og}, \eqref{eq:s:f:infinite:d} and the fact that
$\pk_\rho$ and $\mu_0$ maintain \eqref{eq:rev:spec}, it follows that
\eqref{eq:cond} holds.  Meanwhile, the second requirement in
\eqref{cond:eta:art} is clear.

Regarding \ref{A1}, we set
\begin{align}
 \label{eq:push:it:gauss:2}
F(\br, \bu) = \rho \br + \sqrt{1 - \rho^2} \bu \text{ and }
\nu = \mu_0 = N(0, \cC).
\end{align}
As far as \ref{A4} is concerned, this
condition is fulfilled as a direct consequence of
\eqref{cond:Pot:F:lip:2}, \eqref{cond:Pot:F:lip} and the observation
that, with the Mean Value theorem and \eqref{cond:Pot:F:lip:1}, we
have
\begin{align}
  |\art( F(\br, \bu), \br) -\art( F(\tbr, \bu), \tbr)|
  =& | \exp( - \Pot( \rho \br +\sqrt{1 - \rho^2} \bu))-
     \exp( - \Pot( \rho \tbr +\sqrt{1 - \rho^2} \bu))|
  \notag\\
  \leq& \exp(-M)| \Pot( \rho \br +\sqrt{1 - \rho^2} \bu)
        - \Pot( \rho \tbr +\sqrt{1 - \rho^2} \bu)|,
        \label{eq:weird:lipshitz:1}
\end{align}
for any $\br,\tbr, \bu \in \qsp$.

In order to establish the final condition \ref{A2}, note that under
the conditions given on the covariance operator $\cC$, we can find a
complete orthonormal basis $\{\mathbf{e}_k\}$ for $H$ consisting of
eigenvectors of $\cC$ with corresponding eigenvalues
$\{\lambda_k^2\}$, so that
\begin{align}
    \label{eq:tr:cond}
  \cC \mathbf{e}_k = \lambda_k^2 \mathbf{e}_k
\quad
  \text{ where }
  \sum_{k =1}^\infty \lambda_k^2 < \infty.
\end{align}  
In view of \eqref{eq:tr:cond}, we can select a sequence of
positive, increasing numbers $c_k$ such that moreover
\begin{align}
  \label{eq:Lambda:seq}
  \sum_{k =1}^\infty c_k \lambda_k^2 < \infty.
\end{align}
Form the space
\begin{align}
  \label{eq:comp:g:space:norm}
  Z := \{ \bq \in X | \| \bq\|_{Z} < \infty\}, \quad
\text{
endowed with the norm 
}
  \| \bq\|_{Z}^2 := \sum_{k =1}^\infty 
  c_k\langle \bq,\mathbf{e}_k   \rangle^2.
\end{align}
It is not hard to see that $Z$ is compactly embedded in
$X$.\footnote{Indeed, consider any sequence $\{\bq_j\}$ which is
  bounded in $Z$, namely such that
  $a_0 = \sup_{j} \sum_{k =1}^\infty c_k \langle \bq_j, \mathbf{e}_k
  \rangle^2 < \infty$.  Since $\qsp$ is a separable Hilbert space,
  then $\{\bq_j\}$ is weakly compact in $\qsp$. Thus, there exists a
  subsequence $\{\bq_{l_j}\}_{j \geq 1}$ and limit point
  $\bq_\infty \in Z$ such that $\| \bq_\infty \|_Z^2 \leq a_0$ and
  \begin{align}\label{lim:qj:wk}
    \lim_{j \to \infty} \langle \bq_{l_j}, \mathbf{e}_k \rangle
    = \langle \bq_\infty, \mathbf{e}_k \rangle \quad \mbox{ for every } \,\, k \geq 1,
  \end{align}
  where $\langle \cdot, \cdot \rangle$ denotes the inner product in
  $\qsp$. Hence, denoting by $\| \cdot \|$ the norm in $\qsp$ induced
  by $\langle \cdot, \cdot \rangle$, we have
  \begin{align*}
  	\| \bq_{l_j} - \bq_\infty \|^2  
  	=  \sum_{k=1}^\infty \langle \bq_{l_j} - \bq_\infty, \mathbf{e}_k \rangle^2  
  	&\leq \sum_{k=1}^N \langle \bq_{l_j} - \bq_\infty, \mathbf{e}_k \rangle^2  
  	+ c_{N+1}^{-1} \sum_{k= N+1}^\infty c_k \langle \bq_{l_j} - \bq_\infty, \mathbf{e}_k \rangle^2 
  	\leq \sum_{k=1}^N \langle \bq_{l_j} - \bq_\infty, \mathbf{e}_k \rangle^2   
  	+ c_{N+1}^{-1} 4 a_0.
  \end{align*}
  Invoking \eqref{lim:qj:wk}, it thus follows that
  $\limsup_{j \to \infty} \| \bq_{l_j} - \bq_\infty \|^2 \leq
  c_{N+1}^{-1} 4 a_0$ for all $N \geq 1$. Since the sequence $\{c_k\}$
  is increasing, then
  $\lim_{j \to \infty} \| \bq_{l_j} - \bq_\infty \| = 0$, i.e. this
  subsequence $\{\bq_{l_j}\}$ converges in $\qsp$.}  Furthermore, in
view of the fact we can generate $\bxi \sim \mu_0$ as
$\bxi := \sum_{k =1}^\infty \lambda_k \mathbf{e}_k \xi_k$, where
$\xi_k$ are i.i.d. as $N(0,1)$, it follows that
\begin{align}
  \label{eq:z:mom}
  \int \| \br \|^2_Z \mu_0(d \br)
  = \mathbb{E} \sum_{k =1}^\infty
  c_k \langle \bxi, \mathbf{e}_k\rangle^2 
  = \sum_{k =1}^\infty c_k \lambda_k^2 <\infty.
\end{align}
See e.g. \cite{DPZ2014}.

Given $R > 0$ and $\bbq \in \qsp$, denote as $B_R^Z(\bbq)$ the closed
ball in $Z$ with radius $R$ and centered at $\bbq$.  Note that
$B_R^Z(\bbq)$ is a compact set in $\qsp$, since $Z$ is compactly
embedded in $\qsp$.  Observe that, for any $R > 0$ and $s > 1$, by
H\"older's inequality
\begin{align}
  \int 
  \art(\tbq, \bq_0)  \indFn{B_R^{Z}(\rho \bq_0)^c} (\tbq)\pk_\rho(\bq_0,d\tbq)
  =&   \int 
 \exp(- \Pot(\tbq))  \indFn{B_R^{Z}(\rho \bq_0)^c} (\tbq)
     \pk_\rho(\bq_0,d\tbq)
  \notag\\
  \leq&  \left(\int \exp(-s \Pot(\tbq)) \pk_\rho(\bq_0,d\tbq)\right)^{1/s}
        \pk_\rho(\bq_0 , B_R^{Z}(\rho \bq_0)^c)^{\frac{s-1}{s}}
        \label{int:BRZc:art}
\end{align}
On the other hand, recalling \eqref{eq:Guass:1:prop:pCN} and then using
an infamous inequality variously attributed to Ir\'en\'ee-Jules
Bienaym\'e or to Pafnuty Chebyshev, we have
\begin{align}
  \label{eq:IJBienaymePafnutyChebyshevinequality}
  \pk_\rho(\bq_0 , B_R^{Z}(\rho \bq_0)^c)
  = \mu_0( \sqrt{1 - \rho^2} \| \bxi\|_Z > R)
  \leq \frac{1- \rho^2}{R^2}  \int \| \bxi \|_Z^2 \mu_0(d \bxi).
\end{align}
Combining \eqref{int:BRZc:art},
\eqref{eq:IJBienaymePafnutyChebyshevinequality} and then invoking our
assumption that $\Pot$ is bounded from below by $M$,
\eqref{cond:Pot:F:lip:1}, it follows that
\begin{align}\label{ineq:int:BRZc}
	\int 
	 \art(\tbq, \bq_0) \indFn{B_R^{Z}(\rho \bq_0)^c} (\tbq) \pk(\bq_0,d\tbq)
  \leq e^{-M}
  \left( \int \| \bxi \|_Z^2 \mu_0(d \bxi) \frac{1 - \rho^2}{R^2} \right)^{\frac{s-1}{s}}.
\end{align}
Thus, with \eqref{eq:z:mom}, we deduce that for any $\varepsilon > 0$,
we may choose $R > 0$ sufficiently large so that
\eqref{eq:comp:exterior:demand:MTM} holds, and hence \ref{A2} is
verified.  With this in hand, the proof is complete.

\subsection[Proof of Proposition~\ref{prop:local:MTM:pinf:lim:0}]
{Proof of \cref{prop:local:MTM:pinf:lim:0}}
\label{app:local:local:MTM:pinf:lim:0}

As in the proof of \cref{prop:MTM:glob:p:inf:lim}, we verify conditions
\ref{A3}--\ref{A2} in order to infer the desired result from
\cref{thm:MTM:limit}. 
To prove \ref{A3}, observe that (cf. \eqref{prop:MTM:glob:pf:1})
\begin{align*}
  \int \beta(\tbq, \br) \pk(\br, d \tbq)
  \geq \frac{1}{Z\sqrt{\dmu(\br)}}
  \int \sqrt{\dmu(\tbq)}
   \exp\left( - \frac{1}{ \sigma^2}
   (|\cC^{-1/2} \tbq|^2  + |\cC^{-1/2} \br|^2)  
   \right) d \tbq,
\end{align*}
with $Z = \sqrt{(2 \pi )^{d} \det(\sigma^2\cC)}$.  With
\eqref{eq:MTM:loc:strong:dmu:cond}, this yields the lower bound in
\eqref{inf:int:beta:nu}.  Meanwhile, for the upper bound in
\eqref{inf:int:beta:nu}, we simply note that
$\int \beta(\tbq, \br) \pk(\br, d \tbq) \leq \sqrt{\|
  \dmu\|_\infty/\dmu(\br)}$.  For \eqref{cond:eta:art}, we first note
that \eqref{eq:cond} again holds as observed in the formulation of
\cref{meth:loc:MTM}, so that the first bound in \eqref{cond:eta:art}
is clear.  For the second bound in \eqref{cond:eta:art}, note that
\eqref{eq:MTM:loc:strong:dmu:cond} together with compactness
immediately implies that
\begin{align}
  \label{eq:min:den:cond:1}
  \gamma_R := \inf_{|\bq | \leq R} \dmu(\bq) > 0 \quad
  \text{ for any } R > 0.
\end{align}
The condition \ref{A1} is satisfied precisely as in
\eqref{eq:push:it:gauss:1}.

Regarding the final condition \ref{A4}, we observe that,
for any $R, R' > 0$, given any $\br, \tbr \in B_R$
\begin{align}
  | &\art(F(\br,\bu), \br) - \art(F(\tbr,\bu), \tbr)|
    \leq
    \left|
    \frac{\sqrt{\dmu( \sigma \bu + \br)}}{\sqrt{\dmu(\br)} }
    - \frac{\sqrt{\dmu( \sigma \bu + \tbr)}}{\sqrt{\dmu(\tbr)} } 
    \right| \indFn{B_{R'}}(\bu)
    +  2 \sqrt{\frac{ \|\dmu\|_\infty}{\gamma_R}}
      \indFn{B_{R'}^C}(\bu)
  \notag\\  
    &\leq
      \left(\frac{1}{\sqrt{\gamma_R}}
  | \sqrt{\dmu( \sigma \bu + \br)}
    - \sqrt{\dmu( \sigma \bu + \tbr)}|
      + \frac{\sqrt{\|\dmu\|_\infty}}{\gamma_R}
        | \sqrt{\dmu( \br)}
    - \sqrt{\dmu( \tbr)}|\right)
    \indFn{B_{R'}}(\bu)
    +   2\sqrt{\frac{ \|\dmu\|_\infty}{\gamma_R}}
      \indFn{B_{R'}^C}(\bu)
  \notag\\
          &\leq
            \frac{1}{\sqrt{\gamma_R \gamma_{R + |\sigma| R'}}}
  | \dmu( \sigma \bu + \br)
    - \dmu( \sigma \bu + \tbr)|
      + \sqrt{\frac{\|\dmu\|_\infty}{\gamma_R^3}}
       | \dmu( \br)
            - \dmu( \tbr)|
    +   2\sqrt{\frac{ \|\dmu\|_\infty}{\gamma_R}}
            \indFn{B_{R'}^C}(\bu)
            \label{eq:delta:eps:pushups}
\end{align}
With the bound \eqref{eq:delta:eps:pushups} in hand, we now obtain
\eqref{eq:comp:exterior:demand:MTM} from our assumption in
\eqref{cond:Pot:F:lip:2:loc}, \eqref{cond:Pot:F:lip:loc} as follows.
Given any $R >0$ and any $\epsilon > 0$, first select $R' >0$ large
enough such that
$2\sqrt{ \|\dmu\|_\infty/\gamma_R}\nu(B_{R'}^C) < \epsilon/2$ for
$\nu = N(0,\cC)$.  Next, under \eqref{cond:Pot:F:lip:2:loc},
\eqref{cond:Pot:F:lip:loc}, we may then pick a measurable function
$g_0$ and $\delta_0 > 0$ such that
$| \dmu( \sigma \bu + \br) - \dmu( \sigma \bu + \tbr)| < g_0(\bu)$
whenever $|\br - \tbr| < \delta_0$ and $|\br|, |\tbr| \leq R$ and so
that
$\int g_0(\bu) \nu (d \bu)< \sqrt{\gamma_R \gamma_{R + |\sigma|
    R'}}\epsilon/4$.  Now, according to the Heine–Cantor theorem,
$\dmu$ is uniformly continuous on $B_R$, so that we can find
$\delta \leq \delta_0$ such that
$| \dmu( \br) - \dmu( \tbr)| < \frac{\epsilon}{4}
\sqrt{\frac{\gamma_R^3}{ \|\dmu\|_\infty}}$ for all
$|\br - \tbr| < \delta$ such that $|\br|, |\tbr| \leq R$.  Combining
these elements, we now take
\begin{align*}
  g(\bu) := \frac{1}{\sqrt{\gamma_R \gamma_{R + |\sigma| R'}}} g_0(\bu)
  + \epsilon/4 + 2\sqrt{\frac{ \|\dmu\|_\infty}{\gamma_R}}
            \indFn{B_{R'}^C}(\bu),
\end{align*}
and find that with this $g$ and $\delta >0$, we satisfy the requirements
\eqref{cont:art} and \eqref{eq:g:loc:MTM:cond} commensurate to
our given arbitrary $\epsilon >0$ and compact set $B_R$.

Finally, for \ref{A2}, we simply observe that, for any $\bq_0 \in \RR^d$
and any $R > 0$,
\begin{align*}
\int \art(\tbq, \bq_0) \indFn{B_R^c} (\tbq) Q(\bq_0, d \tbq)
\leq \sqrt{\|\dmu\|_\infty/\dmu(\bq_0)} Q(\bq_0, B_R^c),
\end{align*}
which thus yields \eqref{eq:comp:exterior:demand:MTM}.  The proof of
\cref{prop:local:MTM:pinf:lim:0} is complete.

\subsection[Proof of Proposition~\ref{prop:local:MTM:pinf:lim}]
{Proof of \cref{prop:local:MTM:pinf:lim}}
\label{app:local:local:MTM:pinf:lim}

We once again verify the conditions \ref{A3}--\ref{A2} of
\cref{thm:MTM:limit}.  For  \ref{A3}, we refer back to
\eqref{eq:loc:pCN:beta} and observe that
\begin{align*}
  \int \art(\tbq, \br) \pk_\rho(\br, d \tbq)
  = \int \exp\left( -
  \frac{
  \rho(1 - \rho)}{1 - \rho^2}(\Pot( (\rho \br + \sqrt{1- \rho^2}\bxi)/\rho)-
  \Pot(\br))\right)
  \mu_0( d \bxi)
\end{align*}
A compactness argument very similar to the one given immediately after
\eqref{eq:a3:pushforward:rep:pcond} thus yields the lower bound
required by \eqref{inf:int:beta:nu}, whereas the upper bound follows
from assumption \eqref{cond:Pot:F:lip:1}.  The bounds demanded in
\eqref{cond:eta:art} are clear from the assumed continuity of $\Pot$,
\eqref{eq:RN:eta:etaperp}, \eqref{eq:loc:pCN:beta} and from the fact
that
\begin{align*}
  \frac{d \eta^\perp}{d \eta}(\bq_0, \br)
  \frac{\art(\bq_0,\br)}{\art(\br,\bq_0)}
  = \frac{ \exp( -\Pot(\br) - \frac{\rho(1-\rho)}{1
  -\rho^2}(\Pot(\bq_0/\rho) - \Pot(\br)) ) }
  {\exp( - \Pot(\bq_0)  -
  \frac{\rho(1-\rho)}{1 -\rho^2} (\Pot(\br/\rho) - \Pot(\bq_0)))  }.
\end{align*}
The condition \ref{A1} is maintained under the same $F$ and $\nu$ as
in \eqref{eq:push:it:gauss:2}.  As regards the conditions \eqref{cont:art} and
\eqref{eq:g:loc:MTM:cond} in \ref{A4}, we begin by arguing similarly as
in \eqref{eq:weird:lipshitz:1}. Namely, with the aide of the Mean Value
theorem and our lower bound assumption \eqref{cond:Pot:F:lip:1}, we have that
\begin{align*}
  |\art( F(\br&, \bu), \br) -\art( F(\tbr, \bu), \tbr)|\\
  \leq&  \exp\left(\frac{\rho(1-\rho)}{1
    -\rho^2}\Pot(\br)\right)
    \left|
   \exp\biggl(-\frac{\rho(1-\rho)}{1
    -\rho^2}\Pot\bigl(\br +
         \frac{\sqrt{1 - \rho^2}}{\rho} \bu\bigr)\biggr)
    -
   \exp\biggl(-\frac{\rho(1-\rho)}{1-\rho^2}
    \Pot\bigl(\tbr +
    \frac{\sqrt{1 - \rho^2}}{\rho} \bu\bigr)\biggr)
        \right|\\
       &+ \exp\left(- M \frac{\rho(1-\rho)}{1  -\rho^2}\right)
         \left|
         \exp\left(\frac{\rho(1-\rho)}{1
         -\rho^2}\Pot(\br)\right)-
         \exp\left(\frac{\rho(1-\rho)}{1
    -\rho^2}\Pot(\tbr)\right)
         \right|\\
           \leq&  \exp\left(\frac{\rho(1-\rho)}{1
    -\rho^2}(\Pot(\br) -M)\right)
    \left|\Pot\bigl(\br +
         \frac{\sqrt{1 - \rho^2}}{\rho} \bu\bigr)
    -
    \Pot\bigl(\tbr +
    \frac{\sqrt{1 - \rho^2}}{\rho} \bu\bigr)
        \right|\\
       &+ \exp\left(- M \frac{\rho(1-\rho)}{1  -\rho^2}\right)
         \left|
         \exp\left(\frac{\rho(1-\rho)}{1
         -\rho^2}\Pot(\br)\right)-
         \exp\left(\frac{\rho(1-\rho)}{1
    -\rho^2}\Pot(\tbr)\right)
         \right|,
\end{align*}
for any $\br, \tbr \in \qsp$.  With this upper bound in hand, we now
fix any $\epsilon >0$ and $A \subset \qsp$ compact.  Making another
incantation of the Heine–Cantor theorem, we obtain a $\delta_0 > 0$ so
that
$|e^{\frac{\rho(1-\rho)}{1 -\rho^2}\Pot(\br)} -
e^{\frac{\rho(1-\rho)}{1 -\rho^2}\Pot(\tbr)}| < \epsilon e^{M
  \frac{\rho(1-\rho)}{1 -\rho^2}}/2$ whenever
$|\br - \tbr| < \delta_0$ and $\br,\tbr \in A$.  Next, justified by
our assumption \eqref{eq:some:local:pot}, we can find a
$\delta \leq \delta_0$ and a $g_0: \qsp \to \RR^+$ where
$\int g_0(\bu) \mu_0(d \bu)< \epsilon e^{ -\frac{\rho(1-\rho)}{1
    -\rho^2}(\sup_{\tbq \in A}\Pot(\tbq) -M)}/2$ and such that
$|\Pot(\br + \frac{\sqrt{1 - \rho^2}}{\rho} \bu) - \Pot(\tbr +
\frac{\sqrt{1 - \rho^2}}{\rho} \bu)| < g_0(\bu)$ for any pair with
$|\br - \tbr| < \delta$ $\br,\tbr \in A$.  Hence, the required
relationship for the given $\epsilon$ and $A$ is maintained with the
$\delta >0$ we just found and by taking
$g(\bu) = g_0(\bu) e^{ \frac{\rho(1-\rho)}{1 -\rho^2}(\sup_{\tbq \in
    A}\Pot(\tbq) -M)} + \epsilon/2$.

Finally, to establish \ref{A2}, we argue analogously to
\eqref{int:BRZc:art}--\eqref{ineq:int:BRZc} above.  In our case here,
defining the compact subspace $Z$ exactly as we did previously
in \eqref{eq:comp:g:space:norm}, we have that
\begin{align*}
  &\int 
  \art(\tbq, \bq_0)  \indFn{B_R^{Z}(\rho \bq_0)^c}
  (\tbq)\pk_\rho (\bq_0,d\tbq) \\
  &\qquad\qquad\qquad\qquad\qquad = \exp\left(\frac{\rho(1-\rho)}{1
  -\rho^2}\Pot(\bq_0)\right)
  \int \exp\left(-\frac{\rho(1-\rho)}{1
  -\rho^2}\Pot(\tbq/\rho)\right)
  \indFn{B_R^{Z}(\rho \bq_0)^c} (\tbq) \pk_\rho(\bq_0,d\tbq).
\end{align*}
With this identity in view, one then obtains a bound similar to
\eqref{ineq:int:BRZc}, up to some different constants depending only
on $\bq_0$ and $\rho > 0$, and then immediately infers the condition
\eqref{eq:comp:exterior:demand:MTM}.  The proof of
\cref{prop:local:MTM:pinf:lim} is now complete.

\subsection[Proof of Proposition~\ref{prop:Slingshot:p:lim}]
{Proof of \cref{prop:Slingshot:p:lim}}
\label{app:proof:of:slingshot:lim}

The first case (i) is an obvious consequence of
\cref{cor:large:p:prod}.  Turning to (ii), observe that under our
Gaussian assumption for $\pk$, \eqref{eq:Guass:1:prop:classic}, and in
view of \eqref{eq:ss:gen:beta}, we have that
\begin{align}
  \label{eq:SS:beta:gauss}
  \art(\tbq, \bq_0) = \sqrt{ (2\pi \sigma^2 )^d \det(\cC)}
  \exp\left( \frac{1}{2 \sigma^2} |\cC^{-1/2}( \tbq -
  \bq_0)|^2\right)
  \dmu(\tbq).
\end{align}
Once again we are in the business of verifying the conditions of
\cref{thm:MTM:limit} to establish the desired weak convergence as
$\np \to \infty$.  With \eqref{eq:SS:beta:gauss} and noting,
cf. \eqref{eq:etaperp:eta:SS}, that \eqref{eq:cond} holds, it is clear
that all the conditions for \ref{A3} hold.  Concerning \ref{A1}, we
select $F$ and $\nu$ exactly as above in \eqref{eq:push:it:gauss:1}.
As regards \ref{A4}, we observe that, for any $R, R' > 0$,
\begin{align}
  | \art(F(&\br, \bu), \br) - \art(F(\tbr, \bu), \tbr)|
             = \mathcal{G}(\bu)  |
             \dmu(\br + \sigma \bu )
             - \dmu(\tbr +\sigma \bu) |
             \notag\\
  \leq&  \mathcal{G}(\bu)\left(
        |
        \dmu(\br + \sigma \bu )
        - \dmu(\tbr +\sigma \bu) |
        \indFn{B_{R'}}(\bu)
        + 2 
        \gamma_R( \sigma\bu) \dmu(\sigma \bu )          
        \indFn{B_{R'}^C}(\bu) \right),
          \label{eq:MTM:slingshot:mess}
\end{align}
where
$\mathcal{G}(\bu) = e^{\frac{1}{2}|\cC^{-1/2}\bu|^2} \sqrt{(2\pi \sigma^2)^d
\det(\cC) }$ and $\gamma_R$ is as in the bound
\eqref{eq:uniform:den:bnd}.  We now obtain the conditions
\eqref{cont:art}, \eqref{eq:g:loc:MTM:cond} from
\eqref{eq:MTM:slingshot:mess} by arguing as follows. Given
$\epsilon >0$ and $R> 0$, select $R' > 0$ large enough so that 
$\int_{B_{R'}^c} 2|\sigma|^{-d} \gamma_R( \sigma\bu) \dmu(\sigma
\bu ) < \epsilon/2$.  Having now made this choice of $R' >0$, we invoke
Heine-Cantor yet again to justify the existence of a
$\delta = \delta(R, R', \epsilon) > 0$ such that for all
$\br, \tbr \in B_R$ with $|\br - \tbr| < \delta$ and $\bu \in B_{R'}$,
$|\dmu(\br + \sigma \bu ) - \dmu(\tbr +\sigma \bu) | <
|\sigma|^{d}|B_{R'}|^{-1}\epsilon/2$.  Thus, for the given
$\epsilon >0$, $R >0$ we select the corresponding function
$g(\bu) = \mathcal{G}(\bu)( \frac{1}{2 |B_{R'}|}|\sigma|^{d}\epsilon
\indFn{B_{R'}}(\bu) + 2|\sigma|^{-d} \gamma_R( \sigma\bu)
\dmu(\sigma \bu ) \indFn{B_{R'}^c})$ and the $\delta >0$ we just
found that maintain \eqref{cont:art}, \eqref{eq:g:loc:MTM:cond} .
Finally, noting that
$\int \indFn{B_R^c}\art(\tbq, \bq_0) Q(\bq_0, d \tbq) = \mu(B_R^c)$,
\ref{A2} also follows immediately.  The proof is therefore complete.

\subsection[Proof of Proposition~\ref{prop:Glob:param:Limit}]
{Proof of \cref{prop:Glob:param:Limit}}
\label{app:proof:of:large:sig}

Start with the claims in \eqref{eq:rho:to:zero:Pinf:prop}.  Here observe
that, for any bounded continuous function $\phi: \qsp \to \RR$, we have
\begin{align*}
  \int \phi(\tbq) \mk^\infty_{BB,\rho}(\bq_0, d \tbq)
  = \frac{\int \phi( \tbq) e^{-\Pot(\tbq)} \pk_\rho(\bq_0, d \tbq)}
  {\int e^{-\Pot(\tbq)} \pk_\rho(\bq_0, d \tbq)}
  = \frac{\int \phi( \rho \bq_0 + \sqrt{1- \rho^2}\bxi)
  e^{-\Pot(\rho \bq_0 + \sqrt{1- \rho^2}\bxi)} \mu_0( d\bxi)}
 {\int e^{-\Pot(\rho  \bq_0 + \sqrt{1- \rho^2}\bxi)}
 \mu_0( d\bxi)}
\end{align*}
so that the weak limit
$\lim_{\rho \to 0} \mk^\infty_{BB,\rho}(\bq_0, d \tbq) =\mu(d\tbq)$
follows from the assumption that $\Pot$ is bounded from below and the
Dominated Convergence theorem.

As regards $\mk^\infty_{TJ, \rho}$, we have
\begin{align*}
  \int& \phi(\tbq) \mk^\infty_{TJ,\rho}(\bq_0, d \tbq)
  = \int \frac{\int \phi( \tbq) e^{-\Pot(\tbq)} \pk_\rho(\bbq, d \tbq)}
  {\int e^{-\Pot(\tbq)} \pk_\rho(\bbq, d \tbq)}\pk_\rho(\bq_0, d \bbq)\\
  &=\int \frac{\int \phi( \rho \bbq + \sqrt{1- \rho^2}\bxi)
  e^{-\Pot(\rho \bbq + \sqrt{1- \rho^2}\bxi)} \mu_0( d\bxi)}
 {\int e^{-\Pot(\rho  \bbq + \sqrt{1- \rho^2}\bxi)}
    \mu_0( d\bxi)}\pk_\rho(\bq_0, d \bbq)\\
    &=\int \frac{\int \phi( \rho^2 \bq_0 + \rho\sqrt{1- \rho^2}\tbxi+ \sqrt{1- \rho^2}\bxi)
  e^{-\Pot(\rho^2 \bq_0 + \rho\sqrt{1- \rho^2}\tbxi+\sqrt{1- \rho^2}\bxi)} \mu_0( d\bxi)}
 {\int e^{-\Pot(\rho^2 \bq_0 + \rho\sqrt{1- \rho^2}\tbxi+\sqrt{1- \rho^2}\bxi)}  
    \mu_0( d\bxi)} \mu_0( d\tbxi).
\end{align*}
Thus, once again, the assumption that $\Pot$ is bounded from below
together with the Dominated Convergence theorem imply that
$\lim_{\rho \to 0} \mk^\infty_{TJ,\rho}(\bq_0, d \tbq) =\mu(d\tbq)$.

Finally, regarding  $\mk^\infty_{MT,\rho}$, we obtain by referring back to
\eqref{eq:MTM:inf:ker:BB} that
\begin{align*}
  \int \phi(&\tbq) \mk^\infty_{MT,\rho}(\bq_0, d \tbq)
     = \frac{\int 
  \left[ 1 \wedge \left(
    \frac{\int e^{ -\Pot(\bq')}  Q_\rho(\bq_0, d \bq') }
  { \int e^{-\Pot(\bq')} Q_\rho(\tbq, d \bq')} \right) \right]
 \phi( \tbq) e^{-\Pot(\tbq)} \pk_\rho(\bq_0, d \tbq)}
        {\int  e^{-\Pot(\tbq)} \pk_\rho(\bq_0, d \tbq)}
   \notag\\
      &+ \phi(\bq_0) \frac{
      \int \left(1- 
   1 \wedge \left(
     \frac{\int e^{ -\Pot(\bq')}  Q_\rho(\bq_0, d \bq') }
        { \int e^{ -\Pot(\bq')} Q_\rho(\tbq, d \bq')} \right)
        \right)
        e^{-\Pot(\tbq)} \pk_\rho(\bq_0, d \tbq)
       }{\int_\qsp e^{ -\Pot(\bq')} Q_\rho(\bq_0, d \bq')}
            \notag\\
    =&\frac{\int 
  1 \wedge \left(
    \frac{\int e^{ -\Pot(\rho \bq_0 + \sqrt{1- \rho^2}\bxi')}  \mu_0( d\bxi') }
  { \int e^{ -\Pot(\rho^2 \bq_0 + \rho \sqrt{1- \rho^2}  \bxi + \sqrt{1- \rho^2}\bxi')}\mu_0( d\bxi')  } \right)
\phi( \rho \bq_0 + \sqrt{1- \rho^2}\bxi)
  e^{-\Pot(\rho \bq_0 + \sqrt{1- \rho^2}\bxi)} \mu_0( d\bxi)}
        {\int e^{-\Pot(\rho  \bq_0 + \sqrt{1- \rho^2}\bxi)}
 \mu_0( d\bxi)}
       \notag\\
  &+\phi( \bq_0)\frac{\int \left( 1-
  1 \wedge \left(
    \frac{\int e^{ -\Pot(\rho \bq_0 + \sqrt{1- \rho^2}\bxi')}  \mu_0( d\bxi') }
  { \int e^{ -\Pot(\rho^2 \bq_0 + \rho \sqrt{1- \rho^2}  \bxi + \sqrt{1- \rho^2}\bxi')}\mu_0( d\bxi')  } \right)\right)
  e^{-\Pot(\rho \bq_0 + \sqrt{1- \rho^2}\bxi)} \mu_0( d\bxi)}
        {\int e^{-\Pot(\rho  \bq_0 + \sqrt{1- \rho^2}\bxi)}
 \mu_0( d\bxi)}.
\end{align*}
Thus, another application of
the Dominated Convergence theorem now yields the third claim in
\eqref{eq:rho:to:zero:Pinf:prop} that
$\lim_{\rho \to 0} \mk^\infty_{MT,\rho}(\bq_0, d \tbq) = \mu(d\tbq)$.

Let us now turn to the second set of weak limits recorded in
\eqref{eq:sigma:to:zero:Pinf}. The first limit,
$\lim_{\sigma \to \infty} \mk^\infty_{BB,\sigma}(\bq_0, d \tbq)
=\mu(d\tbq)$, clearly follows from \eqref{eq:weak:conv:iid} and the
Dominated Convergence theorem.  Turning to the second limit,
$\lim_{\sigma \to \infty} \mk^\infty_{TJ,\sigma}(\bq_0, d \tbq)
=\mu(d\tbq)$, let us first fix the convenient notation
\begin{align}\label{not:Cn:Cip}
  \Cn{\tbq} = |\cC^{-1/2} \tbq|, \quad
  \Cip{\tbq}{\bq'} =
  \langle \cC^{-1/2} \tbq, \cC^{-1/2} \bq' \rangle, \quad \tbq, \bq' \in \RR^d.
\end{align}
Then, referring back to \eqref{inf:ker:cond:ind} and applying a
similar logic as in \eqref{eq:weak:conv:iid}, we obtain that
\begin{align}
	&\left| \int \phi(\tbq) \mk^\infty_{TJ,\sigma}(\bq_0, d \tbq) -
	\int\phi(\tbq)\mu(d \tbq)\right|
	\notag\\
	&\qquad \leq \int \left|
	\frac{\int \phi(\bq')
		\dmu( \bq')
		\exp( - \frac{1}{2 \sigma^2} \Cn{\bq' -\tbq}^2) d \bq'}
	{\int \dmu( \bq') \exp( - \frac{1}{2 \sigma^2} \Cn{\bq' -
			\tbq}^2) d \bq'} - \int
	\phi(\bq')  \dmu( \bq')  d \bq'
	\right| \pk_\sigma( \bq_0, d\tbq)
	\notag\\
	&\qquad
	=  \int  2 \| \phi\|_\infty \wedge \left|
	\frac{\int \phi(\bq') \dmu( \bq')
		\exp( - \frac{1}{2 \sigma^2} \Cn{\bq' -\bq_0 - \sigma \bxi}^2) d \bq'}
	{\int \dmu( \bq') \exp( - \frac{1}{2 \sigma^2} \Cn{\bq' - \bq_0 - \sigma \bxi}^2) d \bq'} - \int
	\phi(\bq')  \dmu( \bq')  d \bq'
	\right| \nu_0( d \bxi)
	\notag\\
	&\qquad
	=      \int 2 \| \phi\|_\infty \wedge \biggl|
	\frac{\int \phi(\bq') \dmu( \bq')
		(\exp( - \frac{1}{2 \sigma^2} \Cn{\bq' -\bq_0 - \sigma \bxi}^2) - \exp(- \frac{1}{2} \Cn{\bxi}^2))d \bq'}
	{\int \dmu( \bq')  \exp( - \frac{1}{2 \sigma^2} \Cn{\bq' -\bq_0 - \sigma \bxi}^2)  d \bq'}
	\notag\\
	&\qquad \qquad \quad
	- \int
	\phi(\bq')  \dmu( \bq')  d \bq'  \frac{\int  \dmu( \bq')( \exp(- \frac{1}{2} \Cn{\bxi}^2) -
		\exp( - \frac{1}{2 \sigma^2} \Cn{\bq' -\bq_0 - \sigma \bxi}^2) ) d \bq'  }
	{\int \dmu( \bq') \exp( - \frac{1}{2 \sigma^2} \Cn{\bq' -\bq_0 - \sigma \bxi}^2)   d \bq'}
	\biggr| \nu_0( d \bxi)
	\notag\\
	&\qquad 
	\leq  2 \|\phi\|_\infty \int 1 \wedge
	\left| \frac{\int  \dmu( \bq')| \exp(- \frac{1}{2} \Cn{\bxi}^2) -
		\exp( - \frac{1}{2 \sigma^2} \Cn{\bq' -\bq_0 - \sigma \bxi}^2) | d \bq'  }
	{\int \dmu( \bq') \exp( - \frac{1}{2 \sigma^2} \Cn{\bq' -\bq_0 - \sigma \bxi}^2)    d \bq'}
	\right|\nu_0( d \bxi),
	\label{eq:TJ:sig:inf:lim}
\end{align}
where $\nu_0 = N(0, I)$.  Now, for any fixed $\bxi$, the Dominated
Convergence theorem implies that
\begin{align*}
	\lim_{\sigma \to \infty}\int  \dmu( \bq')| e^{- \frac{1}{2} \Cn{\bxi}^2} -
	e^{ - \frac{1}{2 \sigma^2} \Cn{\bq' -\bq_0 - \sigma \bxi}^2}| d \bq' =0,
	\quad
  \lim_{\sigma \to \infty}\int
  \dmu( \bq')e^{ - \frac{1}{2\sigma^2} \Cn{\bq' -\bq_0 - \sigma \bxi}^2} d \bq'
  = e^{- \frac{1}{2} \Cn{\bxi}^2} .
\end{align*}   
These limits, \eqref{eq:TJ:sig:inf:lim} and a second application of
the Dominated Convergence theorem thus imply the second
claim in  \eqref{eq:sigma:to:zero:Pinf}.

The final claim concerns $\mk^\infty_{MT,\sigma}(\bq_0, d
\tbq)$. Referring back to \eqref{eq:MTM:inf:ker:BB},
\eqref{inf:ker:bb}, and again employing the notation
\eqref{not:Cn:Cip}, we obtain that
\begin{align}
	\biggl| \int& \phi(\tbq) \mk^\infty_{MT,\sigma}(\bq_0, d\tbq)
	- \int \phi(\tbq) \dmu(\tbq) d\tbq\biggr|
	\notag\\
	\leq&   \biggl| \int \!\phi(\tbq) \mk^\infty_{BB,\sigma}(\bq_0, d\tbq)
	- \! \int \!\phi(\tbq) \dmu(\tbq) d\tbq \biggr|
	+ 2\|\phi\|_{\infty} \!\int
	\left|1 -      1 \wedge 
	\frac{\int \dmu(\bq')  \pk_\sigma(\bq_0, d \bq') }
	{ \int \dmu(\bq') \pk_\sigma(\bbq, d \bq')}
	\right|
	\frac{\dmu(\bbq) \pk_\sigma(\bq_0, d \bbq)}
	{\int \dmu(\bq') \pk_\sigma(\bq_0, d \bq')}
	\notag\\
	\leq&\biggl| \int \! \phi(\tbq) \mk^\infty_{BB,\sigma}(\bq_0, d\tbq)
	-\! \int\! \phi(\tbq) \dmu(\tbq) d\tbq \biggr|
	\notag\\
	&\qquad\qquad
	+ 2\|\phi\|_{\infty} \!\int
	\left|1 -  1 \wedge 
	\frac{\int \dmu(\bq')  \exp (- \frac{1}{2 \sigma^2} \Cn{\bq' - \bq_0}^2 ) d \bq'}
	{ \int \dmu(\bq')   \exp (- \frac{1}{2 \sigma^2} \Cn{\bq' - \bbq}^2 ) d \bq'}
	\right|
	\frac{\dmu(\bbq)   \exp (- \frac{1}{2 \sigma^2} \Cn{\bbq - \bq_0}^2 ) }
	{\int \dmu(\bq')  \exp (- \frac{1}{2 \sigma^2} \Cn{\bq' - \bq_0}^2 ) d \bq'} d \bbq,
	\label{eq:MTM:sig:inf:lim}
\end{align}
so that, from the limit
$\lim_{\sigma \to \infty} \mk^\infty_{BB,\sigma}(\bq_0, d \tbq)
=\mu(d\tbq)$ and a few other applications of the Dominated Convergence
theorem on the second term, we deduce the last weak convergence
$\mk^\infty_{MT,\sigma}(\bq_0, d \tbq) = \mu(d \tbq)$ from
\eqref{eq:MTM:sig:inf:lim}.  Thus, the proof of
\cref{prop:Glob:param:Limit} is now complete.

\subsection[Proof of Proposition~\ref{Prop:alg:scalingatinfty:local}]
{Proof of \cref{Prop:alg:scalingatinfty:local}}
\label{app:proof:of:local:scaling}

We start by establishing \eqref{eq:inf:gen:prcond:conv} for the first
generator $\mathcal{L}_\rho$, as given in \eqref{eq:inf:gen:precond}.
Here we observe that
\begin{align}
  \mathcal{L}_\rho \psi(\bq_0)
  \!=& \frac{1+\rho}{4(1-\rho)} \int (\psi(\tbq) \! - \! \psi(\bq_0))
  \bar{\pk}_\rho(\bq_0, d\tbq)
  + \frac{1+\rho}{4(1-\rho)} \int \! (\psi(\tbq) - \psi(\bq_0))
  \left( \frac{d \pk^\infty_\rho(\bq_0, \cdot)}{d \bar{\pk}_\rho(\bq_0, \cdot)}(\tbq) \! - \! 1 \right)
    \! \bar{\pk}_\rho(\bq_0, d\tbq)
  \notag\\
  =& T_1(\rho) + T_2(\rho).
       \label{eq:inf:gen:precond:decomp}
\end{align}
We claim that as $\rho \to 1$, $T_1(\rho)$ converges to the infinitesimal generator of the 
preconditioned Langevin dynamic \eqref{eq:precond:LA:dynam}, as recalled in 
\eqref{eq:inf:gen:pLangevin} and that $T_2(\rho)$ vanishes.
To address the first assertion, referring back to \eqref{eq:pCN:Mala:prop}, and then
Taylor expanding
$ \psi( \rho \bq_0 - (1 - \rho) \cC\nabla \Pot(\bq_0) + \sqrt{1-
  \rho^2}\bxi )$ around $\bq_0$, we find
\begin{align}
  T_1(\rho) =&
  \frac{1+\rho}{4(1-\rho)} \int (
       \psi(
  \rho \bq_0 - (1 - \rho) \cC\nabla \Pot(\bq_0) + \sqrt{1- \rho^2}\bxi
  )
  - \psi(\bq_0))
               \nu(d \bxi)
               \notag\\
    =& \frac{1 + \rho}{4(1-\rho)}
     \int\left( \langle \nabla \psi( \bq_0), \bv(\bxi) \rangle
     + \frac{1}{2} \langle D^2 \psi((1-\theta) \bq_0 + \theta \bv(\bxi)) \bv(\bxi),
     \bv(\bxi) \rangle \right)
     \nu(d \bxi)
     \label{eq:inf:gen:precond:ID1}
\end{align}  
where $\nu = N(0,\cC)$,  $\theta = \theta(\rho) \in [0,1]$, with $\theta(\rho) \to 0$ a.s. as $\rho \to 1$ and
\begin{align}
  \label{eq:my:little:expansion:v}
  \bv(\bxi) = \bv(\bq_0,\rho,\bxi) =- (1- \rho) \bq_0 - (1 - \rho) \cC\nabla \Pot(\bq_0) + \sqrt{1-
  \rho^2}\bxi.
\end{align}
Hence, using that $\nu$ is mean free,
\begin{align}
  T_1(\rho) =& - \frac{1 + \rho}{4}  \langle \nabla \psi( \bq_0),
               \bq_0 + \cC\nabla \Pot(\bq_0) \rangle
                 \notag\\
                 &+
  \frac{(1+\rho)(1- \rho)}{8} \int
   \langle D^2 \psi( (1- \theta)\bq_0 + \theta \bv(\bxi) )(\bq_0 +\cC\nabla \Pot(\bq_0)) ,
  \bq_0 +\cC\nabla \Pot(\bq_0) \rangle
                   \nu(d\bxi)
                                    \notag\\
                 &-
  \frac{(1+\rho)\sqrt{1- \rho^2}}{4} \int
                   \langle D^2 \psi( (1- \theta)\bq_0 + \theta \bv(\bxi) )\bxi,
                   \bq_0 +\cC\nabla \Pot(\bq_0) \rangle
                   \nu(d\bxi)
                                                     \notag\\
                                    &+
  \frac{(1+\rho)^2}{8} \int
                   \langle D^2 \psi( (1- \theta) \bq_0 + \theta \bv(\bxi) )\bxi,
                   \bxi\rangle
  \nu(d\bxi).
   \notag
\end{align}  
Thus, under our assumption that $\| D^2 \psi\|_\infty < \infty$, it clearly follows from these calculations and
the Dominated Convergence theorem that, 
\begin{align}
 \lim_{\rho \to 1} T_1(\rho) =  - \frac{1}{2} \langle \bq + \cC\nabla \Pot(\bq),  \nabla \psi \rangle
  + \frac{1}{2} \mbox{Tr} (\cC D^2 \psi(\bq)),
    \label{eq:T1}
\end{align}
as desired.


We next address the second term $T_2$, in \eqref{eq:inf:gen:precond:decomp}.  Here we proceed 
in two parts via H\"older's inequality.  
To this end we first observe that, 
recalling \eqref{eq:my:little:expansion:v} and \eqref{eq:pCN:Mala:prop}, 
\begin{align}
\int |\psi(\tbq) -& \psi(\bq_0)|^2 \bar{\pk}_\rho(\bq_0, d\tbq)
\leq \|\nabla \psi\|_\infty^2  \int | \bv(\bxi)|^2\nu(d \bxi)
\notag\\
  &= \|\nabla \psi\|_\infty^2\left((1- \rho)^2  |\bq_0 + \cC\nabla \Pot(\bq_0)|^2 +
  (1 - \rho^2) \int |\bxi|^2 \nu( d \bxi)\right)
  \notag\\
  &= (1- \rho)  \|\nabla \psi\|_\infty^2 (|\bq_0 + \cC\nabla \Pot(\bq_0)|^2 +  2 \mbox{Tr} \, \cC).
  \label{ineq:int:phi:barQ:rho}
\end{align}

As regards the second portion of $T_2$ we define, 
\begin{align}
  \label{eq:inf:local:pnc:norm:def}
  \mathcal{I}(\bbq)  & := \!\!
  \int \! \exp \! \left( -
    \frac{\rho}{1 +\rho}(\Pot(\bq'/\rho)-
  \Pot(\bbq))\right) \!
  \pk_\rho(\bbq, d \bq')   = \int e^{ -  \mathcal{A}(\bq', \bq_0, \rho)} 
  \bar{\pk}_\rho(\bq_0, d \bq')
  \notag\\
   &=   \int \! \exp \! \left( - \frac{\rho}{1+\rho}\left[\Pot\left(\bbq +
  \frac{\sqrt{1-\rho^2}}{\rho} \bxi \right)-
  \Pot(\bbq)\right]\right) \!\nu(d \bxi),
\end{align}
where, in view of \eqref{eq:Guass:1:prop:pCN}, \eqref{eq:pCN:Mala:prop},
\begin{align*}
  \mathcal{A}(\tbq, \bq_0, \rho)
  := \frac{1}{1+ \rho}\left( \rho \Pot(\tbq/\rho)-
   \rho  \Pot(\bq_0)
         -\langle \nabla \Pot(\bq_0),
        (\tbq-  \rho \bq_0) \rangle  - \frac{1-\rho}{2} | \cC^{1/2}
   \nabla \Pot(\bq_0)|^2\right).
\end{align*}
Observe that, from  \eqref{eq:pCN:Mala:prop} and \eqref{eq:pcn:loc:uncor:pinf:ker},
\begin{align*}
  \frac{d \pk^\infty_\rho(\bq_0, \cdot)}
  {d \bar{\pk}_\rho(\bq_0,\cdot)}(\tbq)
  =
  \mathcal{I}(\bq_0)^{-1}
  \exp( -  \mathcal{A}(\tbq, \bq_0, \rho)).
\end{align*}
Thus, we have the estimate,
\begin{align}
   \int&
  \left( \frac{d \pk^\infty_\rho(\bq_0, \cdot)}{d \bar{\pk}_\rho(\bq_0, \cdot)}(\tbq)  -  1 \right)^2
     \bar{\pk}_\rho(\bq_0, d\tbq)
     =    \int
  \left( \frac{ e^{ -  \mathcal{A}(\tbq, \bq_0, \rho)}  - \mathcal{I}(\bq_0)}{\mathcal{I}(\bq_0)}\right)^2
     \bar{\pk}_\rho(\bq_0, d\tbq)
     \notag\\
     &\leq e^{ 2\rho \|\Pot\|_{\infty}} \int (e^{ -  \mathcal{A}(\tbq, \bq_0, \rho)}  - \mathcal{I}(\bq_0))^2  \bar{\pk}_\rho(\bq_0, d\tbq)
     \leq 2e^{ 2\rho \|\Pot\|_{\infty}} \int [(e^{ -  \mathcal{A}(\tbq, \bq_0, \rho)}  - 1) ^2 + (1- \mathcal{I}(\bq_0))^2]  \bar{\pk}_\rho(\bq_0, d\tbq)
     \notag\\     
     &\leq 4e^{ 2\rho \|\Pot\|_{\infty}} \int (e^{ -  \mathcal{A}(\tbq, \bq_0, \rho)}  - 1) ^2 \bar{\pk}_\rho(\bq_0, d\tbq)
     \leq 4e^{ 2\rho \|\Pot\|_{\infty}}
       \int  e^{2| \mathcal{A}(\tbq, \bq_0, \rho)|}
  |\mathcal{A}(\tbq, \bq_0, \rho)|^2
  \bar{\pk}_\rho(\bq_0, d\tbq)
  \notag\\
  & \leq
     \left( \int  e^{4| \mathcal{A}(\tbq, \bq_0, \rho)|}
    \bar{\pk}_\rho(\bq_0, d\tbq)\right)^{1/2}
       \left(  \int  |\mathcal{A}(\tbq, \bq_0, \rho)|^4
  \bar{\pk}_\rho(\bq_0, d\tbq)\right)^{1/2},
  \label{eq:yuckyyuckbnd1}
\end{align}
where the penultimate bound follows from the mean value theorem.

For the first term in the upper bound in \eqref{eq:yuckyyuckbnd1}, Fernique's theorem
implies that we can find an $\eta = \eta(\cC) > 0$ such that
$\int e^{2\eta  |\bxi|^2}
\nu(d \bxi)  < \infty$.  
 Thus, with an appropriate use of Young's inequality
\begin{align}
  \int  e^{4| \mathcal{A}(\tbq, \bq_0, \rho)|}
  \bar{\pk}_\rho(\bq_0, d\tbq)
  &\leq e^{ c (\|\Pot\|_{\infty} + \| \nabla \Pot\|^2_\infty )}
      \int  e^{\eta |\tbq-  \rho \bq_0 |^2}
     \bar{\pk}_\rho(\bq_0, d\tbq)
     \notag\\
     &\leq
     e^{ c (\|\Pot\|_{\infty} + \| \nabla \Pot\|^2_\infty )}
      \int e^{2\eta  |\bxi|^2}
     \nu(d \bxi), 
     \label{eq:your:mom:b:1}
\end{align}  
where $c$ depends only on $\cC$ through $\eta$ so that this upper 
bound is independent of $\rho \in (0,1]$.
As far as the second term in the upper bound, we observe that,
taking $\bw = \rho \bq_0 -  (1 - \rho) \cC\nabla \Pot(\bq_0)
+ \sqrt{1-   \rho^2}\bxi$ and applying a secord order Taylor expansion, we obtain 
\begin{align*}
  &\mathcal{A}(\bw, \bq_0, \rho)
  = \frac{1}{1+ \rho}\left( \rho \Pot(\bw/\rho)-
   \rho  \Pot(\bq_0)
         -\langle \nabla \Pot(\bq_0),
        (\bw-  \rho \bq_0) \rangle  - \frac{1-\rho}{2} | \cC^{1/2}
  \nabla \Pot(\bq_0)|^2\right)\\
 &\qquad=
  \frac{1}{1+ \rho}
     \biggl( \rho( \langle \nabla \Pot(\bq_0), \bw/\rho - \bq_0 \rangle
     + \frac{1}{2}\langle \nabla^2 \Pot( (1 - \theta) \bq_0 + \theta \bw/\rho)
     (\bw/\rho - \bq_0), \bw/\rho - \bq_0 \rangle)\\
    &\qquad \qquad \qquad
      -\langle \nabla \Pot(\bq_0),
      (\bw-  \rho \bq_0) \rangle
      - \frac{1-\rho}{2} | \cC^{1/2}\nabla \Pot(\bq_0)|^2
      \biggr)\\
      &\qquad=   \frac{1}{1+ \rho}
         \biggl(\frac{1}{2 \rho}\langle \nabla^2 \Pot( (1 - \theta) \bq_0 + \theta \bw/\rho)
       ((1 - \rho) \cC\nabla \Pot(\bq_0)
         - \sqrt{1-   \rho^2}\bxi),
         (1 - \rho) \cC\nabla \Pot(\bq_0)
          - \sqrt{1-   \rho^2}\bxi\rangle\\
   &\qquad \qquad \qquad
     - \frac{1-\rho}{2} | \cC^{1/2}\nabla \Pot(\bq_0)|^2 \biggr),
\end{align*}
for some $\theta \in [0,1]$. Hence,
\begin{align*}
	|\mathcal{A}(\bw, \bq_0, \rho)|
	&\leq \frac{\rho(1-\rho)}{1 + \rho} \| \nabla^2 \Pot\|_\infty \biggl( (1-\rho) \| \nabla \Pot\|_\infty^2 + (1 + \rho) |\bxi|^2 \biggr) + \frac{1-\rho}{2(1+\rho)} \|\nabla \Pot\|_\infty^2
	\\
	&\leq  \frac{1-\rho}{1+\rho} (1 + \| \nabla^2 \Pot\|_\infty ) (  \| \nabla \Pot\|_\infty^2  + 2|\bxi|^2 ),
\end{align*}
so that
\begin{align}
	  \int  |\mathcal{A}(\tbq, \bq_0, \rho)|^4
	\bar{\pk}_\rho(\bq_0, d\tbq)
	\leq 
	 \left(\frac{1-\rho}{1+\rho} \right)^4(1 + \| \nabla^2 \Pot\|_\infty )^4
	  \int ( \| \nabla \Pot \|_\infty^2  +  2|\bxi|^2 )^4 \nu(d \bxi),
	       \label{eq:your:mom:b:2}
\end{align}
where we again note that $\nu$ has moments of all orders.

We now combine \eqref{ineq:int:phi:barQ:rho} and \eqref{eq:yuckyyuckbnd1} and then incorporate \eqref{eq:your:mom:b:1}, \eqref{eq:your:mom:b:2}
to obtain
\begin{align}
  |T_2(\rho)|^2 \leq& \frac{1}{4(1-\rho)^2} 
\int |\psi(\tbq) - \psi(\bq_0)|^2 \bar{\pk}_\rho(\bq_0, d\tbq)
 \int \left(\frac{d \pk^\infty_\rho(\bq_0, \cdot)}{d \bar{\pk}_\rho(\bq_0, \cdot)}(\tbq)  -  1 \right)^2 \bar{\pk}_\rho(\bq_0, d\tbq)
 \notag\\
 \leq& \frac{1}{4(1-\rho)}  \|\nabla \psi\|_\infty^2 (|\bq_0 + \cC\nabla \Pot(\bq_0)|^2 +  2 \mbox{Tr} \, \cC)  4e^{ 2\rho \|\Pot\|_{\infty}}
       \int  e^{2| \mathcal{A}(\tbq, \bq_0, \rho)|}
  |\mathcal{A}(\tbq, \bq_0, \rho)|^2
  \bar{\pk}_\rho(\bq_0, d\tbq).
  \notag\\
  \leq& c(1- \rho).
     \label{ineq:T2}
\end{align}
Here $c$ is a constant depending only on $\bq_0$, $\cC$,
$\|\nabla \psi\|_\infty$,
$\|\Pot\|_{\infty}, \|\nabla \Pot\|_{\infty}$,
$\| \nabla^2 \Pot\|_{\infty}$ and in particular is independent of
$\rho$.  Hence with \eqref{eq:T1}, \eqref{ineq:T2} we have now
completed the proof of \eqref{eq:inf:gen:prcond:conv} in the case of
$\mathcal{L}_\rho$.

Turn next to $\mathcal{L}_{\rho, lMT}\psi(\bq_0)$.  Having now
established \eqref{eq:inf:gen:prcond:conv} in the first case and given
the decomposition \eqref{eq:inf:gen:precond:MTM} we simply need to
show that the first term in the final equality in
\eqref{eq:inf:gen:precond:MTM} converges to zero as $\rho \to 1$.  To
see this, observe
\begin{align}
&\left|\frac{1+\rho}{4(1-\rho)} \int (\psi(\tbq) - \psi(\bq_0))
  ( \bar{\alpha}^\infty_{\rho}( \tbq, \bq_0) -1)
  \pk^\infty_\rho(\bq_0, d\tbq) \right| \notag\\
  &\quad \leq 
  \frac{1+\rho}{4(1-\rho)} \int |\psi(\tbq) - \psi(\bq_0)|
  \left| \frac{d \pk^\infty_\rho(\bq_0, \cdot)}{d \bar{\pk}_\rho(\bq_0, \cdot)} - 1 \right|
    \bar{\pk}_\rho(\bq_0, d\tbq) + 
          \label{ineq:inf:gen:MTM:FT}\\
  &\quad \qquad
    \frac{1+\rho}{4(1-\rho)} \left(\int |\psi(\tbq) - \psi(\bq_0)|^2
    \bar{\pk}_\rho(\bq_0, d\tbq)\right)^{1/2}
      \left(\int | \bar{\alpha}^\infty_{\rho}( \tbq, \bq_0) -1|^2 \bar{\pk}_\rho(\bq_0, d\tbq)\right)^{1/2}
      \label{ineq:inf:gen:MTM}
\end{align}
The first term, \eqref{ineq:inf:gen:MTM:FT}, is estimated in precisely
same fashion as the term $T_2(\rho)$ in
\eqref{eq:inf:gen:precond:decomp}, so it only remains to estimate the
second term, \eqref{ineq:inf:gen:MTM}.

The first factor in \eqref{ineq:inf:gen:MTM} is bounded as in
\eqref{ineq:int:phi:barQ:rho}.  Regarding the second factor, observe
that, for $y > 0$,
$|1 \wedge y -1| = \max\{(1 - y), 0\} \leq |\log y|$.  Using this
inequality and recalling \eqref{eq:inf:local:pnc:norm:def},
\eqref{eq:pcn:loc:AR:pinf}, we have
\begin{align}
  | \bar{\alpha}^\infty_{\rho}( \tbq, \bq_0) -1|
	&\leq \frac{1}{1 + \rho} \left|  (\Pot(\tbq) - \rho  \Pot(\tbq /\rho)) -  (\Pot(\bq_0) - \rho \Pot(\bq_0 /\rho))  \right|
	+ |\log \mathcal{I}(\bq_0)| + |\log \mathcal{I}(\tbq)|
	\notag\\
	&= \tilde{T}_1 + \tilde{T}_2 + \tilde{T}_3.
	\label{eq:AR:bound:1}
\end{align}
For the first term $\tilde{T}_1$, take $\Psi_\rho(\tbq) = \Pot(\tbq) - \rho  \Pot(\tbq /\rho)$ and observe that, given $\bxi \sim \nu = N(0,\cC)$, 
we have by Taylor expansion that
\begin{align*}
	\Psi_\rho&( \rho \bq_0 - (1 - \rho) \cC\nabla \Pot(\bq_0) + \sqrt{1- \rho^2}\bxi ) - \Psi_\rho(\bq_0)
  = \langle \nabla \Psi_\rho (\bq_0) , \bv  \rangle + \frac12 \langle D^2 \Psi_{\rho}( \tbq_0) \bv, \bv\rangle\\
  =& \langle D^2 \Pot (\bbq_0) (  \bq_0 - \rho^{-1} \bq_0), \bv \rangle+ \langle D^2 \Psi_{\rho}(\tbq_0) \bv, \bv\rangle,
\end{align*}
for some $\tbq_0, \bbq_0 \in X$, and where $\bv = \bv(\bxi)$ is as above in \eqref{eq:my:little:expansion:v}. Thus, 
\begin{align}
	\label{eq:AR:bound:2}
	\int & \tilde{T}_1( \tbq, \bq_0)^2  \bar{Q}_\rho( \bq_0, d \tbq)
	\leq \frac{\|D^2 \Pot \|_{\infty}^2}{(1 + \rho)^2} \int  \left(\frac{1- \rho}{\rho} |\bq_0| |\bv(\bxi)| + |\bv(\bxi)|^2 \right)^2 \nu( d \bxi) 
	\notag\\
	&\leq \frac{9\|D^2 \Pot \|_{\infty}^2}{(1 + \rho)^2} \int  ((1- \rho)^2( (1 + \rho^{-1})|\bq_0|^2  + \| \cC \nabla \Pot\|_\infty^2) +  (1 - \rho)(1+ \rho) |\bxi|^2)^2 \nu (d \bxi)
	\notag\\
	&\leq  9(1-\rho)^2\|D^2 \Pot \|_{\infty}^2 \int  ( (1 + \rho^{-1})|\bq_0|^2  + \| \cC \nabla \Pot\|_\infty^2 +  2 |\bxi|^2)^2 \nu (d \bxi).
\end{align}

Turn next to the estimates for $\tilde{T}_2$, $\tilde{T}_3$ in
\eqref{eq:AR:bound:1}.  Observe that for any $\bbq \in X$ and again
$\bxi \sim \nu$,
\begin{align}
\Pot(\bbq + \rho^{-1}\sqrt{1 - \rho^2} \bxi)-
  \Pot(\bbq) 
  =  \frac{\sqrt{1 - \rho^2}}{\rho} \langle \nabla\Pot(\bbq), \bxi \rangle+ \frac{1 - \rho^2}{2\rho^2}\langle D^2 \Pot(\mathbf{z}(\bbq, \bxi)) \bxi,  \bxi \rangle,
  \label{eq:taylor:push:int}
\end{align}
for some $\bz = \bz(\bbq, \bxi) \in X.$ Meanwhile, noting that for any
$z \in \RR$, $e^{z} = 1 + z + \frac{1}{2} e^{\theta z} z^2$ for some
$\theta = \theta(z) \in [0,1]$, we have from
\eqref{eq:inf:local:pnc:norm:def}, \eqref{eq:taylor:push:int} that
\begin{align}
\label{eq:I:decomp:1}
\mathcal{I}(\bbq) = 1+
	\int (F_\rho(\bbq, \bxi)
	+  F_\rho(\bbq, \bxi)^2 
	\exp( \theta F_\rho(\bbq, \bxi) ))\nu( d \bxi),
\end{align}
where
\begin{align}
	F_\rho(\bbq, \bxi) 
	=  - \frac{\rho}{1 + \rho} \left( \frac{\sqrt{1 -
  \rho^2}}{\rho}
  \langle  \nabla\Pot(\bbq), \bxi \rangle
  + \frac{1 - \rho^2}{2\rho^2}
  \langle D^2 \Pot(\mathbf{z}(\bbq, \bxi)) \bxi,  \bxi \rangle \right).
\end{align}
Now for the first term inside the integral in \eqref{eq:I:decomp:1},
we have a crucial cancellation, resulting in
\begin{align}
	\int F_\rho(\bbq, \bxi)  \nu( d \bxi) 
	=  -\frac{1- \rho}{ 2 \rho} \int \langle D^2 \Pot(\mathbf{z}(\bbq, \bxi)) \bxi,  \bxi \rangle \nu( d \bxi).
	\label{ineq:Frho:cancel}
\end{align}
Moreover, note that
\begin{align}\label{ineq:Frho:rough}
|F_\rho(\bbq, \bxi)|  \leq  \sqrt{1 - \rho^2} \| \nabla
  \Pot\|_{\infty} |\bxi|
  + \frac{1 - \rho}{2 \rho} \| D^2 \Pot\|_{\infty} |\bxi|^2.
\end{align}
Combining \eqref{ineq:Frho:cancel} and \eqref{ineq:Frho:rough}, we find
\begin{align}
	&\left| \int (F_\rho(\bbq, \bxi) 
	+ F_\rho(\bbq, \bxi)^2 
	\exp( \theta F_\rho(\bbq, \bxi) ))\nu( d \bxi) \right|
	\notag\\
	&\leq \frac{1- \rho}{\rho} \int \frac{\| D^2 \Pot\|_{\infty}}{2}  |\bxi|^2 \nu(d \bxi) 
	\notag\\
	&\qquad +  \left(\int (2 (1- \rho^2) \| \nabla \Pot\|_{\infty}^2 |\bxi|^2+ \frac{(1 - \rho)^2}{2 \rho^2} \| D^2 \Pot\|_{\infty}^2 |\bxi|^4)^2 \nu(d \bxi)\right)^{1/2} 
	\left(\int e^{ 2 |F_\rho(\bbq, \bxi)| } \nu(d \bxi)\right)^{1/2}
		\label{ineq:I:bbq}\\
	&\leq (1- \rho) \left( \int \frac{\| D^2 \Pot\|_{\infty}}{2}  |\bxi|^2 \nu(d \bxi)   +
	\left(\int (4\| \nabla \Pot\|_{\infty}^2 |\bxi|^2+ \frac{1}{2 \rho^2} \| D^2 \Pot\|_{\infty}^2 |\bxi|^4)^2 \nu(d \bxi)
	 \int e^{ 2 |F_\rho(\bbq, \bxi)| } \nu(d \bxi)\right)^{1/2} \right).
	\notag
\end{align}
From the upper bound in \eqref{ineq:Frho:rough}, it follows that there
exists $\rho_0 = \rho_0(\text{Tr}\, \cC ) \in (0,1)$ sufficiently
close to $1$ such that the integral
$\int e^{ 2 |F_\rho(\bbq, \bxi)| } \nu(d \bxi)$ is bounded uniformly
in $\rho$ for all $ \rho_0 \leq \rho < 1$, and $\bbq \in \qsp$; see
e.g. \cite[Proposition 2.17]{DPZ2014}.  Hence, from
\eqref{ineq:I:bbq}, we obtain that for all $\rho_0 \leq \rho < 1$
\begin{align}\label{ineq:I:bbq:2}
	\left| \int (F_\rho(\bbq, \bxi) 
	+ F_\rho(\bbq, \bxi)^2 
	\exp( \theta F_\rho(\bbq, \bxi) ))\nu( d \bxi) \right|
	\leq (1-\rho) \kappa,
\end{align}
for some positive constant $\kappa = \kappa (\| \nabla \Pot\|_{\infty}, \| D^2 \Pot\|_{\infty}, \text{Tr} \, \cC)$, independent
of $1 > \rho \geq \rho_0$ and of $\bbq \in \qsp$.

Thus, from \eqref{eq:I:decomp:1} and \eqref{ineq:I:bbq:2}, we deduce
that for all $\rho$ sufficiently close to $1$
\begin{align}
   1- (1- \rho) \kappa \leq \mathcal{I}(\bbq) \leq 1+ (1 - \rho)\kappa,
\end{align}
so that, upon taking the natural logarithm,
\begin{align}
	 \ln (1- (1- \rho) \kappa) \leq \ln (\mathcal{I}(\bbq)) \leq \ln (1+ (1 - \rho)\kappa),
\end{align}
and, consequently,
\begin{align}\label{ineq:ln:I:q}
	\sup_{\bbq \in \qsp} |\ln \cI(\bbq)| \leq \max \{ |\ln (1 - (1-\rho) \kappa)|, \ln (1 + (1-\rho) \kappa) \} 
	\leq 2(1-\rho) \kappa,
\end{align}
for $(1- \rho) \kappa < 1/2$, where we used that $|\ln (1+ x)| \leq 2 |x|$ for all $|x| < 1/2$.

With \eqref{ineq:int:phi:barQ:rho}, \eqref{eq:AR:bound:1},
\eqref{eq:AR:bound:2}, and \eqref{ineq:ln:I:q}, we finally estimate
the last term in the right-hand side of \eqref{ineq:inf:gen:MTM} as
\begin{align*}
	\frac{1+\rho}{4(1-\rho)} \left(\int |\psi(\tbq) - \psi(\bq_0)|^2 \bar{\pk}_\rho(\bq_0, d\tbq)\right)^{1/2}
	\left(\int | \bar{\alpha}^\infty_{\rho}( \tbq, \bq_0) -1|^2 \bar{\pk}_\rho(\bq_0, d\tbq)\right)^{1/2}
	\leq c (1- \rho)^{1/2},
\end{align*}
where $c$ is a positive constant depending only on
$\bq_0, \cC, \|\nabla \psi\|_\infty, \|\nabla \Pot\|_{\infty}, \|
\nabla^2 \Pot\|_{\infty}$.  Therefore, this term also converges to
zero as $\rho \to 1$ and we thus conclude
\eqref{eq:inf:gen:prcond:conv} for $\mathcal{L}_{\rho, lMT}$,
completing the proof \cref{Prop:alg:scalingatinfty:local}.

\subsection[Proof of Theorem~\ref{thm:tv:bound}]
{Proof of \cref{thm:tv:bound}}
\label{app:bounds:proofs}

Working from \eqref{eq:gen:bk:ker}, we observe that
\begin{align}
 \mk^\np&(\bq_0, d \tbq)  
    = \sum_{j=0}^\np \int_{\qsp} \int_{\qsp^\np} \delta_{\bq_j}(d \tbq) \frac{\art(\bq_j, \bq_0)}
    {\art(\bq_0, \bq_0) + p \int_{\qsp} \art(\bq', \bq_0) \pk(\bbq, d \bq')}  
    \prod_{l=1}^\np \pk(\bbq, d \bq_l) \bpk(\bq_0, d \bbq)
    \notag\\
    &- \sum_{j=0}^\np \int_{\qsp} \ \int_{\qsp^\np}  \frac{\delta_{\bq_j}(d \tbq)\art(\bq_j, \bq_0)}
    {\sum_{k=0}^\np \art(\bq_k, \bq_0)} \frac{\tfrac{1}{p}
    \sum_{k=1}^\np (\art(\bq_k, \bq_0) -\int_{\qsp} \art(\bq', \bq_0) \pk(\bbq, d \bq'))}
    {p^{-1}\art(\bq_0, \bq_0) + \int_{\qsp} \art(\tbq, \bq_0) \pk(\bbq, d \tbq)} 
     \prod_{l=1}^\np \pk(\bbq, d \bq_l) \bpk(\bq_0, d \bbq)
    \notag\\
	    =& p \int_{\qsp}  \int_{\qsp}  \frac{\art(\tbq, \bq_0)}
    {\art(\bq_0, \bq_0) + p \int_{\qsp} \art(\bq', \bq_0) \pk(\bbq, d \bq')}  \pk(\bbq, d \tbq) \bpk(\bq_0, d \bbq)
       \notag\\
    &+\delta_{\bq_0}(d \tbq)  \int_{\qsp} \frac{\art(\bq_0, \bq_0)}
      {\art(\bq_0, \bq_0) + p \int_{\qsp} \art(\bq', \bq_0) \pk(\bbq, d \bq')}\bpk(\bq_0, d \bbq)
    \label{int:mk:np:varph:0}\\
    &- \sum_{j=0}^\np \int_{\qsp}\int_{\qsp^\np}    \frac{\delta_{\bq_j}(d \tbq)\art(\bq_j, \bq_0)}
    {\sum_{k=0}^\np \art(\bq_k, \bq_0)} \frac{\tfrac{1}{p}
    \sum_{k=1}^\np (\art(\bq_k, \bq_0) -\int_{\qsp} \art(\bq', \bq_0) \pk(\bbq, d \bq'))}
    {p^{-1}\art(\bq_0, \bq_0) + \int_{\qsp} \art(\bq', \bq_0)\pk(\bbq, d \bq')}
      \prod_{l=1}^\np \pk(\bbq, d \bq_l) \bpk(\bq_0, d \bbq)
	\notag
\end{align}
Thus, with \eqref{eq:p:inf:w:lim:conv:barker} we conclude,
\begin{align}
  \mk^\np&(\bq_0, d \tbq) - P^\infty(\bq_0, d \tbq)
  \notag\\
  =& -  \int_{\qsp} \int_{\qsp} \frac{\art(\bq_0, \bq_0)}
    {\art(\bq_0, \bq_0) + p\int_{\qsp} \art(\bq', \bq_0) \pk(\bbq, d \bq')} 
     \frac{\art(\tbq, \bq_0)}{\int_{\qsp} \art(\bq',
       \bq_0) \pk(\bbq, d \bq')}
       \pk(\bbq, d \tbq) \bpk(\bq_0, d \bbq)
       \notag\\
    &+\delta_{\bq_0}(d \tbq)  \int_{\qsp} \frac{\art(\bq_0, \bq_0)}
      {\art(\bq_0, \bq_0) + p \int_{\qsp} \art(\bq', \bq_0) \pk(\bbq, d \bq')}\bpk(\bq_0, d \bbq) 
        \label{int:mk:np:varphi:1}\\
    &- \sum_{j=0}^\np \int_{\qsp}   \int_{\qsp^\np}\frac{\delta_{\bq_j}(d \tbq)\art(\bq_j, \bq_0)}
    {\sum_{k=0}^\np \art(\bq_k, \bq_0)} \frac{\tfrac{1}{p}
    \sum_{k=1}^\np (\art(\bq_k, \bq_0) -\int_{\qsp} \art(\bq', \bq_0) \pk(\bbq, d \bq'))}
    {p^{-1}\art(\bq_0, \bq_0) + \int_{\qsp} \art(\bq', \bq_0)
      \pk(\bbq, d \bq')} \prod_{l=1}^\np \pk(\bbq, d \bq_l)
      \bpk(\bq_0, d \bbq)
      \notag\\
  =& T_1(\bq_0, d \tbq)  + T_2(\bq_0, d \tbq)  + T_3(\bq_0, d \tbq).
     \notag
\end{align}
Recalling the definition of the TV distance, \eqref{eq:TV:dist:def}, we
now estimate each of the terms $T_1, T_2, T_3$ as follows.  Regarding
$T_1$ we have, cf. \eqref{eq:bart:inf:cond},
\begin{align*}
  \| T_1(\bq_0, d \tbq)  \|_{TV}
  \leq \frac{1}{2} \frac{\art(\bq_0, \bq_0)}
    {\art(\bq_0, \bq_0) + p \bar{\art}( \bq_0)} 
   \int_{\qsp} \int_{\qsp} 
     \frac{\art(\tbq, \bq_0)}{\int_{\qsp} \art(\bq',
       \bq_0) \pk(\bbq, d \bq')}
  \pk(\bbq, d \tbq) \bpk(\bq_0, d \bbq)
  = \frac{1}{2} \frac{\art(\bq_0, \bq_0)}
    {\art(\bq_0, \bq_0) + p \bar{\art}( \bq_0)},
\end{align*}
and similarly for $T_2$
\begin{align*}
  \| T_2(\bq_0, d \tbq)  \|_{TV}
  \leq \frac{1}{2} \frac{\art(\bq_0, \bq_0)}
    {\art(\bq_0, \bq_0) + p \bar{\art}( \bq_0)}.
\end{align*}
Finally, for $T_3$, we estimate
\begin{align}
  \| &T_3(\bq_0, d \tbq)  \|_{TV}\notag\\
  &\leq \frac{1}{2p\bar{\art}( \bq_0)}
    \int_{\qsp}  \int_{\qsp^\np} 
     \left|\sum_{k=1}^\np\bigl(\art(\bq_k, \bq_0) -\int_{\qsp} \art(\bq', \bq_0) \pk(\bbq, d \bq')\bigr) \right|
    \prod_{l=1}^\np \pk(\bbq, d \bq_l)
    \bpk(\bq_0, d \bbq)
    \notag\\
     &\leq \frac{1}{2p\bar{\art}( \bq_0)}
    \int_{\qsp} \left( \int_{\qsp^\np} 
    \left(\sum_{k=1}^\np \art(\bq_k, \bq_0) -\int_{\qsp}
       \art(\bq', \bq_0) \pk(\bbq, d \bq') \right)^2
    \prod_{l=1}^\np \pk(\bbq, d \bq_l)\right)^{1/2}
    \bpk(\bq_0, d \bbq)
       \notag\\
  &=\frac{1}{2\sqrt{p}\bar{\art}( \bq_0)}
    \int_{\qsp}  \left(\int_{\qsp} 
     \left(\art(\bq_1, \bq_0) -\int_{\qsp} \art(\bq', \bq_0) \pk(\bbq, d \bq')\right)^2
   \pk(\bbq, d \bq_1)\right)^{1/2}
    \bpk(\bq_0, d \bbq)
      \label{eq:t3:rate:TV:bnd}
\end{align}
where the final equality follows from the fact the cross terms in the
double sum are zero.  This establishes the first bound
\eqref{eq:ss:tv:bound}.  Noting from \eqref{eq:TV:dist:def} that the
second bound \eqref{eq:ss:tv:bound:msr} follows directly from
\eqref{eq:ss:tv:bound}, the proof is complete.

 \subsection[Proof of Theorem~\ref{thm:RB:1Step}]
 {Proof of \cref{thm:RB:1Step}}
\label{eq:RB:Var:Estimate}

Define
\begin{align}
  X_p^\phi(\bq_0) = X_p^\phi := \frac{1}{p+1}\sum_{j = 0}^p
  \art(\bq_j, \bq_0)\phi(\bq_j),
\end{align}
and take
\begin{align*}
  X_p^{1} =  X_p := \frac{1}{p+1}\sum_{j = 0}^p
  \art(\bq_j, \bq_0).
\end{align*}
In order to obtain \eqref{eq:rb:upper:lower} we
observe, recalling \eqref{eq:cond:var:def} and using elementary properties
of the conditional expectation, that
\begin{align}
  &\Var(\check{\phi}(\bq_0))  = \Var\left(\frac{X_p^\phi}{X_p}\right)
  = \Var\left( \frac{X_p^\phi( \EE (X_p| \bbq_0) -  X_p )}{X_p \cdot \EE (X_p| \bbq_0)}
                               +  \frac{X_p^\phi}{\EE (X_p| \bbq_0)} \right)
  \notag\\
   &\leq 2\Var\left( \frac{X_p^\phi( \EE (X_p| \bbq_0)  -  X_p )}{X_p \cdot \EE (X_p| \bbq_0) }\right)
   +2\Var\left(\frac{X_p^\phi}{\EE (X_p| \bbq_0) } \right) 
  \leq
  2\EE \left( \frac{X_p^\phi( \EE (X_p| \bbq_0)  -  X_p )}{X_p \cdot \EE (X_p| \bbq_0) }\right)^2
    +2\Var\left(\frac{X_p^\phi}{\EE (X_p| \bbq_0) } \right)
  \notag\\
 &\leq 2\|\phi\|^2_\infty\EE\!
 \left( \frac{ \EE (X_p| \bbq_0)  -  X_p }{\EE (X_p| \bbq_0) }\right)^2
   \! \! \!+2\Var \!\left(\frac{X_p^\phi}{\EE (X_p| \bbq_0) } \right)
  \! = \!2\|\phi\|^2_\infty\EE \!
 \left( \frac{\Var(X_p| \bbq_0) }{\EE (X_p| \bbq_0)^2 }\right)
   \!+2\Var\!\left(\frac{X_p^\phi}{\EE (X_p| \bbq_0) } \right).
  \label{eq:upper:var:RB}
\end{align}
Regarding the first term in the upper bound \eqref{eq:upper:var:RB} we
observe that,
\begin{align}
  \Var(X_p|  \bbq_0)
  = \frac{1}{(\np+1)^2}\sum_{j = 0}^\np\Var (\art(\bq_j, \bq_0)| \bbq_0)
  = \frac{p}{(\np+1)^2} \Var (\art(\bq_1, \bq_0)| \bbq_0),
  \label{eq:var:XpID:1}
\end{align}
where we recall the conditionally independent and identically
distributed proposal structure in \eqref{def:vk:Tjelmeland} and note
that $\Var(\bz+ \bw|\cF) = \Var(\bz|\cF) + \Var(\bw|\cF)$ whenever
$\bz, \bw$ are random variables such that
$\EE(\bz \bw|\cF) = \EE(\bz |\cF)\EE(\bw|\cF)$.
Also notice
\begin{align}
  \EE( X_p |\bbq_0) =  \frac{1}{p+1} \art(\bq_0,\bq_0)
    + \frac{p}{p+1} \EE (\art(\bq_1, \bq_0)|\bbq_0),
     \label{eq:mean:num:1:RB}
\end{align}
Thus, with \eqref{eq:var:XpID:1}, \eqref{eq:mean:num:1:RB}
we obtain,
\begin{align}
  \EE\left(\frac{\Var(X_p| \bbq_0)}{\EE(X_p| \bbq_0)^2} \right)
  \leq \frac{1}{\np}
  \EE\left(\frac{\Var (\art(\bq_1, \bq_0)| \bbq_0)}
  {\EE (\art(\bq_1, \bq_0)|\bbq_0)^2} \right).
  \label{eq:RB:1st:upper}
\end{align}

For the second term after the final inequality \eqref{eq:upper:var:RB}
we note, as in \eqref{eq:mean:num:1:RB},
\begin{align}
   \EE (X_p^\phi| \bbq_0)
     = \frac{1}{p+1} \art(\bq_0,\bq_0) \phi(\bq_0)
     + \frac{p}{p+1} \EE  (  \art(\bq_1, \bq_0) \phi(\bq_1)|\bbq_0).
   \label{eq:mean:denom:1:RB}
\end{align}
Thus, \eqref{eq:mean:num:1:RB}, \eqref{eq:mean:denom:1:RB}, another
use of the variance decomposition \eqref{eq:RB:var:decomp},
and the fact that $\Var(\bz\bw|\cF) = \bw^2\Var(\bz|\cF)$ for any random
variables such that $\bw$ is measurable with respect to $\cF$, we
have,
\begin{align}
  &\Var\left( \frac{X^\phi_p}{\EE(X_p| \bbq_0)}\right)
        = \EE\left( \Var\left(\frac{X^\phi_p}{\EE(X_p| \bbq_0)}|  \bbq_0\right)\right)
        + \Var\left(\EE\left(\frac{X^\phi_p}{\EE(X_p| \bbq_0)}|  \bbq_0\right)\right)
        \notag\\
      &= \EE\left( \frac{\Var (X^\phi_p|\bbq_0)}{(\EE(X_p| \bbq_0))^2}\right)
        + \Var\left(\frac{\EE(X^\phi_p|\bbq_0)}{\EE(X_p| \bbq_0)}\right)
  \notag\\
      &= \frac{1}{p}\EE \frac{\Var (\art(\bq_1, \bq_0)\phi(\bq_1)| \bbq_0)}
        {(p^{-1}\art(\bq_0,\bq_0) 
        + \EE  (  \art(\bq_1, \bq_0) |\bbq_0))^2}
        + \Var\left( \frac{\EE  (  \art(\bq_1, \bq_0) \phi(\bq_1)|\bbq_0)
        +p^{-1}\art(\bq_0,\bq_0) \phi(\bq_0) }
        {\EE  (  \art(\bq_1, \bq_0) |\bbq_0)+ p^{-1}\art(\bq_0,\bq_0)} \right)
       \label{eq:RB:2nd:EQ}\\
      &\leq \frac{\|\phi\|_\infty^2}{p}  \EE\left(\frac{\Var (\art(\bq_1, \bq_0)| \bbq_0)}
  {\EE (\art(\bq_1, \bq_0)|\bbq_0)^2}+1 \right)
        + \Var\left( \frac{\EE  (  \art(\bq_1, \bq_0) \phi(\bq_1)|\bbq_0)
        +p^{-1}\art(\bq_0,\bq_0) \phi(\bq_0) }
        {\EE  (  \art(\bq_1, \bq_0) |\bbq_0)+ p^{-1}\art(\bq_0,\bq_0)} \right).
        \label{eq:RB:2nd:upper}
\end{align}
For the final bound note that
\begin{align*}
 \frac{\Var (\art(\bq_1, \bq_0)\phi(\bq_1)| \bbq_0)}
        {(p^{-1}\art(\bq_0,\bq_0) 
  \!+ \EE  (  \art(\bq_1, \bq_0) |\bbq_0))^2}
  \leq
   \frac{\EE (\art(\bq_1, \bq_0)^2\phi(\bq_1)^2| \bbq_0)}
  { [\EE  (  \art(\bq_1, \bq_0) |\bbq_0)]^2}
  \leq
  \|\phi\|^2_\infty \frac{\Var (\art(\bq_1, \bq_0)| \bbq_0) \!+ [\EE (\art(\bq_1, \bq_0)| \bbq_0)]^2}
    { [\EE  (  \art(\bq_1, \bq_0) |\bbq_0)]^2}.
\end{align*}
Thus, combining \eqref{eq:RB:1st:upper}, \eqref{eq:RB:2nd:upper}, with
\eqref{eq:upper:var:RB} we obtain the upper bound in
\eqref{eq:rb:upper:lower}.

Turn now to the lower bound \eqref{eq:rb:upper:lower}.  Here notice that
\begin{align*}
  \Var\left( \frac{X^\phi_p}{\EE(X_p| \bbq_0)}\right)
  &= \Var \left(\frac{X_p^\phi (X_p - \EE(X_p| \bbq_0))}{X_p\EE(X_p| \bbq_0)} \
    +\frac{X^\phi_p}{X_p} \right)
  \leq 2 \|\phi\|_\infty^2
    \EE\left(\frac{\Var(X_p| \bbq_0)}{\EE(X_p| \bbq_0)^2} \right)
    +2\Var(\check{\phi}(\bq_0))
    \notag\\
  &\leq \frac{2  \|\phi\|_\infty^2}{p}  \EE\left(\frac{\Var (\art(\bq_1, \bq_0)| \bbq_0)}
  {\EE (\art(\bq_1, \bq_0)|\bbq_0)^2} \right) + 2\Var(\check{\phi}(\bq_0)),
\end{align*}
where we again used \eqref{eq:RB:1st:upper} in the second
bound. Rearranging,
\begin{align}
   \frac{1}{2}\Var\left( \frac{X^\phi_p}{\EE(X_p| \bbq_0)}\right)
   - \frac{\|\phi\|_\infty^2}{p}   \EE\left(\frac{\Var (\art(\bq_1, \bq_0)| \bbq_0)}
  {\EE (\art(\bq_1, \bq_0)|\bbq_0)^2} \right)
   \leq \Var(\check{\phi}(\bq_0)).
       \label{eq:lower:var:RB}
\end{align}
Thus, combining \eqref{eq:lower:var:RB} with \eqref{eq:RB:2nd:EQ}
yields the desired lower bound, completing the proof.

\subsection[Proof of Theorem~\ref{thm:large:p:MH:hmc}]
{Proof of \cref{thm:large:p:MH:hmc}}
\label{app:proof:large:p:MH:hmc}

We start by showing the first limit in \eqref{lim:mkmh:mkB}, namely
$\mkmhinf (\bq, d \tbq) = \lim_{\np \to \infty} \mkmh (\bq, d \tbq)$,
with $\mkmhinf$ as given in \eqref{def:mkmhinf}. Fix $\bq \in \RR^\dm$
and a continuous and bounded function $\varphi: \RR^{\dm} \to
\RR$. From \eqref{def:mk:mh:hmc}, we have
\begin{align}\label{eq:phi:mkmh}
	\int_{\RR^\dm} \varphi(\tbq) \mkmh(\bq, d \tbq)
  &=  \frac{1}{\np} \sum_{j=1}^\np \int_{\RR^\dm}
    \left[ 1 \wedge \exp \left( \Ham(\bq, \bv) - \Ham(R
    \hS^\np_{jT/\np}(\bq, \bv)) \right) \right]
    \varphi(\Pi_1 \hS^\np_{jT/\np}(\bq, \bv)) \vk(\bq, d \bv) \nonumber \\
  &\qquad + \varphi(\bq) \int_{\RR^\dm}
    \left\{ 1 - \frac{1}{\np} \sum_{j=1}^\np
    \left[ 1 \wedge \exp \left( \Ham(\bq, \bv) - \Ham(R
    \hS^\np_{jT/\np}(\bq, \bv))
    \right) \right] \right\} \vk(\bq, d \bv),
\end{align}
where $\vk(\bq, d \bv) = e^{-\VPot(\bq, \bv)} d \bv$. 

Since $x \in \RR \mapsto 1 \wedge e^x$ is a Lipschitz function with
Lipschitz constant $1$, we may estimate the integrand of the second
term in the right-hand side of \eqref{eq:phi:mkmh} as
\begin{align}\label{intnd:rej:term}
  1 - \frac{1}{\np} \sum_{j=1}^\np
  \left[ 1 \wedge \exp \left( \Ham(\bq, \bv) - \Ham(R \hS^\np_{jT/\np}(\bq, \bv)) \right) \right]
  &= \frac{1}{\np} \sum_{j=1}^\np
    \left\{ 1 -  \left[ 1 \wedge \exp \left( \Ham(\bq, \bv) -
    \Ham(R \hS^\np_{jT/\np}(\bq, \bv)) \right) \right] \right\} \nonumber \\
  &\leq \frac{1}{\np} \sum_{j=1}^\np | \Ham(\bq, \bv) - \Ham(R \hS^\np_{jT/\np}(\bq, \bv)) |.
\end{align}

Note that the assumption of symmetry of $\VPot$ in $\bv$, namely
\eqref{VPot:sym:v}, implies that $\Ham \circ R = \Ham$, where we
recall that $R(\bq, \bv) = (\bq, -\bv)$. With this fact and the
invariance of $\Ham$ in \eqref{Ham:inv:hSt}, we can rewrite
\eqref{intnd:rej:term} and further estimate it as
\begin{align}\label{ineq:intnd:rej:term}
	\frac{1}{\np} \sum_{j=1}^\np | \Ham(\hS_{jT/\np}(\bq, \bv)) - \Ham(\hS_{jT/\np}^\np (\bq, \bv)) |
	\leq \sup_{t \in [0,T]} | \Ham(\hS_t (\bq, \bv)) - \Ham(\hS_t^\np (\bq, \bv)) |.
\end{align}

Due to the continuity of $t \mapsto \hS_t(\bq,\bv)$ on the compact
interval $[0,T]$, it follows that there exists $M = M(\bq,\bv) > 0$
such that $\sup_{t \in [0,T]} \| \hS_t(\bq,\bv)\| \leq M(\bq, \bv)$
for each fixed $(\bq, \bv) \in \RR^{2\dm}$. Then, assumption
\eqref{ass:hSt:hSnpt} implies that, for $\np = \np(\bq,\bv)$
sufficiently large,
$\sup_{t \in [0,T]} \|\hS_t^\np (\bq,\bv)\| \leq 2 M(\bq, \bv)$. Since
$\Ham$ is a continuous mapping, and hence uniformly continuous on the
closed ball of radius $2 M(\bq,\bv)$ in $\RR^{2\dm}$, then
\eqref{ass:hSt:hSnpt} together with the latter two facts yield that
the right-hand side of \eqref{ineq:intnd:rej:term} converges to zero
as $\np \to \infty$ for each $(\bq, \bv) \in \RR^{2 \dm}$. Moreover,
note that the integrand of the second term in the right-hand side of
\eqref{eq:phi:mkmh} is bounded above by the constant $2$. Therefore,
by the Dominated Convergence theorem, we obtain that
\begin{align}\label{lim:p:rej:term}
  \lim_{\np \to \infty}  \int_{\RR^\dm}
  \left\{ 1 - \frac{1}{\np} \sum_{j=1}^\np
  \left[ 1 \wedge \exp \left( \Ham(\bq, \bv) -
  \Ham(R \hS^\np_{jT/\np}(\bq, \bv)) \right) \right] \right\} \vk(\bq, d \bv) = 0.
\end{align}

Regarding the first term in the right-hand side of
\eqref{eq:phi:mkmh}, we first rewrite it as
\begin{multline*}
  \frac{1}{\np} \sum_{j=1}^\np \int_{\RR^\dm}
  \left\{  \left[ 1 \wedge \exp \left( \Ham(\bq, \bv) -
        \Ham(R \hS^\np_{jT/\np}(\bq, \bv)) \right) \right]  -1
       \right\}
  \varphi(\Pi_1 \hS^\np_{jT/\np}(\bq, \bv)) \vk(\bq, d \bv) \\
  + \frac{1}{\np} \sum_{j=1}^\np \int_{\RR^\dm}
  \varphi(\Pi_1 \hS^\np_{jT/\np}(\bq, \bv)) \vk(\bq, d \bv) 
	= (I) + (II).
\end{multline*}

Since $\varphi$ is bounded, we may estimate the integrand in $(I)$ as
\begin{multline*}
  \frac{1}{\np} \sum_{j=1}^\np
  \left\{  \left[ 1 \wedge \exp \left( \Ham(\bq, \bv) -
        \Ham(R \hS^\np_{jT/\np}(\bq, \bv)) \right) \right]  -1 \right\} \varphi(\Pi_1 \hS^\np_{jT/\np}(\bq, \bv)) \\
	\leq \frac{\| \varphi\|_\infty}{\np} \sum_{j=1}^\np |\Ham(\bq, \bv) - \Ham(R \hS_{jT/\np}^\np (\bq, \bv))|,
\end{multline*}
where $\|\varphi\|_\infty = \sup_{\bq \in \RR^\dm}
|\varphi(\bq)|$. Thus, proceeding analogously as with the analysis of
the previous term, we arrive at
\begin{align}\label{lim:p:I}
	\lim_{\np \to \infty} (I) = 0.
\end{align}

For the term $(II)$, we first split it as
\begin{align*}
  (II) &= \frac{1}{\np} \sum_{j=1}^\np \int_{\RR^\dm}
         \left[ \varphi(\Pi_1 \hS_{jT/\np}^\np (\bq, \bv)) - \varphi(\Pi_1 \hS_{jT/\np} (\bq, \bv)) \right] \vk(\bq, d \bv) 
	+ \frac{1}{\np} \sum_{j=1}^\np \int_{\RR^\dm} \varphi(\Pi_1 \hS_{jT/\np} (\bq, \bv)) \vk(\bq, d \bv) \\
	&= (II_a) + (II_b).
\end{align*}
The integrand in $(II_a)$ can be estimated as
\begin{align*}
	\frac{1}{\np} \sum_{j=1}^\np \left[ \varphi(\Pi_1 \hS_{jT/\np}^\np (\bq, \bv)) - \varphi(\Pi_1 \hS_{jT/\np} (\bq, \bv)) \right] 
	\leq \sup_{t \in [0,T]} |\varphi(\Pi_1 \hS_t^\np (\bq, \bv)) - \varphi(\Pi_1 \hS_t (\bq, \bv))|.
\end{align*}
Since $\varphi \circ \Pi_1: \RR^{2 \dm} \to \RR$ is a continuous and
bounded function, we again proceed similarly as before to deduce from
assumption \eqref{ass:hSt:hSnpt} and the Dominated Convergence theorem
that
\begin{align}\label{lim:p:IIa}
	\lim_{\np \to \infty} (II_a) = 0.
\end{align}

Finally, regarding the term $(II_b)$, we first note that
\begin{align}\label{lim:Riemann:int}
	\frac{1}{\np} \sum_{j=1}^\np \varphi(\Pi_1 \hS_{jT/\np} (\bq,\bv)) = \frac{1}{T} \frac{T}{\np} \sum_{j=1}^\np \varphi(\Pi_1 \hS_{jT/\np} (\bq,\bv))  
	\longrightarrow \frac{1}{T} \int_0^T \varphi(\Pi_1 \hS_t (\bq, \bv)) \, dt \quad \mbox{as } \np \to \infty,
\end{align}
for every $(\bq, \bv) \in \RR^{2\dm}$. Hence, it follows by applying once again the Dominated Convergence theorem that
\begin{align}\label{lim:p:IIb}
	\lim_{\np \to \infty} (II_b) = \frac{1}{T} \int_{\RR^\dm} \int_0^T \varphi(\Pi_1 \hS_t (\bq, \bv)) \, dt \, \vk(\bq, d \bv).
\end{align}

Therefore, from \eqref{eq:phi:mkmh}, \eqref{lim:p:rej:term}-\eqref{lim:p:IIb}, we deduce that
\begin{align*}
	\lim_{\np \to \infty} \int_{\RR^\dm} \varphi(\tbq) \mkmh(\bq,
  d \tbq)
  = \frac{1}{T} \int_{\RR^\dm} \int_0^T \varphi(\Pi_1 \hS_t (\bq,
  \bv)) \, dt \, \vk(\bq, d \bv)
  = \int_{\RR^\dm} \varphi(\tbq) \mkmhinf (\bq, d \tbq)
\end{align*}
for all $\bq \in \RR^\dm$, with $\mkmhinf$ as defined in
\eqref{def:mkmhinf}. This concludes the proof of the first limit in
\eqref{lim:mkmh:mkB}.

Regarding the second limit in \eqref{lim:mkmh:mkB}, let us again fix
$\bq \in \RR^\dm$ and a continuous and bounded function
$\varphi: \RR^\dm \to \RR$. From the definition of the kernel $\mkB$
in \eqref{def:mkB:hmc}, we have

\begin{align}\label{eq:phi:mkB}
	\int_{\RR^\dm} \varphi(\tbq) \mkB(\bq, d \tbq) 
  = \sum_{j=0}^\np \int_{\RR^\dm}
  \frac{\exp(- \Ham(S_j(\bq, \bv)))}{\sum_{k=0}^\np
  \exp(-\Ham (S_k(\bq,\bv)))} \varphi (\Pi_1 \hS^\np_{jT/\np}(\bq, \bv)) \vk(\bq, d \bv).
\end{align}

We rewrite the integrand in \eqref{eq:phi:mkB} as
\begin{align}
	&\sum_{j=0}^\np \frac{\exp(- \Ham(S_j(\bq, \bv)))}{\sum_{k=0}^\np \exp(-\Ham (S_k(\bq,\bv)))} \varphi (\Pi_1 \hS^\np_{jT/\np}(\bq, \bv)) \notag \\
	&\qquad = \frac{1}{\np + 1} \sum_{j=0}^\np \left( \frac{\exp(- \Ham(S_j(\bq, \bv)))}{\frac{1}{\np + 1} \sum_{k=0}^\np \exp(-\Ham (S_k(\bq,\bv)))} -1 \right) \varphi (\Pi_1 \hS^\np_{jT/\np}(\bq, \bv))
	+ \frac{1}{\np + 1}  \sum_{j=0}^\np \varphi (\Pi_1 \hS^\np_{jT/\np}(\bq, \bv)) \notag \\
	&\qquad = (I) + (II).
	\label{eq:I:II}
\end{align}

For the first term, (I), we have
\begin{align*}
	(I) 
	&\leq \frac{\| \varphi\|_\infty}{\np + 1} \sum_{j=0}^\np \left| \frac{\exp( \Ham(\bq, \bv) - \Ham(S_j(\bq, \bv)))}{\frac{1}{\np + 1} \sum_{k=0}^\np \exp(  \Ham(\bq, \bv) -\Ham (S_k(\bq,\bv)))} - 1 \right| \\
	&\leq \| \varphi\|_\infty \frac{\max_{j=0, \ldots, \np} | \exp( \Ham(\bq, \bv) - \Ham(S_j(\bq, \bv))) -1 | + \left| 1 - \frac{1}{\np + 1} \sum_{k=0}^\np \exp(  \Ham(\bq, \bv) -\Ham (S_k(\bq,\bv))) \right| }{\frac{1}{\np + 1} \sum_{k=0}^\np \exp(  \Ham(\bq, \bv) -\Ham (S_k(\bq,\bv)))}
	\\
	&\leq 2  \| \varphi\|_\infty \frac{ \max_{j=0, \ldots, \np} | \exp( \Ham(\bq, \bv) - \Ham(S_j(\bq, \bv))) -1 | }{\min_{j=0,\ldots,\np} \exp(  \Ham(\bq, \bv) -\Ham (S_j(\bq,\bv)))}.
\end{align*}

Invoking the symmetry of $\VPot$, \eqref{VPot:sym:v}, which implies
$\Ham \circ R = \Ham$, and also the invariance of $\hS_t$,
$t \in [0,T]$, under $\Ham$, namely \eqref{Ham:inv:hSt}, it follows
that

\begin{align*}
	(I) &\leq 2  \| \varphi\|_\infty \frac{ \max_{j=0, \ldots, \np} | \exp( \Ham(\hS_{jT/\np} (\bq, \bv)) - \Ham(\hS_{jT/\np}^\np (\bq, \bv))) -1 | }{\min_{k=0,\ldots,\np} \exp(  \Ham(\hS_{kT/\np}( \bq, \bv)) -\Ham (\hS_{kT/\np}^\np (\bq,\bv)))} \\
	&\leq  2  \| \varphi\|_\infty \frac{ \sup_{t \in [0,T]} | \exp( \Ham(\hS_t (\bq, \bv)) - \Ham(\hS_t^\np (\bq, \bv))) -1 | }{\inf_{t \in [0,T]} \exp(  \Ham(\hS_t( \bq, \bv)) -\Ham (\hS_t^\np (\bq,\bv)))}.
\end{align*}

Due to the continuity of $\Ham$ and assumption \eqref{ass:hSt:hSnpt},
it is not difficult to show that
\begin{align*}
	 \lim_{\np \to \infty} \sup_{t \in [0,T]} | \exp( \Ham(\hS_t (\bq, \bv)) - \Ham(\hS_t^\np (\bq, \bv))) -1 |  = 0,
\end{align*}
and
\begin{align*}
	 \lim_{\np \to \infty} \inf_{t \in [0,T]} \exp(  \Ham(\hS_t( \bq, \bv)) -\Ham (\hS_t^\np (\bq,\bv))) = 1,
\end{align*}
so that $\lim_{\np \to \infty} (I) = 0$ for all $(\bq, \bv) \in \RR^{2 \dm}$.

Moreover, for the term (II) in \eqref{eq:I:II}, we have similarly as
in \eqref{lim:Riemann:int} that
\begin{align*}
	(II) = \frac{\np}{T(\np + 1)} \frac{T}{\np}  \sum_{j=0}^\np \varphi (\Pi_1 \hS^\np_{jT/\np}(\bq, \bv)) \longrightarrow  \frac{1}{T} \int_0^T \varphi(\Pi_1 \hS_t (\bq, \bv)) \, dt \quad \mbox{as } \np \to \infty,
\end{align*}
for all $(\bq, \bv) \in \RR^{2 \dm}$.

Therefore, it follows from \eqref{eq:phi:mkB}-\eqref{eq:I:II} and the
Dominated Convergence theorem that
\begin{align*}
	\lim_{\np \to \infty} \int_{\RR^\dm} \varphi(\tbq) \mkB(\bq, d \tbq) = \frac{1}{T} \int_{\RR^\dm} \int_0^T \varphi(\Pi_1 \hS_t (\bq, \bv)) \, dt \, \vk(\bq, d \bv) = \int_{\RR^\dm} \varphi(\tbq) \mkmhinf (\bq, d \tbq)
\end{align*}
for all $\bq \in \RR^\dm$, as desired.

\section{Additional Proposal Tuning Experiments}\label{app:tuning}

\subsection{Convergence of MCMC Results to Target}
This section supplements \cref{sec:numexp:mcmc:proposal} and presents
additional numerical experiment related to the tuning of the proposal
distributions to optimize MCMC convergence. \cref{fig:tune:triMix}
shows the results of MCMC sampling by target acceptance rate for a
target distribution consisting of a mixture of three Gaussian
distributions, given by
\begin{align}\label{eq:trimix}
	\dps(x) = \frac{1}{Z} \left[ 
	0.3 \exp\left(-\frac{1}{2\tau^2}(x+7)^2\right)  +
	0.5 \exp\left(-\frac{1}{2\tau^2}(x+2)^2\right) +
	\exp\left(-\frac{1}{2\tau^2}(x-2)^2\right)
	\right]
\end{align}
for $\tau = 1/\sqrt{8}$. \cref{fig:tune:triMix} is the analogue of
\cref{fig:tune:tiltbanana}, which shows similar results for the tilted
banana distribution given in \eqref{eq:tiltedbanana}. Each point on
the plot represents the results of a chain of 20,000 MCMC samples. As
in \cref{fig:tune:tiltbanana}, the left plot of \cref{fig:tune:triMix}
shows the proposal standard deviation $\sigma$ for each target
acceptance rate, the middle plots shows that the target acceptance
rate was largely achieved by the selected $\sigma$, and the right plot
shows loss -- defined to be the sum of the relative error in the
first, second, and fourth moments -- by target acceptance rate. As in
\cref{fig:tune:tiltbanana}, the loss is largely robust across a wide
range of target acceptance rates.

\begin{figure}[htbp]
	\centering
	\includegraphics[width=\textwidth]{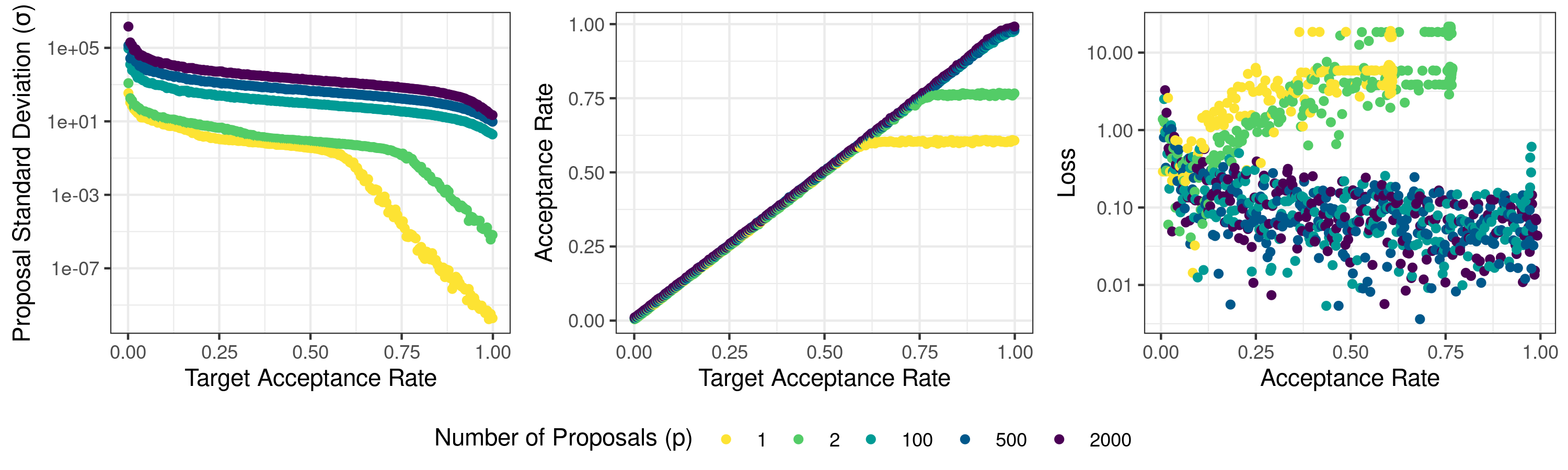}
	\caption{Tuning $\sigma_\dpk$ for the mixture of Gaussians, by
          number of proposals $\np$. Each point represents a chain of
          20,000 MCMC samples with $\sigma_\dpk$ adaptively tuned to a
          given target acceptance rate. Left: The choice of
          $\sigma_\dpk$ by target acceptance rate. Middle: The
          acceptance rate by target acceptance rate, showing that the
          target acceptance rate was achieved up to the limits of the
          Barker formulation. Right: Loss by acceptance rate, showing
          accurate approximation across a broad range of acceptance
          rates.}
	\label{fig:tune:triMix}
\end{figure}

\end{document}